\documentclass[a4paper,11pt]{article}
\pdfoutput=1 

\usepackage{jcappub} 

\usepackage[T1]{fontenc} 

\usepackage[dvipsnames]{xcolor}

\usepackage{placeins}
\usepackage{enumitem}

\usepackage[nameinlink]{cleveref}
\crefname{figure}{figure}{figures}
\crefname{section}{section}{sections}
\crefname{equation}{equation}{}

\usepackage{hyperref}

\usepackage{soul}
\usepackage{lineno}

\def\bdE{{\bf E }}

\def\bdX{{\bf X }}

\def\bdw{{\bf w }}

\def\bda{{\bf a }}
\def\bdb{{\bf b }}
\def\bdB{{\bf B }}

\def\bdb{{\bf b }}

\def\tC{\boldsymbol{\rm C}}

\def\tI{\boldsymbol{\rm I}}

\def\tP{\boldsymbol{\rm P}}

\def\setsymbol#1#2{\expandafter\def\csname #1\endcsname{#2}}
\def\getsymbol#1{\csname #1\endcsname}

\newbox\tablebox    \newdimen\tablewidth
\def\leaderfil{\leaders\hbox to 5pt{\hss.\hss}\hfil}
\def\tablenote#1 #2\par{\begingroup \parindent=0.8em
    \abovedisplayshortskip=0pt\belowdisplayshortskip=0pt
    \noindent
    $$\hss\vbox{\hsize\tablewidth \hangindent=\parindent \hangafter=1 \noindent
    \hbox to \parindent{$^#1$\hss}\strut#2\strut\par}\hss$$
    \endgroup}

\def\L2{\ifmmode L_2\else $L_2$\fi}

\def\DeltaT{\ifmmode \Delta T\else $\Delta T$\fi}
\def\deltat{\ifmmode \Delta t\else $\Delta t$\fi}
\def\fknee{\ifmmode f_{\rm knee}\else $f_{\rm knee}$\fi}
\def\Fmax{\ifmmode F_{\rm max}\else $F_{\rm max}$\fi}
\def\solar{\ifmmode{\rm M}_{\mathord\odot}\else${\rm M}_{\mathord\odot}$\fi}
\def\Msolar{\ifmmode{\rm M}_{\mathord\odot}\else${\rm M}_{\mathord\odot}$\fi}
\def\Lsolar{\ifmmode{\rm L}_{\mathord\odot}\else${\rm L}_{\mathord\odot}$\fi}
\def\inv{\ifmmode^{-1}\else$^{-1}$\fi}
\def\mo{\ifmmode^{-1}\else$^{-1}$\fi}
\def\sup#1{\ifmmode ^{\rm #1}\else $^{\rm #1}$\fi}
\def\expo#1{\ifmmode \times 10^{#1}\else $\times 10^{#1}$\fi}
\def\,{\thinspace}
\def\lsim{\mathrel{\raise .4ex\hbox{\rlap{$<$}\lower 1.2ex\hbox{$\sim$}}}}
\def\gsim{\mathrel{\raise .4ex\hbox{\rlap{$>$}\lower 1.2ex\hbox{$\sim$}}}}

\def\simprop{\mathrel{\raise .4ex\hbox{\rlap{$\propto$}\lower 1.2ex\hbox{$\sim$}}}}
\def\deg{\ifmmode^\circ\else$^\circ$\fi}
\def\pdeg{\ifmmode $\setbox0=\hbox{$^{\circ}$}\rlap{\hskip.11\wd0 .}$^{\circ}
          \else \setbox0=\hbox{$^{\circ}$}\rlap{\hskip.11\wd0 .}$^{\circ}$\fi}
\def\arcs{\ifmmode {^{\scriptstyle\prime\prime}}
          \else $^{\scriptstyle\prime\prime}$\fi}
\def\arcm{\ifmmode {^{\scriptstyle\prime}}
          \else $^{\scriptstyle\prime}$\fi}
\newdimen\sa  \newdimen\sb
\def\parcs{\sa=.07em \sb=.03em
     \ifmmode \hbox{\rlap{.}}^{\scriptstyle\prime\kern -\sb\prime}\hbox{\kern -\sa}
     \else \rlap{.}$^{\scriptstyle\prime\kern -\sb\prime}$\kern -\sa\fi}
\def\parcm{\sa=.08em \sb=.03em
     \ifmmode \hbox{\rlap{.}\kern\sa}^{\scriptstyle\prime}\hbox{\kern-\sb}
     \else \rlap{.}\kern\sa$^{\scriptstyle\prime}$\kern-\sb\fi}
\def\ra[#1 #2 #3.#4]{#1\sup{h}#2\sup{m}#3\sup{s}\llap.#4}
\def\dec[#1 #2 #3.#4]{#1\deg#2\arcm#3\arcs\llap.#4}
\def\deco[#1 #2 #3]{#1\deg#2\arcm#3\arcs}
\def\rra[#1 #2]{#1\sup{h}#2\sup{m}}

\def\dots{\relax\ifmmode \ldots\else $\ldots$\fi}
\def\WHzsr{\ifmmode $W\,Hz\mo\,sr\mo$\else W\,Hz\mo\,sr\mo\fi}
\def\mHz{\ifmmode $\,mHz$\else \,mHz\fi}
\def\GHz{\ifmmode $\,GHz$\else \,GHz\fi}
\def\mKs{\ifmmode $\,mK\,s$^{1/2}\else \,mK\,s$^{1/2}$\fi}
\def\muKs{\ifmmode \,\mu$K\,s$^{1/2}\else \,$\mu$K\,s$^{1/2}$\fi}
\def\muKRJs{\ifmmode \,\mu$K$_{\rm RJ}$\,s$^{1/2}\else \,$\mu$K$_{\rm RJ}$\,s$^{1/2}$\fi}
\def\muKHz{\ifmmode \,\mu$K\,Hz$^{-1/2}\else \,$\mu$K\,Hz$^{-1/2}$\fi}
\def\MJysr{\ifmmode \,$MJy\,sr\mo$\else \,MJy\,sr\mo\fi}
\def\MJysrmK{\ifmmode \,$MJy\,sr\mo$\,mK$_{\rm CMB}\mo\else \,MJy\,sr\mo\,mK$_{\rm CMB}\mo$\fi}
\def\microns{\ifmmode \,\mu$m$\else \,$\mu$m\fi}

\def\muK{\ifmmode \,\mu$K$\else \,$\mu$\hbox{K}\fi}
\def\microK{\ifmmode \,\mu$K$\else \,$\mu$\hbox{K}\fi}
\def\muW{\ifmmode \,\mu$W$\else \,$\mu$\hbox{W}\fi}
\def\kms{\ifmmode $\,km\,s$^{-1}\else \,km\,s$^{-1}$\fi}
\def\kmsMpc{\ifmmode $\,\kms\,Mpc\mo$\else \,\kms\,Mpc\mo\fi}
\providecommand{\sorthelp}[1]{}

\def\lb{\textit{LiteBIRD}}
\def\pl{\textit{Planck}}
\def\wm{\textit{WMAP}}

\def\creff@jnl#1{{\rm#1\/}}

\def\aj{\creff@jnl{AJ}}                  
\def\araa{\creff@jnl{ARA\&A}}            
\def\apj{\creff@jnl{ApJ}}                
\def\apjl{\creff@jnl{ApJ}}               
\def\apjs{\creff@jnl{ApJS}}              
\def\ao{\creff@jnl{Appl.Optics}}         
\def\apss{\creff@jnl{Ap\&SS}}            
\def\aap{\creff@jnl{A\&A}}               
\def\aapr{\creff@jnl{A\&A~Rev.}}         
\def\aaps{\creff@jnl{A\&AS}}             
\def\azh{\creff@jnl{AZh}}                        
\def\baas{\creff@jnl{BAAS}}              
\def\jcap{\creff@jnl{JCAP}}              
\def\jrasc{\creff@jnl{JRASC}}            
\def\memras{\creff@jnl{MmRAS}}           
\def\mnras{\creff@jnl{MNRAS}}            
\def\pra{\creff@jnl{Phys.Rev.A}}         
\def\prb{\creff@jnl{Phys.Rev.B}}         
\def\prc{\creff@jnl{Phys.Rev.C}}         
\def\prd{\creff@jnl{Phys.Rev.D}}         
\def\prl{\creff@jnl{Phys.Rev.Lett}}      
\def\physrep{\creff@jnl{Phys.Rep.}}      
\def\pasp{\creff@jnl{PASP}}              
\def\pasj{\creff@jnl{PASJ}}              
\def\qjras{\creff@jnl{QJRAS}}            
\def\skytel{\creff@jnl{S\&T}}            
\def\solphys{\creff@jnl{Solar~Phys.}}    
\def\sovast{\creff@jnl{Soviet~Ast.}}     
 \def\ssr{\creff@jnl{Space~Sci.Rev.}}    
\def\zap{\creff@jnl{ZAp}}                
\def\nat{\creff@jnl{Nature}}             

\title{\boldmath Field-level constraints on cosmic birefringence from hybrid ILC maps combining E- and B-mode channels}

\author{Mathieu Remazeilles}
\affiliation{Instituto de Física de Cantabria (CSIC-UC), \\ Avda. de los Castros s/n, 39005 Santander, Spain}

\emailAdd{remazeilles@ifca.unican.es}

\abstract{
Cosmic birefringence, arising from a potential parity-violating interaction between cosmic microwave background (CMB) photons and evolving pseudo-scalar fields such as axion-like particles, can rotate the CMB polarization plane and induce an effective correlation between the CMB $E$- and $B$-mode polarization. In this work, we introduce a hybrid internal linear combination (ILC) method that combines both $E$- and $B$-mode frequency maps into the component separation pipeline, enabling the disentanglement of correlated and uncorrelated components of CMB polarization in the presence of cosmic birefringence and instrumental polarization angle miscalibration. We derive an analytic linear relation connecting the birefringence-induced correlated component of the CMB $E$- (or $B$-) mode field to the full CMB $B$- (or $E$-) mode field convolved with a modulating field. By performing linear regression between these fields across multiple sky patches, we directly estimate the birefringence angle at the field level. This allows us to distinguish cosmic birefringence from polarization angle miscalibration and foreground contamination, as the ILC responds differently to achromatic cosmic birefringence and chromatic systematic effects, with its weights projecting spatial or harmonic dependence only onto the latter. This non-parametric, field-level approach provides a novel way to probe cosmic birefringence directly in real space. When applied to realistic simulations of the forthcoming \lb\ satellite mission, our method yields constraints that are competitive with, and complementary to, existing power spectrum-based analyses. When applied to \pl\ Release 4 (PR4) data, we find a birefringence angle of $\beta = 0.32^\circ \pm 0.12^\circ$, a $2.7\sigma$ detection that remains robust against varying sky fractions.
}

\begin{document}
\maketitle
\flushbottom

\section{Introduction}
\label{sec:intro}

Cosmic birefringence is a potential signature of parity-violating physics beyond the Standard Model of particle physics and cosmology~\cite{Carroll1990}. It is predicted to induce a rotation of the plane of linear polarization of the cosmic microwave background (CMB) radiation as it propagates over cosmological distances. This effect leads to mixing between the CMB polarization $E$- and $B$-modes, resulting in observable imprints such as non-zero $EB$ and $TB$ cross-correlations (see \cite{Komatsu2022} for a review). A detection of such a signal would provide unique insights into couplings between photons and time-dependent pseudo-scalar fields, such as axion-like particles (ALPs)\cite{Marsh2016,Nakagawa2021,Obata2022}, which are compelling dark matter candidates, or dark energy quintessence fields~\cite{Carroll1998,Liu2006,Li2008}.

In many extensions of the Standard Model, pseudo-scalar fields couple to electromagnetism via a Chern-Simons interaction of the form $\mathcal{L}_{\phi\gamma} = -\frac{1}{4} g_{\phi\gamma} \, \phi \, F_{\mu\nu} \tilde{F}^{\mu\nu}$, where $\phi$ is the pseudo-scalar field, such as an ALP, $g_{\phi\gamma}$ is the coupling constant, $F_{\mu\nu}$ is the electromagnetic field strength tensor, and $\tilde{F}^{\mu\nu}$ is its dual. This interaction violates parity and causes a rotation of the linear polarization of photons as they travel through regions where $\phi$ varies in time or space~\cite{Harari1992,Finelli2009,Pospelov2009}. The net rotation angle, commonly referred to as the cosmic birefringence angle $\beta$, is given by $\beta = \frac{1}{2} g_{\phi\gamma} \left[\phi(t_{\mathrm{obs}}) - \phi(t_{\mathrm{em}})\right]$, where $\phi(t_{\mathrm{obs}})$ and $\phi(t_{\mathrm{em}})$ are the field values at the time of observation and at the last-scattering surface, respectively. Constraints on $\beta$ therefore translate into bounds on either the dynamics of the pseudo-scalar field or its coupling to photons.

The cumulative rotation imprinted on the CMB polarization results in a non-zero, parity-odd $EB$ cross-power spectrum, which depends on $\beta$ as $C_\ell^{EB} \simeq 2\beta (C_\ell^{EE} - C_\ell^{BB})$~\cite{Lue1999,Feng2005,Liu2006}. Searches for parity violation in CMB data and constraints on the cosmic birefringence angle $\beta$ have thus primarily relied on detecting such $EB$ correlations at the power-spectrum level.

Instrumental systematics, especially miscalibration of detector polarization angles, can mimic similar $E$-$B$ mixing~\cite{Hu2003,Shimon2008,Miller2009,Yadav2009,Keating2013}. However, as shown by~\cite{Minami2019,Minami2020}, although both cosmic birefringence and instrumental miscalibration generate $EB$ correlations, they differ in their frequency dependence and in how they affect the CMB versus Galactic foregrounds. This distinction enables multifrequency observations to disentangle the physical birefringence signal from instrumental systematics. Using this approach, tentative evidence for isotropic cosmic birefringence has been reported in \pl\ and \wm\ data, with measured rotation angles of $\beta \sim 0.3^\circ$ detected at $2.4$ to $3.6\sigma$ significance~\cite{Minami2020b,Diego2022,Eskilt2022,Eskilt2022b}. Future CMB experiments such as \lb~\cite{LiteBIRD-CB2025}, the Simons Observatory~\cite{SO-LAT2025}, and CMB-S4-like~\cite{CMB-S4_2016} are expected to achieve substantially improved sensitivity to cosmic birefringence due to enhanced polarization sensitivity, wider frequency coverage, and improved control of systematics. However, the robustness of current constraints remains limited by the reliability of foreground models and assumptions in the yet poorly-known $EB$ cross-spectrum of Galactic dust and synchrotron emissions~\cite{Clark2021,Diego2023,Vacher2023,Hervias2025}.

So far, the most stringent constraints on the cosmic birefringence angle $\beta$ have come from power-spectrum analyses \citep{Minami2020b,Diego2022,Eskilt2022,Eskilt2022b,delaHoz2022,Jost2023}. While powerful, these approaches aggregate information across the sky, potentially washing out spatial variations in $\beta$ or correlations that are more naturally expressed at the map (or field) level. Alternative map-based approaches, such as stacking $B$-mode polarization maps at extrema of the temperature or $E$-mode maps  \citep{Komatsu2011,Contreras2017,Jow2019}, have also been explored, but have generally yielded lower significance than power-spectrum-based methods \citep{planck2014-a23,Sullivan2025,LiteBIRD-CB2025}. In this context, developing robust \emph{field-level} estimators that can disentangle physical birefringence from instrumental systematics and foregrounds is therefore essential. The non-parametric component separation method proposed in this work provides such a framework, leveraging $EB$ correlations and multifrequency information to extract birefringence signatures directly from the CMB polarization maps, without relying on assumptions about the foregrounds.

We present a \emph{field-level} approach to infer $\beta$ directly from CMB polarization maps. Our method constructs a CMB $E$-mode map using a \emph{hybrid} Internal Linear Combination (ILC) technique that combines both $E$- and $B$-mode frequency maps with frequency-dependent weights. This novel Hybrid ILC enables a decomposition of the CMB $E$-mode field into components that are respectively \emph{correlated} and \emph{uncorrelated} with the $B$-modes, while also exploiting the distinct frequency signatures of cosmic birefringence and instrumental miscalibration to discriminate between them.

From this decomposition, we define two CMB $E$-mode maps: one reconstructed from a standard ILC (containing both correlated and uncorrelated components), and another from the Hybrid ILC (containing only the uncorrelated component). Their difference isolates the $E$-mode component induced by cosmic birefringence, which, in the small-angle limit, is linearly related to the $B$-mode field. This relation allows a map-level regression analysis to estimate $\beta$,  potentially yielding tighter constraints than those derived from power-spectrum methods.

The proposed approach is particularly relevant for future experiments like \lb~\cite{LiteBIRD2023}, which aim to achieve high sensitivity in polarization across multiple frequency channels with tightly controlled systematics. By combining $E$- and $B$-mode information at the map level and performing field-level inference, our method offers a powerful and complementary alternative to traditional birefringence searches.

This paper is organised as follows. Section~\ref{sec:theory} provides a brief overview of the theoretical background related to cosmic birefringence and instrumental miscalibration effects on CMB polarization. In Section~\ref{sec:hybridilc}, we describe the Hybrid ILC formalism and its role in isolating uncorrelated CMB $E$- and $B$-mode components. Section~\ref{sec:linreg} introduces our linear regression framework, with Section~\ref{subsec:formula} presenting the theoretical basis for the linear relationship between correlated $E$- and $B$-modes, and Section~\ref{subsec:regression} describing our map-level regression method for estimating $\beta$. Section~\ref{sec:results} presents forecasts based on simulated observations with \lb-like specifications (Sections~\ref{subsec:sims}--\ref{subsec:beta_linreg}), followed by an application of the method to current \pl\ PR4 data (Section~\ref{subsec:pr4}).
We conclude in Section~\ref{sec:conclusion}. A summary of the notations used throughout the paper is provided in Appendix~\ref{sec:notation}.

\section{Theoretical background}
\label{sec:theory}

Cosmic birefringence \citep{Carroll1990} and miscalibrated instrumental polarization angles \citep{Hu2003,Shimon2008,Miller2009,Yadav2009,Keating2013} can both induce mixing between $E$- and $B$-mode polarization in multi-frequency sky observations. However, as highlighted by \cite{Minami2019,Minami2020}, these two effects impact the cosmic microwave background (CMB) radiation and Galactic foreground emission in distinct ways:
\begin{align}
\label{eq:rotation}
   \left(\begin{array}{c}
        E_\nu(\hat{n}) \\
        B_\nu(\hat{n})
    \end{array}\right)
= R\left(\alpha_\nu + \beta\right) 
       \left(\begin{array}{c}
         a_\nu E^{\rm CMB}(\hat{n}) \\
        a_\nu B^{\rm CMB}(\hat{n}) 
    \end{array}\right)
    + R\left(\alpha_\nu\right) 
      \left(\begin{array}{c}
         E_\nu^{\rm FG}(\hat{n}) \\
        B_\nu^{\rm FG}(\hat{n}) 
    \end{array}\right)
    + \left(\begin{array}{c}
         E_\nu^{\rm N}(\hat{n}) \\
        B_\nu^{\rm N}(\hat{n}) 
    \end{array}\right)\,,
\end{align}
where $a_\nu$ is the spectral energy distribution (SED) of the CMB across frequencies $\nu$\footnote{In thermodynamic temperature units, $a_\nu = 1$ across all frequencies $\nu$.} and $E_\nu(\hat{n})$, $B_\nu(\hat{n})$ are the observed $E$- and $B$-mode polarization signals at frequency $\nu$ and sky direction $\hat{n}$. These receive contributions from the CMB anisotropies $E^{\rm CMB}(\hat{n})$ and $B^{\rm CMB}(\hat{n})$, Galactic foreground components $E_\nu^{\rm FG}(\hat{n})$ and $B_\nu^{\rm FG}(\hat{n})$, and instrumental noise $E_\nu^{\rm N}(\hat{n})$ and $B_\nu^{\rm N}(\hat{n})$. Here, $\alpha_\nu$ denotes the frequency-dependent polarization angle miscalibration from the instrument, $\beta$ is the cosmic birefringence angle, and
\begin{align}
\label{eq:rotation_matrix}
R\left(\theta\right) = \left(\begin{array}{cc}
         \cos 2\theta & - \sin 2\theta \\
       \sin 2\theta &  \cos 2\theta
   \end{array}\right)
\end{align}
is the rotation matrix by an angle $2\theta$.

The immediate consequence of these rotations is an effective cross-correlation between the observed $E$- and $B$-modes for any pair of frequency channels $(\nu,\nu')$. Hereafter, we compute the resulting $EB$ angular cross-power spectrum using two complementary formulations: one that distinguishes contributions from the CMB and foregrounds (Section~\ref{sec:eb1}), and another that separates cosmological and instrumental contributions (Section~\ref{sec:eb2}). Both formulations will be used later to derive key properties of the Hybrid ILC method (Section~\ref{sec:hybridilc}).

\subsection{CMB and foreground contributions to $EB$ cross-power spectrum}
\label{sec:eb1}

The angular cross-power spectrum between the observed $E$- and $B$-modes for any pair of frequency channels $(\nu,\nu')$ can be directly derived from Equation~\eqref{eq:rotation} by expressing the rotated polarization fields in harmonic space using spherical harmonic coefficients. Assuming no intrinsic $EB$ correlation in the CMB, foregrounds, or instrumental noise,\footnote{Although standard $\Lambda\text{CDM}$ cosmology predicts no primordial $EB$ correlation, a nonzero foreground $EB$ signal could arise from the filamentary structure of dust and synchrotron emission \citep{Clark2021}. This would introduce additional terms, $\cos 4\alpha_\nu\, C_\ell^{E_\nu B_{\nu},{\rm FG}}$ and $C_\ell^{E_\nu B_{\nu},{\rm FG}} / \cos 4\alpha_\nu$, in Equations~\eqref{eq:ceb_nunu} and~\eqref{eq:ceb2_nunu}, respectively. However, we omit these terms for two reasons: (i) current measurements indicate that the foreground $EB$ signal is statistically consistent with zero \citep{planck2016-l11A,Martire2022}, and (ii) non-parametric component separation methods, such as the Hybrid ILC employed in this work, do not require any explicit modelling or parameterization of the foreground $EB$ contribution.} the resulting expression is:
\begin{align}
\label{eq:ceb}
C^{E_\nu B_{\nu'}}_\ell &= \frac{\sin\left(4\beta +2\left(\alpha_\nu+\alpha_{\nu'}\right)\right)}{2}a_\nu a_{\nu'}\left(C_\ell^{EE,{\rm CMB}} - C_\ell^{BB,{\rm CMB}}\right)\cr
				&  - \frac{\sin\left(2\left(\alpha_\nu-\alpha_{\nu'}\right)\right)}{2}a_\nu a_{\nu'}\left(C_\ell^{EE,{\rm CMB}} + C_\ell^{BB,{\rm CMB}}\right)\cr
				& + \frac{\sin\left(2\left(\alpha_\nu+\alpha_{\nu'}\right)\right)}{2}\left(C_\ell^{E_\nu E_{\nu'},{\rm FG}} - C_\ell^{B_\nu B_{\nu'},{\rm FG}}\right)\cr
				&  - \frac{\sin\left(2\left(\alpha_\nu-\alpha_{\nu'}\right)\right)}{2}\left(C_\ell^{E_\nu E_{\nu'},{\rm FG}} + C_\ell^{B_\nu B_{\nu'},{\rm FG}}\right)\,,
\end{align}
For a pair of identical frequencies, i.e. $\nu=\nu'$, this expression reduces to:
\begin{align}
\label{eq:ceb_nunu}
C^{E_\nu B_{\nu}}_\ell &= \frac{\sin\left(4\beta +4\alpha_\nu\right)}{2}a_\nu a_{\nu}\left(C_\ell^{EE,{\rm CMB}} - C_\ell^{BB,{\rm CMB}}\right)\cr
				& + \frac{\sin\left(4\alpha_\nu\right)}{2}\left(C_\ell^{E_\nu E_{\nu},{\rm FG}} - C_\ell^{B_\nu B_{\nu},{\rm FG}}\right)\,.
\end{align}

This result can be further simplified in specific limits. First, in the absence of cosmic birefringence ($\beta = 0$), Equation~\eqref{eq:ceb_nunu} becomes:
\begin{align}
\label{eq:ceb_zerobeta_nunu}
C^{E_\nu B_{\nu}}_\ell &\simeq \frac{\sin\left(4\alpha_\nu\right)}{2}a_\nu a_{\nu}\left(C_\ell^{EE,{\rm CMB}} - C_\ell^{BB,{\rm CMB}}\right)\cr
				& + \frac{\sin\left(4\alpha_\nu\right)}{2}\left(C_\ell^{E_\nu E_{\nu},{\rm FG}} - C_\ell^{B_\nu B_{\nu},{\rm FG}}\right)\,.
\end{align}
Conversely, in the presence of cosmic birefringence but relatively small instrumental miscalibration ($\alpha_\nu \ll \beta$), Equation~\eqref{eq:ceb_nunu} reduces to:
\begin{align}
\label{eq:ceb_zeroalpha_nunu}
C^{E_\nu B_{\nu}}_\ell &\simeq \frac{\sin\left(4\beta\right)}{2}a_\nu a_{\nu}\left(C_\ell^{EE,{\rm CMB}} - C_\ell^{BB,{\rm CMB}}\right)\cr
				& + \frac{\sin\left(4\alpha_\nu\right)}{2}\left(C_\ell^{E_\nu E_{\nu},{\rm FG}} - C_\ell^{B_\nu B_{\nu},{\rm FG}}\right)\,.
\end{align}
The strong asymmetry $C_\ell^{EE,{\rm CMB}} \gg C_\ell^{BB,{\rm CMB}}$ for the CMB and the moderate asymmetry $C_\ell^{E_\nu E_{\nu},{\rm FG}} \gtrsim 2 C_\ell^{B_\nu B_{\nu},{\rm FG}}$ for Galactic foregrounds thus makes the $EB$ cross-power spectrum a sensitive probe of cosmic birefringence through the CMB contribution, and of polarization angle miscalibration through the Galactic foreground contribution (Equations~\ref{eq:ceb_nunu} and \ref{eq:ceb_zeroalpha_nunu}).

\subsection{Cosmological and instrumental contributions to $EB$ cross-power spectrum}
\label{sec:eb2}

While Equations~\eqref{eq:ceb}-\eqref{eq:ceb_nunu} distinguish between contributions from the CMB and foregrounds, an alternative formulation of the observed $EB$ cross-power spectrum can be derived that instead separates cosmological ($\beta$) and instrumental ($\alpha_\nu$) contributions. To begin, we consider how the observed $EE$ and $BB$ auto-power spectra for a pair of frequency channels $(\nu,\nu')$ are modified by the polarization rotation:
\begin{subequations}
\begin{flalign}
\label{eq:cee}
C^{E_\nu E_{\nu'}}_\ell &= \frac{\cos\left(4\beta +2\left(\alpha_\nu+\alpha_{\nu'}\right)\right)}{2}a_\nu a_{\nu'}\left(C_\ell^{EE,{\rm CMB}} - C_\ell^{BB,{\rm CMB}}\right)\cr
				&  + \frac{\cos\left(2\left(\alpha_\nu-\alpha_{\nu'}\right)\right)}{2}a_\nu a_{\nu'}\left(C_\ell^{EE,{\rm CMB}} + C_\ell^{BB,{\rm CMB}}\right)\cr
				& + \frac{\cos\left(2\left(\alpha_\nu+\alpha_{\nu'}\right)\right)}{2}\left(C_\ell^{E_\nu E_{\nu'},{\rm FG}} - C_\ell^{B_\nu B_{\nu'},{\rm FG}}\right)\cr
				&  + \frac{\cos\left(2\left(\alpha_\nu-\alpha_{\nu'}\right)\right)}{2}\left(C_\ell^{E_\nu E_{\nu'},{\rm FG}} + C_\ell^{B_\nu B_{\nu'},{\rm FG}}\right)\cr
				& + C_\ell^{E_\nu E_{\nu'},{\rm N}}\delta_\nu^{\nu'}\,,\\
\cr
\label{eq:cbb}
C^{B_\nu B_{\nu'}}_\ell &= \frac{\cos\left(4\beta +2\left(\alpha_\nu+\alpha_{\nu'}\right)\right)}{2}a_\nu a_{\nu'}\left(C_\ell^{BB,{\rm CMB}} - C_\ell^{EE,{\rm CMB}}\right)\cr
				&  + \frac{\cos\left(2\left(\alpha_\nu-\alpha_{\nu'}\right)\right)}{2}a_\nu a_{\nu'}\left(C_\ell^{BB,{\rm CMB}} + C_\ell^{EE,{\rm CMB}}\right)\cr
				& + \frac{\cos\left(2\left(\alpha_\nu+\alpha_{\nu'}\right)\right)}{2}\left(C_\ell^{B_\nu B_{\nu'},{\rm FG}} - C_\ell^{E_\nu E_{\nu'},{\rm FG}}\right)\cr
				&  + \frac{\cos\left(2\left(\alpha_\nu-\alpha_{\nu'}\right)\right)}{2}\left(C_\ell^{B_\nu B_{\nu'},{\rm FG}} + C_\ell^{E_\nu E_{\nu'},{\rm FG}}\right)\cr
				& + C_\ell^{B_\nu B_{\nu'},{\rm N}}\delta_\nu^{\nu'}\,.
\end{flalign}
\end{subequations}

Given that the noise polarization angles are randomly oriented, we have $C_\ell^{E_\nu E_\nu,{\rm N}} \simeq C_\ell^{B_\nu B_\nu,{\rm N}}$, so that the difference between the $EE$ and $BB$ spectra can be expressed as:
\begin{align}
\label{eq:cee-cbb}
C^{E_\nu E_{\nu'}}_\ell - C^{B_\nu B_{\nu'}}_\ell &= \cos\left(4\beta +2\left(\alpha_\nu+\alpha_{\nu'}\right)\right)a_\nu a_{\nu'}\left(C_\ell^{EE,{\rm CMB}} - C_\ell^{BB,{\rm CMB}}\right)\cr
				& + \cos\left(2\left(\alpha_\nu+\alpha_{\nu'}\right)\right)\left(C_\ell^{E_\nu E_{\nu'},{\rm FG}} - C_\ell^{B_\nu B_{\nu'},{\rm FG}}\right)\,.
\end{align}
Multiplying both sides by $\tan(2(\alpha_\nu + \alpha_{\nu'}))/2$ yields
\begin{align}
\label{eq:cee-cbb2}
 &\frac{\tan\left(2\left(\alpha_\nu+\alpha_{\nu'}\right)\right)}{2}\left(C^{E_\nu E_{\nu'}}_\ell - C^{B_\nu B_{\nu'}}_\ell\right) \cr
& = \frac{\sin\left(2\left(\alpha_\nu+\alpha_{\nu'}\right)\right)\cos\left(4\beta +2\left(\alpha_\nu+\alpha_{\nu'}\right)\right)}{2\cos\left(2\left(\alpha_\nu+\alpha_{\nu'}\right)\right)}a_\nu a_{\nu'}\left(C_\ell^{EE,{\rm CMB}} - C_\ell^{BB,{\rm CMB}}\right)\cr
				& + \frac{\sin\left(2\left(\alpha_\nu+\alpha_{\nu'}\right)\right)}{2}\left(C_\ell^{E_\nu E_{\nu'},{\rm FG}} - C_\ell^{B_\nu B_{\nu'},{\rm FG}}\right)\cr
				& = \frac{\sin\left(4\beta +2\left(\alpha_\nu+\alpha_{\nu'}\right)\right)}{2}a_\nu a_{\nu'}\left(C_\ell^{EE,{\rm CMB}} - C_\ell^{BB,{\rm CMB}}\right)\cr
				& - \frac{\sin\left(4\beta\right)}{2\cos\left(2\left(\alpha_\nu+\alpha_{\nu'}\right)\right)}a_\nu a_{\nu'}\left(C_\ell^{EE,{\rm CMB}} - C_\ell^{BB,{\rm CMB}}\right)\cr
				& + \frac{\sin\left(2\left(\alpha_\nu+\alpha_{\nu'}\right)\right)}{2}\left(C_\ell^{E_\nu E_{\nu'},{\rm FG}} - C_\ell^{B_\nu B_{\nu'},{\rm FG}}\right)\,,
\end{align}
where the trigonometric identity $\cos x \sin y =  \sin x \cos y - \sin\left(x-y\right)$, with $x=4\beta +2\left(\alpha_\nu+\alpha_{\nu'}\right)$ and ${y=2\left(\alpha_\nu+\alpha_{\nu'}\right)}$, has been used for the last equality. 

Therefore, Equation~\eqref{eq:ceb} can be equivalently expressed as:
\begin{align}
\label{eq:ceb2}
				C^{E_\nu B_{\nu'}}_\ell & = \frac{\tan\left(2\left(\alpha_\nu+\alpha_{\nu'}\right)\right)}{2}\left(C^{E_\nu E_{\nu'}}_\ell - C^{B_\nu B_{\nu'}}_\ell\right) \cr
				&+ \frac{\sin\left(4\beta\right)}{2\cos\left(2\left(\alpha_\nu+\alpha_{\nu'}\right)\right)}a_\nu a_{\nu'}\left(C_\ell^{EE,{\rm CMB}} - C_\ell^{BB,{\rm CMB}}\right)\cr
				&  - \frac{\sin\left(2\left(\alpha_\nu-\alpha_{\nu'}\right)\right)}{2}a_\nu a_{\nu'}\left(C_\ell^{EE,{\rm CMB}} + C_\ell^{BB,{\rm CMB}}\right)\cr
				&  - \frac{\sin\left(2\left(\alpha_\nu-\alpha_{\nu'}\right)\right)}{2}\left(C_\ell^{E_\nu E_{\nu'},{\rm FG}} + C_\ell^{B_\nu B_{\nu'},{\rm FG}}\right)\,,
\end{align}
providing an alternative expression to Equation~\eqref{eq:ceb} that explicitly separates contributions from $\beta$ and $\alpha_\nu$, in the small-angle limit where $\cos(2(\alpha_\nu + \alpha_{\nu'})) \simeq 1$.
For identical frequency channels $\nu = \nu'$, Equation~\eqref{eq:ceb2} simplifies to: 
\begin{align}
\label{eq:ceb2_nunu}
C^{E_\nu B_{\nu}}_\ell & = \frac{\tan\left(4\alpha_\nu\right)}{2}\left(C^{E_\nu E_{\nu}}_\ell - C^{B_\nu B_{\nu}}_\ell\right) + \frac{\sin\left(4\beta\right)}{2\cos\left(4\alpha_\nu\right)}a_\nu a_{\nu}\left(C_\ell^{EE,{\rm CMB}} - C_\ell^{BB,{\rm CMB}}\right)\,,
\end{align}
 which corresponds to the expression commonly adopted in recent CMB polarization analyses of cosmic birefringence \citep{Minami2019,Minami2020,Diego2022,Eskilt2022,Eskilt2022b,Komatsu2022,Diego2023}.

Again, this alternative expression can be further simplified in specific cases. In the absence of cosmic birefringence ($\beta = 0$), Equation~\eqref{eq:ceb2_nunu} reduces to:
\begin{align}
\label{eq:ceb2_zerobeta_nunu}
				C^{E_\nu B_{\nu}}_\ell & \simeq \frac{\tan\left(4\alpha_\nu\right)}{2}\left(C^{E_\nu E_{\nu}}_\ell - C^{B_\nu B_{\nu}}_\ell\right)\,,
\end{align}
which is consitent with the former Equation~\eqref{eq:ceb_zerobeta_nunu} when considering the limit $\beta\to 0$ in Equation~\eqref{eq:cee-cbb}. Conversely, in the presence of cosmic birefringence but with relatively small instrumental miscalibration ($\alpha_\nu \ll \beta$),  Equation~\eqref{eq:ceb2_nunu} becomes: 
\begin{align}
\label{eq:ceb2_zeroalpha_nunu}
				C^{E_\nu B_{\nu}}_\ell & \simeq \frac{\sin\left(4\beta\right)}{2}a_\nu a_{\nu}\left(C_\ell^{EE,{\rm CMB}} - C_\ell^{BB,{\rm CMB}}\right)\cr
								& +  \frac{\tan\left(4\alpha_\nu\right)}{2}\left(C^{E_\nu E_{\nu}}_\ell - C^{B_\nu B_{\nu}}_\ell\right)\,.
\end{align}

\section{A hybrid ILC CMB map from combined E- and B-mode channels}
\label{sec:hybridilc}

We introduce a novel approach for constructing a \emph{hybrid} CMB $E$-mode map, $\widehat{E}^{\rm CMB}(\hat{n})$, which retains only the \emph{uncorrelated} component to $B$-modes, by combining and weighting both $E$- and $B$-mode frequency maps in an Internal Linear Combination (ILC). We refer to this new method as the Hybrid ILC. In parallel, we build the full CMB $E$-mode map, $\widetilde{E}^{\rm CMB}(\hat{n})$, using a standard ILC \citep{WMAP2003,Tegmark2003,Eriksen2004,Delabrouille2009} that combines only $E$-mode frequency maps. The difference between the two, $\widetilde{E}^{\rm CMB}(\hat{n}) - \widehat{E}^{\rm CMB}(\hat{n})$, isolates the \emph{correlated} component of the CMB $E$-modes that is sensitive to cosmic birefringence, while significantly reducing residual foreground contamination.

Importantly, since instrumental miscalibration varies with frequency, it is naturally downweighted by the ILC across both sky position and angular scale, leaving an anisotropic residual. In contrast, the cosmological birefringence angle $\beta$, being achromatic, is consistently preserved across the entire map.

This method aims to place tighter constraints on the birefringence angle $\beta$ by leveraging spatial correlations between $E$- and $B$-modes at the map level, rather than relying solely on power-spectrum-based summary statistics. Such a field-level approach has the potential to achieve greater sensitivity than traditional $EB$ cross-spectrum analyses.

\subsection{Formulation of the component separation problem}
\label{subsec:compsep}

Equation~\eqref{eq:rotation} can be recast as a component separation problem:
\begin{align}
\label{eq:rotation2}
E_\nu(\hat{n})  = a_\nu E_\nu^{\,\prime\,\rm CMB}(\hat{n}) +  E_\nu^{\rm FG+N}(\hat{n})\,\\
\label{eq:rotation2b}
B_\nu(\hat{n})  = a_\nu B_\nu^{\,\prime\,\rm CMB}(\hat{n}) +  B_\nu^{\rm FG+N}(\hat{n})\,
\end{align}
where $E_\nu^{\,\prime\,\rm CMB}(\hat{n})$ and $B_\nu^{\,\prime\,\rm CMB}(\hat{n})$ are the rotated CMB $E$- and $B$-mode components, and $E_\nu^{\rm FG+N}(\hat{n})$, $B_\nu^{\rm FG+N}(\hat{n})$ represent the overall contamination from foregrounds and noise in each polarization channel. These nuisance terms are left unparameterized within our blind ILC framework,\footnote{In this non-parametric approach, the nuisance terms $E_\nu^{\rm FG+N}(\hat{n})$ and $B_\nu^{\rm FG+N}(\hat{n})$ may include intrinsic foreground $EB$ correlations.} allowing for maximal model independence.

This formulation highlights effective, \emph{frequency-dependent} CMB polarization anisotropies:
\begin{align}
\label{eq:effcmb}
E_\nu^{\,\prime\,\rm CMB}(\hat{n}) = \cos(2\beta + 2\alpha_\nu) E^{\rm CMB}(\hat{n}) -  \sin(2\beta + 2\alpha_\nu) B^{\rm CMB}(\hat{n})\,,\\
\label{eq:effcmbb}
B_\nu^{\,\prime\,\rm CMB}(\hat{n}) = \cos(2\beta + 2\alpha_\nu) B^{\rm CMB}(\hat{n}) +  \sin(2\beta + 2\alpha_\nu) E^{\rm CMB}(\hat{n})\,,
\end{align}
with non-zero effective cross-correlation (see Equation~\ref{eq:ceb}):
\begin{align}
\label{eq:cmb_eb}
C^{E_\nu^{\,\prime\,} B_{\nu'}^{\,\prime\,}, {\rm CMB}}_\ell &= \frac{\sin\left(4\beta +2\left(\alpha_\nu+\alpha_{\nu'}\right)\right)}{2}\left(C_\ell^{EE,{\rm CMB}} - C_\ell^{BB,{\rm CMB}}\right)\cr
				&  - \frac{\sin\left(2\left(\alpha_\nu-\alpha_{\nu'}\right)\right)}{2}\left(C_\ell^{EE,{\rm CMB}} + C_\ell^{BB,{\rm CMB}}\right)\,.
\end{align}
Here $E^{\rm CMB}(\hat{n})$ and $B^{\rm CMB}(\hat{n})$ denote the intrinsic, achromatic, and uncorrelated CMB polarization fields predicted by $\Lambda$CDM, prior to any rotation, with corresponding power spectra  $C_\ell^{EE,{\rm CMB}}$ and $C_\ell^{BB,{\rm CMB}}$, respectively.

Under the small-angle approximation ($\alpha_\nu \ll 1$, $\beta \ll 1$), the system can be expressed in an $n_f$-dimensional vector form, where $n_f$ is the number of frequency channels:
\begin{subequations}
\begin{align}
\label{eq:rotation3e}
\bdE(\hat{n})  = \bda E^{\rm CMB}(\hat{n}) -  2\beta \bda B^{\rm CMB}(\hat{n}) - 2\boldsymbol{\alpha} B^{\rm CMB}(\hat{n}) +  \bdE^{\rm FG+N}(\hat{n})\,,\\
\label{eq:rotation3b}
\bdB(\hat{n})  = \bda B^{\rm CMB}(\hat{n}) +  2\beta \bda E^{\rm CMB}(\hat{n}) + 2\boldsymbol{\alpha} E^{\rm CMB}(\hat{n}) +  \bdB^{\rm FG+N}(\hat{n})\,.
\end{align}
\end{subequations}
Here  $\bda$ is the $n_f$-dimensional vector of CMB SED coefficients $a_\nu$ across frequency channels, and $\boldsymbol{\alpha}$ is the corresponding vector of instrumental miscalibration angles $\alpha_\nu$.\footnote{Technically, the entries of $\boldsymbol{\alpha}$ are the products $a_\nu \alpha_\nu$ between CMB SED coefficients and miscalibration angles; however, under the assumption of thermodynamic temperature units, where $a_\nu=1$ for all $\nu$, they reduce to the miscalibration angles themselves.} The vectors $\bdE(\hat{n})$ and $\bdB(\hat{n})$ contains the observed $E$- and $B$-mode signals at each frequency, while the vectors $\bdE^{\rm FG+N}(\hat{n})$ and $\bdB^{\rm FG+N}(\hat{n})$ capture the combined contributions from foregrounds and instrumental noise in the respective polarization modes.

In the following, we denote the covariance matrices of the observed $E$- and $B$-mode frequency maps as $\tC^{EE} = \langle \bdE\bdE^\top\rangle$ and $\tC^{BB} = \langle \bdB\bdB^\top\rangle$, where $\bdE^\top$ and $\bdB^\top$ are the transposes of the vectors $\bdE$ and $\bdB$, respectively. These $(n_f \times n_f)$ matrices may be computed in either pixel or harmonic space, depending on the specific ILC implementation; hence, we retain a general notation that is agnostic to representation. Similarly, the ILC weights $w_\nu$ across the $n_f$ frequency channels are assembled into the $n_f$-dimensional $\bdw$, which may vary with position or angular scale (for instance, pixel-by-pixel in map space \citep{Eriksen2004}, multipole-by-multipole in harmonic space \citep{Tegmark2003}, or both in a localized frame such as needlets \citep{Delabrouille2009}). 

\subsection{ILC-based discrimination between cosmic birefringence and miscalibration} 
\label{subsec:discrimination}

Applying the standard ILC weights \cite{WMAP2003,Tegmark2003,Eriksen2004,Delabrouille2009} 
\begin{align} 
\label{eq:stdilcweights}
\widetilde{\bdw}_{E} &= \frac{ (\tC^{EE})^{-1} \bda }{ \bda^\top (\tC^{EE})^{-1} \bda }\,, \quad \widetilde{\bdw}_{B} = \frac{ (\tC^{BB})^{-1} \bda }{ \bda^\top (\tC^{BB})^{-1} \bda }\,, 
\end{align}
separately to the observed $E$- and $B$-mode frequency maps (Equations~\ref{eq:rotation3e}--\ref{eq:rotation3b}) yields the following CMB estimates:
\begin{subequations}
\begin{align}
\label{eq:example_ilce}
\widetilde{E}^{\rm CMB} = \widetilde{\bdw}_{E}^\top \bdE = E^{\rm CMB} - 2\left(\beta +  \widetilde{\bdw}_{E}^\top\boldsymbol{\alpha}\right) B^{\rm CMB}  + \widetilde{\bdw}_{E}^\top  \bdE^{\rm FG+N}\\
\label{eq:example_ilcb}
\widetilde{B}^{\rm CMB} = \widetilde{\bdw}_{B}^\top \bdB = B^{\rm CMB} + 2\left(\beta +  \widetilde{\bdw}_{B}^\top\boldsymbol{\alpha}\right) E^{\rm CMB}  + \widetilde{\bdw}_{B}^\top  \bdB^{\rm FG+N}
\end{align}
\end{subequations}
where we have used the fact that the standard ILC weights are constrained to offer unit response to the CMB SED, i.e., $\widetilde{\bdw}_{E}^\top \bda = 1$ and $\widetilde{\bdw}_{B}^\top \bda = 1$, but are agnostic to the miscalibration angles $\alpha_\nu$, which remain unknown. Consequently, the extracted CMB maps inherit weighted averages of these angles, denoted 
\begin{align}
\label{eq:projected_alpha}
\overline{\alpha}^{\,(E)} =  \widetilde{\bdw}_{E}^\top\boldsymbol{\alpha}\,,\quad \overline{\alpha}^{\,(B)} =  \widetilde{\bdw}_{B}^\top\boldsymbol{\alpha}\,,
\end{align} 
which vary across the sky or multipoles (e.g., $\overline{\alpha}^{\,(E)}(\hat{n})$ or $\overline{\alpha}^{\,(E)}_{\ell m}$). Unlike the constant birefringence angle $\beta$, these variations arise from the spatial and spectral dependence of the ILC weights, which adapt to the local variance of the foreground contamination.
As a result, the ILC responds differently to achromatic (cosmic birefringence) and chromatic (instrumental) rotation effects, offering a pathway to discriminate between them based on their distinct spatial signatures after component separation, as illustrated below.

In the regime where the reconstructed CMB dominates over residual foreground and noise contamination, the $EB$ cross-power spectrum between the standard ILC CMB $E$- and $B$-mode maps (Equations~\ref{eq:example_ilce}--\ref{eq:example_ilcb}) is given by
\begin{align} 
\label{eq:eb_stdilc}
C_\ell^{\widetilde{E}\widetilde{B},{\rm CMB}} &= \left\langle \widetilde{E}_{\ell m}^{\rm CMB} \widetilde{B}_{\ell m}^{*\rm CMB}\right\rangle =  2\left(\beta + \overline{\alpha}^{\,(B)}_\ell \right) C_\ell^{EE,{\rm CMB}} - 2\left(\beta + \overline{\alpha}^{\,(E)}_\ell\right) C_\ell^{BB,{\rm CMB}}\,,
\end{align}
where, for illustration, we assume an ILC implementation in harmonic space, resulting in $\ell$-dependent weights $\widetilde{\bdw}_{E}(\ell)$ and $\widetilde{\bdw}_{B}(\ell)$, and hence $\ell$-dependent projected miscalibration angles $\overline{\alpha}^{\,(E)}_{\ell}$ and $\overline{\alpha}^{\,(B)}_{\ell}$.

A key insight from Equation~\eqref{eq:eb_stdilc} is that the ratio
\begin{align} 
\label{eq:magicratio}
\mathcal{R}_\ell = \frac{ C_\ell^{\widetilde{E}\widetilde{B},{\rm CMB}} } { C_\ell^{EE,{\rm CMB}}  - C_\ell^{BB,{\rm CMB}} } = 2 \frac{\left(\beta + \overline{\alpha}^{\,(B)}_\ell \right) C_\ell^{EE,{\rm CMB}} - \left(\beta + \overline{\alpha}^{\,(E)}_\ell\right) C_\ell^{BB,{\rm CMB}} } { C_\ell^{EE,{\rm CMB}}  - C_\ell^{BB,{\rm CMB}} }
\end{align}
acts as a diagnostic for birefringence: in the absence of cosmic birefringence ($\beta = 0$), $\mathcal{R}_\ell$ exhibits multipole-dependent variations driven by the miscalibration terms $\overline{\alpha}^{\,(E)}_\ell$ and $\overline{\alpha}^{\,(B)}_\ell$:
\begin{align} 
\label{eq:magicratio_alpha}
\mathcal{R}_\ell \underset{\beta = 0}{\simeq} 2 \frac{ \overline{\alpha}^{\,(B)}_\ell C_\ell^{EE,{\rm CMB}} - \overline{\alpha}^{\,(E)}_\ell C_\ell^{BB,{\rm CMB}} } { C_\ell^{EE,{\rm CMB}}  - C_\ell^{BB,{\rm CMB}} }\,;
\end{align}
whereas in the presence of significant birefringence ($\beta \gg \langle \alpha_{E} \rangle_\ell,  \langle \alpha_{B} \rangle_\ell$), the ratio becomes approximately flat, with 
\begin{align}
\label{eq:magicratio_beta}
\mathcal{R}_\ell \underset{\beta \gg \alpha_\nu}{\simeq} 2\beta = \text{constant}\,.
\end{align}
 This makes $\mathcal{R}_\ell$ a useful and easily computable indicator to distinguish between instrumental miscalibration and genuine cosmic birefringence, an approach we will adopt in the forthcoming data analysis. 

\subsection{Disentangling correlated and uncorrelated CMB components: a toy model} 
\label{subsec:toy}

To build some intuition, we consider a simplified case with two polarization channels but only a single, perfectly calibrated frequency channel, i.e., $\alpha_\nu = 0$ for that channel $\nu$. In this simplified, single-frequency scenario, and assuming a high signal-to-noise regime where noise and foregrounds can be neglected (e.g., at intermediate multipoles and high Galactic latitudes), the standard ILC CMB maps reduce to the observed $E$- and $B$-mode maps, i.e. $\widetilde{E}^{\rm CMB} = E \simeq E^{\rm CMB} - 2\beta B^{\rm CMB}$ and $\widetilde{B}^{\rm CMB} = B \simeq B^{\rm CMB} + 2\beta E^{\rm CMB}$. 
Although these two fields are correlated through cosmic birefringence, there is no direct linear regression between them, as each contains contributions from both intrinsic $E$- and $B$-modes. 

The motivation behind our Hybrid ILC approach is to isolate precisely the component of the CMB $E$-mode field that does linearly regress with the CMB $B$-modes, namely, a field of the form $\widehat{E}^{\rm CMB} = -2 \beta K B^{\rm CMB}$, where $K$ is a deterministic kernel. In practice, this involves reconstructing only the portion of the CMB $E$-mode signal that is correlated with the $B$-modes, while projecting out the uncorrelated component.

This separation proceeds in two steps. First, we reconstruct the \emph{uncorrelated} component of the CMB $E$-mode field, denoted $\widehat{E}^{\rm CMB}$, via a Hybrid ILC that linearly combines and weights the observed $E$- and $B$-mode maps to eliminate any correlation with CMB $B$-modes. Second, we subtract this uncorrelated component from the full $E$-mode map obtained via the standard ILC, $\widetilde{E}^{\rm CMB}$, thus leaving only the \emph{correlated} component of the CMB $E$-modes which linearly regresses with CMB $B$-modes. 

In our toy model, we thus seek a combination of the form $\widehat{E}^{\rm CMB} = E + w B$ that is uncorrelated with the CMB $B$-mode field $\widetilde{B}^{\rm CMB} = B$. Enforcing this decorrelation $\langle \widehat{E}^{\rm CMB}, \widetilde{B}^{\rm CMB}\rangle = 0$ leads to the condition:
\begin{align}
\left\langle E + w B, B \right\rangle = 0\,,
\end{align}
which, in harmonic space, gives the optimal weight:
\begin{align}
w = - \frac{C_\ell^{EB}}{C_\ell^{BB}}\,.
\end{align}

With this choice of $w$, the uncorrelated component of the $E$-mode field is given by:
\begin{align}
\widehat{E}^{\rm CMB}_{\ell m} = E_{\ell m} - \frac{C_\ell^{EB}}{C_\ell^{BB}}B_{\ell m}\,.
\end{align}
Since the standard ILC reduces to $\widetilde{E}^{\rm CMB} = E$ in this single-frequency case, the difference between the standard and Hybrid ILC maps becomes:
\begin{align}
\widetilde{E}^{\rm CMB}_{\ell m} - \widehat{E}^{\rm CMB}_{\ell m} = \frac{C_\ell^{EB}}{C_\ell^{BB}}B_{\ell m}\,.
\end{align}
Replacing the observed $E$ and $B$ fields with the standard ILC estimates, $\widetilde{E}^{\rm CMB}$ and $\widetilde{B}^{\rm CMB}$, and using the expression for the $EB$ cross-power spectrum from Equation~\eqref{eq:eb_stdilc}, we obtain:
\begin{align}
\widetilde{E}^{\rm CMB}_{\ell m} - \widehat{E}^{\rm CMB}_{\ell m} =2\beta \frac{C_\ell^{EE,{\rm CMB}} - C_\ell^{BB,{\rm CMB}}}{C_\ell^{\widetilde{B}\widetilde{B},{\rm CMB}}}\widetilde{B}^{\rm CMB}_{\ell m}\,.
\end{align}
This expression isolates the correlated, first-order component of the CMB $E$-mode field that linearly regresses with the CMB $B$-modes. Moreover, the difference $\widetilde{E}^{\rm CMB} - \widehat{E}^{\rm CMB}$ between the two reconstructed CMB $E$-mode maps is also expected to substantially suppress residual foreground contamination, as both maps share similar residuals that tend to cancel in the subtraction. 
A full and rigorous derivation of the Hybrid ILC and the associated linear regression in the general case is presented in the following sections.

\subsection{Full derivation of the Hybrid ILC}
\label{subsec:hybridweights}

In this work, we aim to extract the \emph{uncorrelated} component of the CMB $E$-mode anisotropies, denoted $\widehat{E}^{\rm CMB}(\hat{n})$, by combining both $E$ and $B$ polarization channels within the ILC framework. To this end, we seek $2n_f$ specific weights $\bdw^\top = \begin{pmatrix} \bdw_{E}^\top\  \bdw_{B}^\top \end{pmatrix}$ to be assigned to the $2n_f$-dimensional vector  $\bdX^\top  = \begin{pmatrix} \bdE^\top\  \bdB^\top \end{pmatrix}$, which is composed of the $n_f$ frequency maps observed in $E$-mode and the $n_f$ frequency maps observed in $B$-mode.
The resulting Hybrid ILC estimate of the CMB $E$-modes is given by:
\begin{align}
\label{eq:ehat}
\widehat{E}^{\rm CMB}(\hat{n}) &= \bdw^\top \bdX(\hat{n}) \cr
        & =  \begin{pmatrix}
         \bdw_{E}^\top\ \bdw_{B}^\top \end{pmatrix}  \begin{pmatrix}
         \bdE(\hat{n})\\ \bdB(\hat{n}) \end{pmatrix} \cr 
         & = \bdw_{E}^\top \bdE(\hat{n}) +  \bdw_{B}^\top \bdB(\hat{n}) \cr
         & = \sum_{\nu=1}^{n_f}w_{E,\nu}\,  E_{\nu}(\hat{n}) + \sum_{\nu=1}^{n_f}w_{B,\nu}\,  B_{\nu}(\hat{n})\,,
\end{align}
where the weights $\bdw^\top = \begin{pmatrix} \bdw_{E}^\top\ \bdw_{B}^\top \end{pmatrix}$ are optimized to minimize the variance of $\widehat{E}^{\rm CMB}(\hat{n})$, while providing unit response to the CMB in the $E$-mode channels only:
\begin{align}
 \bdw_{E}^\top \bda = 1\,,
\end{align}
with $\bda$ representing the $n_f$-dimensional vector encoding the CMB SED across the $E$-mode channels.

Denoting the full $(2n_f \times 2n_f)$ covariance matrix of the joint $E$- and $B$-mode data as
\begin{align}
\label{eq:covmatrix}
\tC =  \left\langle \bdX \bdX^\top \right\rangle = \begin{pmatrix} \tC^{EE}\  \tC^{EB} \\ \tC^{BE}\  \tC^{BB} \end{pmatrix}\,,
\end{align}
where $\tC^{BE} = \left(\tC^{EB}\right)^\top$,\footnote{Note that $\tC^{BE} \neq \tC^{EB}$ because Equations~\eqref{eq:ceb} and \eqref{eq:ceb2} are not symmetric under the exchange $\nu \leftrightarrow \nu'$, unless all polarization angles are perfectly calibrated, i.e., $\alpha_\nu = 0$ for all $\nu$.} the solution for the Hybrid ILC weights $\bdw$ is obtained by minimizing the Lagrangian:
\begin{eqnarray}
\label{eq:lagrangian}
\mathcal{L} &=& \bdw^\top  \tC \bdw + \lambda\left(1 - \bdw^\top \bdb\right)\cr
\cr
&=&  \bdw_{E}^\top  \tC^{EE} \bdw_{E} + \bdw_{B}^\top  \tC^{BB} \bdw_{B} + \bdw_{E}^\top  \tC^{EB} \bdw_{B} + \bdw_{B}^\top  \tC^{BE} \bdw_{E} \cr
\cr
& & +\, \lambda\left(1 - \bdw_{E}^\top  \bda \right)\,,
\end{eqnarray}
which ensures overall variance minimization while preserving the targeted CMB $E$-mode signal.
Here, the $2n_f$-dimensional vector $\bdb^\top = \begin{pmatrix} \bda^\top & \boldsymbol{0}^\top \end{pmatrix}$ encodes the SED coefficients of the CMB $E$-mode across the $E$- and $B$-mode frequency channels, with the $n_f$-dimensional subvector $\bda^\top = \begin{pmatrix} 1 & \cdots & 1 \end{pmatrix}$ providing unit response across the $E$-mode channels (in thermodynamic units) and the $n_f$-dimensional vector $\boldsymbol{0}^\top = \begin{pmatrix} 0 & \cdots & 0 \end{pmatrix}$ ensuring a null response along the $B$-mode channels. The Lagrange multiplier $\lambda$ enforces the constraint that the $E$-mode component of the CMB is preserved during variance minimization. 
However, as we will see, due to the effective cross-correlation between the CMB $E$- and $B$-mode signals (Equation~\ref{eq:cmb_eb}), this constraint does not preserve the full CMB $E$-mode field, but only its component that is uncorrelated with the $B$-modes, which is precisely the target of our Hybrid ILC approach.

The gradient of the Lagrangian must vanish at the ILC weight solution:
\begin{align}
\label{eq:gradient_lagragian}
\nabla_{\bdw^\top} \mathcal{L} = \begin{pmatrix} \nabla_{\bdw_{E}^\top} \mathcal{L} \\  \nabla_{\bdw_{B}^\top} \mathcal{L} \end{pmatrix} =  \begin{pmatrix} 0\\ 0\end{pmatrix}\,.
\end{align}
Using the form of the Lagrangian from the first line of Equation~\eqref{eq:lagrangian}, the saddle point satisfying this condition is given by:
\begin{align}
\label{eq:weights0}
\bdw = \frac{  \tC^{-1}\bdb }{  \bdb^\top \tC^{-1}\bdb }\,.
\end{align}
This expression defines the general form of the Hybrid ILC weights, which we implement in practice in our analysis. However, to gain further insight into how these weights respond differently to $E$- and $B$-mode contributions, we re-derive the solution by solving the gradient condition in Equation~\eqref{eq:gradient_lagragian}, using the expanded form of the Lagrangian given in the second line of Equation~\eqref{eq:lagrangian}. This yields the following system of equations:
\begin{align}
  \begin{cases}
\tC^{EE}\bdw_{E}  +  \tC^{EB} \bdw_{B} &= \lambda \bda \\
\tC^{BB}\bdw_{B} +  \tC^{BE} \bdw_{E} &= 0
\end{cases}\,.
\end{align}
Solving for the weights, we obtain:
\begin{align}
\label{eq:relation}
  \begin{cases}
  \bdw_{E} + \left(\tC^{EE}\right)^{-1}\tC^{EB} \bdw_{B} = \lambda  \left(\tC^{EE}\right)^{-1}\bda \\
\bdw_{B} =   - \left(\tC^{BB}\right)^{-1}\tC^{BE} \bdw_{E} 
\end{cases}\,.
\end{align}
Substituting the second equation of the system into the first yields:
\begin{eqnarray}
\begin{cases}
 \left(\tI - \left(\tC^{EE}\right)^{-1}\tC^{EB} \left(\tC^{BB}\right)^{-1}\tC^{BE}\right)\bdw_{E}  = \lambda  \left(\tC^{EE}\right)^{-1}\bda \\
\bdw_{B} =   -\left(\tC^{BB}\right)^{-1}\tC^{BE} \bdw_{E} 
\end{cases}\,,
\end{eqnarray}
where $\tI$ denotes the identity matrix. Solving for $\bdw_E$ gives:
\begin{eqnarray}
\begin{cases}
 \bdw_{E}  = \lambda  \left(\tI - \left(\tC^{EE}\right)^{-1}\tC^{EB} \left(\tC^{BB}\right)^{-1}\tC^{BE}\right)^{-1}\left(\tC^{EE}\right)^{-1}\bda \\
\bdw_{B} =   -\left(\tC^{BB}\right)^{-1}\tC^{BE} \bdw_{E} 
\end{cases}\,.
\end{eqnarray}
Multiplying from the left by $\bda^\top$ and applying the constraint $\bda^\top \bdw_E = 1$ to preserve the CMB $E$-mode component, we find:
\begin{align}
  \lambda = \frac{1}{ \bda^\top \left(\tI - \left(\tC^{EE}\right)^{-1}\tC^{EB} \left(\tC^{BB}\right)^{-1}\tC^{BE}\right)^{-1}\left(\tC^{EE}\right)^{-1}\bda }\,.
\end{align}
Therefore, the Hybrid ILC weights across the $E$ and $B$ channels are given by:
\begin{eqnarray}
\label{eq:weights}
\begin{cases}
 \bdw_{E}  = \frac{  \left(\tI - \tP\right)^{-1}\left(\tC^{EE}\right)^{-1}\bda }{  \bda^\top \left(\tI - \tP\right)^{-1}\left(\tC^{EE}\right)^{-1}\bda } \\
\bdw_{B} =   - \left(\tC^{BB}\right)^{-1}\tC^{BE} \bdw_{E} 
\end{cases}\,,
\end{eqnarray}
where the matrix $\tP$ represents the effective squared cross-correlation between the observed $E$- and $B$-mode channels:
\begin{align}
\label{eq:pearson}
\tP = \left(\tC^{EE}\right)^{-1}\tC^{EB} \left(\tC^{BB}\right)^{-1}\tC^{BE}\,.
\end{align}
Note that in the absence of $EB$ correlations, such as those induced by cosmic birefringence or instrumental miscalibration, i.e., when $\tC^{EB} = 0$ so that $\tP = 0$, the expression for $\bdw_E$ reduces to the standard ILC weights $\widetilde{\bdw}_{E}  = ( \bda^\top ( \tC^{EE} )^{-1} \bda)^{-1}( \tC^{EE} )^{-1} \bda$.

Our hybrid ILC approach, which combines both $E$- and $B$-mode frequency channels, yields an estimate $\widehat{E}^{\rm CMB}(\hat{n})$ of the \emph{uncorrelated} component of the CMB $E$-mode anisotropies, since the cross-correlation --- whether evaluated via the cross-power spectrum in harmonic space or the two-point correlation in map space --- between the Hybrid ILC estimate $\widehat{E}^{\rm CMB}(\hat{n})$ and the standard ILC estimate $\widetilde{B}^{\rm CMB}(\hat{n})$ of the full CMB $B$-mode signal vanishes:
 \begin{align}
\label{eq:nulleb}
C^{\widehat{E} \widetilde{B}, {\rm CMB}} & = \left\langle \widehat{E}^{\rm CMB},  \widetilde{B}^{\rm CMB}\right\rangle \cr
							     & = \left\langle \bdw_{E}^\top\bdE + \bdw_{B}^\top\bdB,  \widetilde{\bdw}_{B}^\top\bdB \right\rangle \cr
							     & =  \bdw_{E}^\top \left\langle\bdE\bdB^\top \right\rangle \widetilde{\bdw}_{B} + \bdw_{B}^\top \left\langle \bdB \bdB^\top \right\rangle \widetilde{\bdw}_{B}\cr
							     & =  \bdw_{E}^\top \tC^{EB} \widetilde{\bdw}_{B} + \bdw_{B}^\top \tC^{BB} \widetilde{\bdw}_{B}\cr
							     & =  \bdw_{E}^\top \tC^{EB} \widetilde{\bdw}_{B} -  \bdw_{E}^\top \tC^{EB} \left(\tC^{BB}\right)^{-1} \tC^{BB} \widetilde{\bdw}_{B}\cr
							     & =  \bdw_{E}^\top \tC^{EB} \widetilde{\bdw}_{B} -  \bdw_{E}^\top \tC^{EB} \widetilde{\bdw}_{B}\cr
							     & = 0\,,
\end{align}
where in the antepenultimate line we used the expression $\bdw_{B} =   - \left(\tC^{BB}\right)^{-1}\tC^{BE} \bdw_{E}$  (Equation~\ref{eq:weights}) and the identity $\left(\tC^{BE}\right)^\top = \tC^{EB}$. 
Equation~\eqref{eq:nulleb} demonstrates the effective \emph{decorrelation} between the Hybrid ILC CMB $E$-mode map and the standard ILC CMB $B$-mode map, in stark contrast to Equation~\eqref{eq:eb_stdilc}, where the standard ILC $E$- and $B$-mode maps exhibit non-zero correlation in the presence of cosmic birefringence and polarization angle miscalibration.

Before closing this section, we note that using $\bdb^\top = \begin{pmatrix}  \boldsymbol{0}^\top & \bda^\top \end{pmatrix}$ instead of $\begin{pmatrix} \bda^\top & \boldsymbol{0}^\top \end{pmatrix}$ in Equation~\eqref{eq:weights0}, which defines the Hybrid ILC weights, or, equivalently, swapping $E$ and $B$ in Equation~\eqref{eq:weights}, enables the reconstruction of the \emph{uncorrelated} component of the CMB $B$-mode anisotropies, $\widehat{B}^{\rm CMB}(\hat{n})$. In this case, the resulting map is uncorrelated with the standard ILC CMB $E$-mode map, i.e., $C^{\widehat{B} \widetilde{E}, {\rm CMB}} = \langle \widehat{B}^{\rm CMB},  \widetilde{E}^{\rm CMB}\rangle = 0$.

\section{Estimation of the birefringence angle via field-level linear regression}
\label{sec:linreg}

With three ILC maps at our disposal, namely, the full CMB $E$- and $B$-mode maps, $\widetilde{E}^{\rm CMB}(\hat{n})$ and $\widetilde{B}^{\rm CMB}(\hat{n})$, obtained via the standard ILC method (Equation~\ref{eq:stdilcweights}), and the \emph{uncorrelated} CMB $E$-mode (or $B$-mode) map, $\widehat{E}^{\rm CMB}(\hat{n})$ (or $\widehat{B}^{\rm CMB}(\hat{n})$), derived from our Hybrid ILC approach (Equation~\ref{eq:weights}), we are now in a position to constrain the cosmic birefringence angle $\beta$ at the map level (i.e., through field-level inference, as opposed to inference based on summary statistics), as we demonstrate in the following section.

\subsection{Linear regression formula}
\label{subsec:formula}

In Equation~\eqref{eq:pearson}, the matrix $\tP$ contributes only at second order in $\beta$ and $\boldsymbol{\alpha}$ due to its quadratic dependence on $\tC^{EB}$. Therefore, to linear order in the small angles $\boldsymbol{\alpha}$ and $\beta$, the Hybrid ILC weights (Equation~\ref{eq:weights}) take the form:
\begin{eqnarray}
\label{eq:weights_firstorder}
\begin{cases}
 \bdw_{E}  \simeq \widetilde{\bdw}_{E} = \frac{  \left(\tC^{EE}\right)^{-1}\bda }{  \bda^\top \left(\tC^{EE}\right)^{-1}\bda } \\
\bdw_{B} \simeq  - \left(\tC^{BB}\right)^{-1}\tC^{BE} \widetilde{\bdw}_{E} 
\end{cases}\,.
\end{eqnarray}
The Hybrid ILC CMB $E$-mode map can thus be expressed, at linear order, as:
\begin{align}
\label{eq:theorem00}
\widehat{E}^{\rm CMB}(\hat{n}) &= \bdw_{E}^\top\bdE(\hat{n}) + \bdw_{B}^\top\bdB(\hat{n}) \cr
& \simeq \widetilde{\bdw}_{E}^\top\bdE(\hat{n}) + \left(\bdw_{B}^\top \bda\right) B^{\rm CMB}(\hat{n}) +  \bdw_{B}^\top\bdB^{\rm FG+N}(\hat{n})\cr
& \simeq \widetilde{E}^{\rm CMB}(\hat{n}) +  \left(\bdw_{B}^\top \bda\right) \widetilde{B}^{\rm CMB}(\hat{n}) +  \left[\bdw_{B}^\top -  \left(\bdw_{B}^\top \bda\right) \widetilde{\bdw}_{B}^\top\right]\bdB^{\rm FG+N}(\hat{n})\,,
\end{align}
where we have used $\widetilde{E}^{\rm CMB} = \widetilde{\bdw}_{E}^\top\bdE$ and $\widetilde{B}^{\rm CMB} =  \widetilde{\bdw}_{B}^\top\bdB \simeq B^{\rm CMB} + \widetilde{\bdw}_{B}^\top\bdB^{\rm FG+N}(\hat{n})$, while neglecting terms of order $\beta E^{\rm CMB}$ in $\bdB$, since the prefactor $\bdw_{B}$ is already first order.

Consequently, the difference between the standard and hybrid ILC CMB $E$-mode maps isolates the component of the CMB $E$-modes that is linearly correlated with the CMB $B$-modes:
 \begin{align}
\label{eq:theorem0}
\widetilde{E}^{\rm CMB}(\hat{n}) - \widehat{E}^{\rm CMB}(\hat{n}) &\simeq \frac{  \bda^\top \left(\tC^{EE}\right)^{-1} \tC^{EB} \left(\tC^{BB}\right)^{-1} \bda}{  \bda^\top \left(\tC^{EE}\right)^{-1}\bda }\widetilde{B}^{\rm CMB}(\hat{n}) + \varepsilon(\hat{n})\,,
\end{align}
with an error term:
\begin{align}
\label{eq:error}
\varepsilon(\hat{n}) = \left[ \bdw_{B}^\top -  \left(\bdw_{B}^\top \bda \right) \widetilde{\bdw}_{B}^\top \right]\bdB^{\rm FG+N}(\hat{n})\,,
\end{align}
which is uncorrelated with $\widetilde{B}^{\rm CMB}(\hat{n})$ and independent of the cosmic birefringence angle $\beta$, as we show next.

Using the expressions of the standard ILC weights from Equation~\eqref{eq:stdilcweights}, the regression can be reformulated as:
 \begin{align}
\label{eq:theorem0}
\widetilde{E}^{\rm CMB}(\hat{n}) - \widehat{E}^{\rm CMB}(\hat{n}) &\simeq \frac{\widetilde{\bdw}_{E}^\top \tC^{EB} \widetilde{\bdw}_{B}}{\widetilde{\bdw}_{B}^\top \tC^{BB} \widetilde{\bdw}_{B}}\widetilde{B}^{\rm CMB}(\hat{n}) + \varepsilon(\hat{n})\cr
												   &\simeq \frac{\widetilde{\bdw}_{E}^\top \left\langle \bdE \bdB^\top \right\rangle \widetilde{\bdw}_{B}}{\widetilde{\bdw}_{B}^\top \left\langle \bdB \bdB^\top \right\rangle  \widetilde{\bdw}_{B}}\widetilde{B}^{\rm CMB}(\hat{n}) + \varepsilon(\hat{n})\cr												   						
												   &\simeq \frac{ \left\langle \widetilde{E}^{\rm CMB}  \widetilde{B}^{\rm CMB} \right\rangle}{ \left\langle \widetilde{B}^{\rm CMB} \widetilde{B}^{\rm CMB} \right\rangle} \widetilde{B}^{\rm CMB}(\hat{n}) + \varepsilon(\hat{n})\,.
\end{align}
By cross-correlating both sides of Equation~\eqref{eq:theorem0} with $\widetilde{B}^{\rm CMB}$, we obtain that the error term $\varepsilon(\hat{n})$ is uncorrelated with the field $\widetilde{B}^{\rm CMB}(\hat{n})$:
\begin{align}
\label{eq:uncorr_error}
 \left\langle \varepsilon(\hat{n}) \widetilde{B}^{\rm CMB}(\hat{n}) \right\rangle = 0\,,
\end{align}
since the Hybrid ILC field $\widehat{E}^{\rm CMB}$ and the standard ILC field $\widetilde{B}^{\rm CMB}$ are uncorrelated by construction (Equation~\ref{eq:nulleb}). 

The error term $\varepsilon(\hat{n}) $ is also independent of $\beta$, since it vanishes in the absence of polarization angle miscalibration. Indeed,  if $\alpha_\nu = 0$, the observed $EB$ cross-correlation from Equation~\eqref{eq:ceb2} simplifies to
\begin{align}
\label{eq:ceb2_firstorder0}
\tC^{E B}_\ell  \underset{\alpha_\nu = 0}{\simeq} \beta f_\ell\bda \bda^\top\,,
\end{align}
with the scalar function $f_\ell$ given by 
\begin{align}
\label{eq:fl}
f_\ell  =  2\left(C_\ell^{EE,{\rm CMB}} - C_\ell^{BB,{\rm CMB}}\right)\,.
\end{align}
It follows that both $\bdw_B^\top$ and $(\bdw_B^\top\bda)\widetilde{\bdw}_B^\top$ reduce to the same expression:
\begin{align}
\label{eq:error_terms}
\bdw_{B}^\top   & \underset{\alpha_\nu = 0}{=} -  \beta f_\ell  \bdw_{E}^\top\bda \bda^\top \left(\tC^{BB}\right)^{-1} = -  \beta f_\ell  \bda^\top \left(\tC^{BB}\right)^{-1}\,, \\ 		
\left(\bdw_{B}^\top \bda\right) \widetilde{\bdw}_{B}^\top   &  \underset{\alpha_\nu = 0}{=} \left(-  \beta f_\ell \bda^\top \left(\tC^{BB}\right)^{-1}\bda\right) \frac{\bda^\top \left(\tC^{BB}\right)^{-1}}{\bda^\top \left(\tC^{BB}\right)^{-1}\bda} = -  \beta f_\ell  \bda^\top \left(\tC^{BB}\right)^{-1}\,,      
\end{align}
which leads to the cancellation of the error term:
\begin{align}
\label{eq:error_perfectcalib}
\varepsilon(\hat{n}) = \left[ \bdw_{B}^\top -  \left(\bdw_{B}^\top \bda \right) \widetilde{\bdw}_{B}^\top \right]\bdB^{\rm FG+N}(\hat{n}) \underset{\alpha_\nu = 0}{=} 0\,. 	      
\end{align}

The regression in Equation~\eqref{eq:theorem0} applies in pixel, harmonic, or needlet space. Without any loss of generality, assuming harmonic domain weights, the regression relation becomes:
\begin{align}
\label{eq:theorem0_lm}
\widetilde{E}^{\rm CMB}_{\ell m} - \widehat{E}^{\rm CMB}_{\ell m} &\simeq \left(\frac{ C_\ell^{\widetilde{E}\widetilde{B},{\rm CMB}} }{ C_\ell^{\widetilde{B}\widetilde{B},{\rm CMB}} }\right) \widetilde{B}^{\rm CMB}_{\ell m} + \varepsilon_{\ell m} \,.
\end{align}
Using the expression for the standard ILC $EB$ cross-power spectrum (Equation~\ref{eq:eb_stdilc}), we obtain:
\begin{align}
 \label{eq:theorem_lm}
\widetilde{E}^{\rm CMB}_{\ell m} - \widehat{E}^{\rm CMB}_{\ell m} &\simeq \left(\beta + \overline{\alpha}_\ell\right)F_\ell  \widetilde{B}^{\rm CMB}_{\ell m} + \varepsilon_{\ell m}\,,
\end{align}
with
\begin{align}
\label{eq:modulation}
F_\ell &= 2\frac{C_\ell^{EE,{\rm CMB}} - C_\ell^{BB,{\rm CMB}}}{C^{\widetilde{B} \widetilde{B}, {\rm CMB}}_\ell}\,,\\
\label{eq:offset}
\overline{\alpha}_\ell & =  \frac{  \overline{\alpha}^{\,(B)}_\ell C_\ell^{EE,{\rm CMB}} -  \overline{\alpha}^{\,(E)}_\ell C_\ell^{BB,{\rm CMB}} }{ C_\ell^{EE,{\rm CMB}} - C_\ell^{BB,{\rm CMB}} }\,,
\end{align}
where $F_\ell$ modulates the $B$-mode field, and $\overline{\alpha}_\ell$ is a residual offset angle related to the weighted averages of miscalibrated polarization angles across frequency channels, $\overline{\alpha}^{\,(E)}_\ell\equiv \widetilde{\bdw}_{E}(\ell)^\top\boldsymbol{\alpha}$ and $\overline{\alpha}^{\,(B)}_\ell \equiv \widetilde{\bdw}_{B}(\ell)^\top\boldsymbol{\alpha}$, as a result of component separation (see Section~\ref{subsec:discrimination}). 
Unlike $\beta$, the offset angle $\overline{\alpha}_\ell$ inherits $\ell$-dependence from harmonic ILC weights due to the chromatic nature of polarization angle miscalibration. 
If the ILC is implemented in pixel or needlet space rather than the harmonic domain, $\overline{\alpha}_\ell$ may also exhibit  $(\ell,m)$-dependence or spatial variations due to the anisotropy of the weights.
In these expressions, $C_\ell^{EE,{\rm CMB}}$ and $C_\ell^{BB,{\rm CMB}}$ denote the intrinsic  (i.e., pre-rotation) CMB $E$- and $B$-mode power spectra, typically taken from the best-fit $\Lambda$CDM model, while $C^{\widetilde{B} \widetilde{B}, {\rm CMB}}_\ell$ is the power spectrum of the standard ILC CMB $B$-mode map.

Equations~\eqref{eq:theorem_lm}--\eqref{eq:offset} constitute the main result of this work, revealing a linear relationship at the map level between the \emph{correlated} component of the CMB $E$-mode anisotropies (defined as the difference $\widetilde{E}^{\rm CMB}(\hat{n}) - \widehat{E}^{\rm CMB}(\hat{n})$ between the standard and hybrid ILC CMB $E$-mode maps) and the \emph{modulated} standard ILC CMB $B$-mode anisotropies, $( F \ast \widetilde{B}^{\rm CMB})(\hat{n})$, where $\ast$ denotes convolution in real space. This linear relation arises as a direct consequence of cosmic birefringence.

In the absence of polarization angle miscalibration (i.e., $\alpha_\nu=0$), both the residual offset and the error term vanish, $\overline{\alpha}_\ell  = 0$ and $\varepsilon_{\ell m} = 0$, respectively. In this case, Equation~\eqref{eq:theorem_lm} reduces to a clean linear spatial regression:
\begin{align}
\label{eq:theorem_pixel}
\widetilde{E}^{\rm CMB}(\hat{n}) - \widehat{E}^{\rm CMB}(\hat{n}) & \simeq \beta \left( F \ast \widetilde{B}^{\rm CMB} \right)(\hat{n})\,,
\end{align}
enabling unbiased estimation of the cosmic birefringence angle $\beta$ across different regions of the sky. However, when $\alpha_\nu \neq 0$, the resulting residual offset $\overline{\alpha}(\hat{n})$ may introduce a small, spatially varying bias in the estimation of $\beta$ when using ordinary least squares (OLS) regression (e.g., \cite{Isobe1990}). To address this, in Section~\ref{subsec:regression} we introduce an Instrumental Variable Two-Stage Least Squares (IV-2SLS) regression method \citep{IV-2SLS}, which treats the residual offset as an endogenous error term and mitigates this bias.

Lastly, a symmetric expression exists for the $B$-mode map difference, i.e., a regression of $\widetilde{B}^{\rm CMB} - \widehat{B}^{\rm CMB}$ on $\widetilde{E}^{\rm CMB}$, by simply exchanging $E$ and $B$ in Equations~\eqref{eq:theorem_lm}--\eqref{eq:offset}.

\subsection{Linear regression approach}
\label{subsec:regression}

The linear regression problem defined in Equations~\eqref{eq:theorem_lm}--\eqref{eq:offset} can be recast in map space in the standard form:
\begin{align}
\label{eq:textbook}
Y(\hat{n}) = \beta X(\hat{n}) + e(\hat{n})\,,
\end{align}
where the dependent variable $Y$, the explanatory variable $X$, and the error term $e$ are given by:
\begin{align}
\label{eq:y}
Y(\hat{n}) &= \widetilde{E}^{\rm CMB}(\hat{n}) - \widehat{E}^{\rm CMB}(\hat{n})\,,\\
\label{eq:x}
X(\hat{n}) &= (F \ast \widetilde{B}^{\rm CMB})(\hat{n})\,,\\
\label{eq:e}
e(\hat{n}) &= (\overline{\alpha} \ast X)(\hat{n}) + \varepsilon(\hat{n})\,.
\end{align} 
Although $\varepsilon(\hat{n})$ is uncorrelated with $X(\hat{n})$ (see Equation~\ref{eq:uncorr_error}), the full error term $e(\hat{n})$ may still be correlated with $X(\hat{n})$, due to the offset term $(\overline{\alpha} \ast X)(\hat{n})$ being itself a function of this signal. This may introduce endogeneity, potentially biasing ordinary least squares (OLS) estimates of $\beta$:
\begin{align}
\widehat{\beta}^{\,\rm OLS} = \frac{\left\langle X , Y\right\rangle}{\left\langle X , X\right\rangle} = \frac{\left\langle X , \beta X + e \right\rangle}{\left\langle X , X\right\rangle} =  \beta + \frac{\left\langle X , e \right\rangle}{\left\langle X , X\right\rangle}\,.
\end{align}

To address potential endogeneity, we employ an Instrumental Variable (IV) approach using Two-Stage Least Squares (2SLS) estimation \citep{IV-2SLS}, which requires identifying an instrumental variable $Z$ that is strongly correlated with $X$ but uncorrelated with the error $e$ (exogeneity condition). The IV-2SLS method proceeds in two stages. First, we regress $X$ on $Z$ to estimate the relationship $X = \gamma Z$:
\begin{align}
\widehat{\gamma} = \frac{\left\langle Z , X\right\rangle}{\left\langle Z , Z\right\rangle}\,,
\end{align}
and then we regress $Y$ on the predicted variable $\widehat{X} = \widehat{\gamma} Z$, yielding an unbiased estimate of $\beta$:
\begin{align}
\widehat{\beta}^{\,\rm IV\text{-}2SLS} = \frac{\left\langle \widehat{X} , Y\right\rangle}{\left\langle \widehat{X} , X\right\rangle} = \frac{\left\langle Z , Y\right\rangle}{\left\langle Z , X\right\rangle}= \frac{\left\langle Z , \beta X + e \right\rangle}{\left\langle Z , X\right\rangle} = \beta\,,
\end{align}
since $\left\langle Z , e\right\rangle = 0$ by construction.

When regressing $Y = \widetilde{E}^{\rm CMB}(\hat{n}) - \widehat{E}^{\rm CMB}(\hat{n})$ on $X = (F \ast \widetilde{B}^{\rm CMB})(\hat{n})$, we find that the error $e$ is correlated with $X$ by about $15\,\%$ on average across the sky regions used in our simulations (Section~\ref{sec:results}), indicating substantial endogeneity. In contrast, when regressing $Y = \widetilde{B}^{\rm CMB}(\hat{n}) - \widehat{B}^{\rm CMB}(\hat{n})$ on $X = (G \ast\widetilde{E}^{\rm CMB})(\hat{n})$, where $G(\hat{n})$ is the modulation function $F(\hat{n})$ with $E$ and $B$ swapped in Equation~\eqref{eq:modulation}, the correlation between $e$ and $X$ drops to around $3\,\%$. Although endogeneity appears minimal in the latter case, the true correlation between $e$ and $X$ is data-dependent and difficult to assess in practice, as it relies on unknown values of $\alpha_\nu$. To be conservative, we adopt the IV-2SLS method in all cases to avoid bias in $\beta$, accepting a slight increase in variance compared to OLS.

For the regression of $Y = \widetilde{E}^{\rm CMB}(\hat{n}) - \widehat{E}^{\rm CMB}(\hat{n})$ on $X = (F \ast \widetilde{B}^{\rm CMB})(\hat{n})$, using $Z = \widetilde{B}^{\rm CMB}(\hat{n})$ as an instrumental variable provides strong correlation with the explanatory variable ($\langle Z, X\rangle \sim 62\,\%$) and near-zero correlation with the error term ($\langle Z, e\rangle \sim 1\,\%$), on average across the simulated sky regions. Similarly, for the regression of $Y = \widetilde{B}^{\rm CMB}(\hat{n}) - \widehat{B}^{\rm CMB}(\hat{n})$ on $X = (G \ast\widetilde{E}^{\rm CMB})(\hat{n})$, choosing $Z = \widetilde{E}^{\rm CMB}(\hat{n})$ yields $\langle Z, X\rangle \sim 78\,\%$ and $\langle Z, e\rangle \sim 2\,\%$. The effectiveness of our IV-2SLS approach in recovering unbiased estimates of $\beta$ is demonstrated in Section~\ref{sec:results}.

\begin{figure}[tbp]
\centering 
\includegraphics[width=0.8\textwidth,clip]{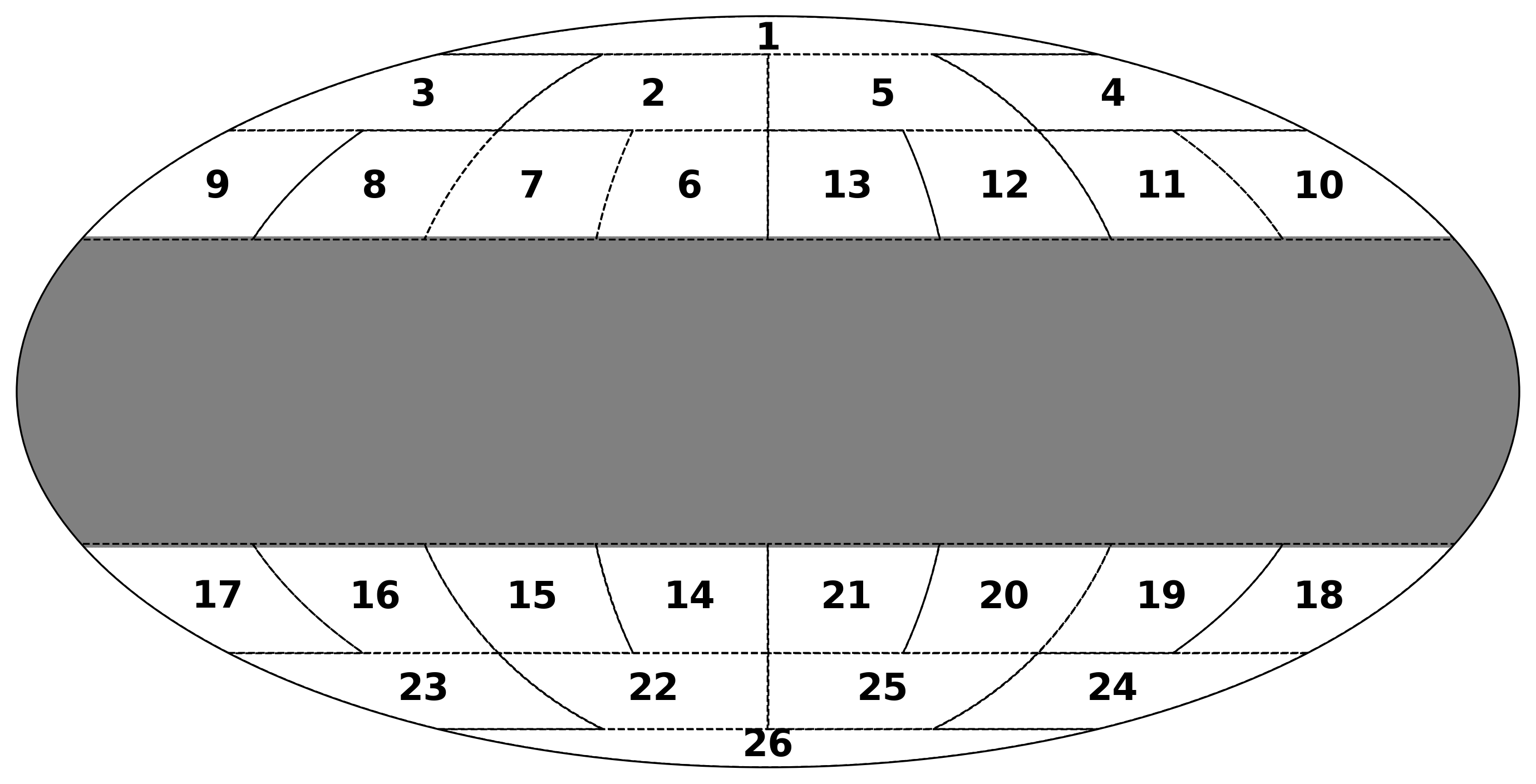}~
\hfill
\caption{\label{fig:igloo} Sky partitioning into 26 equal-area patches, each covering about $2\,\%$ of the sky. The total observed sky fraction is $f_{\rm sky}=50\,\%$, excluding the grey area.}
\end{figure}

\section{Data analysis}
\label{sec:results}

\subsection{LiteBIRD sky simulations}
\label{subsec:sims}

LiteBIRD is designed to observe the full sky with three telescopes, covering 22 channels of observation across 15 frequency bands \citep{LiteBIRD2023}. Overlapping frequency bands are not fully independent, as they probe the same sky signal, but their instrumental noise is independent. Reproducing the approach of \cite{LiteBIRD-CB2025}, we generate sky simulations for the \lb\ experiment across its $22$ frequency channels, spanning $40$--$402$\,GHz, including contributions from the CMB, Galactic foregrounds (synchrotron and thermal dust), and instrumental noise, all in Stokes $Q$ and $U$ maps at \texttt{HEALPix}\footnote{\url{https://healpix.jpl.nasa.gov/}} \citep{gorski2005} resolution $N_{\rm side}=512$, under four different rotation scenarios: 
\begin{enumerate}[label=(\roman*)]
\item$\beta = 0$, $\alpha_\nu = 0$ (no birefringence, no miscalibration), 
\item $\beta = 0$, $\alpha_\nu \neq 0$  (miscalibration only),
\item $\beta = 0.3^\circ$, $\alpha_\nu = 0$ (birefringence only),
\item $\beta = 0.3^\circ$, $\alpha_\nu \neq 0$ (both birefringence and miscalibration).
\end{enumerate}
The choice of birefringence angle $\beta = 0.3^\circ$ is informed by current observational constraints \citep{Diego2022}.
The miscalibration angles $\alpha_\nu$ are independently drawn per frequency from a uniform distribution centred at zero with width $[-\sigma_{\alpha_\nu}, \sigma_{\alpha_\nu}]$, using $\sigma_{\alpha_\nu}$ values from the \lb\ instrumental requirements given in the Table~3 of \cite{Vielva2022} (case~2.3). For each scenario, the appropriate rotation is applied to the CMB and foreground $Q,U$ maps according to Equations~\eqref{eq:rotation}–\eqref{eq:rotation_matrix}.

Following \cite{LiteBIRD-CB2025}, we simulate Galactic foregrounds using the \texttt{PySM} \texttt{d1s1} model \citep{pysm,pysm3,Panex2025}, which includes spatially varying SEDs consisting of, prior to any rotation, a modified blackbody for thermal dust and a power-law for synchrotron. After applying the respective rotations, CMB, dust, and synchrotron maps are coadded per frequency channel and smoothed with a Gaussian beam using the full-width at half-maximum (FWHM) values listed in Table~3 of \cite{LiteBIRD2023}. We assume Dirac delta-function bandpasses for all channels. Instrumental noise is simulated as Gaussian white noise per channel, with per-pixel RMS given by the sensitivities quoted in Table~3 of \cite{LiteBIRD2023}. This noise is added to the coadded component maps, yielding a set of $22$ \lb\ full-sky $Q$ and $U$ maps per rotation scenario.

From these simulations, we reconstruct full-sky CMB $E$- and $B$-mode maps using the standard ILC method (Equation~\ref{eq:stdilcweights}) applied to $E$- or $B$-channel inputs, producing the estimated fields $\widetilde{E}^{\rm CMB}(\hat{n})$ and $\widetilde{B}^{\rm CMB}(\hat{n})$ at $30'$ angular resolution. In addition, we reconstruct the \emph{uncorrelated} CMB $E$-mode field, $\widehat{E}^{\rm CMB}(\hat{n})$, at the same angular resolution using the Hybrid ILC method (Equation~\ref{eq:weights0} or \ref{eq:weights}) described in Section~\ref{subsec:hybridweights}, applied jointly to the $E$ and $B$ channels.
We then compute the difference $\widetilde{E}^{\rm CMB}(\hat{n}) - \widehat{E}^{\rm CMB}(\hat{n})$, representing the \emph{correlated} CMB $E$-mode field, and perform a linear regression with the modulated CMB $B$-mode field $(F \ast \widetilde{B}^{\rm CMB})(\hat{n})$, using Equations~\eqref{eq:theorem_lm}–\eqref{eq:offset}. This spatial regression is carried out over the $26$ sky patches defined in Figure~\ref{fig:igloo}, after applying a harmonic-domain top-hat filter retaining multipoles $\ell \in [250, 500]$ to enhance both the signal-to-noise ratio and the degree of correlation between the fields. 

We likewise reconstruct the uncorrelated CMB $B$-mode field, $\widehat{B}^{\rm CMB}(\hat{n})$, using the Hybrid ILC, allowing for the alternative regression of the correlated $B$-mode component, $\widetilde{B}^{\rm CMB}(\hat{n}) - \widehat{B}^{\rm CMB}(\hat{n})$, onto the modulated $E$-mode field $(G \ast \widetilde{E}^{\rm CMB})(\hat{n})$.

We implement both the standard and Hybrid ILC methods in a needlet frame \citep{Narcowich2006,Marinucci2008}, whose localization properties in both harmonic and pixel space allow the ILC weights to adapt to local conditions of foreground and noise contamination across the sky and angular scales \citep{Delabrouille2009}. We therefore refer to these implementations as standard NILC and Hybrid NILC, respectively. Both methods are applied using the same needlet configuration, as defined by the equation~3.3 and the figure~3 of \cite{Remazeilles2024}.

\begin{figure}[tbp]
\centering 
\includegraphics[width=0.5\textwidth,clip]{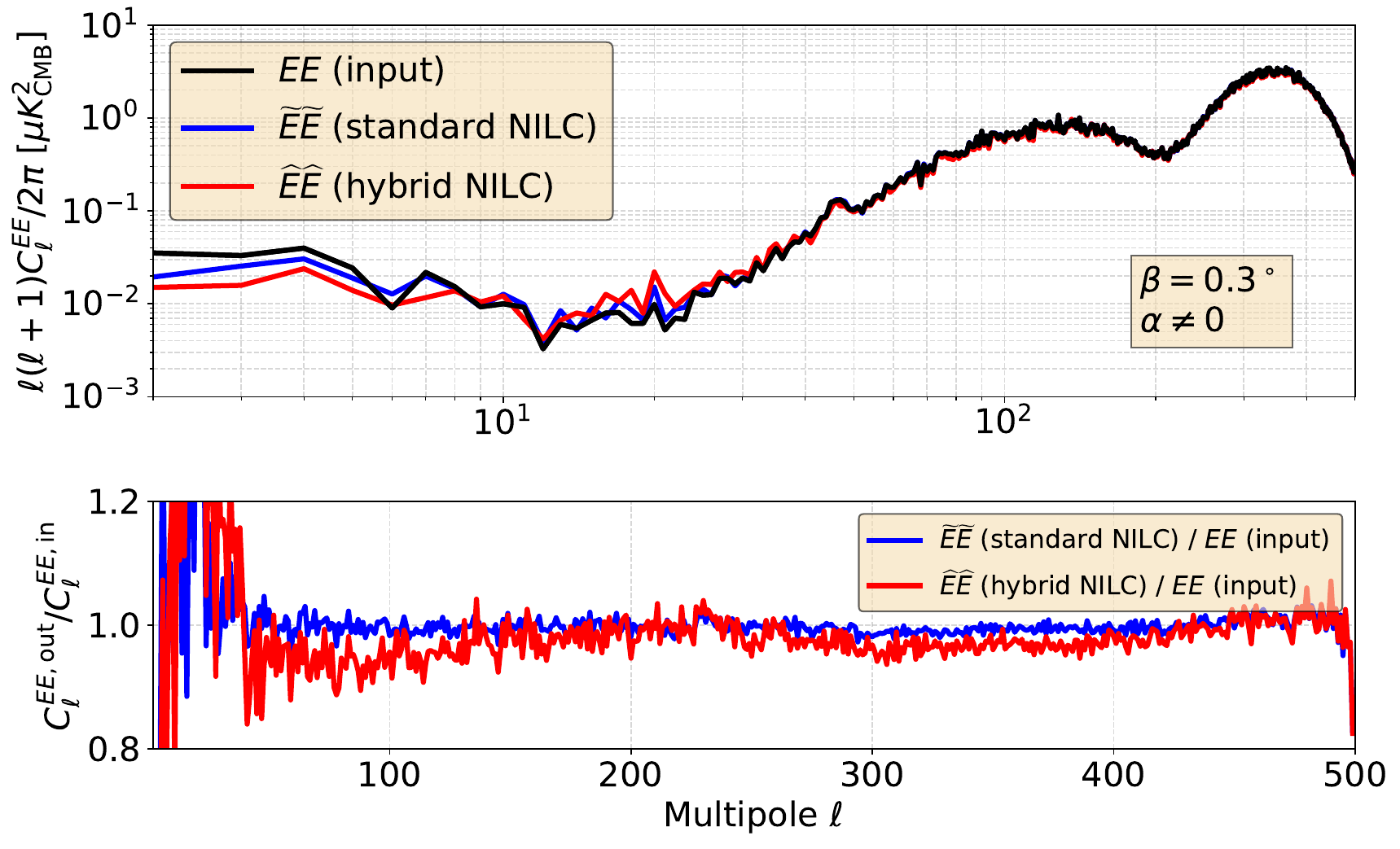}~
\includegraphics[width=0.5\textwidth,clip]{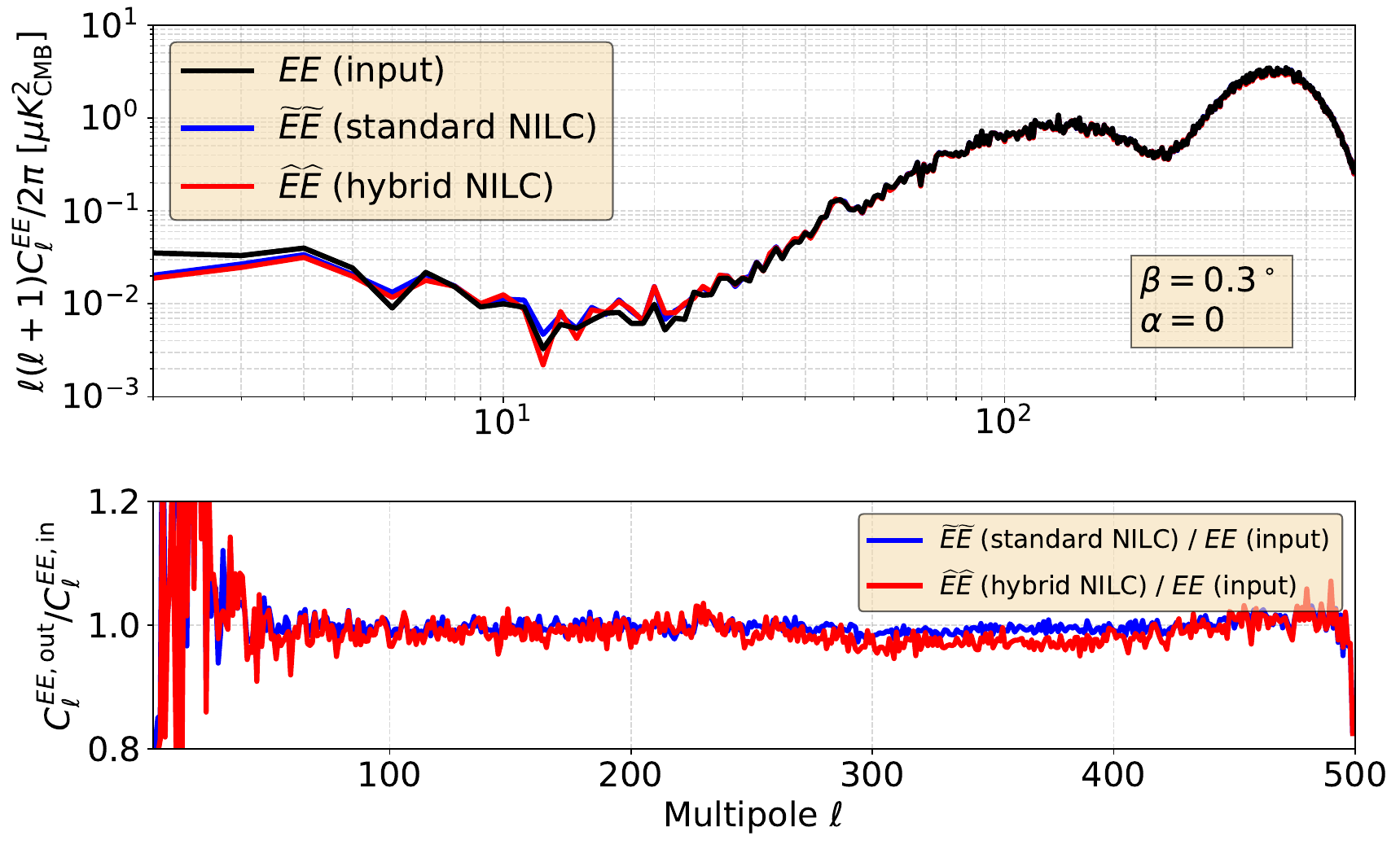}\\
\includegraphics[width=0.5\textwidth,clip]{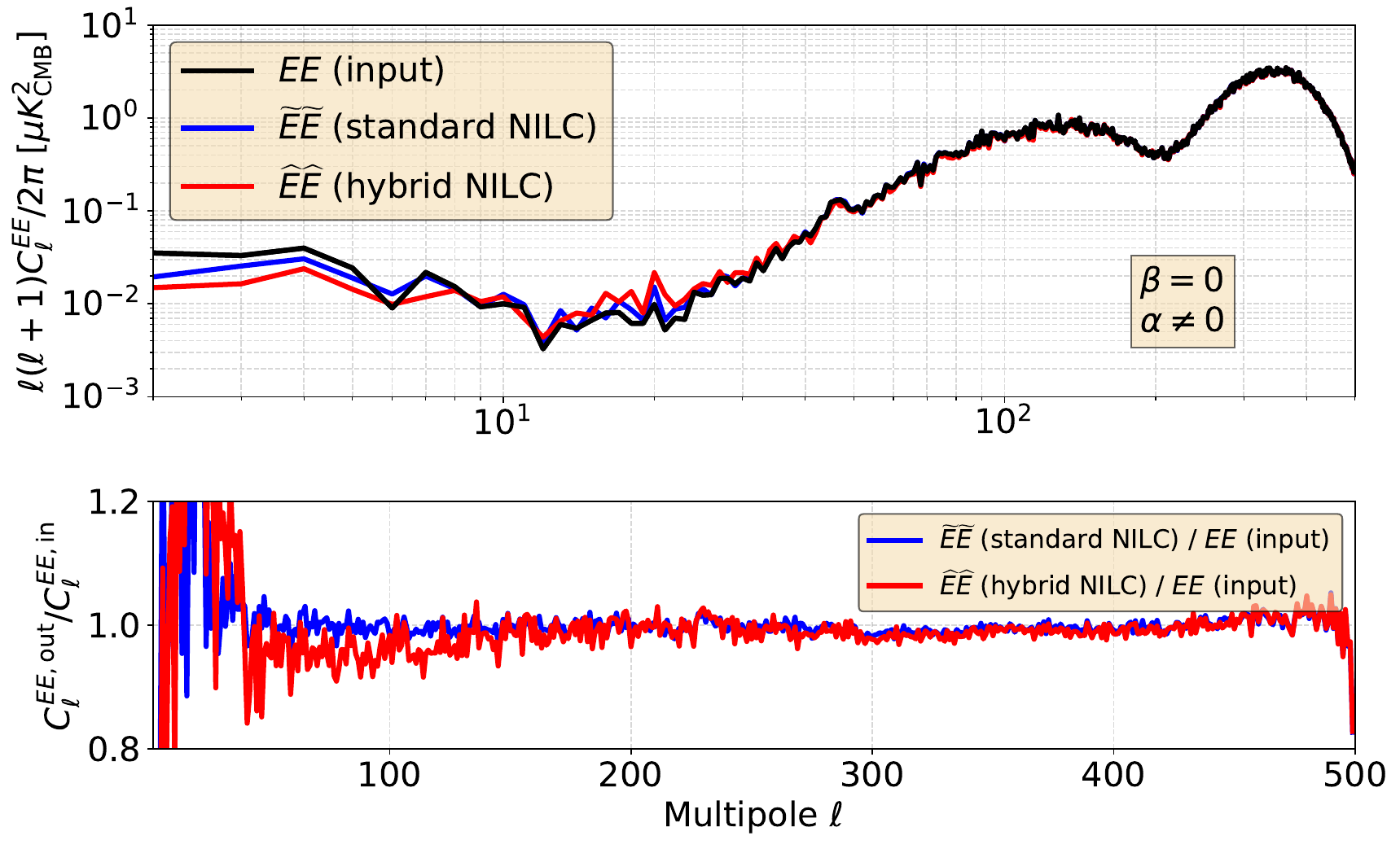}~
\includegraphics[width=0.5\textwidth,clip]{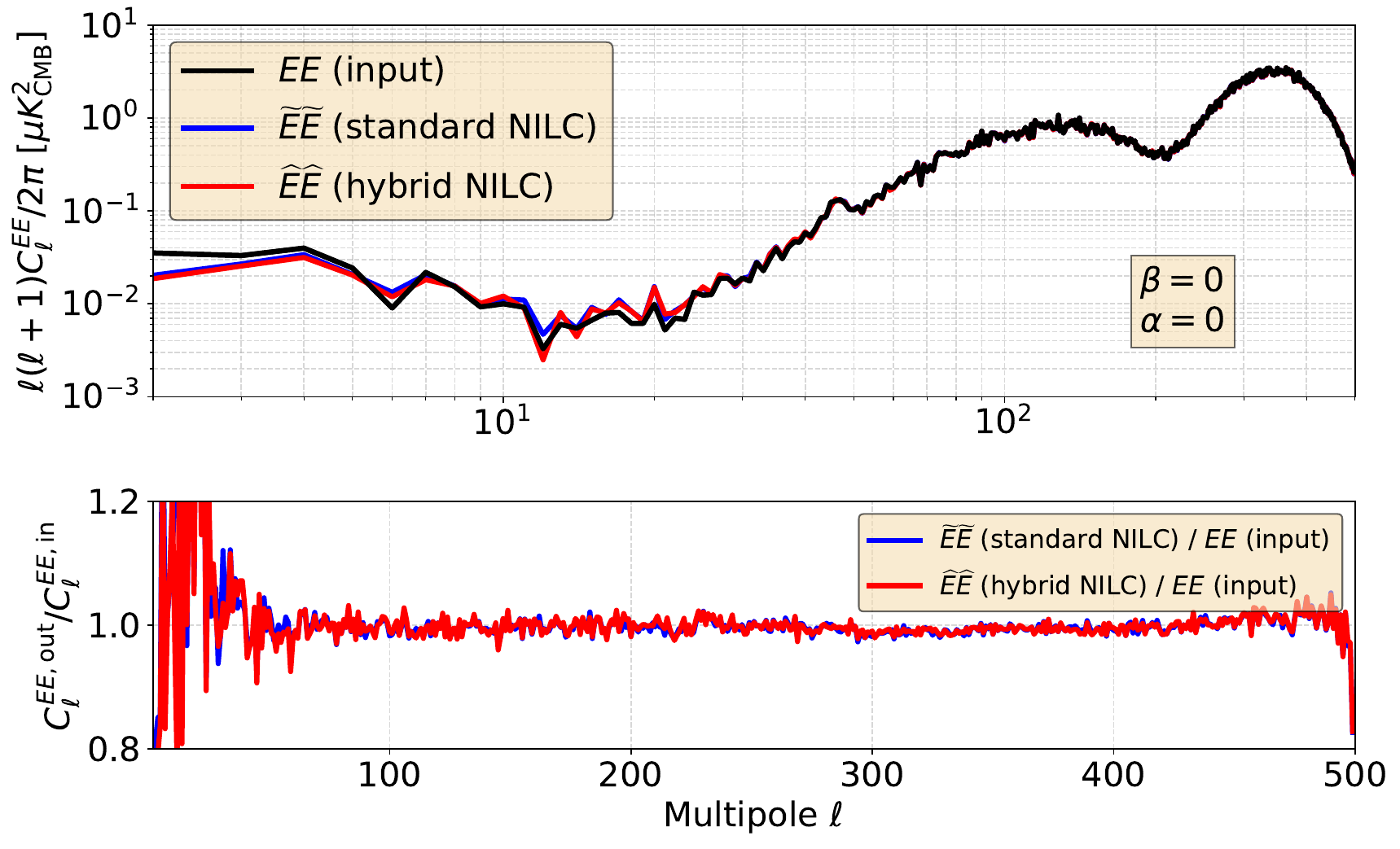}
\hfill
\caption{\label{fig:eeps} $EE$ power spectra from the input CMB $E$-mode map (\emph{black}), the standard NILC CMB $E$-mode map (\emph{blue}), and the hybrid NILC CMB $E$-mode map (\emph{red}), computed over $f_{\rm sky}=50\,\%$ of the sky. \emph{Top left}: Case $\beta = 0.3^\circ$, $\alpha \neq 0$.   \emph{Top right}: Case $\beta = 0.3^\circ$, $\alpha = 0$. \emph{Bottom left}: Case $\beta = 0$, $\alpha \neq 0$.  \emph{Bottom right}: Case $\beta = 0$, $\alpha = 0$. 
A clear power deficit appears in the Hybrid NILC CMB $E$-mode spectra for cases with either $\beta \neq 0$ (top panels) or $\alpha \neq 0$ (left panels), consistent with the method extracting only the \emph{uncorrelated} component of the CMB $E$-modes. Suppression from instrumental polarization angle miscalibration ($\alpha \neq 0$) is most evident at low multipoles ($\ell < 150$), where residual foreground contamination is stronger, while suppression from cosmic birefringence ($\beta \neq 0$) is most prominent at intermediate multipoles ($\ell \simeq 250$--$450$).}
\end{figure}

\subsection{CMB EE and EB power spectra from standard and hybrid NILC}
\label{subsec:power_spectra}

Figure~\ref{fig:eeps} presents the recovered CMB $EE$ power spectrum from the standard ($\widetilde{E}\widetilde{E}$, blue line) and hybrid ($\widehat{E}\widehat{E}$, red line) NILC maps, compared to the input CMB $EE$ power spectrum (black line), over a sky fraction of $f_{\rm sky} = 50\,\%$ (excluding the grey region in Figure~\ref{fig:igloo}), for the four rotation scenarios considered. In each panel, the lower subplot shows the ratio between the recovered and input $EE$ power spectra for both ILC methods. 
As expected, the Hybrid NILC reconstruction (red) shows a power deficit across multipoles when either $\beta \neq 0$ (top panels) or $\alpha_\nu \neq 0$ (left panels), reflecting the fact that the Hybrid NILC removes the component of the CMB $E$-modes correlated with $B$-modes, preserving only the uncorrelated part. 

Instrumental miscalibration ($\alpha_\nu \neq 0$, bottom left) primarily affects low multipoles ($\ell < 150$), where foreground contamination is more significant, while cosmic birefringence ($\beta \neq 0$, top right) induces power suppression at intermediate multipoles ($\ell \simeq 250$--$450$). These distinct spectral signatures in the Hybrid NILC $EE$ power spectrum offer a first diagnostic to distinguish between instrumental miscalibration, cosmic birefringence, or a combination of both, and also help identify the optimal multipole range, here $\ell \simeq 250$--$450$, for constraining cosmic birefringence. Finally, in the absence of any rotation ($\beta = 0$, $\alpha_\nu = 0$, bottom right), no significant power suppression is observed, as the Hybrid NILC reduces to the standard NILC in this case. 

\begin{figure}[tbp]
\centering 
\includegraphics[width=0.5\textwidth,clip]{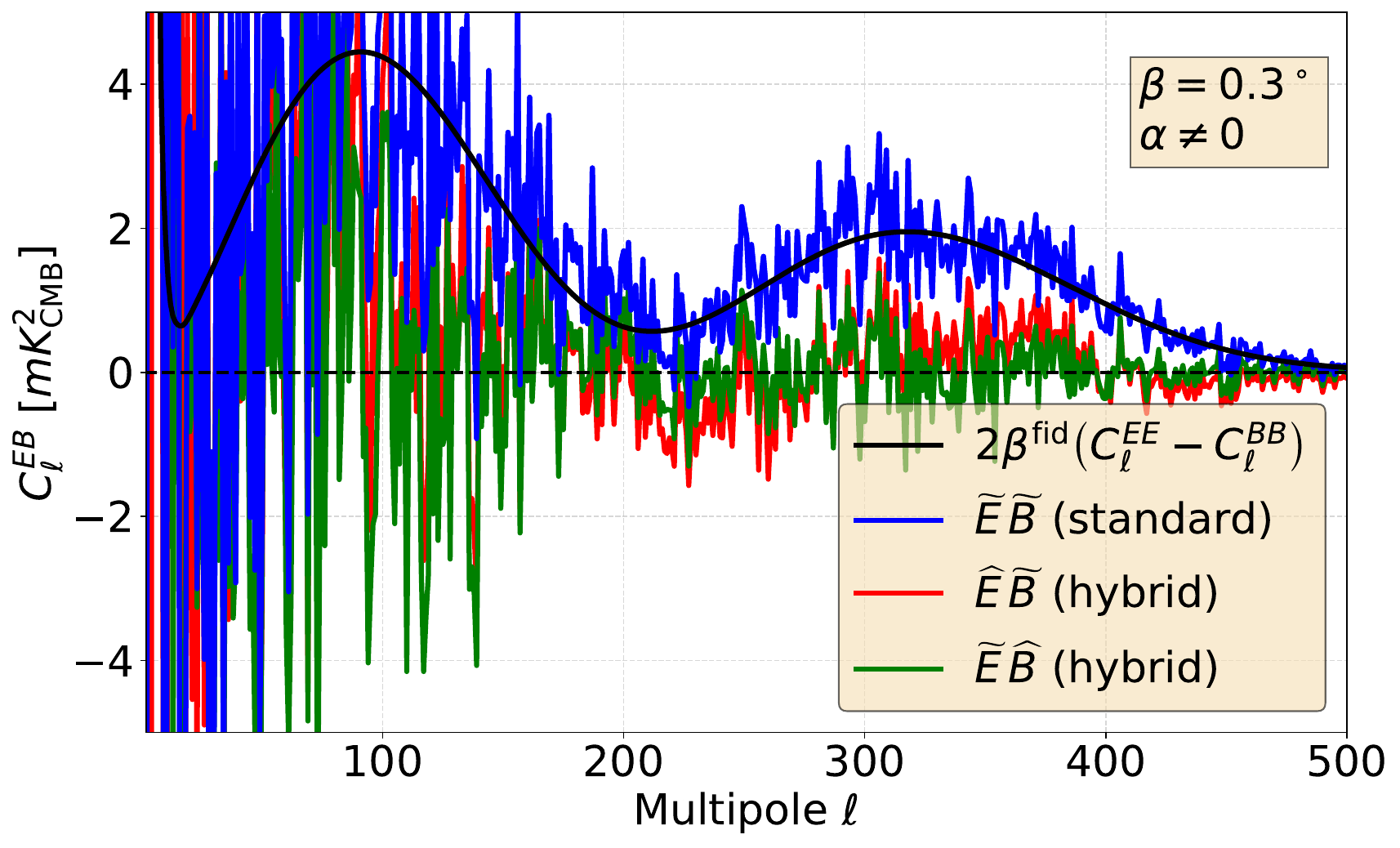}~
\includegraphics[width=0.5\textwidth,clip]{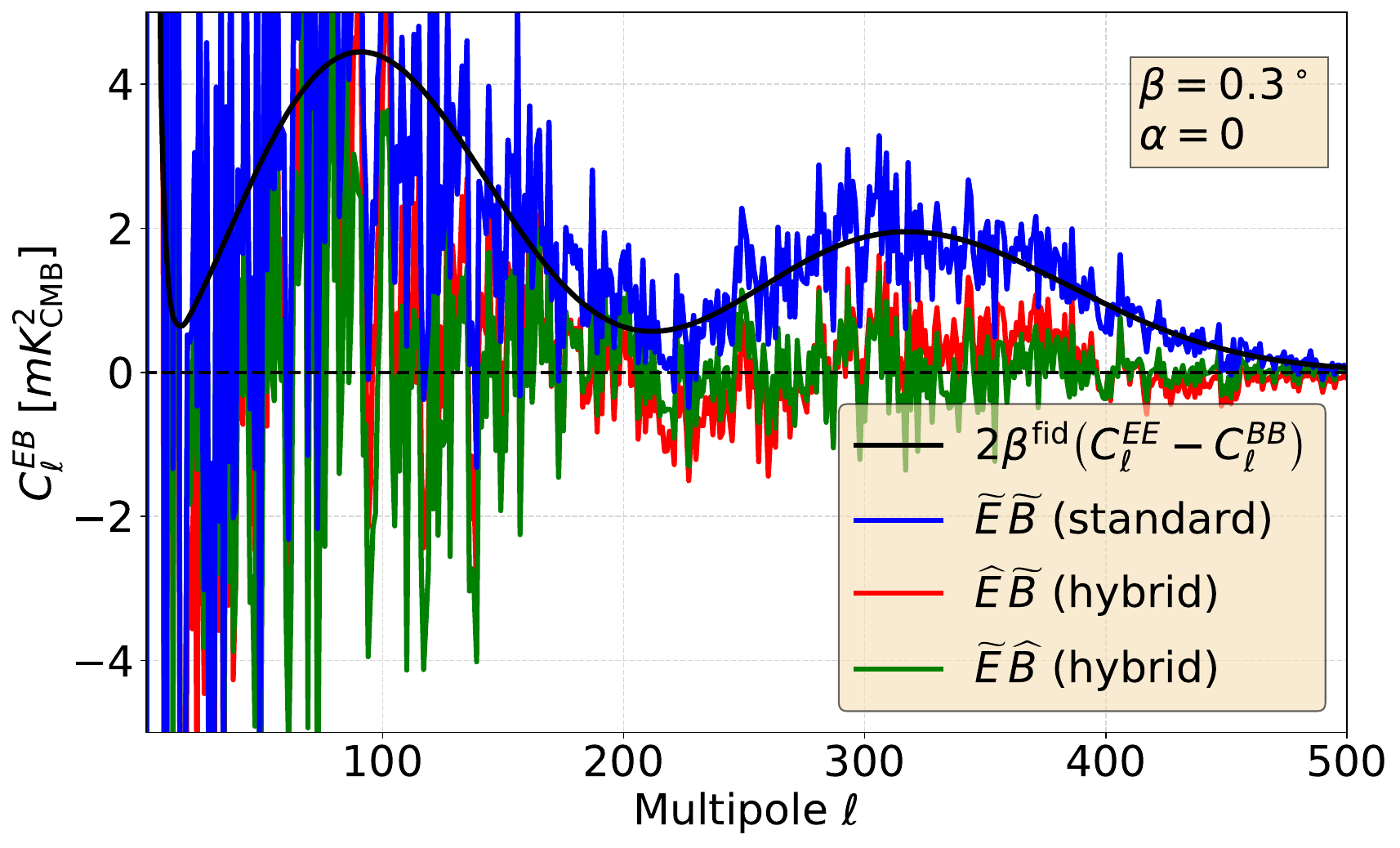}\\
\includegraphics[width=0.5\textwidth,clip]{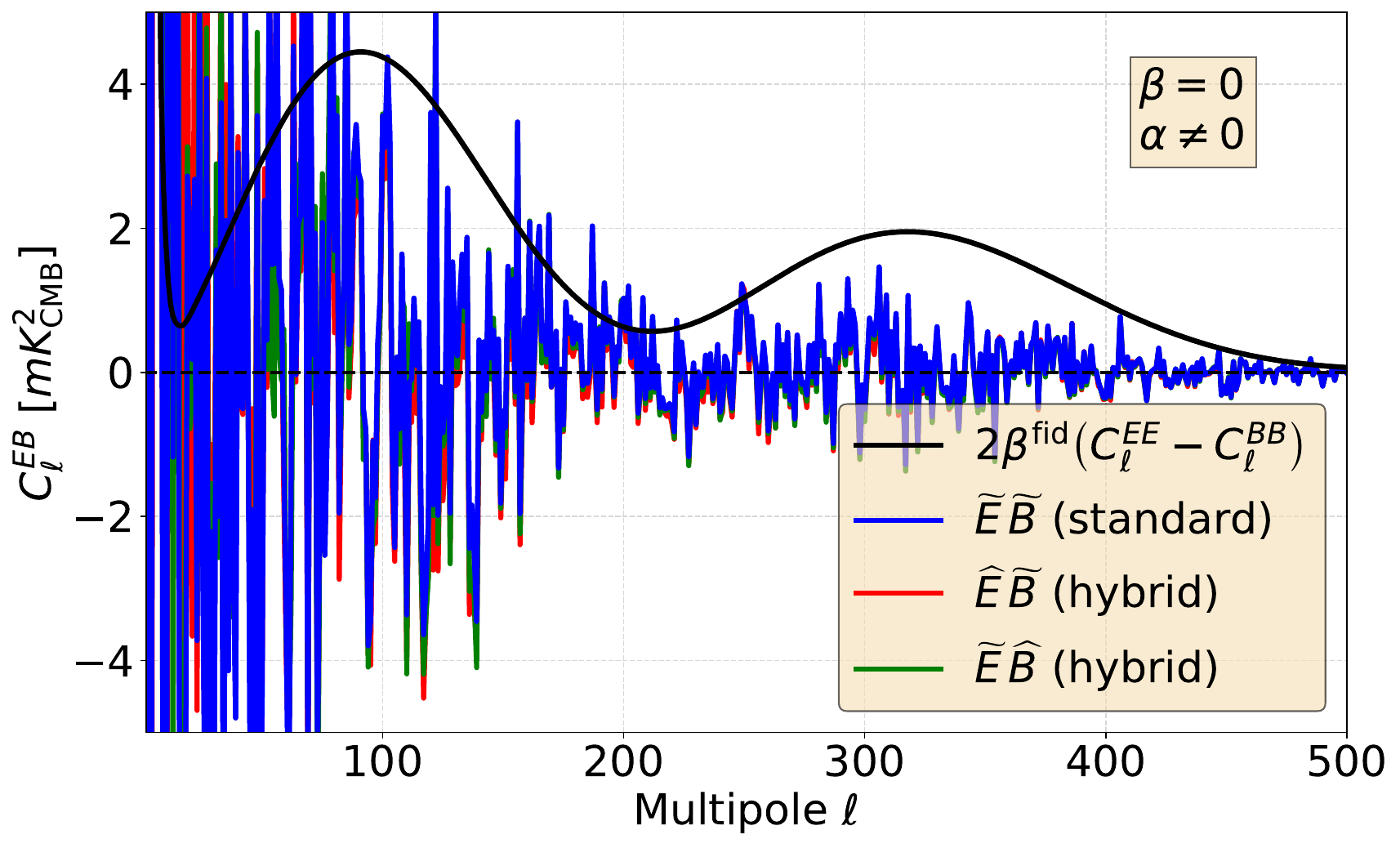}~
\includegraphics[width=0.5\textwidth,clip]{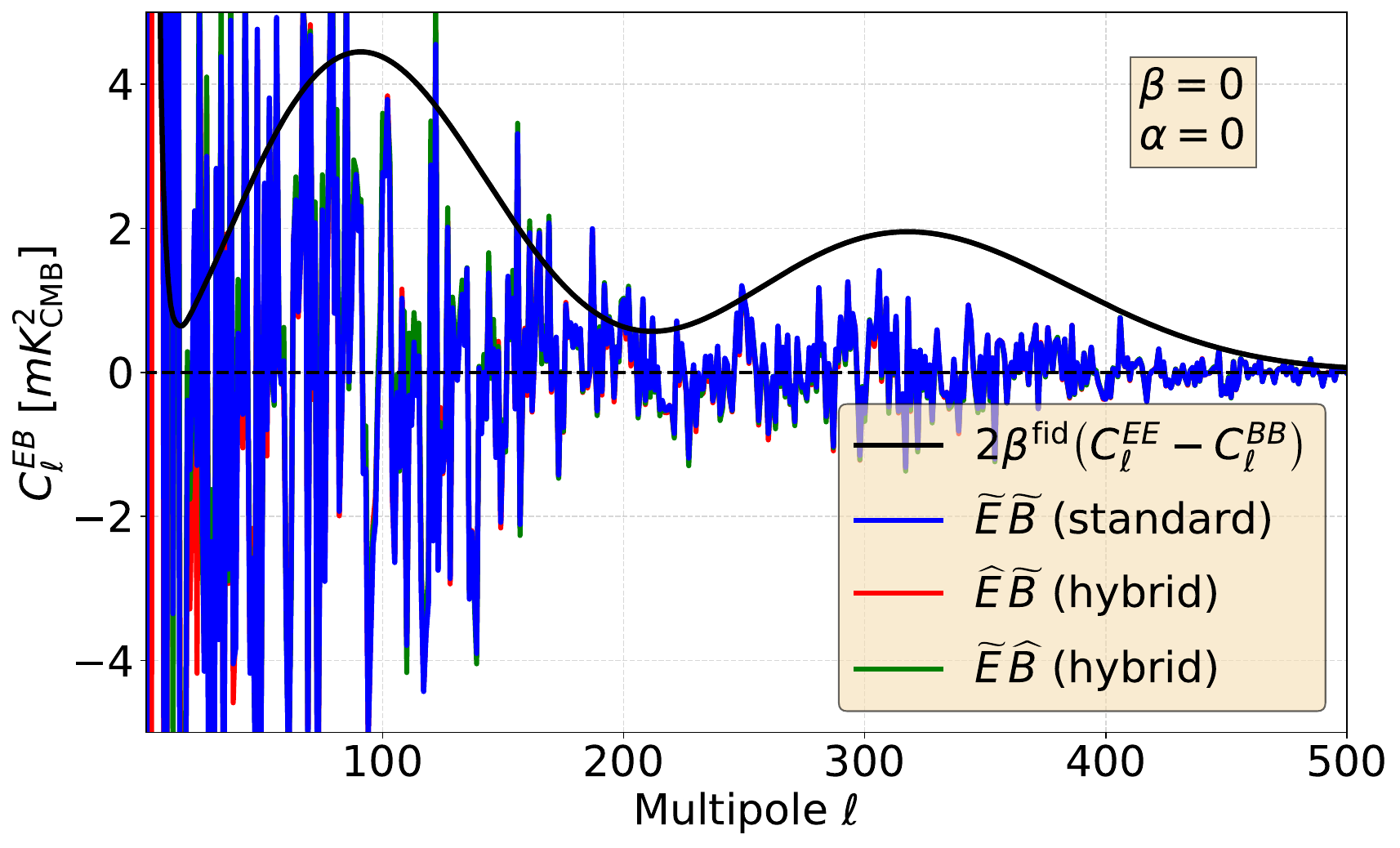}
\hfill
\caption{\label{fig:ebps} $EB$ cross-power spectra computed over $f_{\rm sky} = 50\,\%$ of the sky between the standard NILC CMB $E$- and $B$-mode maps (\emph{blue}), the Hybrid NILC CMB $E$-mode map and the standard NILC CMB $B$-mode map (\emph{red}),  and the Hybrid NILC CMB $B$-mode map and the standard NILC CMB $E$-mode map (\emph{green}), along with the theory prediction for the effective $EB$ cross-spectrum, $2\beta^{\rm fid}(C_\ell^{EE} - C_\ell^{BB})$, for a fiducial birefringence angle $\beta^{\rm fid} = 0.3^\circ$ (\emph{black}). \emph{Top left}: Case $\beta = 0.3^\circ$, $\alpha \neq 0$. \emph{Top right}: Case $\beta = 0.3^\circ$, $\alpha = 0$.  \emph{Bottom left}: Case $\beta = 0$, $\alpha \neq 0$.  \emph{Bottom right}: Case $\beta = 0$, $\alpha = 0$. When $\beta \neq 0$ (top panels), the standard NILC $E$- and $B$-mode cross-spectrum (blue) agrees with the theoretical prediction (black), while the Hybrid NILC $E$-mode map (red) shows minimal correlation, aside from residuals, with the standard NILC $B$-mode map, as expected from Equation~\eqref{eq:nulleb}.}
\end{figure}

Figure~\ref{fig:ebps} shows the recovered $EB$ cross-power spectrum between the standard NILC CMB $E$- and $B$-mode maps ($\widetilde{E}\widetilde{B}$, blue) across the four rotation scenarios, along with the theoretical prediction for the effective $EB$ spectrum, $2\beta^{\rm fid}(C_\ell^{EE,\rm{CMB}} - C_\ell^{BB,\rm{CMB}})$ (black line), assuming a fiducial birefringence angle $\beta^{\rm fid} = 0.3^\circ$. The standard NILC $EB$ cross-spectrum agrees with the theoretical expectation only when cosmic birefringence is present ($\beta = 0.3^\circ$, top panels). 

In contrast, the cross-spectra $\widehat{E}\widetilde{B}$ (red) and $\widetilde{E}\widehat{B}$ (green), involving hybrid and standard NILC maps, remain consistent with zero in all scenarios. This behaviour, consistent with Equation~\eqref{eq:nulleb}, reflects the construction of the Hybrid ILC, which isolates the \emph{uncorrelated} components of the CMB $E$- and $B$-modes.

Although the standard NILC $EB$ cross-spectrum (blue) could, in principle, be used to constrain $\beta$ by fitting the theoretical model (black), as commonly done in the literature, we instead propose leveraging the Hybrid-ILC separation of correlated and uncorrelated CMB components to infer $\beta$ directly at the map level via spatial linear regression (Equations~\ref{eq:textbook}--\ref{eq:e}), aiming for tighter constraints. Nonetheless, in Section~\ref{subsec:diagnostic}, we still make use of the standard NILC $EB$ cross-spectrum to compute the ratio $\mathcal{R}_\ell$ defined in Equation~\eqref{eq:magicratio}, a diagnostic tool developed in Section~\ref{subsec:discrimination} to help distinguish between cosmic birefringence and instrumental miscalibration.

\subsection{Diagnosing cosmic birefringence vs. polarization angle miscalibration}
\label{subsec:diagnostic}

The ratio $\mathcal{R}_\ell=C_\ell^{\widetilde{E}\widetilde{B}, {\rm CMB}}/(C_\ell^{EE, {\rm CMB}} - C_\ell^{BB, {\rm CMB}})$, introduced in Equation~\eqref{eq:magicratio}, compares the shape of standard NILC CMB $EB$ cross-spectrum to that of the difference between the $\Lambda$CDM $EE$ and $BB$ auto-spectra. As shown analytically in Section~\ref{subsec:discrimination}, this ratio serves as a diagnostic tool to distinguish between the signatures of cosmic birefringence and instrumental polarization angle miscalibration.

The key idea is that cosmic birefringence induces an achromatic (i.e., frequency-independent) rotation, which remains unaltered by the ILC weights and thus produces a ratio $\mathcal{R}_\ell$ that is constant across multipoles and sky regions. In contrast, polarization angle miscalibration is chromatic, with frequency-dependent angles projected through the NILC weights, introducing multipole and spatial variations in $\mathcal{R}_\ell$. The spectral shape of $\mathcal{R}_\ell$ therefore reflects the origin of the observed $EB$ correlation, providing a way to break the degeneracy between cosmological and instrumental effects.

\begin{figure}[tbp]
\centering 
\includegraphics[width=0.5\textwidth,clip]{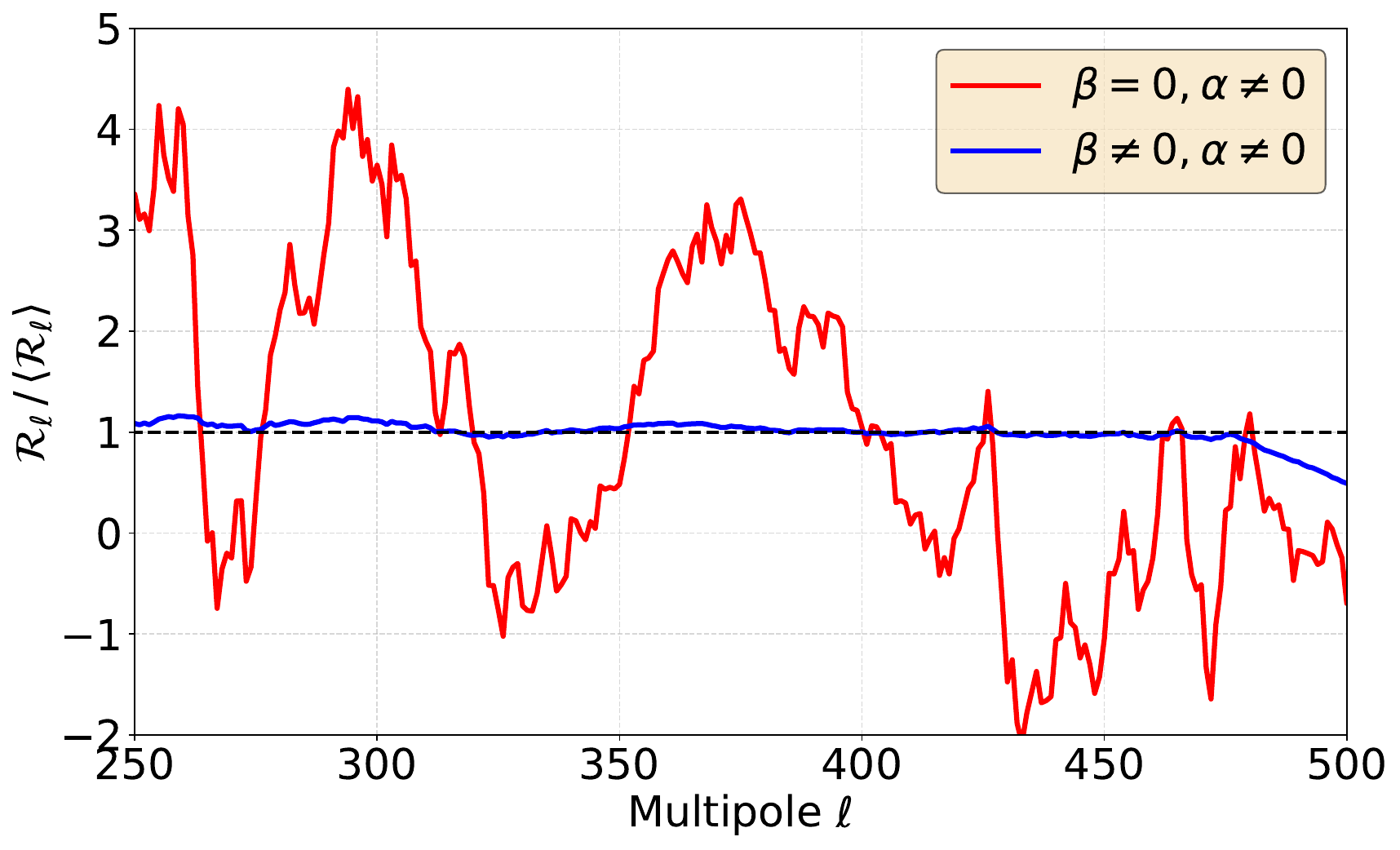}~
\includegraphics[width=0.5\textwidth,clip]{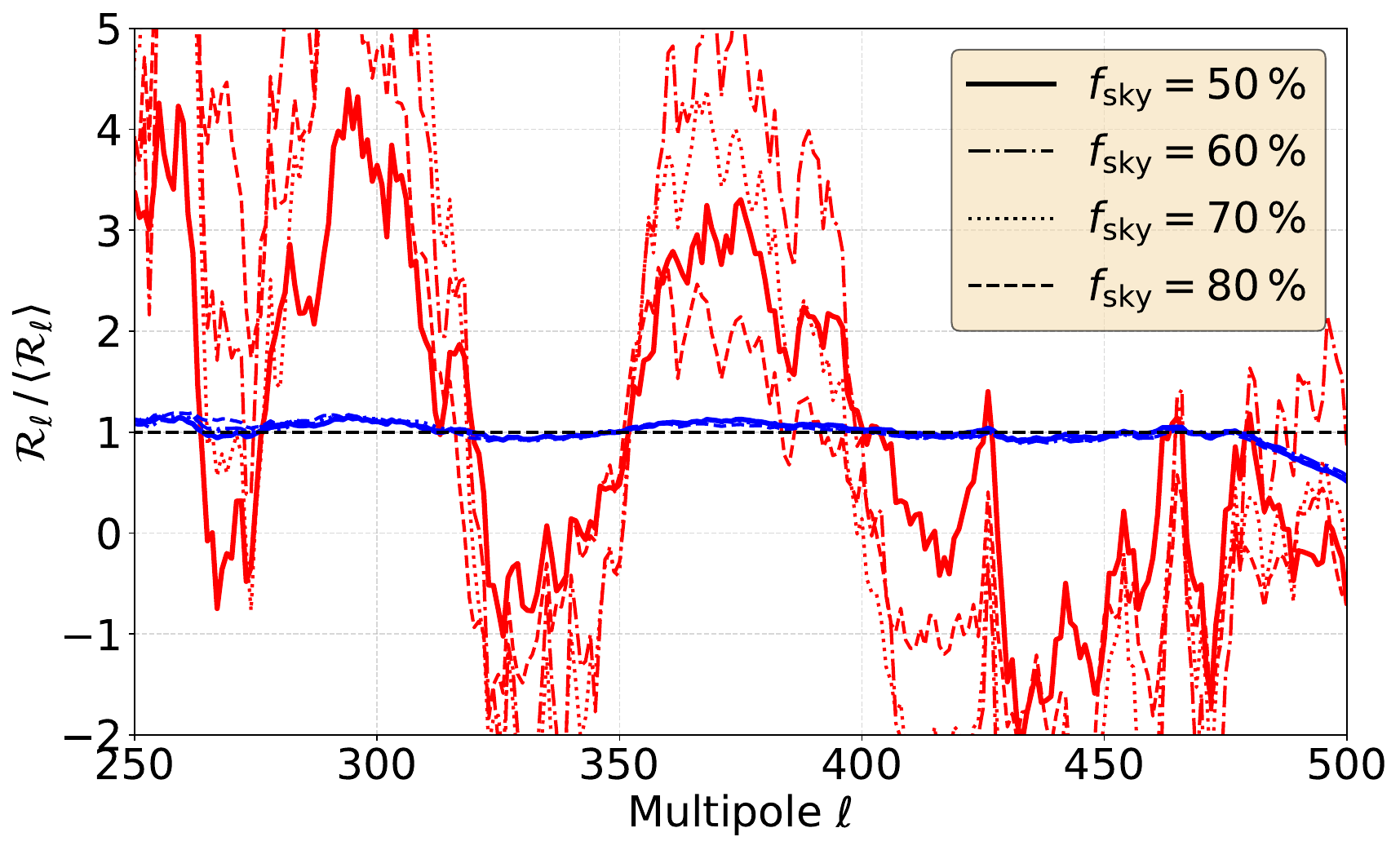}~
\hfill
\caption{\label{fig:degeneracy_breaking} 
\emph{Left}: Ratio $\mathcal{R}_\ell = C^{\widetilde{E} \widetilde{B}, {\rm CMB}}_\ell/(C^{EE, {\rm CMB}}_\ell - C^{BB, {\rm CMB}}_\ell)$ (Equation~\ref{eq:magicratio}) of the standard NILC $EB$ cross-power spectrum over $f_{\rm sky}=50\,\%$ of the sky to the difference between the $\Lambda\text{CDM}$ $EE$ and $BB$ auto-power spectra, further normalised by its mean $\langle \mathcal{R}_\ell\rangle$ over $\ell \in [250,500]$. 
In the presence of cosmic birefringence, this ratio remains relatively constant across multipoles (\emph{blue line}), with $\mathcal{R}_\ell \simeq \langle \mathcal{R}_\ell\rangle$, as predicted by Equation~\eqref{eq:magicratio_beta}. In contrast, it exhibits significant distortion when polarization angle miscalibration is the primary source of rotation (\emph{red line}), as predicted by Equation~\eqref{eq:magicratio_alpha}. This observable allows us to distinguish between cosmic birefringence and instrumental miscalibration, complementing the field-level inference of the birefringence angle developed in this work. \emph{Right}: Same observable as in the left panel, but computed over varying sky fractions. Under achromatic cosmic birefringence (\emph{blue}), the ratio remains stable across sky areas. In contrast, chromatic miscalibration (\emph{red}) induces spatial distortions in this observable, inherited from anisotropies in the NILC weights. Sensitivity to sky coverage thus provides an additional diagnostic for disentangling cosmic birefringence from instrumental miscalibration.} 
\end{figure}

This behaviour is illustrated in the left panel of Figure~\ref{fig:degeneracy_breaking}, which shows $\mathcal{R}_\ell$ normalised by its mean over the multipole range $\ell \in [250, 500]$, i.e., $\mathcal{R}_\ell/\langle \mathcal{R}_\ell \rangle$, for two distinct scenarios:\footnote{For visual rendering, the spectrum $\mathcal{R}_\ell/\langle \mathcal{R}_\ell \rangle$ in Figure~\ref{fig:degeneracy_breaking} has been smoothed using a sliding window with relative width $\Delta\ell /\ell = 0.1$.} instrumental miscalibration without cosmic birefringence ($\alpha_\nu \neq 0$, $\beta = 0$, red), and both cosmic birefringence and instrumental miscalibration ($\alpha_\nu \neq 0$, $\beta \neq 0$, blue). In the presence of cosmic birefringence (blue), we recover a flat, constant ratio $\mathcal{R}_\ell\simeq \langle \mathcal{R}_\ell \rangle$. Conversely, when the observed $EB$ signal is purely due to  polarization angle miscalibration (red), the ratio shows clear spectral distortions with significant variations across multipoles.   

The right panel of Figure~\ref{fig:degeneracy_breaking}  shows the same observable computed over varying sky fractions. Again, the ratio remains relatively stable and insensitive to the sky fraction under achromatic cosmic birefringence (blue), but varies with sky coverage under chromatic instrumental miscalibration (red) due to spatial structure inherited from the NILC weights. The non-monotonic variation of the ratio with sky fraction arises from the properties of the foreground $E/B$ scalar maps, which are signed and non-local fields. The $Q/U \to E/B$ transformation produces large regions of opposite polarity whose contributions can partially cancel in the harmonic coefficients as the mask changes to include more sky. Consequently, the measured foreground power --- and, in turn, the NILC $E$ and $B$ weights involved in the ratio through Equations~\eqref{eq:projected_alpha}--\eqref{eq:magicratio} --- do not vary strictly monotonically with Galactic latitude.

Together, these behaviours make $\mathcal{R}_\ell$ a practical and interpretable diagnostic for real data analysis: flatness supports the presence of cosmic birefringence, while distortions point to polarization angle miscalibration.

\subsection{Cosmic birefringence angle estimation via spatial linear regression}
\label{subsec:beta_linreg}

\begin{figure}[tbp]
\centering 
\includegraphics[width=0.505\textwidth,clip]{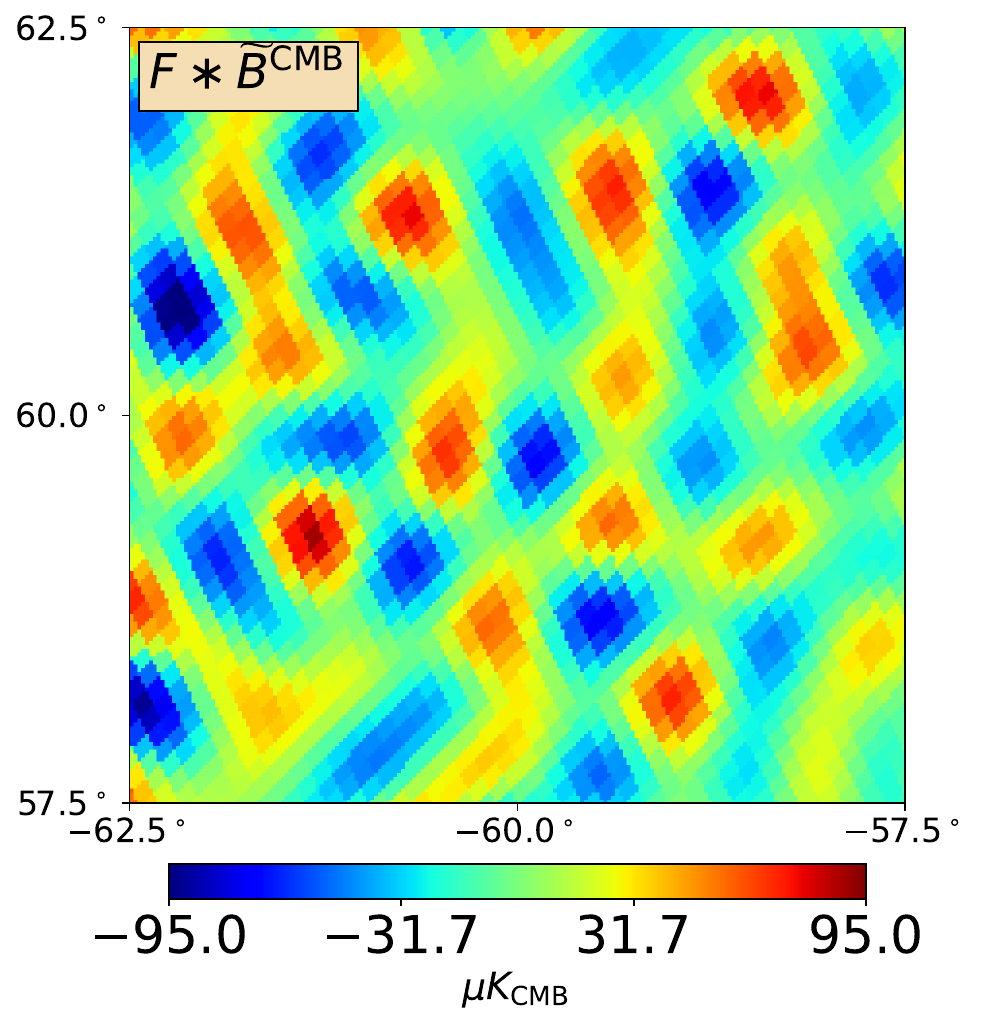}~
\includegraphics[width=0.495\textwidth,clip]{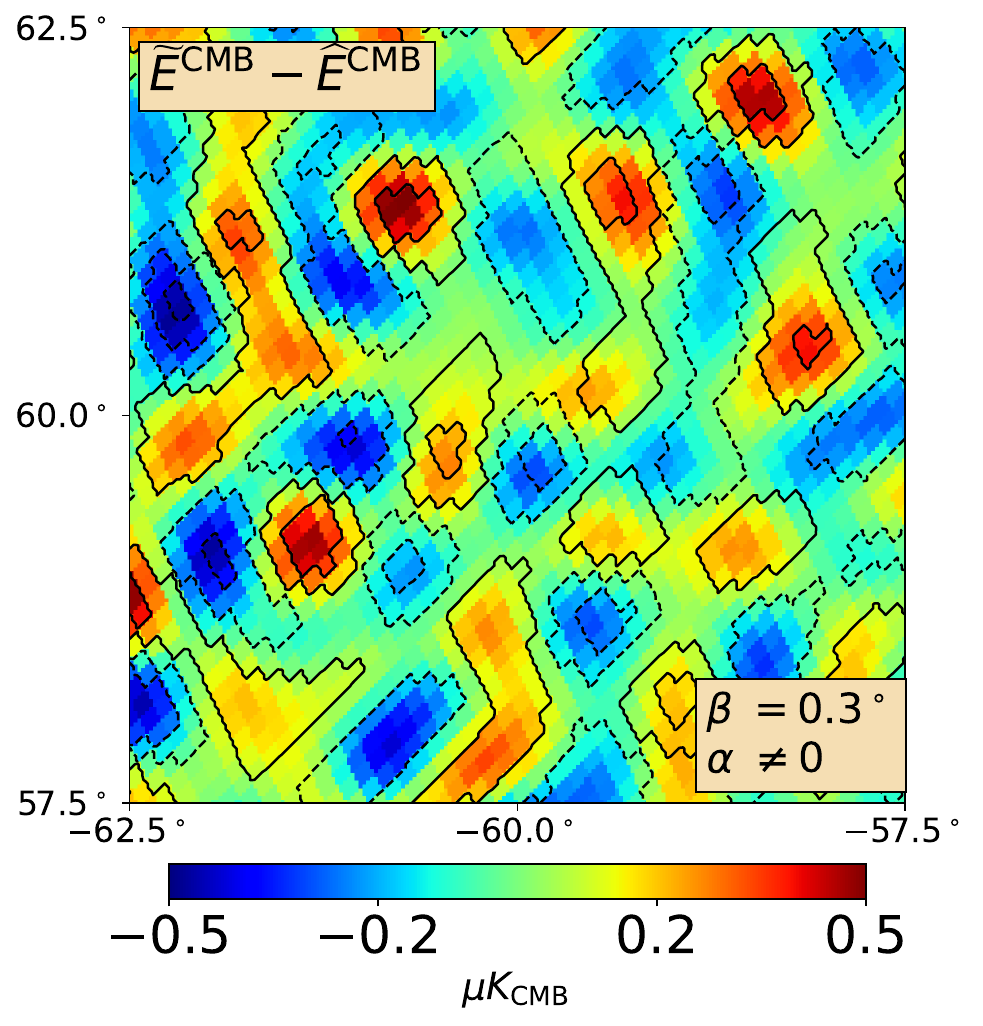}\\
\includegraphics[width=0.495\textwidth,clip]{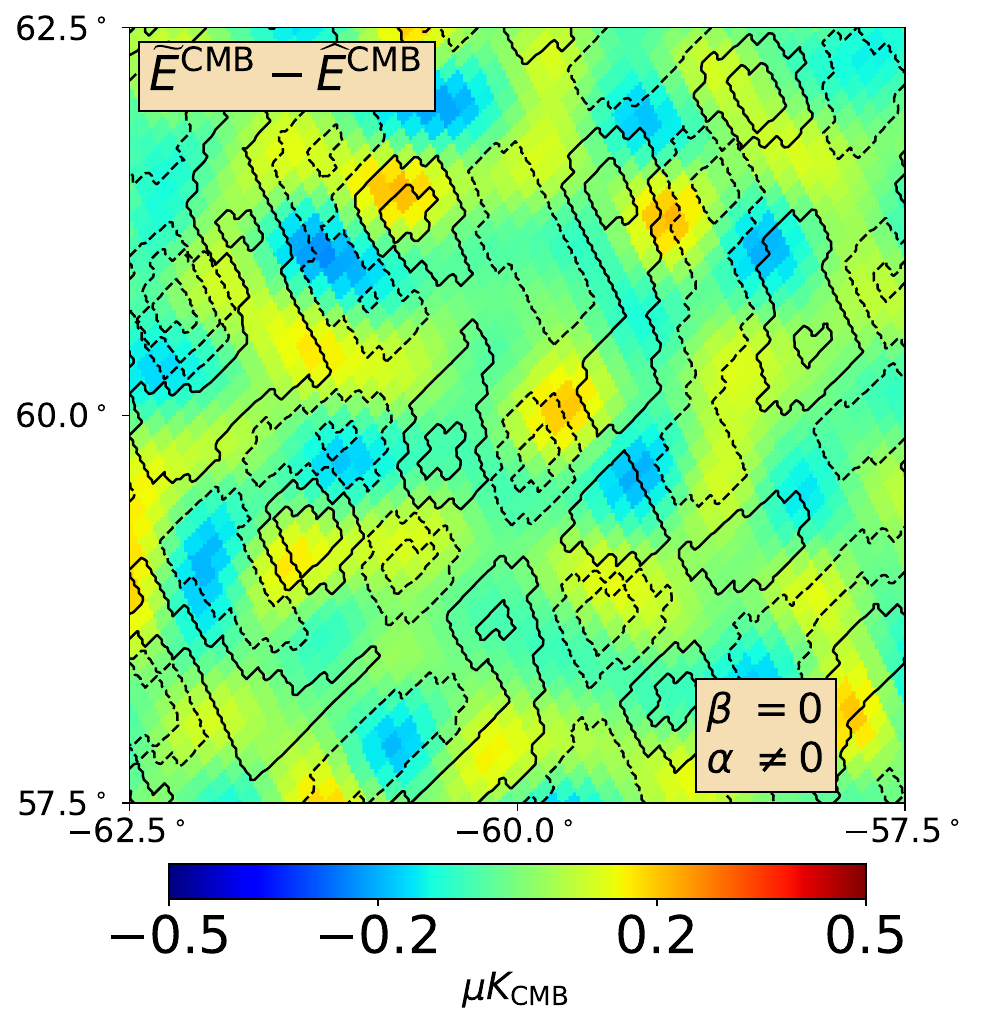}~
\includegraphics[width=0.495\textwidth,clip]{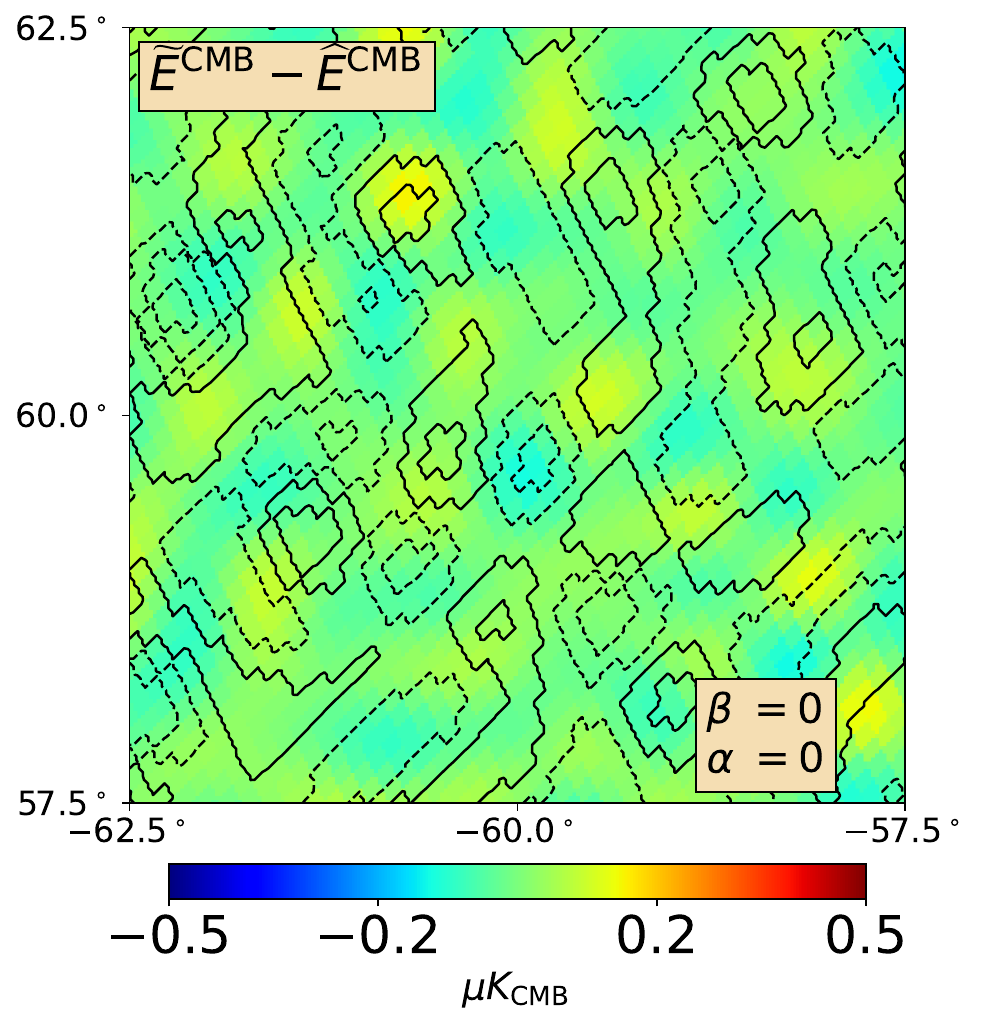}
\hfill
\caption{\label{fig:maps} 
Spatial correlation between the modulated CMB $B$-mode field $F \ast \widetilde{B}^{\rm CMB}$ (\emph{top left}) and the difference $\widetilde{E}^{\rm CMB} - \widehat{E}^{\rm CMB}$ (remaining panels) in a $5^\circ \times 5^\circ$ patch centred at Galactic coordinates $(\ell, b) = (-60^\circ, 60^\circ)$, i.e., within the sky patch 5 of Figure~\ref{fig:igloo}. Both fields were reconstructed using standard and hybrid NILC methods, and all maps in this figure have been filtered in harmonic space with a top-hat bandpass retaining multipoles $\ell \in [250, 500]$. Black isocontours of the modulated $B$-mode field $F \ast \widetilde{B}^{\rm CMB}$ are overlaid on the $E$-mode difference field $\widetilde{E}^{\rm CMB} - \widehat{E}^{\rm CMB}$ to highlight spatial correlation.
\emph{Top right}: When $\beta = 0.3^\circ$ and $\alpha_\nu \ne 0$, strong correlation ($86\,\%$ in the patch, $83\,\%$ over $f_{\rm sky} = 50\,\%$ of the sky) is observed, as expected from Equation~\eqref{eq:theorem_pixel}. In the ideal case with perfect calibration, i.e., $\beta = 0.3^\circ$ and $\alpha_\nu = 0$ (not shown), the correlation reaches $94\,\%$/$92\,\%$.
In the absence of birefringence ($\beta = 0$, bottom panels), correlation drops to $24\,\%$/$7\,\%$ for $\alpha_\nu \ne 0$ (\emph{bottom left}), and to $37\,\%$/$5\,\%$ for $\alpha_\nu = 0$ (\emph{bottom right}).
}
\end{figure}

\begin{figure}[tbp]
\centering 
\includegraphics[width=0.505\textwidth,clip]{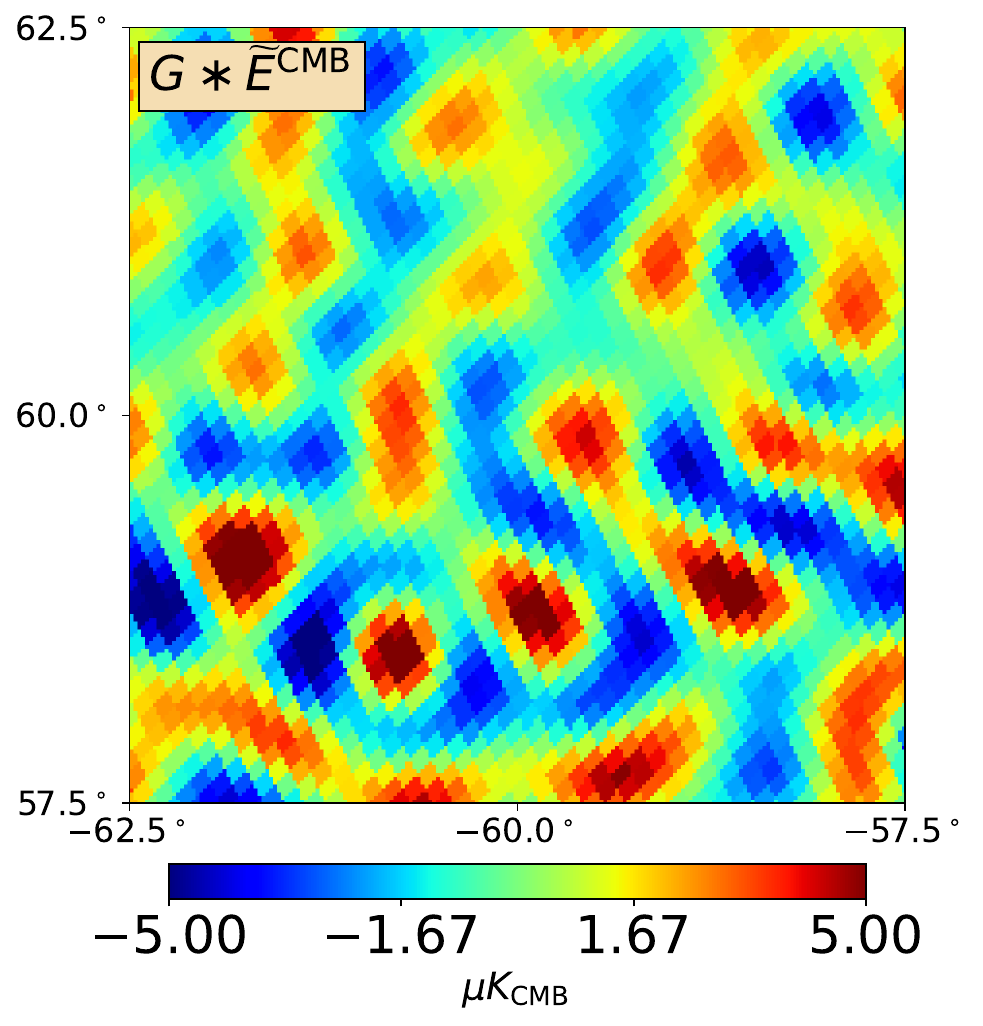}~
\includegraphics[width=0.495\textwidth,clip]{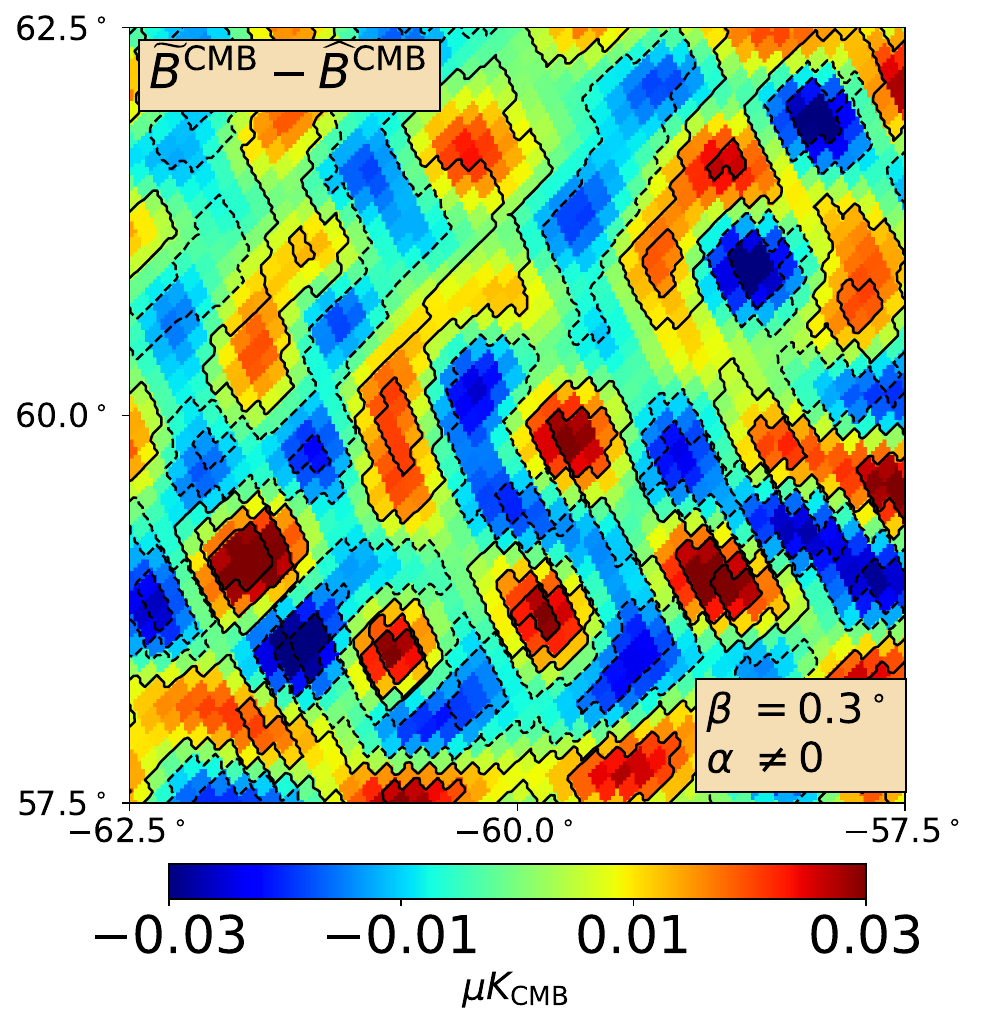}\\
\includegraphics[width=0.495\textwidth,clip]{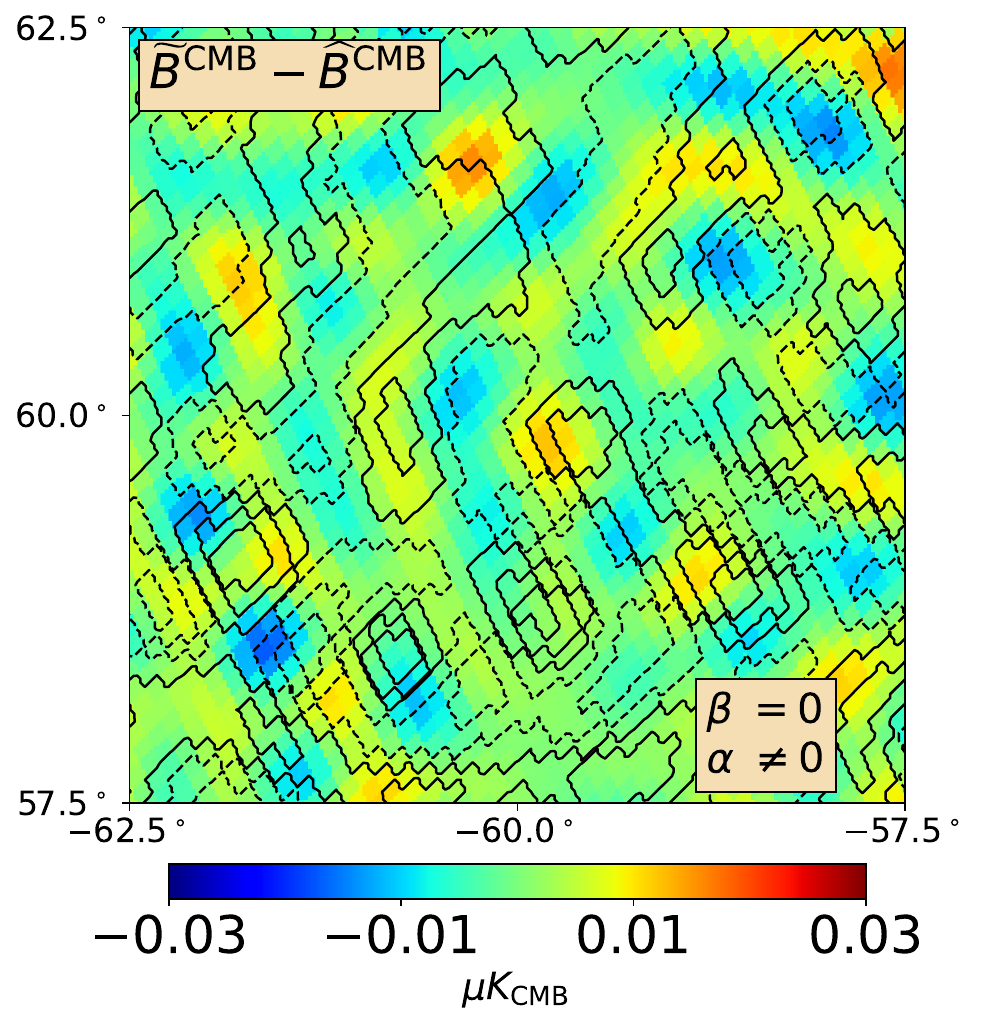}~
\includegraphics[width=0.495\textwidth,clip]{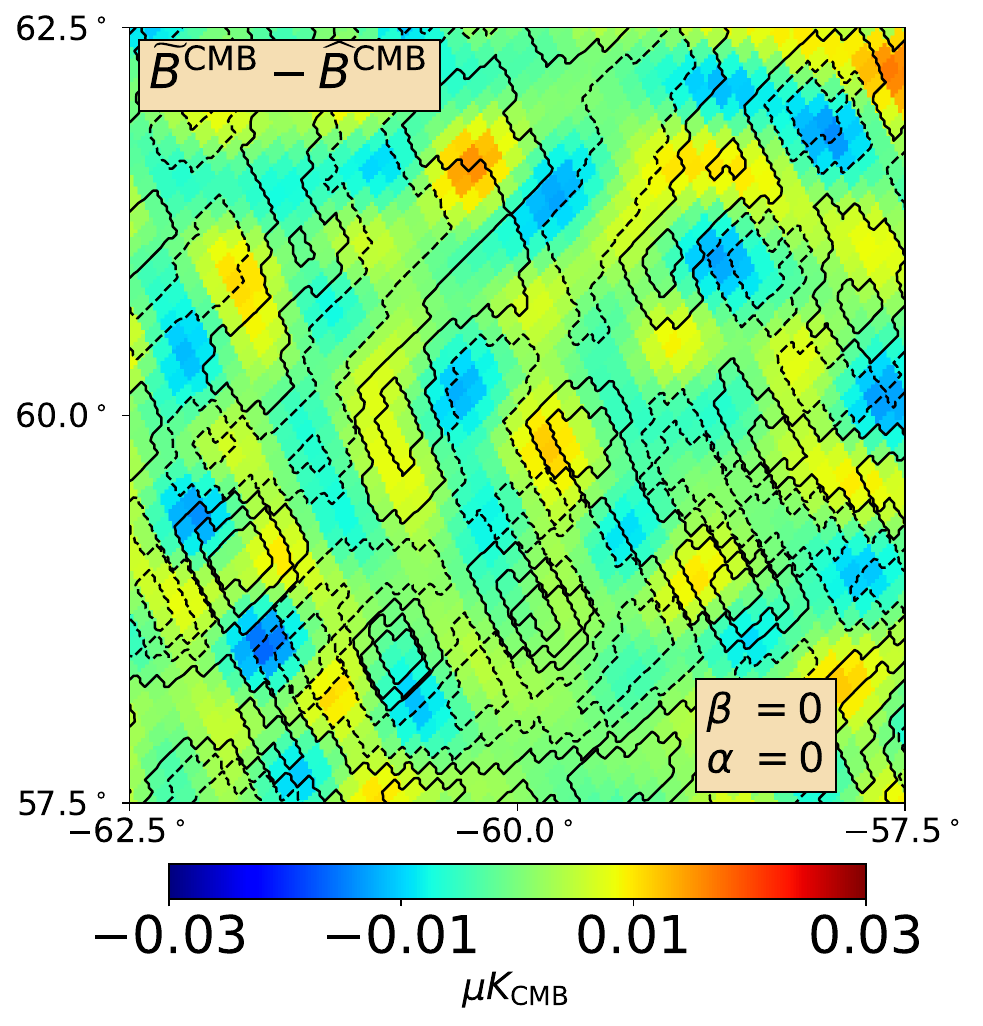}
\hfill
\caption{\label{fig:maps_b} 
Same as Figure~\ref{fig:maps}, but for the modulated CMB $E$-mode field $G \ast \widetilde{E}^{\rm CMB}$ (\emph{top left}) and the difference $\widetilde{B}^{\rm CMB} - \widehat{B}^{\rm CMB}$ (remaining panels).
\emph{Top right}: When $\beta = 0.3^\circ$ and $\alpha_\nu \ne 0$, the spatial correlation between the two fields reaches $93\,\%$ in the patch and $92\,\%$ over $f_{\rm sky} = 50\,\%$ of the sky.
In the absence of birefringence (bottom panels), the correlation falls to $32\,\%$/$8\,\%$ for $\alpha_\nu \ne 0$ (\emph{bottom left}), and to $29\,\%$/$3\,\%$ for $\alpha_\nu = 0$ (\emph{bottom right}).
}
\end{figure}

In the presence of cosmic birefringence, the difference between the standard and hybrid NILC maps, $\widetilde{E}^{\rm CMB}(\hat{n}) - \widehat{E}^{\rm CMB}(\hat{n})$, which captures the correlated component of the CMB $E$-modes, is expected to exhibit strong spatial correlation with the modulated standard NILC CMB $B$-mode field $(F \ast \widetilde{B}^{\rm CMB})(\hat{n})$, as predicted by the linear relation in Equation~\eqref{eq:theorem_pixel}. This expectation is visually confirmed in Figures~\ref{fig:maps} and~\ref{fig:maps_b}.

Figure~\ref{fig:maps} compares these two fields in a $5^\circ \times 5^\circ$ sky patch (patch 5 of Figure~\ref{fig:igloo}), filtered in harmonic space with a top-hat bandpass retaining multipoles $\ell \in [250, 500]$ to enhance signal-to-noise. The comparison is shown for three different scenarios: (i) $\beta = 0.3^\circ$ and $\alpha_\nu\neq 0$ (birefringence and miscalibration; top right), (ii) $\beta = 0$ and $\alpha_\nu\neq 0$ (only miscalibration; bottom left), and (iii) $\beta = 0$ and $\alpha_\nu = 0$ (neither effect; bottom right). 
When birefringence is present (top right), a strong spatial correlation emerges between the two fields, highlighted by isocontours of the modulated $B$-mode map $F\ast \widetilde{B}^{\rm CMB}$ overlaid on the difference map $\widetilde{E}^{\rm CMB} - \widehat{E}^{\rm CMB}$. The Pearson correlation coefficient reaches $86\,\%$ in the patch and $83\,\%$ across $f_{\rm sky} = 50\,\%$ of the sky.\footnote{The Pearson correlation coefficient, defined as $\rho = \mathrm{cov}(X,Y)/\sqrt{\mathrm{cov}(X,X),\mathrm{cov}(Y,Y)}$, where $\mathrm{cov}(X,Y)$ is the covariance between two vector fields $X(\hat{n})$ and $Y(\hat{n})$ and $\mathrm{cov}(X,X)$ and $\mathrm{cov}(Y,Y)$ are their respective variances, quantifies the linear correlation between the two fields $X(\hat{n})$ and $Y(\hat{n})$.} In the idealized case of perfect angle calibration ($\alpha_\nu = 0$ and $\beta=0.3^\circ$, not shown), the correlation would rise to $94\,\%$ and $92\,\%$ over $50\,\%$ of the sky.

In the absence of cosmic birefringence ($\beta = 0$, bottom panels), the visible correlation between the two fields disappears due to spatial variations in the projected miscalibration angles, inherited from those of the NILC weights, which in turn disrupt the perfect linear relation described by Equation~\eqref{eq:theorem_pixel}, leading to a transition to the more general form in Equation~\eqref{eq:theorem_lm}. In this case, the Pearson correlation coefficient drops to $24\,\%$ in the patch ($7\,\%$ over $50\,\%$ of the sky) for $\alpha_\nu \neq 0$ (bottom left), and to $37\,\%$ ($5\,\%$ over $ 50\,\%$ of the sky) for $\alpha_\nu = 0$ (bottom right).

Figure~\ref{fig:maps_b} presents a complementary analysis: the difference between standard and hybrid NILC $B$-mode maps, $\widetilde{B}^{\rm CMB}(\hat{n}) - \widehat{B}^{\rm CMB}(\hat{n})$, is compared to the modulated $E$-mode field $(G \ast \widetilde{E}^{\rm CMB})(\hat{n})$ for the same three scenarios. This alternative regression yields even stronger correlation  in the birefringence case ($\beta = 0.3^\circ$, $\alpha_\nu\neq 0$), reaching $93\,\%$ in the patch and $92\,\%$ over $f_{\rm sky} = 50\,\%$ of the sky, approaching the level seen in the perfectly calibrated case. This improvement is attributed to the higher signal-to-noise in the modulated $E$-mode map $G\ast \widetilde{E}^{\rm CMB}$ compared to the modulated $B$-mode map $F\ast \widetilde{B}^{\rm CMB}$. 

Figures~\ref{fig:ttplot} and~\ref{fig:ttplot_zerobeta_nonzeroalpha} in Appendix~\ref{sec:ttplots} display the correlation plots ("T-T plots") between the fields $Y(\hat{n})=\widetilde{E}^{\rm CMB}(\hat{n}) - \widehat{E}^{\rm CMB}(\hat{n})$ and $X(\hat{n})=F \ast \widetilde{B}^{\rm CMB}$ across the 26 sky patches defined in Figure~\ref{fig:igloo}, for two distinct scenarios: with cosmic birefringence ($\beta=0.3^\circ$, $\alpha_\nu\neq 0$; Figure~\ref{fig:ttplot}) and without cosmic birefringence ($\beta=0$, $\alpha_\nu\neq 0$; Figure~\ref{fig:ttplot_zerobeta_nonzeroalpha}). In the presence of cosmic birefringence, we observe a consistent and pronounced correlation between the two fields across all sky patches. In contrast, when birefringence is absent but polarization angle miscalibration is still present, no clear correlation emerges.

\begin{table}
    \centering
    \resizebox{\columnwidth}{!}{%
    \begin{tabular}{|c|c|}
    \hline
    & ($\beta = 0.3^\circ$, $\alpha \neq 0$) \\        
    Sky patch & $\widehat{\beta}$ (degrees) \\
    & \\
    \hline
        1 & $0.330^\circ \pm 0.002^\circ$ \\
    2 & $0.307^\circ \pm 0.002^\circ$ \\
    3 & $0.302^\circ \pm 0.002^\circ$ \\
    4 & $0.331^\circ \pm 0.002^\circ$ \\
    5 & $0.327^\circ \pm 0.002^\circ$ \\
    6 & $0.301^\circ \pm 0.002^\circ$ \\
    7 & $0.296^\circ \pm 0.002^\circ$ \\
    8 & $0.291^\circ \pm 0.002^\circ$ \\
    9 & $0.299^\circ \pm 0.002^\circ$ \\
    10 & $0.313^\circ \pm 0.002^\circ$ \\
    11 & $0.315^\circ \pm 0.002^\circ$ \\
    12 & $0.323^\circ \pm 0.002^\circ$ \\
    13 & $0.291^\circ \pm 0.002^\circ$ \\
    14 & $0.271^\circ \pm 0.002^\circ$ \\
    15 & $0.277^\circ \pm 0.002^\circ$ \\
    16 & $0.284^\circ \pm 0.002^\circ$ \\
    17 & $0.274^\circ \pm 0.002^\circ$ \\
    18 & $0.273^\circ \pm 0.002^\circ$ \\
    19 & $0.290^\circ \pm 0.002^\circ$ \\
    20 & $0.300^\circ \pm 0.002^\circ$ \\
    21 & $0.282^\circ \pm 0.002^\circ$ \\
    22 & $0.283^\circ \pm 0.002^\circ$ \\
    23 & $0.274^\circ \pm 0.002^\circ$ \\
    24 & $0.276^\circ \pm 0.002^\circ$ \\
    25 & $0.307^\circ \pm 0.002^\circ$ \\
    26 & $0.288^\circ \pm 0.002^\circ$ \\
    \hline
    & \\
    All patches & $\langle\,\widehat{\beta}\,\rangle$ (degrees) \\
    & \\
    \hline
        1--26 & $ 0.296^\circ \pm 0.018^\circ$\\
    \hline
    \end{tabular}%
        \begin{tabular}{|c|}
    \hline
    ($\beta = 0.3^\circ$, $\alpha = 0$) \\        
    $\widehat{\beta}$ (degrees) \\
     \\
    \hline
        $0.324^\circ \pm 0.002^\circ$ \\
    $0.301^\circ \pm 0.001^\circ$ \\
    $0.300^\circ \pm 0.001^\circ$ \\
    $0.330^\circ \pm 0.002^\circ$ \\
    $0.322^\circ \pm 0.002^\circ$ \\
    $0.292^\circ \pm 0.002^\circ$ \\
    $0.292^\circ \pm 0.001^\circ$ \\
    $0.282^\circ \pm 0.001^\circ$ \\
    $0.294^\circ \pm 0.001^\circ$ \\
    $0.308^\circ \pm 0.001^\circ$ \\
    $0.310^\circ \pm 0.001^\circ$ \\
    $0.318^\circ \pm 0.002^\circ$ \\
    $0.283^\circ \pm 0.001^\circ$ \\
    $0.261^\circ \pm 0.001^\circ$ \\
    $0.271^\circ \pm 0.001^\circ$ \\
    $0.272^\circ \pm 0.001^\circ$ \\
    $0.268^\circ \pm 0.001^\circ$ \\
    $0.266^\circ \pm 0.001^\circ$ \\
    $0.282^\circ \pm 0.001^\circ$ \\
    $0.292^\circ \pm 0.002^\circ$ \\
    $0.274^\circ \pm 0.001^\circ$ \\
    $0.277^\circ \pm 0.001^\circ$ \\
    $0.271^\circ \pm 0.001^\circ$ \\
    $0.269^\circ \pm 0.001^\circ$ \\
    $0.299^\circ \pm 0.002^\circ$ \\
    $0.284^\circ \pm 0.001^\circ$ \\
    \hline
    \\
    $\langle\,\widehat{\beta}\,\rangle$ (degrees) \\
    \\
    \hline
        $ 0.290^\circ \pm 0.019^\circ$\\
    \hline
    \end{tabular}%
    \begin{tabular}{|c|}
    \hline
    ($\beta = 0$, $\alpha \neq 0$) \\        
    $\widehat{\beta}$ (degrees) \\
     \\
    \hline
        $0.031^\circ \pm 0.001^\circ$ \\
    $0.012^\circ \pm 0.001^\circ$ \\
    $0.013^\circ \pm 0.001^\circ$ \\
    $0.035^\circ \pm 0.001^\circ$ \\
    $0.042^\circ \pm 0.001^\circ$ \\
    $0.022^\circ \pm 0.001^\circ$ \\
    $0.007^\circ \pm 0.001^\circ$ \\
    $0.010^\circ \pm 0.001^\circ$ \\
    $0.016^\circ \pm 0.001^\circ$ \\
    $0.024^\circ \pm 0.001^\circ$ \\
    $0.034^\circ \pm 0.001^\circ$ \\
    $0.046^\circ \pm 0.001^\circ$ \\
    $0.020^\circ \pm 0.001^\circ$ \\
    $-0.006^\circ \pm 0.001^\circ$ \\
    $-0.006^\circ \pm 0.001^\circ$ \\
    $0.005^\circ \pm 0.001^\circ$ \\
    $-0.009^\circ \pm 0.001^\circ$ \\
    $-0.013^\circ \pm 0.001^\circ$ \\
    $0.009^\circ \pm 0.001^\circ$ \\
    $0.016^\circ \pm 0.001^\circ$ \\
    $0.004^\circ \pm 0.001^\circ$ \\
    $-0.003^\circ \pm 0.001^\circ$ \\
    $-0.013^\circ \pm 0.001^\circ$ \\
    $-0.008^\circ \pm 0.001^\circ$ \\
    $0.014^\circ \pm 0.001^\circ$ \\
    $-0.006^\circ \pm 0.001^\circ$ \\
    \hline
    \\
    $\langle\,\widehat{\beta}\,\rangle$ (degrees) \\
    \\
    \hline
        $ 0.011^\circ \pm 0.017^\circ$\\
    \hline
    \end{tabular}%
    \begin{tabular}{|c|}
    \hline
    ($\beta = 0$, $\alpha = 0$) \\        
    $\widehat{\beta}$ (degrees) \\
     \\
    \hline
        $0.024^\circ \pm 0.001^\circ$ \\
    $0.006^\circ \pm 0.001^\circ$ \\
    $0.010^\circ \pm 0.001^\circ$ \\
    $0.032^\circ \pm 0.001^\circ$ \\
    $0.036^\circ \pm 0.001^\circ$ \\
    $0.011^\circ \pm 0.001^\circ$ \\
    $0.002^\circ \pm 0.001^\circ$ \\
    $-0.000^\circ \pm 0.001^\circ$ \\
    $0.010^\circ \pm 0.001^\circ$ \\
    $0.018^\circ \pm 0.001^\circ$ \\
    $0.029^\circ \pm 0.001^\circ$ \\
    $0.040^\circ \pm 0.001^\circ$ \\
    $0.011^\circ \pm 0.001^\circ$ \\
    $-0.017^\circ \pm 0.001^\circ$ \\
    $-0.014^\circ \pm 0.001^\circ$ \\
    $-0.007^\circ \pm 0.001^\circ$ \\
    $-0.016^\circ \pm 0.001^\circ$ \\
    $-0.021^\circ \pm 0.001^\circ$ \\
    $0.000^\circ \pm 0.001^\circ$ \\
    $0.008^\circ \pm 0.001^\circ$ \\
    $-0.004^\circ \pm 0.001^\circ$ \\
    $-0.010^\circ \pm 0.001^\circ$ \\
    $-0.017^\circ \pm 0.001^\circ$ \\
    $-0.016^\circ \pm 0.001^\circ$ \\
    $0.005^\circ \pm 0.001^\circ$ \\
    $-0.011^\circ \pm 0.001^\circ$ \\
    \hline
    \\
    $\langle\,\widehat{\beta}\,\rangle$ (degrees) \\
    \\
    \hline
        $ 0.004^\circ \pm 0.017^\circ$\\
    \hline
    \end{tabular}%
    }
    \caption{Estimates $\widehat{\beta}$ of the cosmic birefringence angle in the 26 sky patches shown in Figure~\ref{fig:igloo}, obtained from linear regression between the NILC fields $\widetilde{E}^{\rm CMB} - \widehat{E}^{\rm CMB}$ and $F \ast \widetilde{B}^{\rm CMB}$ (see Equations~\ref{eq:textbook}--\ref{eq:e}), for four different rotation scenarios. The average and standard deviation across all sky patches are given in the last row, yielding a recovered cosmic birefringence angle of $\widehat{\beta} = 0.296^\circ \pm 0.018^\circ$ for $\beta = 0.3^\circ$, $\alpha \neq 0$ (\emph{first column}); $\widehat{\beta} = 0.290^\circ \pm 0.019^\circ$ for $\beta = 0.3^\circ$, $\alpha = 0$ (\emph{second column}); $\widehat{\beta} = 0.011^\circ \pm 0.017^\circ$ for $\beta = 0$, $\alpha \neq 0$ (\emph{third column}); and $\widehat{\beta} = 0.004^\circ \pm 0.017^\circ$ for $\beta = 0$, $\alpha = 0$ (\emph{fourth column}).\label{tab:beta}}
    \end{table}

\begin{table}
    \centering
    \resizebox{\columnwidth}{!}{%
    \begin{tabular}{|c|c|}
    \hline
    & ($\beta = 0.3^\circ$, $\alpha \neq 0$) \\        
    Sky patch & $\widehat{\beta}$ (degrees) \\
    & \\
    \hline
        1 & $0.322^\circ \pm 0.001^\circ$ \\
    2 & $0.312^\circ \pm 0.001^\circ$ \\
    3 & $0.312^\circ \pm 0.001^\circ$ \\
    4 & $0.331^\circ \pm 0.001^\circ$ \\
    5 & $0.336^\circ \pm 0.001^\circ$ \\
    6 & $0.308^\circ \pm 0.001^\circ$ \\
    7 & $0.307^\circ \pm 0.001^\circ$ \\
    8 & $0.300^\circ \pm 0.001^\circ$ \\
    9 & $0.311^\circ \pm 0.001^\circ$ \\
    10 & $0.318^\circ \pm 0.001^\circ$ \\
    11 & $0.327^\circ \pm 0.001^\circ$ \\
    12 & $0.341^\circ \pm 0.001^\circ$ \\
    13 & $0.310^\circ \pm 0.001^\circ$ \\
    14 & $0.290^\circ \pm 0.001^\circ$ \\
    15 & $0.292^\circ \pm 0.001^\circ$ \\
    16 & $0.298^\circ \pm 0.001^\circ$ \\
    17 & $0.289^\circ \pm 0.001^\circ$ \\
    18 & $0.279^\circ \pm 0.001^\circ$ \\
    19 & $0.295^\circ \pm 0.001^\circ$ \\
    20 & $0.307^\circ \pm 0.001^\circ$ \\
    21 & $0.294^\circ \pm 0.001^\circ$ \\
    22 & $0.292^\circ \pm 0.001^\circ$ \\
    23 & $0.279^\circ \pm 0.001^\circ$ \\
    24 & $0.289^\circ \pm 0.001^\circ$ \\
    25 & $0.303^\circ \pm 0.001^\circ$ \\
    26 & $0.288^\circ \pm 0.001^\circ$ \\
    \hline
    & \\
    All patches & $\langle\,\widehat{\beta}\,\rangle$ (degrees) \\
    & \\
    \hline
        1--26 & $ 0.305^\circ \pm 0.016^\circ$\\
    \hline
    \end{tabular}%
        \begin{tabular}{|c|}
    \hline
    ($\beta = 0.3^\circ$, $\alpha = 0$) \\        
    $\widehat{\beta}$ (degrees) \\
     \\
    \hline
        $0.318^\circ \pm 0.001^\circ$ \\
    $0.308^\circ \pm 0.001^\circ$ \\
    $0.307^\circ \pm 0.001^\circ$ \\
    $0.327^\circ \pm 0.001^\circ$ \\
    $0.334^\circ \pm 0.001^\circ$ \\
    $0.304^\circ \pm 0.001^\circ$ \\
    $0.300^\circ \pm 0.001^\circ$ \\
    $0.296^\circ \pm 0.001^\circ$ \\
    $0.306^\circ \pm 0.001^\circ$ \\
    $0.313^\circ \pm 0.001^\circ$ \\
    $0.323^\circ \pm 0.001^\circ$ \\
    $0.338^\circ \pm 0.001^\circ$ \\
    $0.306^\circ \pm 0.001^\circ$ \\
    $0.283^\circ \pm 0.001^\circ$ \\
    $0.285^\circ \pm 0.001^\circ$ \\
    $0.293^\circ \pm 0.001^\circ$ \\
    $0.283^\circ \pm 0.001^\circ$ \\
    $0.274^\circ \pm 0.001^\circ$ \\
    $0.291^\circ \pm 0.001^\circ$ \\
    $0.302^\circ \pm 0.001^\circ$ \\
    $0.288^\circ \pm 0.001^\circ$ \\
    $0.285^\circ \pm 0.001^\circ$ \\
    $0.274^\circ \pm 0.001^\circ$ \\
    $0.285^\circ \pm 0.001^\circ$ \\
    $0.298^\circ \pm 0.001^\circ$ \\
    $0.282^\circ \pm 0.001^\circ$ \\
    \hline
    \\
    $\langle\,\widehat{\beta}\,\rangle$ (degrees) \\
    \\
    \hline
        $ 0.300^\circ \pm 0.017^\circ$\\
    \hline
    \end{tabular}%
    \begin{tabular}{|c|}
    \hline
    ($\beta = 0$, $\alpha \neq 0$) \\        
    $\widehat{\beta}$ (degrees) \\
     \\
    \hline
        $0.026^\circ \pm 0.001^\circ$ \\
    $0.017^\circ \pm 0.001^\circ$ \\
    $0.017^\circ \pm 0.001^\circ$ \\
    $0.036^\circ \pm 0.001^\circ$ \\
    $0.042^\circ \pm 0.001^\circ$ \\
    $0.014^\circ \pm 0.001^\circ$ \\
    $0.012^\circ \pm 0.001^\circ$ \\
    $0.005^\circ \pm 0.001^\circ$ \\
    $0.016^\circ \pm 0.001^\circ$ \\
    $0.023^\circ \pm 0.001^\circ$ \\
    $0.033^\circ \pm 0.001^\circ$ \\
    $0.045^\circ \pm 0.001^\circ$ \\
    $0.016^\circ \pm 0.001^\circ$ \\
    $-0.006^\circ \pm 0.001^\circ$ \\
    $-0.003^\circ \pm 0.001^\circ$ \\
    $0.004^\circ \pm 0.001^\circ$ \\
    $-0.006^\circ \pm 0.001^\circ$ \\
    $-0.016^\circ \pm 0.001^\circ$ \\
    $0.001^\circ \pm 0.001^\circ$ \\
    $0.012^\circ \pm 0.001^\circ$ \\
    $-0.000^\circ \pm 0.001^\circ$ \\
    $-0.003^\circ \pm 0.001^\circ$ \\
    $-0.014^\circ \pm 0.001^\circ$ \\
    $-0.006^\circ \pm 0.001^\circ$ \\
    $0.007^\circ \pm 0.001^\circ$ \\
    $-0.006^\circ \pm 0.001^\circ$ \\
    \hline
    \\
    $\langle\,\widehat{\beta}\,\rangle$ (degrees) \\
    \\
    \hline
        $ 0.010^\circ \pm 0.016^\circ$\\
    \hline
    \end{tabular}%
    \begin{tabular}{|c|}
    \hline
    ($\beta = 0$, $\alpha = 0$) \\        
    $\widehat{\beta}$ (degrees) \\
     \\
    \hline
        $0.022^\circ \pm 0.001^\circ$ \\
    $0.012^\circ \pm 0.001^\circ$ \\
    $0.012^\circ \pm 0.001^\circ$ \\
    $0.031^\circ \pm 0.001^\circ$ \\
    $0.038^\circ \pm 0.001^\circ$ \\
    $0.009^\circ \pm 0.001^\circ$ \\
    $0.005^\circ \pm 0.001^\circ$ \\
    $-0.001^\circ \pm 0.001^\circ$ \\
    $0.010^\circ \pm 0.001^\circ$ \\
    $0.018^\circ \pm 0.001^\circ$ \\
    $0.027^\circ \pm 0.001^\circ$ \\
    $0.041^\circ \pm 0.001^\circ$ \\
    $0.011^\circ \pm 0.001^\circ$ \\
    $-0.013^\circ \pm 0.001^\circ$ \\
    $-0.010^\circ \pm 0.001^\circ$ \\
    $-0.003^\circ \pm 0.001^\circ$ \\
    $-0.013^\circ \pm 0.001^\circ$ \\
    $-0.022^\circ \pm 0.001^\circ$ \\
    $-0.006^\circ \pm 0.001^\circ$ \\
    $0.006^\circ \pm 0.001^\circ$ \\
    $-0.008^\circ \pm 0.001^\circ$ \\
    $-0.011^\circ \pm 0.001^\circ$ \\
    $-0.021^\circ \pm 0.001^\circ$ \\
    $-0.011^\circ \pm 0.001^\circ$ \\
    $0.001^\circ \pm 0.001^\circ$ \\
    $-0.013^\circ \pm 0.001^\circ$ \\
    \hline
    \\
    $\langle\,\widehat{\beta}\,\rangle$ (degrees) \\
    \\
    \hline
        $ 0.004^\circ \pm 0.017^\circ$\\
    \hline
    \end{tabular}%
    }
    \caption{Same as Table~\ref{tab:beta}, but for the linear regression between the NILC fields $\widetilde{B}^{\rm CMB} - \widehat{B}^{\rm CMB}$ and $G \ast \widetilde{E}^{\rm CMB}$. In this case, the recovered cosmic birefringence angle is $\widehat{\beta} = 0.305^\circ \pm 0.016^\circ$ for $\beta = 0.3^\circ$, $\alpha \neq 0$ (\emph{first column}); $\widehat{\beta} = 0.300^\circ \pm 0.017^\circ$ for $\beta = 0.3^\circ$, $\alpha = 0$ (\emph{second column}); $\widehat{\beta} = 0.010^\circ \pm 0.016^\circ$ for $\beta = 0$, $\alpha \neq 0$ (\emph{third column}); and $\widehat{\beta} = 0.004^\circ \pm 0.017^\circ$ for $\beta = 0$, $\alpha = 0$ (\emph{fourth column}).\label{tab:beta_b}}
    \end{table}

We estimate the cosmic birefringence angle in each sky patch by applying the IV-2SLS linear regression method described in Section~\ref{subsec:regression}, regressing $Y(\hat{n})=\widetilde{E}^{\rm CMB}(\hat{n}) - \widehat{E}^{\rm CMB}(\hat{n})$ on $X(\hat{n})=F \ast \widetilde{B}^{\rm CMB}$. The slope obtained from each regression yields a local estimate of the birefringence angle, and the results are reported in Table~\ref{tab:beta} for the four rotation scenarios introduced in Section~\ref{subsec:sims}.

The isotropic birefringence angle is then estimated as the mean of the patch-wise $\beta$ values. To quantify the associated uncertainty, we compute the standard deviation across the 26 patch estimates, providing a conservative error estimate that captures both statistical and systematic contributions, including anisotropic foreground residuals and spatial fluctuations in the projected miscalibration angles. This approach avoids the potential underestimation of the uncertainty that would arise from assuming independent patches, as done in standard bootstrapping or when computing the standard error on the mean, which may fail to account for coherent contamination across patches.

For the case with cosmic birefringence and miscalibration ($\beta=0.3^\circ$, $\alpha_\nu\neq 0$), the mean estimated angle across all patches is $\widehat{\beta} = 0.296^\circ\pm0.018^\circ$, corresponding to a highly significant detection of $\beta=0.3^\circ$ at the $16\sigma$ level, with a negligible bias of less than $0.3\sigma$. In the absence of birefringence ($\beta=0$, $\alpha_\nu\neq 0$) the recovered mean angle is $\widehat{\beta} = 0.011^\circ\pm0.017^\circ$, fully consistent with zero within $1\sigma$,  demonstrating the robustness of the field-level inference against false detections due to instrumental systematics. 

Figures~\ref{fig:ttplot_b} and~\ref{fig:ttplot_zerobeta_nonzeroalpha_b} in Appendix~\ref{sec:ttplots} present the alternative regression analysis between the fields $Y(\hat{n})=\widetilde{B}^{\rm CMB}(\hat{n}) - \widehat{B}^{\rm CMB}(\hat{n})$ and $X(\hat{n})=G \ast \widetilde{E}^{\rm CMB}$, evaluated over the 26 sky patches defined in Figure~\ref{fig:igloo}, for the two scenarios with and without cosmic birefringence. In the case with birefringence ($\beta=0.3^\circ$, $\alpha_\nu\neq 0$; Figure~\ref{fig:ttplot_b}), this regression reveals an even stronger and consistent correlation between the two fields across all patches, compared to the previous approach. 

Using the IV-2SLS regression method as before, we estimate the birefringence angle in each patch, with results summarized in Table~\ref{tab:beta_b}. The mean estimated angle across all patches in the birefringent case is $\widehat{\beta} = 0.305^\circ \pm 0.016^\circ$, yielding an unbiased detection of $\beta=0.3^\circ$ at approximately $19\sigma$ significance. Conversely, in the non-birefringent scenario ($\beta=0$, $\alpha_\nu\neq 0$), the recovered angle is $\widehat{\beta} = -0.010^\circ \pm 0.016^\circ$, fully consistent with zero. 

The enhanced detection significance in this alternative regression stems from the higher signal-to-noise ratio in the modulated standard ILC CMB $E$-mode field ($G \ast \widetilde{E}^{\rm CMB}$) compared to the modulated standard ILC CMB $B$-mode field ($F \ast \widetilde{B}^{\rm CMB}$), leading to more precise angle estimates.

These results demonstrate the robustness and effectiveness of spatial linear regression for estimating the cosmic birefringence angle from CMB polarization data processed via standard and hybrid ILC component separation. Both regression strategies, whether based on reconstructed $E$-mode or $B$-mode fields, achieve highly significant detections of birefringence when present, and consistently return null results when absent, even in the presence of instrumental polarization angle miscalibration.

\subsection{Application to Planck Release 4 data}
\label{subsec:pr4}

In this final section, we apply the same analysis pipeline used for \lb\ simulations to the latest \pl\ Release 4 (PR4) data \citep{planck2020-LVII}, which include seven frequency channels spanning $30$ to $353$\,GHz from both the Low Frequency Instrument (LFI) and High Frequency Instrument (HFI). We first reconstruct full-sky PR4 CMB $E$- and $B$-mode maps using the standard NILC approach (Equation~\ref{eq:stdilcweights}) applied independently to the seven frequency maps for either $E$- or $B$-mode polarization, yielding the estimated fields $\widetilde{E}^{\rm CMB}$ and $\widetilde{B}^{\rm CMB}$ at $15'$ angular resolution. We then apply the Hybrid NILC method (Equations~\ref{eq:weights0} or~\ref{eq:weights}) jointly to the fourteen $E$- and $B$-mode PR4 channel maps to reconstruct the \emph{uncorrelated} CMB $B$-mode field, $\widehat{B}^{\rm CMB}$, over the full sky at the same resolution.

\begin{figure}[tbp]
\centering 
\includegraphics[width=0.8\textwidth,clip]{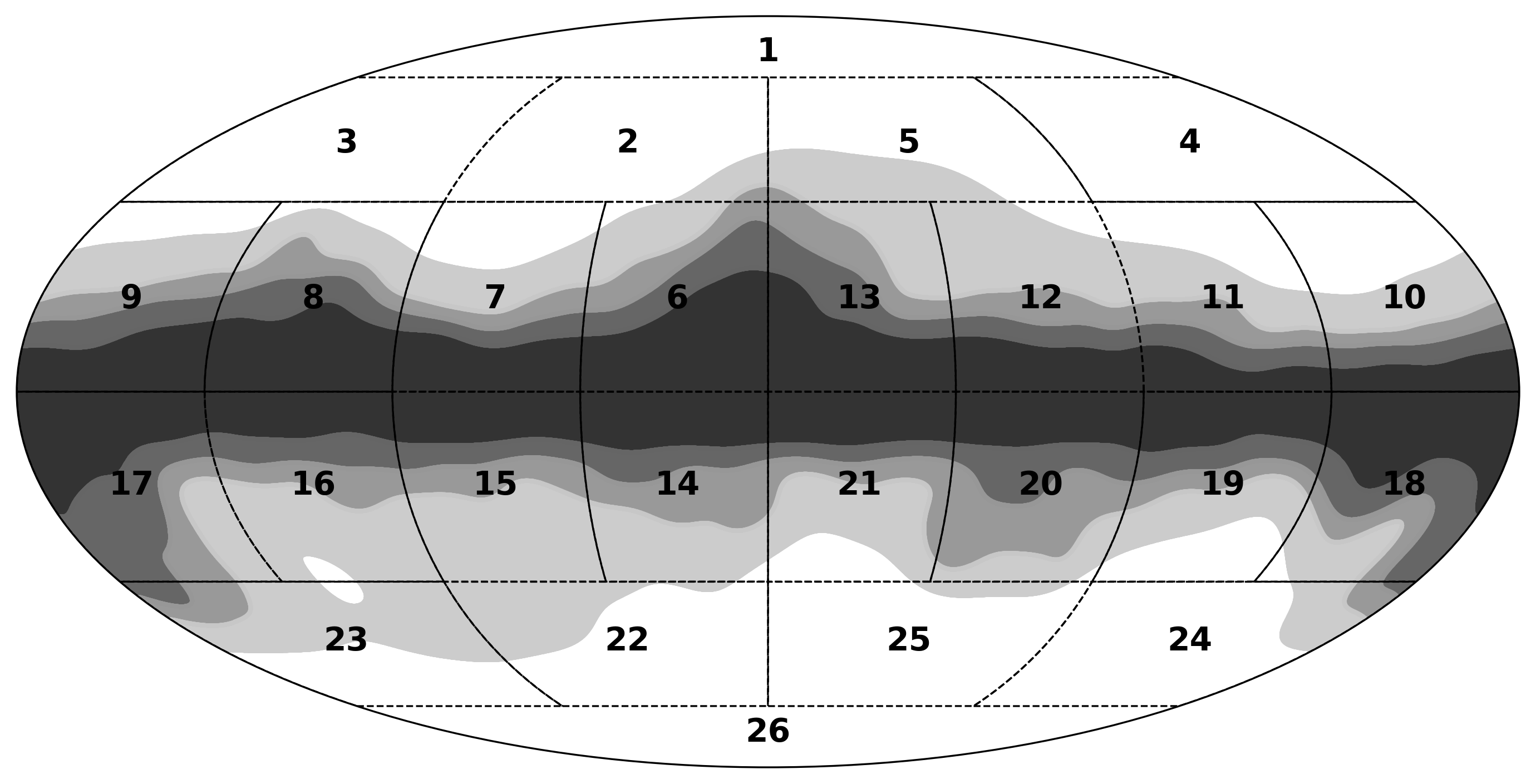}~
\hfill
\caption{\label{fig:igloopr4} Sky partitioning into 26 equal-area patches, each covering about $3.8\,\%$ of the sky, used for \pl\ PR4 data analysis. The total observed sky fractions are $f_{\rm sky}=40$, $60$, $70$, and $80\,\%$, excluding the grey areas from lightest to darkest shading, respectively.}
\end{figure}

The difference $\widetilde{B}^{\rm CMB} - \widehat{B}^{\rm CMB}$ isolates the \emph{correlated} CMB $B$-mode component, which we regress against the modulated $E$-mode field $G \ast \widetilde{E}^{\rm CMB}$, following Equations~\eqref{eq:theorem_lm}--\eqref{eq:offset}. We adopt this regression, rather than using $\widetilde{E}^{\rm CMB} - \widehat{E}^{\rm CMB}$ versus $F \ast \widetilde{B}^{\rm CMB}$, because it yields tighter constraints on the birefringence angle $\beta$, as shown in simulations (Section~\ref{subsec:beta_linreg}). The regression is carried out independently in each of the 26 sky patches defined in Figure~\ref{fig:igloopr4}, after applying a top-hat filter in harmonic space over $\ell \in [350, 500]$\footnote{The selected multipole range is narrower for \pl\ than for \lb, as \pl\ provides a lower signal-to-noise ratio on the reconstructed CMB polarization over the low-multipole range.} to enhance the signal-to-noise ratio of the reconstructed fields and their correlation.

\begin{figure}[tbp]
\centering 
\includegraphics[width=0.8\textwidth,clip]{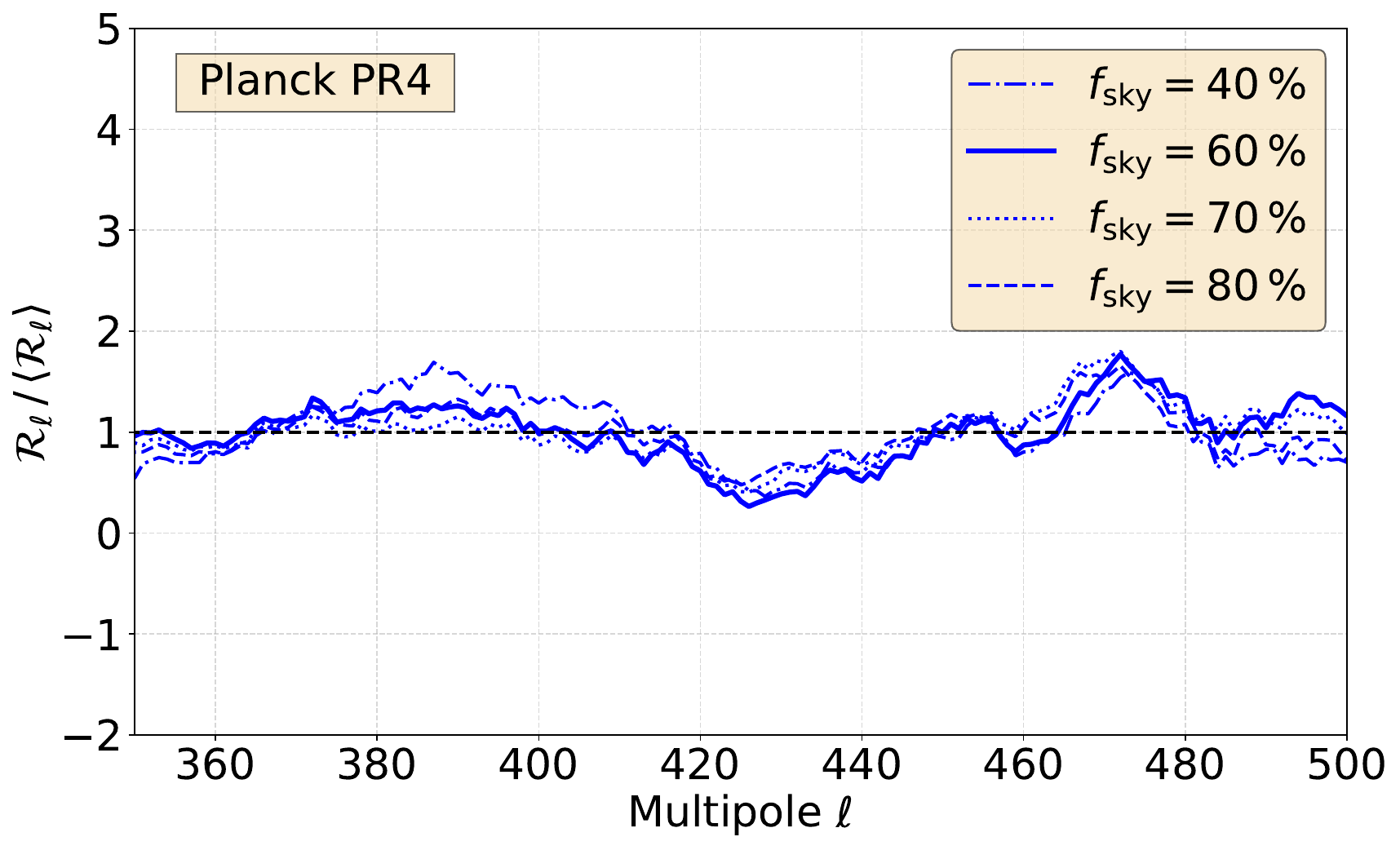}~
\hfill
\caption{\label{fig:degeneracy_breaking_pr4} 
Same as Figure~\ref{fig:degeneracy_breaking}, but for \pl\ PR4 data. The absence of large distortions in $\mathcal{R}_\ell$ relative to its mean, across both multipoles and sky regions, indicates that the effective $EB$ cross-spectrum from the \pl\ PR4 NILC CMB maps is consistent with achromatic cosmic birefringence rather than chromatic miscalibration.} 
\end{figure}

\begin{figure}[tbp]
\centering 
\includegraphics[width=0.8\textwidth,clip]{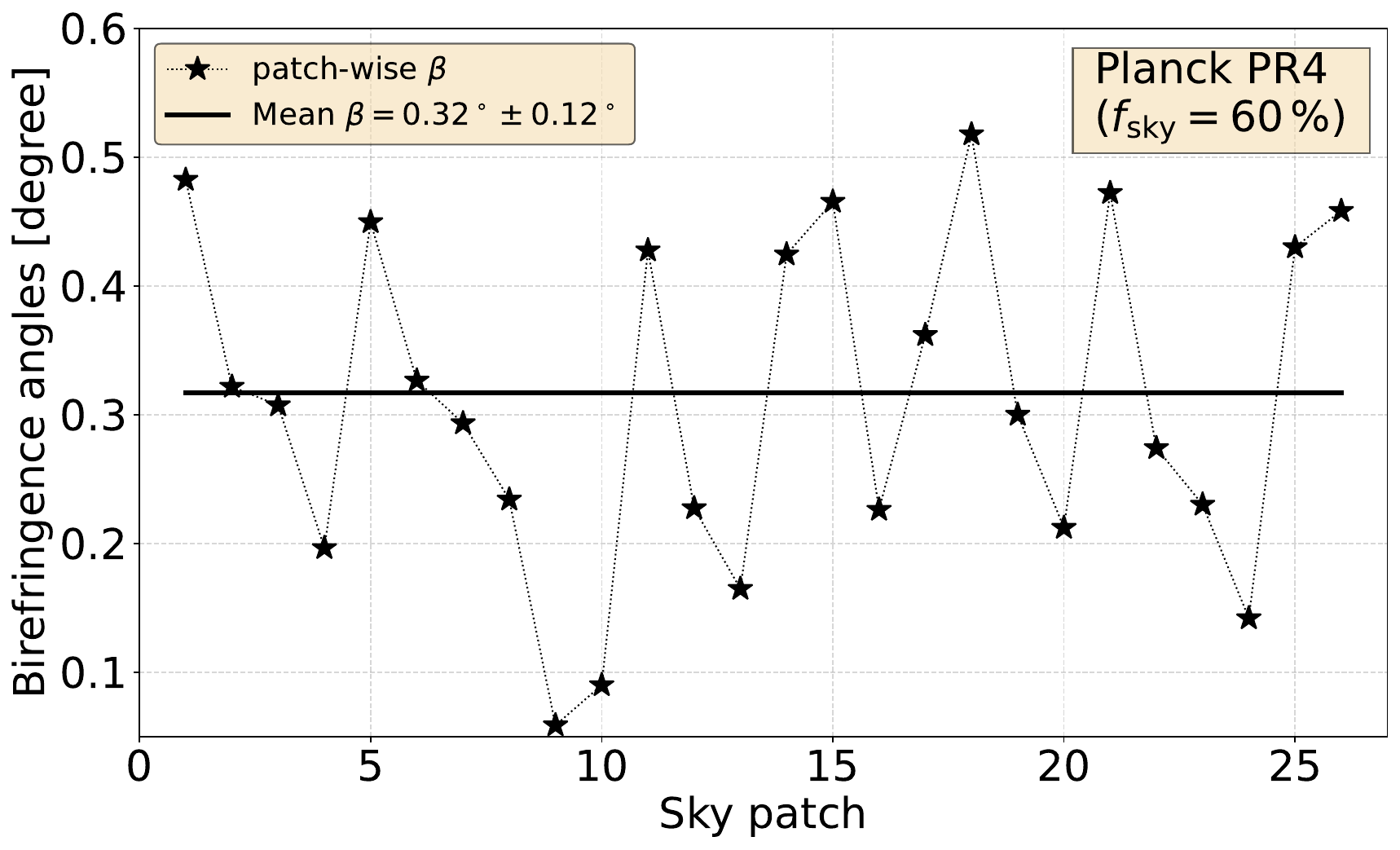}~\\
\includegraphics[width=0.8\textwidth,clip]{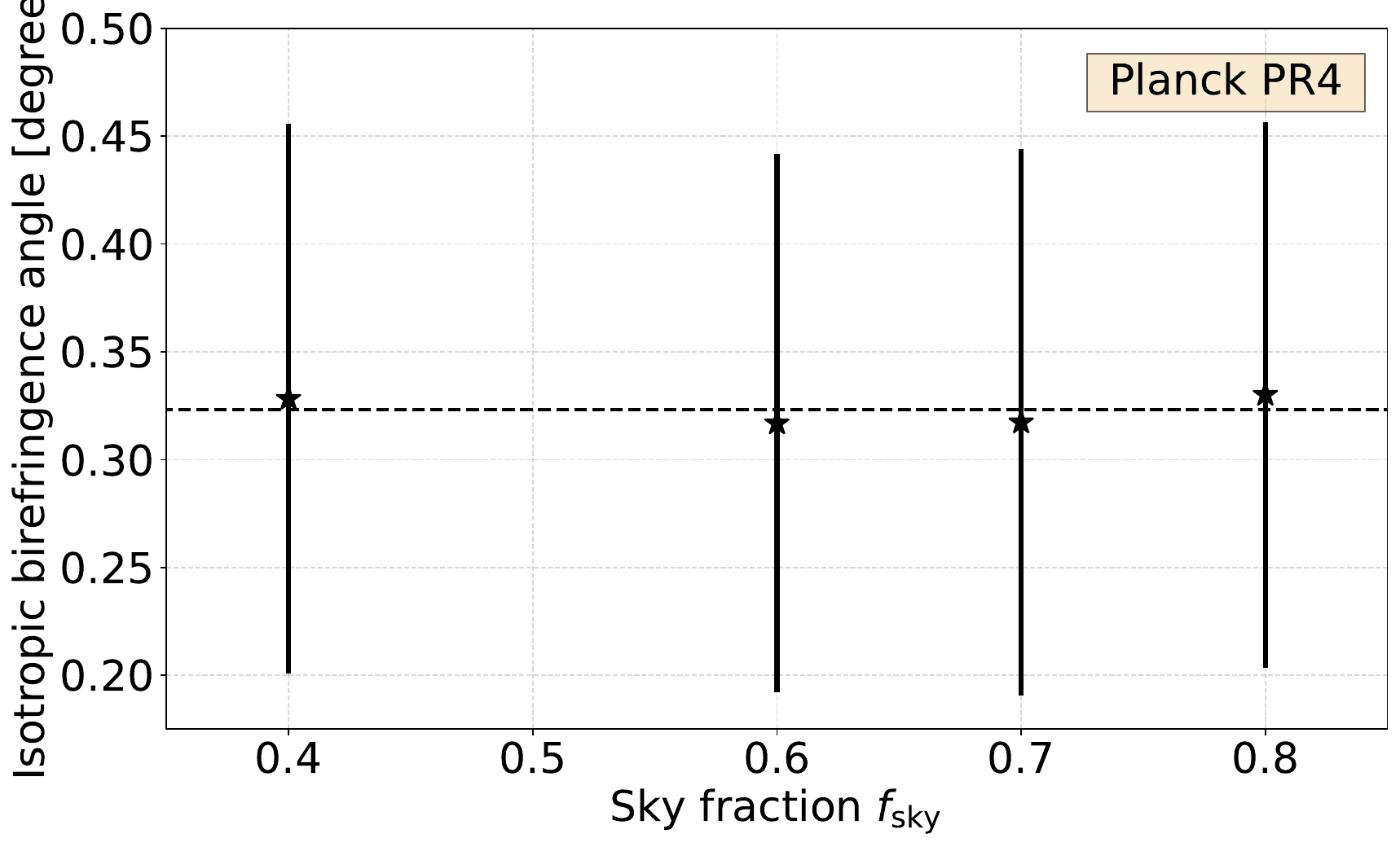}~
\hfill
\caption{\label{fig:beta_pr4} 
\emph{Top}: Estimates of the \pl\ PR4 cosmic birefringence angle in the 26 sky patches from Figure~\ref{fig:igloopr4}, with a collective area of $f_{\rm sky}=60\,\%$ of the sky. These are obtained via linear regression between the reconstructed PR4 fields $\widetilde{B}^{\rm CMB} - \widehat{B}^{\rm CMB}$ and $G \ast \widetilde{E}^{\rm CMB}$ (see Section~\ref{sec:linreg}). The area-weighted mean and standard deviation across patches yield an isotropic birefringence angle of $\widehat{\beta} = 0.32^\circ \pm 0.12^\circ$ (solid black line). \emph{Bottom}: Isotropic birefringence angle estimate from PR4 for varying sky fractions, demonstrating the stability of the result across sky coverage and supporting its robustness.} 
\end{figure}

Before presenting the linear regression results, we first apply the diagnostic analysis of Section~\ref{subsec:diagnostic}, previously performed on simulations (Figure~\ref{fig:degeneracy_breaking}), to the \pl\ PR4 data. Specifically, we compute the ratio $\mathcal{R}_\ell=C_\ell^{\widetilde{E}\widetilde{B}, {\rm CMB}}/(C_\ell^{EE, {\rm CMB}} - C_\ell^{BB, {\rm CMB}})$ (Equation~\ref{eq:magicratio}), which compares the shape of the PR4 NILC CMB $EB$ cross-spectrum to that of the difference between the $\Lambda$CDM $EE$ and $BB$ auto-spectra. Figure~\ref{fig:degeneracy_breaking_pr4} shows this ratio, normalised by its mean $\langle \mathcal{R}_\ell \rangle$ over $\ell = 350$--500, for various sky fractions.
The absence of significant distortions in $\mathcal{R}_\ell$ across both multipoles and sky coverages, especially compared to the red curves in Figure~\ref{fig:degeneracy_breaking}, suggests that the observed $EB$ correlation in the PR4 NILC CMB maps is predominantly due to achromatic cosmic birefringence, rather than chromatic instrumental miscalibration, as the ratio does not inherit the multipole or spatial dependence of the NILC weights.

The top panel of Figure~\ref{fig:beta_pr4} shows estimates of the cosmic birefringence angle in the 26 sky patches defined in Figure~\ref{fig:igloopr4}, covering $f_{\rm sky} = 60\,\%$ of the sky after masking the Galactic plane. These patchwise $\beta$ values are obtained via spatial linear regression between the PR4 fields $Y(\hat{n}) = \widetilde{B}^{\rm CMB}(\hat{n}) - \widehat{B}^{\rm CMB}(\hat{n})$ and $X(\hat{n}) = (G \ast \widetilde{E}^{\rm CMB})(\hat{n})$, as described in Section~\ref{subsec:regression}. Because the \pl\ $f_{\rm sky} = 60\,\%$ Galactic mask partially overlaps with some patches, the effective area per patch is not strictly uniform. The global isotropic birefringence angle is therefore computed as a weighted mean of the patchwise estimates, with weights given by the effective number of unmasked pixels in each patch. The associated uncertainty is given by the weighted standard deviation across the 26 patchwise estimates.\footnote{Patches 9 and 10, located near the edge of the \pl\ Galactic mask around the Galactic anti-centre, show slightly outlying $\beta$ values due to residual foreground $E/B$ leakage near the mask boundary. Since the \pl\ masks are defined from the local $Q/U$ polarization fields, while our analysis operates on the non-local $E/B$ maps, foreground contamination can delocalize and extend beyond the mask in $E/B$ space.}  This yields an isotropic birefringence angle of $\widehat{\beta} = 0.32^\circ \pm 0.12^\circ$ from PR4 data over $f_{\rm sky} = 60\,\%$ of the sky, corresponding to a $2.7\sigma$ detection. This map-based result is consistent with recent PR4-based birefringence constraints \citep{Diego2022,Eskilt2022}, which instead relied on a parametric, power-spectrum-based approach \citep{Minami2019,Minami2020}.

\begin{figure}[tbp]
\centering 
\includegraphics[width=0.8\textwidth,clip]{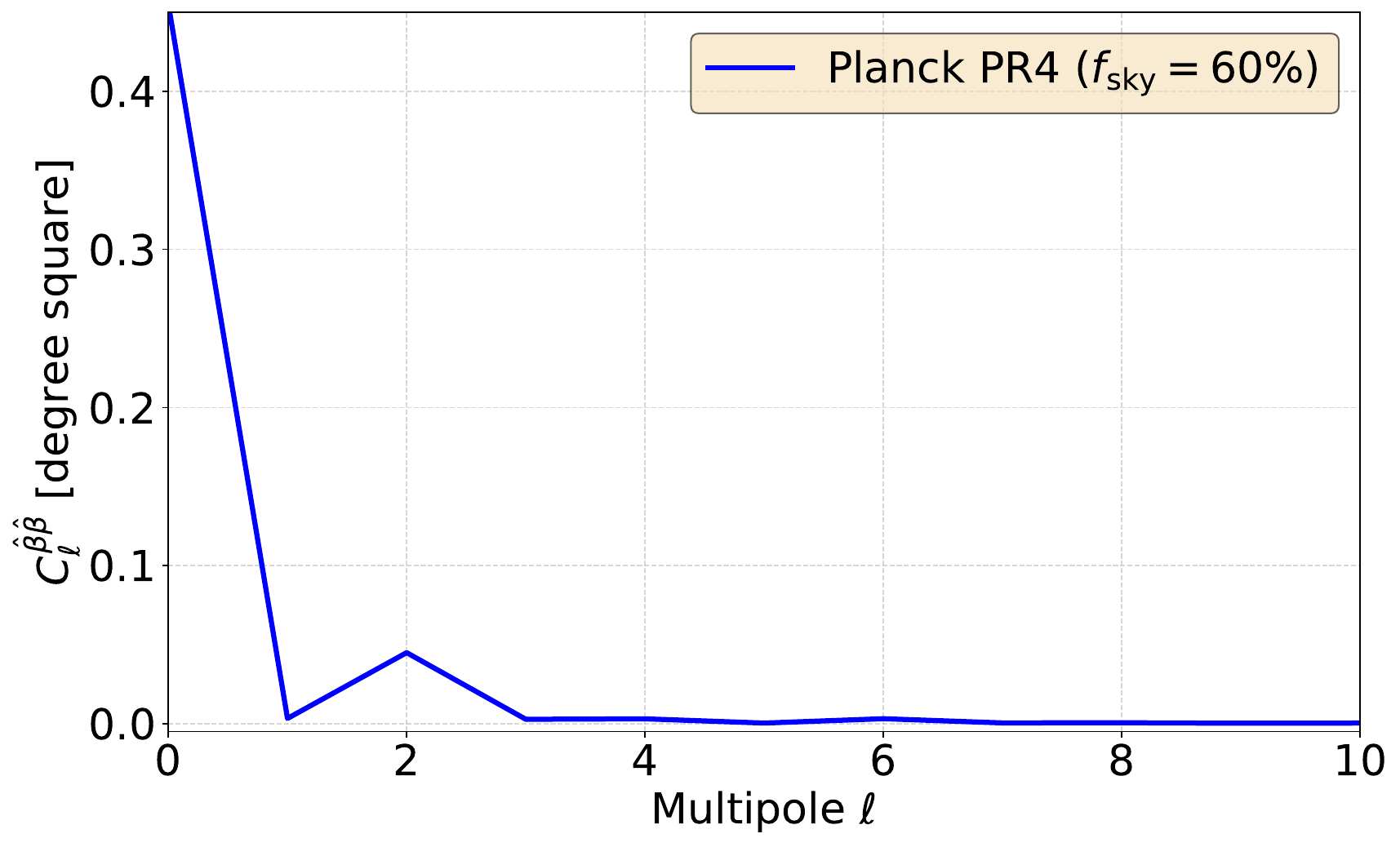}~\\
\includegraphics[width=0.8\textwidth,clip]{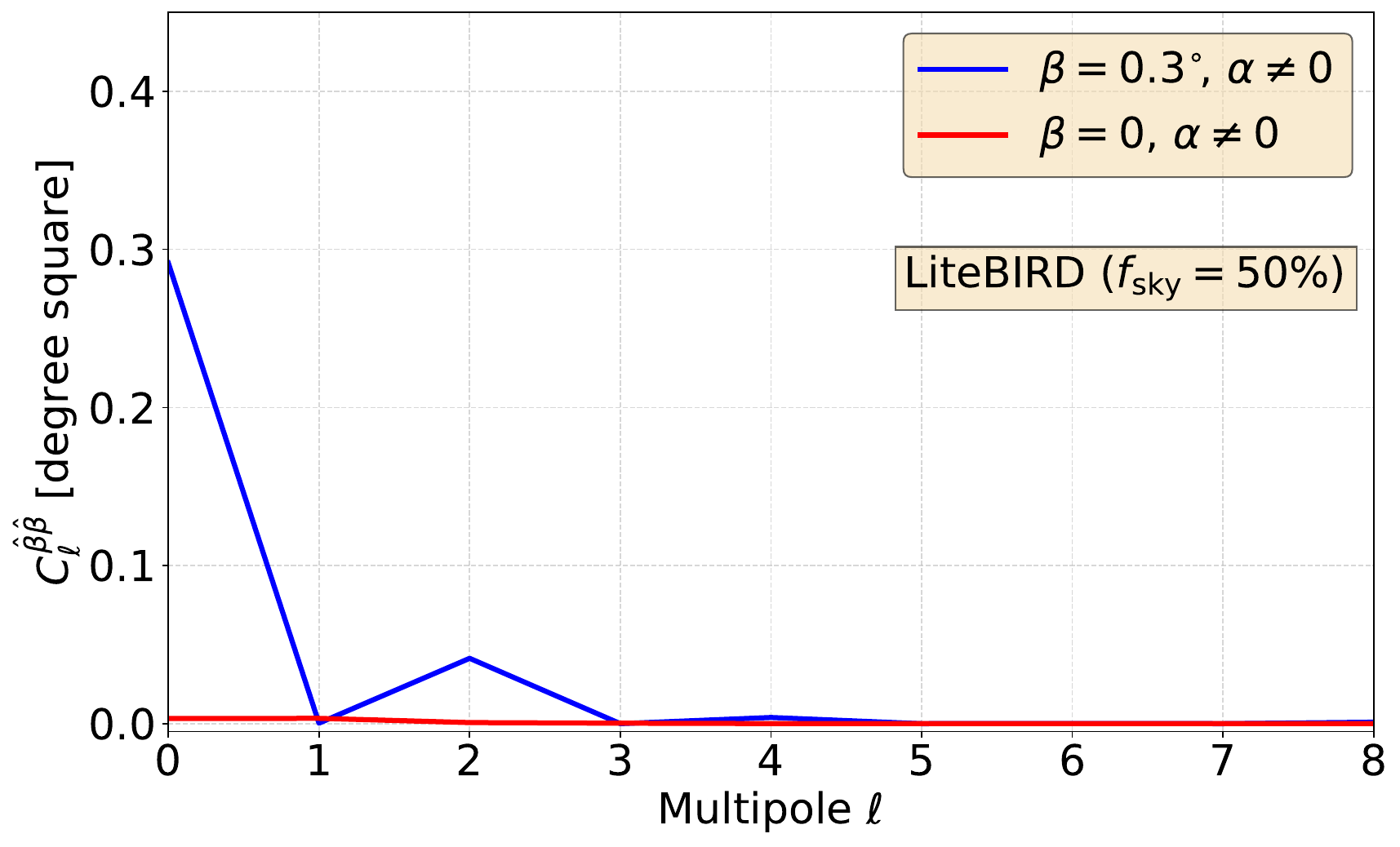}~
\hfill
\caption{\label{fig:cl_beta} 
\emph{Top}: Angular power spectrum of the estimated \pl\ PR4 map of the cosmic birefringence angle $\widehat{\beta}(\hat{n})$, derived from the spatial regression analysis over $f_{\rm sky}=60\,\%$ of the sky (figure~\ref{fig:igloopr4}). The spectrum, computed without mask deconvolution, shows a dominant monopole component, which for this sky fraction corresponds to a mean angle $\langle\widehat{\beta}(\hat{n})\rangle = 0.32^\circ$, with only minor leakage to higher multipoles caused by masking. 
\emph{Bottom}: Same spectrum for \lb\ simulations over $f_{\rm sky}=50\,\%$ of the sky (Figure~\ref{fig:igloo}). In the absence of birefringence (\emph{red}), the monopole power is negligible, while in the presence of isotropic birefringence (\emph{blue}) it is consistent with the injected value $\beta = 0.3^\circ$.}
\end{figure}

 As a complement to the top panel of Figure~\ref{fig:beta_pr4}, the top panel of Figure~\ref{fig:cl_beta} shows the angular power spectrum of the \pl\ PR4 $\widehat{\beta}$ map estimated from the spatial regression over $f_{\rm sky}=60\,\%$ of the sky. The spectrum, computed without mask deconvolution, is consistent with a monopole-dominated signal, with $C_{\ell = 0}^{\widehat{\beta}\widehat{\beta}} = 4\pi f_{\rm sky}^2\langle\widehat{\beta}(\hat{n}) \rangle^2 \simeq 0.45$ degree square, corresponding to a mean birefringence angle of $\langle\widehat{\beta}(\hat{n}) \rangle \simeq 0.32^\circ$. It shows only minor leakage to higher multipoles due to masking.\footnote{The monopole power leaks primarily into even multipoles due to the mask's approximate symmetry in latitude.} This result also supports recent evidence for the scale-independence of cosmic birefringence from \pl\ data \citep{Ballardini2025}.
For reference, the bottom panel of Figure~\ref{fig:cl_beta} shows the same spectrum for the \lb\ simulation analysis over $f_{\rm sky}=50\,\%$ of the sky (see Section~\ref{subsec:beta_linreg}), where the monopole power for this sky fraction is consistent with the injected isotropic birefringence angle, either $\beta = 0.3^\circ$ (blue) or $\beta = 0$ (red).

Finally, the bottom panel of Figure~\ref{fig:beta_pr4} shows our estimates of the isotropic birefringence angle for different sky coverages, from $f_{\rm sky} = 40$ to $80\,\%$, using the \pl\ Galactic masks\footnote{\href{http://pla.esac.esa.int/pla/aio/product-action?MAP.MAP_ID=HFI_Mask_GalPlane-apo0_2048_R2.00.fits}{http://pla.esac.esa.int/pla/aio/product-action?MAP.MAP\_ID=HFI\_Mask\_GalPlane-apo0\_2048\_R2.00.fits}} in Figure~\ref{fig:igloopr4}. The inferred birefringence angle remains stable across sky fractions, highlighting the robustness of our blind, map-level approach, which makes no assumptions about foregrounds. This observed consistency also supports a cosmological origin for the signal, as instrumental miscalibration and foreground contamination, being chromatic, would be modulated by the NILC weights and thus expected to vary with sky coverage.

\section{Conclusion}
\label{sec:conclusion}

In this work, we introduced a novel, non-parametric, and data-driven approach to constrain cosmic birefringence in map space. By jointly incorporating both $E$- and $B$-mode frequency maps into the component separation process, our Hybrid NILC method enables, for the first time, to disentangle the correlated and uncorrelated components of the CMB polarization field over the sky (Section~\ref{sec:hybridilc} and Figure~\ref{fig:ebps}). This decomposition allows for spatial linear regression between the correlated component of the CMB $E$-mode (or $B$-mode) field and the full component of the CMB $B$-mode (or $E$-mode) field modulated by an analytic kernel proportional to $\beta$ (Section~\ref{sec:linreg}), enabling a direct, field-level estimation of the birefringence angle (Section~\ref{subsec:beta_linreg}).

We also demonstrated the discriminating power of this NILC-based framework in distinguishing cosmic birefringence from polarization angle miscalibration (Sections~\ref{subsec:discrimination} and~\ref{subsec:diagnostic}; Figures~\ref{fig:degeneracy_breaking} and~\ref{fig:degeneracy_breaking_pr4}). While frequency-dependent effects such as instrumental miscalibration and foreground contamination are modulated by the NILC weights, thus inheriting their multipole and spatial variations, cosmic birefringence, being achromatic, remains unaltered and uniform across both the sky and multipoles in the reconstructed CMB map.

We forecast that the forthcoming CMB space mission \lb, a JAXA-led mission with contributions from Europe, Canada, and the United States, will achieve a $16$--$19\sigma$ detection of a birefringence angle $\beta = 0.3^\circ$ with this method (Section~\ref{subsec:beta_linreg}; Figures~\ref{fig:ttplot} and~\ref{fig:ttplot_b}; Tables~\ref{tab:beta} and~\ref{tab:beta_b}), assuming the mission's target specifications for polarization angle calibration are met \cite{Vielva2022}. This makes the Hybrid NILC and regression pipeline a powerful, blind, map-based technique that is both complementary to and competitive with existing parametric and power-spectrum-based methods in the literature \citep{LiteBIRD-CB2025}.

When applied to \pl\ PR4 data, our method yields an isotropic birefringence angle of $\beta = 0.32^\circ \pm 0.12^\circ$ with $2.7\sigma$ significance (Section~\ref{subsec:pr4}, Figure~\ref{fig:beta_pr4}). This result is consistent with the tightest existing constraints \citep{Diego2022,Eskilt2022}, while also exhibiting robustness to variations in sky cuts, thus giving further support to its cosmological origin.

Looking ahead, the Hybrid NILC method could benefit from additional refinements, such as moment deprojection \citep{Remazeilles2021,Carones2024} and multi-clustering techniques \citep{Carones2023}, to further enhance sensitivity to cosmic birefringence. It would also be interesting to apply this method to other CMB datasets, such as those from the Atacama Cosmology Telescope \citep{ACT2025} and the Simons Observatory \citep{SO-LAT2025}, or their combination with \pl\ data, so as to check the consistency of the birefringence angle across independent instruments with different calibration strategies and uncorrelated systematics.

Finally, the Hybrid ILC method developed in this work also marks the first demonstration of the benefits of incorporating both $E$- and $B$-mode channels into the ILC framework to disentangle correlated and uncorrelated components. This opens up several promising avenues for future component separation strategies, including:
(i) Combining both temperature and polarization channels in the Hybrid ILC to separate spectrally degenerate signals, such as CMB and kinetic Sunyaev-Zeldovich (kSZ) effect, or thermal dust and cosmic infrared background (CIB), in cases where only one of the two components is polarized. In such cases, the intrinsic $TE$ correlation of the polarized component (e.g., CMB or thermal dust) can help the ILC reduce sample variance contamination from that component in the reconstruction of the unpolarized signal (e.g., kSZ or CIB), thereby partially breaking the spectral degeneracy. 
(ii) Applying the Hybrid ILC jointly to temperature and polarization channels to enhance the cleaning of Galactic foreground contamination in unpolarized extragalactic signals, such as the thermal SZ effect and 21-cm line intensity maps. These applications are currently under active investigation and will be presented in future work.

\acknowledgments
We thank Patricia Diego-Palazuelos for useful discussions during the early stages of this work. This work was supported by the Spanish Ministry of Science and Innovation (MCIN) and the Agencia Estatal de Investigación (AEI) through the project grants PID2022-139223OB-C21 and PID2022-140670NA-I00. Some of the results in this paper were derived using the \texttt{healpy} \citep{healpy} and \texttt{HEALPix} \citep{gorski2005} packages.

\bibliographystyle{JHEP}
\bibliography{main,Planck_bib}

\providecommand{\href}[2]{#2}\begingroup\raggedright\begin{thebibliography}{10}

\bibitem{Carroll1990}
S.M.~{Carroll}, G.B.~{Field} and R.~{Jackiw}, \emph{{Limits on a Lorentz- and
  parity-violating modification of electrodynamics}},
  \href{https://doi.org/10.1103/PhysRevD.41.1231}{\emph{\prd} {\bfseries 41}
  (1990) 1231}.

\bibitem{Komatsu2022}
E.~{Komatsu}, \emph{{New physics from the polarized light of the cosmic
  microwave background}},
  \href{https://doi.org/10.1038/s42254-022-00452-4}{\emph{Nature Reviews
  Physics} {\bfseries 4} (2022) 452}
  [\href{https://arxiv.org/abs/2202.13919}{{\ttfamily 2202.13919}}].

\bibitem{Marsh2016}
D.J.E.~{Marsh}, \emph{{Axion cosmology}},
  \href{https://doi.org/10.1016/j.physrep.2016.06.005}{\emph{\physrep}
  {\bfseries 643} (2016) 1}.

\bibitem{Nakagawa2021}
S.~{Nakagawa}, F.~{Takahashi} and M.~{Yamada}, \emph{{Cosmic Birefringence
  Triggered by Dark Matter Domination}},
  \href{https://doi.org/10.1103/PhysRevLett.127.181103}{\emph{\prl} {\bfseries
  127} (2021) 181103} [\href{https://arxiv.org/abs/2103.08153}{{\ttfamily
  2103.08153}}].

\bibitem{Obata2022}
I.~{Obata}, \emph{{Implications of the cosmic birefringence measurement for the
  axion dark matter search}},
  \href{https://doi.org/10.1088/1475-7516/2022/09/062}{\emph{\jcap} {\bfseries
  2022} (2022) 062} [\href{https://arxiv.org/abs/2108.02150}{{\ttfamily
  2108.02150}}].

\bibitem{Carroll1998}
S.M.~{Carroll}, \emph{{Quintessence and the Rest of the World: Suppressing
  Long-Range Interactions}},
  \href{https://doi.org/10.1103/PhysRevLett.81.3067}{\emph{\prl} {\bfseries 81}
  (1998) 3067} [\href{https://arxiv.org/abs/astro-ph/9806099}{{\ttfamily
  astro-ph/9806099}}].

\bibitem{Liu2006}
G.-C.~{Liu}, S.~{Lee} and K.-W.~{Ng}, \emph{{Effect on Cosmic Microwave
  Background Polarization of Coupling of Quintessence to Pseudoscalar Formed
  from the Electromagnetic Field and its Dual}},
  \href{https://doi.org/10.1103/PhysRevLett.97.161303}{\emph{\prl} {\bfseries
  97} (2006) 161303} [\href{https://arxiv.org/abs/astro-ph/0606248}{{\ttfamily
  astro-ph/0606248}}].

\bibitem{Li2008}
M.~{Li} and X.~{Zhang}, \emph{{Cosmological CPT violating effect on CMB
  polarization}}, \href{https://doi.org/10.1103/PhysRevD.78.103516}{\emph{\prd}
  {\bfseries 78} (2008) 103516}
  [\href{https://arxiv.org/abs/0810.0403}{{\ttfamily 0810.0403}}].

\bibitem{Harari1992}
D.~{Harari} and P.~{Sikivie}, \emph{{Effects of a Nambu-Goldstone boson on the
  polarization of radio galaxies and the cosmic microwave background}},
  \href{https://doi.org/10.1016/0370-2693(92)91363-E}{\emph{Physics Letters B}
  {\bfseries 289} (1992) 67}.

\bibitem{Finelli2009}
F.~{Finelli} and M.~{Galaverni}, \emph{{Rotation of linear polarization plane
  and circular polarization from cosmological pseudoscalar fields}},
  \href{https://doi.org/10.1103/PhysRevD.79.063002}{\emph{\prd} {\bfseries 79}
  (2009) 063002} [\href{https://arxiv.org/abs/0802.4210}{{\ttfamily
  0802.4210}}].

\bibitem{Pospelov2009}
M.~{Pospelov}, A.~{Ritz} and C.~{Skordis}, \emph{{Pseudoscalar Perturbations
  and Polarization of the Cosmic Microwave Background}},
  \href{https://doi.org/10.1103/PhysRevLett.103.051302}{\emph{\prl} {\bfseries
  103} (2009) 051302} [\href{https://arxiv.org/abs/0808.0673}{{\ttfamily
  0808.0673}}].

\bibitem{Lue1999}
A.~{Lue}, L.~{Wang} and M.~{Kamionkowski}, \emph{{Cosmological Signature of New
  Parity-Violating Interactions}},
  \href{https://doi.org/10.1103/PhysRevLett.83.1506}{\emph{\prl} {\bfseries 83}
  (1999) 1506} [\href{https://arxiv.org/abs/astro-ph/9812088}{{\ttfamily
  astro-ph/9812088}}].

\bibitem{Feng2005}
B.~{Feng}, H.~{Li}, M.~{Li} and X.~{Zhang}, \emph{{Gravitational leptogenesis
  and its signatures in CMB [rapid communication]}},
  \href{https://doi.org/10.1016/j.physletb.2005.06.009}{\emph{Physics Letters
  B} {\bfseries 620} (2005) 27}
  [\href{https://arxiv.org/abs/hep-ph/0406269}{{\ttfamily hep-ph/0406269}}].

\bibitem{Hu2003}
W.~{Hu}, M.M.~{Hedman} and M.~{Zaldarriaga}, \emph{{Benchmark parameters for
  CMB polarization experiments}},
  \href{https://doi.org/10.1103/PhysRevD.67.043004}{\emph{\prd} {\bfseries 67}
  (2003) 043004} [\href{https://arxiv.org/abs/astro-ph/0210096}{{\ttfamily
  astro-ph/0210096}}].

\bibitem{Shimon2008}
M.~{Shimon}, B.~{Keating}, N.~{Ponthieu} and E.~{Hivon}, \emph{{CMB
  polarization systematics due to beam asymmetry: Impact on inflationary
  science}}, \href{https://doi.org/10.1103/PhysRevD.77.083003}{\emph{\prd}
  {\bfseries 77} (2008) 083003}
  [\href{https://arxiv.org/abs/0709.1513}{{\ttfamily 0709.1513}}].

\bibitem{Miller2009}
N.J.~{Miller}, M.~{Shimon} and B.G.~{Keating}, \emph{{CMB polarization
  systematics due to beam asymmetry: Impact on cosmological birefringence}},
  \href{https://doi.org/10.1103/PhysRevD.79.103002}{\emph{\prd} {\bfseries 79}
  (2009) 103002} [\href{https://arxiv.org/abs/0903.1116}{{\ttfamily
  0903.1116}}].

\bibitem{Yadav2009}
A.P.S.~{Yadav}, R.~{Biswas}, M.~{Su} and M.~{Zaldarriaga}, \emph{{Constraining
  a spatially dependent rotation of the cosmic microwave background
  polarization}}, \href{https://doi.org/10.1103/PhysRevD.79.123009}{\emph{\prd}
  {\bfseries 79} (2009) 123009}
  [\href{https://arxiv.org/abs/0902.4466}{{\ttfamily 0902.4466}}].

\bibitem{Keating2013}
B.G.~{Keating}, M.~{Shimon} and A.P.S.~{Yadav}, \emph{{Self-calibration of
  Cosmic Microwave Background Polarization Experiments}},
  \href{https://doi.org/10.1088/2041-8205/762/2/L23}{\emph{\apjl} {\bfseries
  762} (2013) L23} [\href{https://arxiv.org/abs/1211.5734}{{\ttfamily
  1211.5734}}].

\bibitem{Minami2019}
Y.~{Minami}, H.~{Ochi}, K.~{Ichiki}, N.~{Katayama}, E.~{Komatsu} and
  T.~{Matsumura}, \emph{{Simultaneous determination of the cosmic birefringence
  and miscalibrated polarization angles from CMB experiments}},
  \href{https://doi.org/10.1093/ptep/ptz079}{\emph{Progress of Theoretical and
  Experimental Physics} {\bfseries 2019} (2019) 083E02}.

\bibitem{Minami2020}
Y.~{Minami} and E.~{Komatsu}, \emph{{Simultaneous determination of the cosmic
  birefringence and miscalibrated polarization angles II: Including
  cross-frequency spectra}},
  \href{https://doi.org/10.1093/ptep/ptaa130}{\emph{Progress of Theoretical and
  Experimental Physics} {\bfseries 2020} (2020) 103E02}
  [\href{https://arxiv.org/abs/2006.15982}{{\ttfamily 2006.15982}}].

\bibitem{Minami2020b}
Y.~{Minami} and E.~{Komatsu}, \emph{{New Extraction of the Cosmic Birefringence
  from the Planck 2018 Polarization Data}},
  \href{https://doi.org/10.1103/PhysRevLett.125.221301}{\emph{\prl} {\bfseries
  125} (2020) 221301} [\href{https://arxiv.org/abs/2011.11254}{{\ttfamily
  2011.11254}}].

\bibitem{Diego2022}
P.~{Diego-Palazuelos}, J.R.~{Eskilt}, Y.~{Minami}, M.~{Tristram},
  R.M.~{Sullivan}, A.J.~{Banday} et~al., \emph{{Cosmic Birefringence from the
  Planck Data Release 4}},
  \href{https://doi.org/10.1103/PhysRevLett.128.091302}{\emph{\prl} {\bfseries
  128} (2022) 091302} [\href{https://arxiv.org/abs/2201.07682}{{\ttfamily
  2201.07682}}].

\bibitem{Eskilt2022}
J.R.~{Eskilt} and E.~{Komatsu}, \emph{{Improved constraints on cosmic
  birefringence from the WMAP and Planck cosmic microwave background
  polarization data}},
  \href{https://doi.org/10.1103/PhysRevD.106.063503}{\emph{\prd} {\bfseries
  106} (2022) 063503} [\href{https://arxiv.org/abs/2205.13962}{{\ttfamily
  2205.13962}}].

\bibitem{Eskilt2022b}
J.R.~{Eskilt}, \emph{{Frequency-dependent constraints on cosmic birefringence
  from the LFI and HFI Planck Data Release 4}},
  \href{https://doi.org/10.1051/0004-6361/202243269}{\emph{\aap} {\bfseries
  662} (2022) A10} [\href{https://arxiv.org/abs/2201.13347}{{\ttfamily
  2201.13347}}].

\bibitem{LiteBIRD-CB2025}
E.~{de la Hoz}, P.~{Diego-Palazuelos}, J.~{Errard}, A.~{Gruppuso}, B.~{Jost},
  R.M.~{Sullivan} et~al., \emph{{LiteBIRD Science Goals and Forecasts:
  constraining isotropic cosmic birefringence}},
  \href{https://doi.org/10.48550/arXiv.2503.22322}{\emph{arXiv e-prints} (2025)
  arXiv:2503.22322} [\href{https://arxiv.org/abs/2503.22322}{{\ttfamily
  2503.22322}}].

\bibitem{SO-LAT2025}
{The Simons Observatory Collaboration}, M.~{Abitbol}, I.~{Abril-Cabezas},
  S.~{Adachi}, P.~{Ade}, A.E.~{Adler} et~al., \emph{{The Simons Observatory:
  Science Goals and Forecasts for the Enhanced Large Aperture Telescope}},
  \href{https://doi.org/10.48550/arXiv.2503.00636}{\emph{arXiv e-prints} (2025)
  arXiv:2503.00636} [\href{https://arxiv.org/abs/2503.00636}{{\ttfamily
  2503.00636}}].

\bibitem{CMB-S4_2016}
K.N.~{Abazajian}, P.~{Adshead}, Z.~{Ahmed}, S.W.~{Allen}, D.~{Alonso},
  K.S.~{Arnold} et~al., \emph{{CMB-S4 Science Book, First Edition}},
  \href{https://doi.org/10.48550/arXiv.1610.02743}{\emph{arXiv e-prints} (2016)
  arXiv:1610.02743} [\href{https://arxiv.org/abs/1610.02743}{{\ttfamily
  1610.02743}}].

\bibitem{Clark2021}
S.E.~{Clark}, C.-G.~{Kim}, J.C.~{Hill} and B.S.~{Hensley}, \emph{{The Origin of
  Parity Violation in Polarized Dust Emission and Implications for Cosmic
  Birefringence}}, \href{https://doi.org/10.3847/1538-4357/ac0e35}{\emph{\apj}
  {\bfseries 919} (2021) 53}
  [\href{https://arxiv.org/abs/2105.00120}{{\ttfamily 2105.00120}}].

\bibitem{Diego2023}
P.~{Diego-Palazuelos}, E.~{Mart{\'\i}nez-Gonz{\'a}lez}, P.~{Vielva},
  R.B.~{Barreiro}, M.~{Tristram}, E.~{de la Hoz} et~al., \emph{{Robustness of
  cosmic birefringence measurement against Galactic foreground emission and
  instrumental systematics}},
  \href{https://doi.org/10.1088/1475-7516/2023/01/044}{\emph{\jcap} {\bfseries
  2023} (2023) 044} [\href{https://arxiv.org/abs/2210.07655}{{\ttfamily
  2210.07655}}].

\bibitem{Vacher2023}
L.~{Vacher}, J.~{Aumont}, F.~{Boulanger}, L.~{Montier}, V.~{Guillet},
  A.~{Ritacco} et~al., \emph{{Frequency dependence of the thermal dust E/B
  ratio and EB correlation: Insights from the spin-moment expansion}},
  \href{https://doi.org/10.1051/0004-6361/202245292}{\emph{\aap} {\bfseries
  672} (2023) A146} [\href{https://arxiv.org/abs/2210.14768}{{\ttfamily
  2210.14768}}].

\bibitem{Hervias2025}
C.~{Herv{\'\i}as-Caimapo}, A.J.~{Cukierman}, P.~{Diego-Palazuelos},
  K.M.~{Huffenberger} and S.E.~{Clark}, \emph{{Modeling parity-violating
  spectra in Galactic dust polarization with filaments and its applications to
  cosmic birefringence searches}},
  \href{https://doi.org/10.1103/PhysRevD.111.083532}{\emph{\prd} {\bfseries
  111} (2025) 083532} [\href{https://arxiv.org/abs/2408.06214}{{\ttfamily
  2408.06214}}].

\bibitem{delaHoz2022}
E.~{de la Hoz}, P.~{Diego-Palazuelos}, E.~{Mart{\'\i}nez-Gonz{\'a}lez},
  P.~{Vielva}, R.B.~{Barreiro} and J.D.~{Bilbao-Ahedo}, \emph{{Determination of
  polarization angles in CMB experiments and application to CMB component
  separation analyses}},
  \href{https://doi.org/10.1088/1475-7516/2022/03/032}{\emph{\jcap} {\bfseries
  2022} (2022) 032} [\href{https://arxiv.org/abs/2110.14328}{{\ttfamily
  2110.14328}}].

\bibitem{Jost2023}
B.~{Jost}, J.~{Errard} and R.~{Stompor}, \emph{{Characterizing cosmic
  birefringence in the presence of Galactic foregrounds and instrumental
  systematic effects}},
  \href{https://doi.org/10.1103/PhysRevD.108.082005}{\emph{\prd} {\bfseries
  108} (2023) 082005} [\href{https://arxiv.org/abs/2212.08007}{{\ttfamily
  2212.08007}}].

\bibitem{Komatsu2011}
E.~{Komatsu}, K.M.~{Smith}, J.~{Dunkley}, C.L.~{Bennett}, B.~{Gold},
  G.~{Hinshaw} et~al., \emph{{Seven-year Wilkinson Microwave Anisotropy Probe
  (WMAP) Observations: Cosmological Interpretation}},
  \href{https://doi.org/10.1088/0067-0049/192/2/18}{\emph{\apjs} {\bfseries
  192} (2011) 18} [\href{https://arxiv.org/abs/1001.4538}{{\ttfamily
  1001.4538}}].

\bibitem{Contreras2017}
D.~{Contreras}, P.~{Boubel} and D.~{Scott}, \emph{{Constraints on
  direction-dependent cosmic birefringence from Planck polarization data}},
  \href{https://doi.org/10.1088/1475-7516/2017/12/046}{\emph{\jcap} {\bfseries
  2017} (2017) 046} [\href{https://arxiv.org/abs/1705.06387}{{\ttfamily
  1705.06387}}].

\bibitem{Jow2019}
D.L.~{Jow}, D.~{Contreras}, D.~{Scott} and E.F.~{Bunn}, \emph{{Taller in the
  saddle: constraining CMB physics using saddle points}},
  \href{https://doi.org/10.1088/1475-7516/2019/03/031}{\emph{\jcap} {\bfseries
  2019} (2019) 031} [\href{https://arxiv.org/abs/1811.05629}{{\ttfamily
  1811.05629}}].

\bibitem{planck2014-a23}
{\sorthelp{Planck Collaboration IntZX}}{Planck Collaboration Int. XLIX},
  \emph{{\textit{Planck} intermediate results. XLIX. Parity-violation
  constraints from polarization data}},
  \href{https://doi.org/10.1051/0004-6361/201629018}{\emph{\aap} {\bfseries
  596} (2016) A110} [\href{https://arxiv.org/abs/1605.08633}{{\ttfamily
  1605.08633}}].

\bibitem{Sullivan2025}
R.M.~{Sullivan}, A.~{Abghari}, P.~{Diego-Palazuelos}, L.T.~{Hergt} and
  D.~{Scott}, \emph{{Planck PR4 (NPIPE) map-space cosmic birefringence}},
  \href{https://doi.org/10.1088/1475-7516/2025/06/025}{\emph{\jcap} {\bfseries
  2025} (2025) 025} [\href{https://arxiv.org/abs/2502.07654}{{\ttfamily
  2502.07654}}].

\bibitem{LiteBIRD2023}
{LiteBIRD Collaboration}, E.~{Allys}, K.~{Arnold}, J.~{Aumont}, R.~{Aurlien},
  S.~{Azzoni} et~al., \emph{{Probing cosmic inflation with the LiteBIRD cosmic
  microwave background polarization survey}},
  \href{https://doi.org/10.1093/ptep/ptac150}{\emph{Progress of Theoretical and
  Experimental Physics} {\bfseries 2023} (2023) 042F01}
  [\href{https://arxiv.org/abs/2202.02773}{{\ttfamily 2202.02773}}].

\bibitem{planck2016-l11A}
{\sorthelp{Planck Collaboration 2018K}}{Planck Collaboration XI},
  \emph{{\textit{Planck} 2018 results. XI. Polarized dust foregrounds}},
  \href{https://doi.org/10.1051/0004-6361/201832618}{\emph{\aap} {\bfseries
  641} (2020) A11} [\href{https://arxiv.org/abs/1801.04945}{{\ttfamily
  1801.04945}}].

\bibitem{Martire2022}
F.A.~{Martire}, R.B.~{Barreiro} and E.~{Mart{\'\i}nez-Gonz{\'a}lez},
  \emph{{Characterization of the polarized synchrotron emission from Planck and
  WMAP data}},
  \href{https://doi.org/10.1088/1475-7516/2022/04/003}{\emph{\jcap} {\bfseries
  2022} (2022) 003} [\href{https://arxiv.org/abs/2110.12803}{{\ttfamily
  2110.12803}}].

\bibitem{WMAP2003}
C.L.~{Bennett}, R.S.~{Hill}, G.~{Hinshaw}, M.R.~{Nolta}, N.~{Odegard},
  L.~{Page} et~al., \emph{{First-Year Wilkinson Microwave Anisotropy Probe
  (WMAP) Observations: Foreground Emission}},
  \href{https://doi.org/10.1086/377252}{\emph{\apjs} {\bfseries 148} (2003) 97}
  [\href{https://arxiv.org/abs/astro-ph/0302208}{{\ttfamily
  astro-ph/0302208}}].

\bibitem{Tegmark2003}
M.~{Tegmark}, A.~{de Oliveira-Costa} and A.J.~{Hamilton}, \emph{{High
  resolution foreground cleaned CMB map from WMAP}},
  \href{https://doi.org/10.1103/PhysRevD.68.123523}{\emph{\prd} {\bfseries 68}
  (2003) 123523} [\href{https://arxiv.org/abs/astro-ph/0302496}{{\ttfamily
  astro-ph/0302496}}].

\bibitem{Eriksen2004}
H.K.~{Eriksen}, A.J.~{Banday}, K.M.~{G{\'o}rski} and P.B.~{Lilje}, \emph{{On
  Foreground Removal from the Wilkinson Microwave Anisotropy Probe Data by an
  Internal Linear Combination Method: Limitations and Implications}},
  \href{https://doi.org/10.1086/422807}{\emph{\apj} {\bfseries 612} (2004) 633}
  [\href{https://arxiv.org/abs/astro-ph/0403098}{{\ttfamily
  astro-ph/0403098}}].

\bibitem{Delabrouille2009}
J.~{Delabrouille}, J.~{Cardoso}, M.~{Le Jeune}, M.~{Betoule}, G.~{Fay} and
  F.~{Guilloux}, \emph{{A full sky, low foreground, high resolution CMB map
  from WMAP}}, \href{https://doi.org/10.1051/0004-6361:200810514}{\emph{\aap}
  {\bfseries 493} (2009) 835}
  [\href{https://arxiv.org/abs/0807.0773}{{\ttfamily 0807.0773}}].

\bibitem{Isobe1990}
T.~{Isobe}, E.D.~{Feigelson}, M.G.~{Akritas} and G.J.~{Babu}, \emph{{Linear
  Regression in Astronomy. I.}},
  \href{https://doi.org/10.1086/169390}{\emph{\apj} {\bfseries 364} (1990)
  104}.

\bibitem{IV-2SLS}
R.L.~Basmann, \emph{A generalized classical method of linear estimation of
  coefficients in a structural equation},
  \href{https://doi.org/10.2307/1907743}{\emph{Econometrica} {\bfseries 25}
  (1957) 77}.

\bibitem{gorski2005}
K.M.~{G{\'o}rski}, E.~{Hivon}, A.J.~{Banday}, B.D.~{Wandelt}, F.K.~{Hansen},
  M.~{Reinecke} et~al., \emph{{HEALPix: A Framework for High-Resolution
  Discretization and Fast Analysis of Data Distributed on the Sphere}},
  \href{https://doi.org/10.1086/427976}{\emph{\apj} {\bfseries 622} (2005) 759}
  [\href{https://arxiv.org/abs/astro-ph/0409513}{{\ttfamily
  astro-ph/0409513}}].

\bibitem{Vielva2022}
P.~{Vielva}, E.~{Mart{\'\i}nez-Gonz{\'a}lez}, F.J.~{Casas}, T.~{Matsumura},
  S.~{Henrot-Versill{\'e}}, E.~{Komatsu} et~al., \emph{{Polarization angle
  requirements for CMB B-mode experiments. Application to the LiteBIRD
  satellite}},
  \href{https://doi.org/10.1088/1475-7516/2022/04/029}{\emph{\jcap} {\bfseries
  2022} (2022) 029} [\href{https://arxiv.org/abs/2202.01324}{{\ttfamily
  2202.01324}}].

\bibitem{pysm}
B.~{Thorne}, J.~{Dunkley}, D.~{Alonso} and S.~{N{\ae}ss}, \emph{{The Python Sky
  Model: software for simulating the Galactic microwave sky}},
  \href{https://doi.org/10.1093/mnras/stx949}{\emph{\mnras} {\bfseries 469}
  (2017) 2821} [\href{https://arxiv.org/abs/1608.02841}{{\ttfamily
  1608.02841}}].

\bibitem{pysm3}
A.~{Zonca}, B.~{Thorne}, N.~{Krachmalnicoff} and J.~{Borrill}, \emph{{The
  Python Sky Model 3 software}},
  \href{https://doi.org/10.21105/joss.03783}{\emph{The Journal of Open Source
  Software} {\bfseries 6} (2021) 3783}
  [\href{https://arxiv.org/abs/2108.01444}{{\ttfamily 2108.01444}}].

\bibitem{Panex2025}
{The Pan-Experiment Galactic Science Group}, {:}, J.~{Borrill}, S.E.~{Clark},
  J.~{Delabrouille}, A.V.~{Frolov} et~al., \emph{{Full-sky Models of Galactic
  Microwave Emission and Polarization at Sub-arcminute Scales for the Python
  Sky Model}}, \href{https://doi.org/10.48550/arXiv.2502.20452}{\emph{arXiv
  e-prints} (2025) arXiv:2502.20452}
  [\href{https://arxiv.org/abs/2502.20452}{{\ttfamily 2502.20452}}].

\bibitem{Narcowich2006}
F.J.~Narcowich, P.~Petrushev and J.D.~Ward, \emph{Localized tight frames on
  spheres}, \href{https://doi.org/10.1137/040614359}{\emph{SIAM Journal on
  Mathematical Analysis} {\bfseries 38} (2006) 574}
  [\href{https://arxiv.org/abs/https://doi.org/10.1137/040614359}{{\ttfamily
  https://doi.org/10.1137/040614359}}].

\bibitem{Marinucci2008}
D.~{Marinucci}, D.~{Pietrobon}, A.~{Balbi}, P.~{Baldi}, P.~{Cabella},
  G.~{Kerkyacharian} et~al., \emph{{Spherical needlets for cosmic microwave
  background data analysis}},
  \href{https://doi.org/10.1111/j.1365-2966.2007.12550.x}{\emph{\mnras}
  {\bfseries 383} (2008) 539}
  [\href{https://arxiv.org/abs/0707.0844}{{\ttfamily 0707.0844}}].

\bibitem{Remazeilles2024}
M.~{Remazeilles}, M.~{Douspis}, J.A.~{Rubi{\~n}o-Mart{\'\i}n}, A.J.~{Banday},
  J.~{Chluba}, P.~{de Bernardis} et~al., \emph{{LiteBIRD science goals and
  forecasts. Mapping the hot gas in the Universe}},
  \href{https://doi.org/10.1088/1475-7516/2024/12/026}{\emph{\jcap} {\bfseries
  2024} (2024) 026} [\href{https://arxiv.org/abs/2407.17555}{{\ttfamily
  2407.17555}}].

\bibitem{planck2020-LVII}
{\sorthelp{Planck Collaboration IntZZG}}{Planck Collaboration Int. LVII},
  \emph{{\textit{Planck} intermediate results. LVII. NPIPE: Joint \Planck\ LFI
  and HFI data processing}},
  \href{https://doi.org/10.1051/0004-6361/202038073}{\emph{\aap} {\bfseries
  643} (2020) 42} [\href{https://arxiv.org/abs/2007.04997}{{\ttfamily
  2007.04997}}].

\bibitem{Ballardini2025}
M.~{Ballardini}, A.~{Gruppuso}, S.~{Paradiso}, S.S.~{Sirletti} and P.~{Natoli},
  \emph{{Planck constraints on the scale dependence of isotropic cosmic
  birefringence}},
  \href{https://doi.org/10.1088/1475-7516/2025/09/075}{\emph{\jcap} {\bfseries
  2025} (2025) 075} [\href{https://arxiv.org/abs/2507.16714}{{\ttfamily
  2507.16714}}].

\bibitem{Remazeilles2021}
M.~{Remazeilles}, A.~{Rotti} and J.~{Chluba}, \emph{{Peeling off foregrounds
  with the constrained moment ILC method to unveil primordial CMB B modes}},
  \href{https://doi.org/10.1093/mnras/stab648}{\emph{\mnras} {\bfseries 503}
  (2021) 2478} [\href{https://arxiv.org/abs/2006.08628}{{\ttfamily
  2006.08628}}].

\bibitem{Carones2024}
A.~{Carones} and M.~{Remazeilles}, \emph{{Optimization of foreground moment
  deprojection for semi-blind CMB polarization reconstruction}},
  \href{https://doi.org/10.1088/1475-7516/2024/06/018}{\emph{\jcap} {\bfseries
  2024} (2024) 018} [\href{https://arxiv.org/abs/2402.17579}{{\ttfamily
  2402.17579}}].

\bibitem{Carones2023}
A.~{Carones}, M.~{Migliaccio}, G.~{Puglisi}, C.~{Baccigalupi}, D.~{Marinucci},
  N.~{Vittorio} et~al., \emph{{Multiclustering needlet ILC for CMB B-mode
  component separation}},
  \href{https://doi.org/10.1093/mnras/stad2423}{\emph{\mnras} {\bfseries 525}
  (2023) 3117} [\href{https://arxiv.org/abs/2212.04456}{{\ttfamily
  2212.04456}}].

\bibitem{ACT2025}
T.~{Louis}, A.~{La Posta}, Z.~{Atkins}, H.T.~{Jense}, I.~{Abril-Cabezas},
  G.E.~{Addison} et~al., \emph{{The Atacama Cosmology Telescope: DR6 Power
  Spectra, Likelihoods and $\varLambda$CDM Parameters}},
  \href{https://doi.org/10.48550/arXiv.2503.14452}{\emph{arXiv e-prints} (2025)
  arXiv:2503.14452} [\href{https://arxiv.org/abs/2503.14452}{{\ttfamily
  2503.14452}}].

\bibitem{healpy}
A.~{Zonca}, L.~{Singer}, D.~{Lenz}, M.~{Reinecke}, C.~{Rosset}, E.~{Hivon}
  et~al., \emph{{healpy: equal area pixelization and spherical harmonics
  transforms for data on the sphere in Python}},
  \href{https://doi.org/10.21105/joss.01298}{\emph{The Journal of Open Source
  Software} {\bfseries 4} (2019) 1298}.

\end{thebibliography}\endgroup


\appendix

\section{Notation summary}\label{sec:notation}

For reference, we provide in Table~\ref{tab:notation} a summary of the main symbols and conventions used throughout the paper. Subscripted quantities indexed by $\nu$ (e.g., $E_\nu(\hat{n})$) denote field values at a specific frequency channel $\nu$ and sky position $\hat{n}$, while bold symbols (e.g., $\bdw_{E}$ and $\mathbf{C}^{EE}$) represent vectors and matrices across all frequency channels. A prime (e.g., $E^{\,\prime\,\rm CMB}_\nu$) indicates a rotated polarization field, incorporating both cosmic birefringence and instrumental miscalibration. Tildes (e.g., $\widetilde{E}^{\rm CMB}$) denote standard ILC reconstructions, and hats (e.g., $\widehat{E}^{\rm CMB}$) refer to Hybrid ILC estimates that isolate uncorrelated CMB components. These conventions are used consistently in both pixel-space and harmonic-space formulations.

\begin{table}[t]
\centering
    \resizebox{\columnwidth}{!}{
\renewcommand{\arraystretch}{1.3}
\begin{tabular}{ll}
\hline
\textbf{Symbol} & \textbf{Description} \\
\hline
$E_\nu(\hat{n})$, $B_\nu(\hat{n})$ & Observed $E$- and $B$-mode channel maps at frequency $\nu$ (Eq.~\ref{eq:rotation})\\
$E^{\rm CMB}(\hat{n})$, $B^{\rm CMB}(\hat{n})$ & Unrotated (primordial) CMB $E$- and $B$-mode fields (Eq.~\ref{eq:rotation})\\
$E^{\,\prime\,\rm CMB}_\nu(\hat{n})$, $B^{\,\prime\,\rm CMB}_\nu(\hat{n})$ & Rotated CMB polarization components at frequency $\nu$ (Eqs.~\ref{eq:effcmb}--\ref{eq:effcmbb})\\
$E_\nu^{\rm FG+N}(\hat{n})$, $B_\nu^{\rm FG+N}(\hat{n})$ & Foreground + noise contamination at frequency $\nu$  (Eqs.~\ref{eq:rotation2}--\ref{eq:rotation2b})\\
$a_\nu$ & Spectral energy distribution (SED) of the CMB at frequency $\nu$ (Eq.~\ref{eq:rotation})\\
$\alpha_\nu$ & Instrumental polarization angle miscalibration at frequency $\nu$ (Eq.~\ref{eq:rotation})\\
$\overline{\alpha}^{\,(E)}_\ell$, $\overline{\alpha}^{\,(B)}_\ell$ & Projected miscalibration angles from ILC weights (Eq.~\ref{eq:projected_alpha})\\
$\beta$ & Cosmic birefringence angle (achromatic) (Eq.~\ref{eq:rotation})\\
$R(\theta)$ & Polarization rotation matrix by angle $\theta$ (Eq.~\ref{eq:rotation_matrix})\\
\hline
$\widetilde{E}^{\rm CMB}(\hat{n})$, $\widetilde{B}^{\rm CMB}(\hat{n})$ & Standard ILC estimates of CMB $E$- and $B$-modes  (Eqs.~\ref{eq:example_ilce}--\ref{eq:example_ilcb})\\
$\widehat{E}^{\rm CMB}(\hat{n})$, $\widehat{B}^{\rm CMB}(\hat{n})$ & Hybrid ILC estimates of uncorrelated CMB $E$- and $B$-modes  (Eq.~\ref{eq:ehat})\\
$\widetilde{\bdw}_{E}$, $\widetilde{\bdw}_{B}$ & Standard ILC weights across frequencies for $E$-/$B$-mode channels (Eq.~\ref{eq:stdilcweights})\\
$\bdw_{E}$, $\bdw_{B}$ & Hybrid ILC weights across frequencies for $E$-/$B$-mode channels (Eq.~\ref{eq:weights0}, Eq.~\ref{eq:weights})\\
$\mathbf{C}^{EE}$, $\mathbf{C}^{BB}$, $\mathbf{C}^{EB}$ & Covariance matrices across frequencies for observed $E$/$B$ channel maps (Eq.~\ref{eq:covmatrix}) \\
\hline
$F_\ell$, $G_\ell$ & Multipole filters modulating standard ILC $B$- and $E$-modes in regression (Eq.~\ref{eq:modulation})\\
$X(\hat{n})$, $Y(\hat{n})$ & Regression variables in map-space estimation of $\beta$ (Eqs.~\ref{eq:y}--\ref{eq:x})\\
$\varepsilon(\hat{n})$, $e(\hat{n})$ & Residual error terms in field-level regression (Eq.~\ref{eq:error}, Eq.~\ref{eq:e})\\
$\mathcal{R}_\ell$ & Diagnostic ratio for distinguishing $\beta$ from miscalibration (Eq.~\ref{eq:magicratio})\\
\hline
\end{tabular}}
\caption{Summary of symbols used throughout the paper. Bold symbols denote vectors and matrices over frequency channels. Hat and tilde quantities refer to Hybrid ILC and standard ILC reconstructions, respectively.}
\label{tab:notation}
\end{table}

\section{Correlation plots of reconstructed CMB fields for spatial regression}\label{sec:ttplots}

\begin{figure}[tbp]
\centering 
\includegraphics[width=0.25\textwidth,clip]{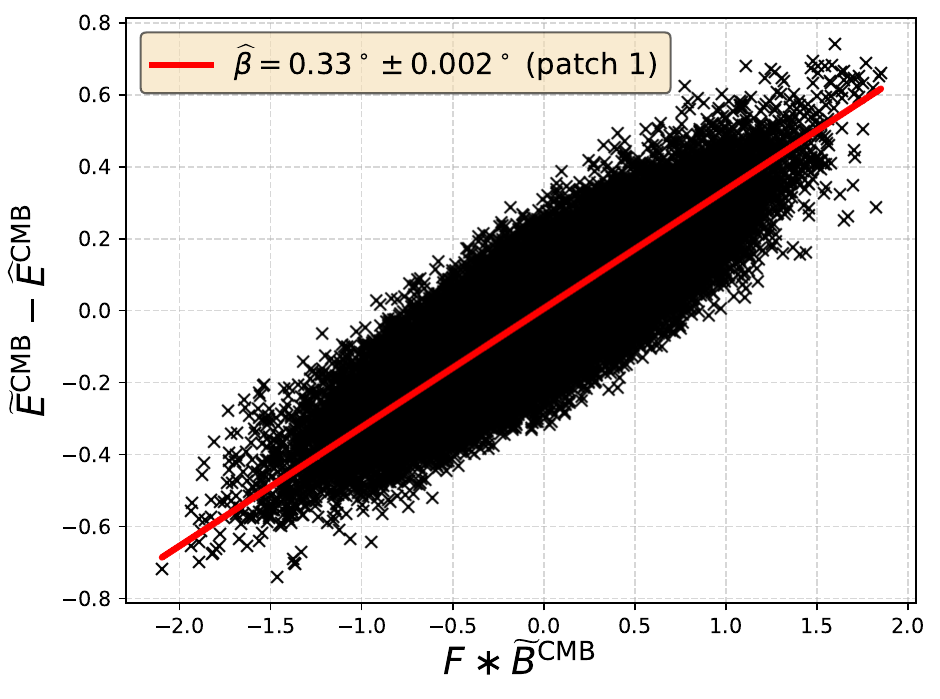}~
\includegraphics[width=0.25\textwidth,clip]{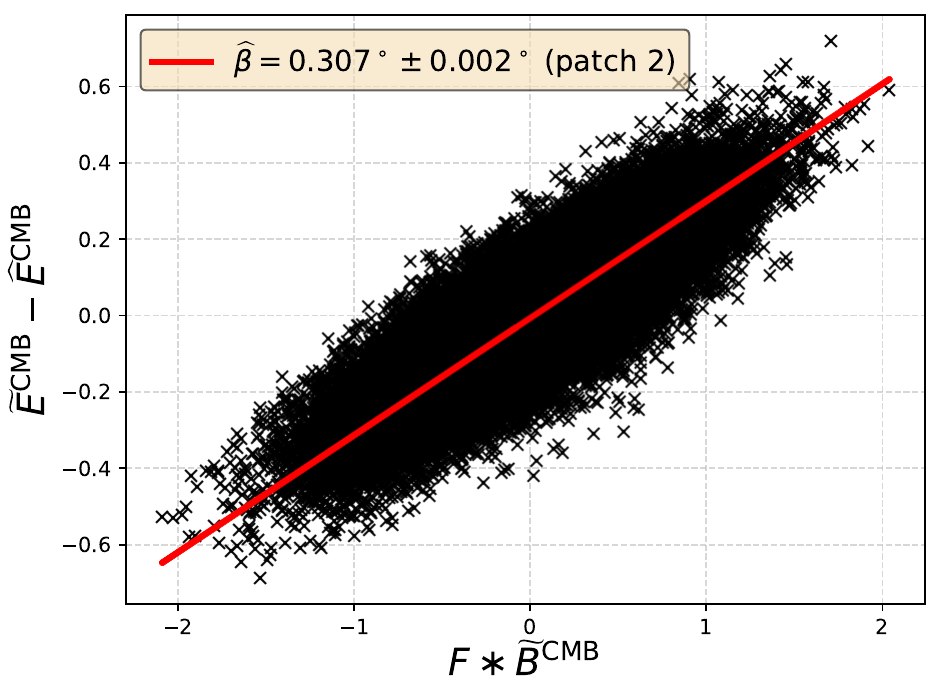}~
\includegraphics[width=0.25\textwidth,clip]{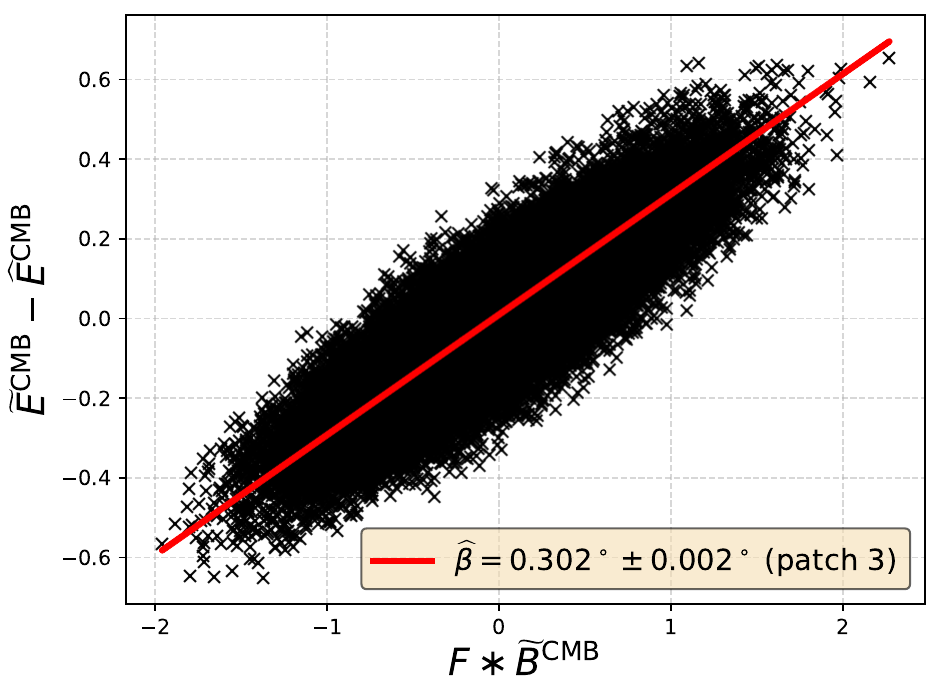}~
\includegraphics[width=0.25\textwidth,clip]{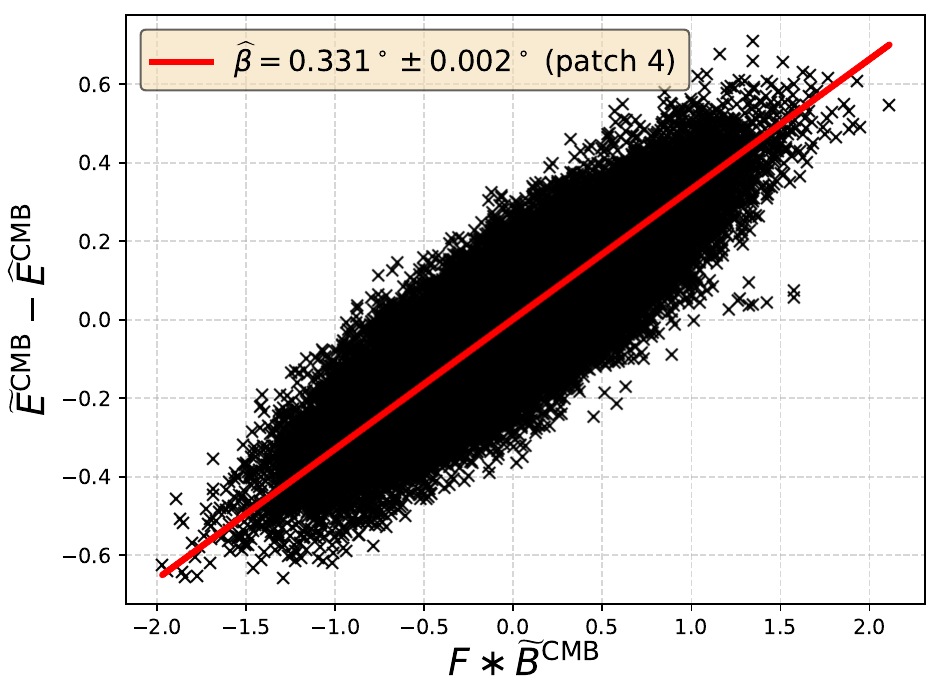}\\
\includegraphics[width=0.25\textwidth,clip]{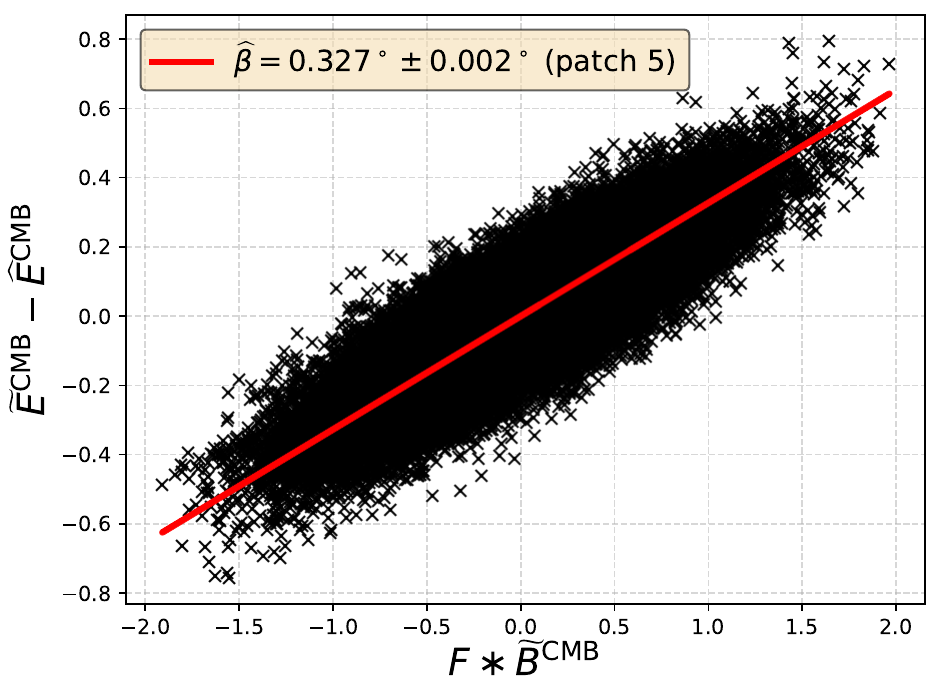}~
\includegraphics[width=0.25\textwidth,clip]{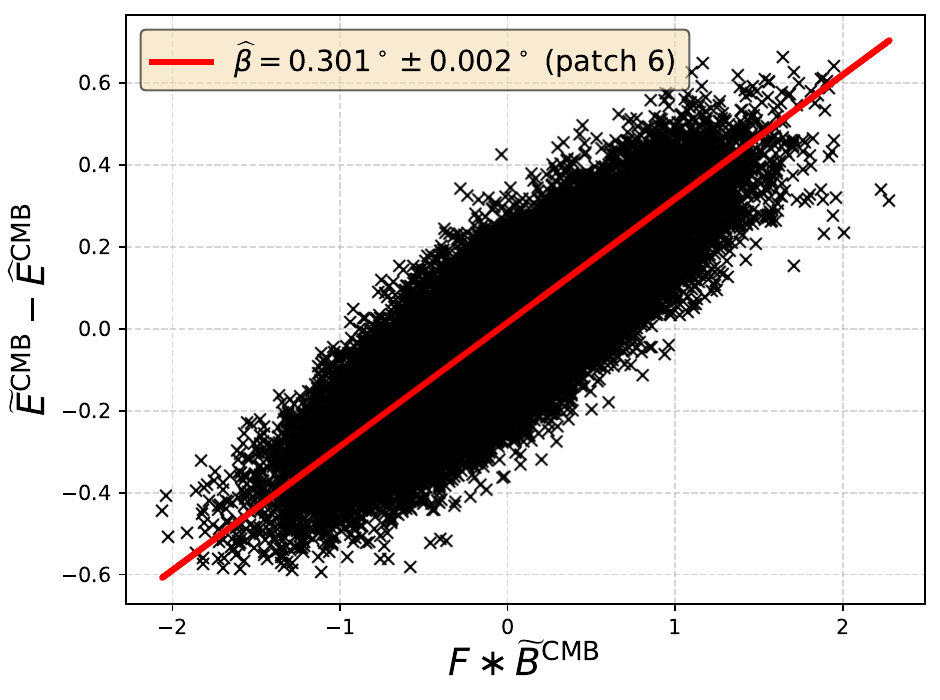}~
\includegraphics[width=0.25\textwidth,clip]{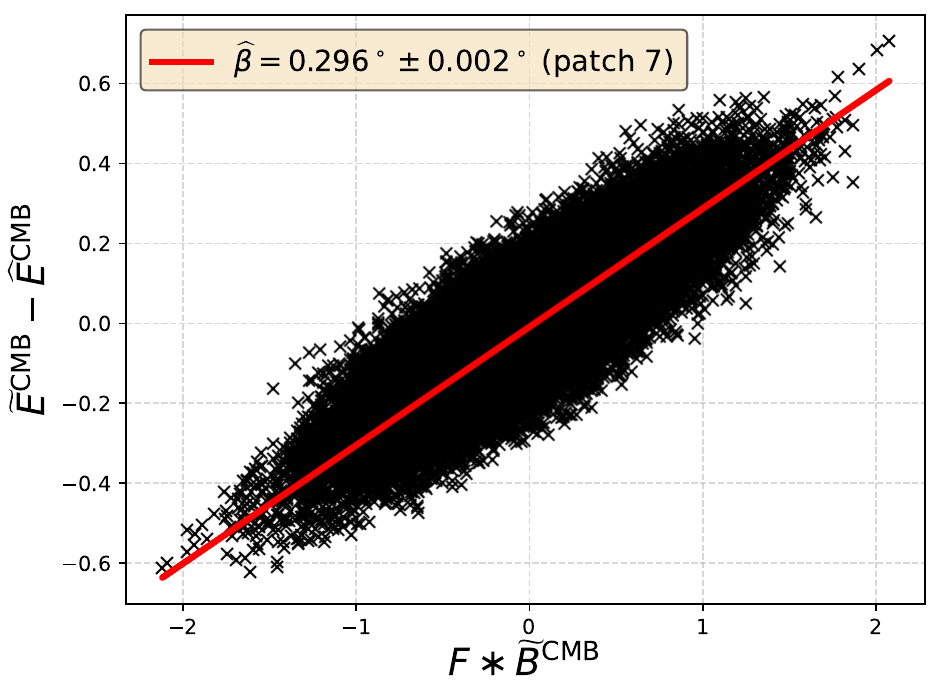}~
\includegraphics[width=0.25\textwidth,clip]{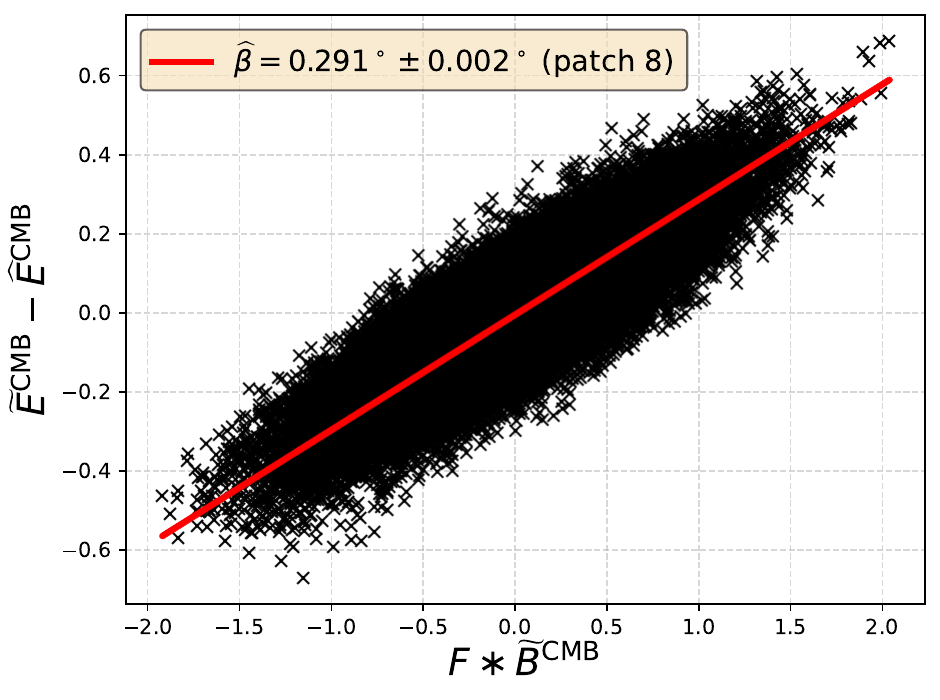}\\
\includegraphics[width=0.25\textwidth,clip]{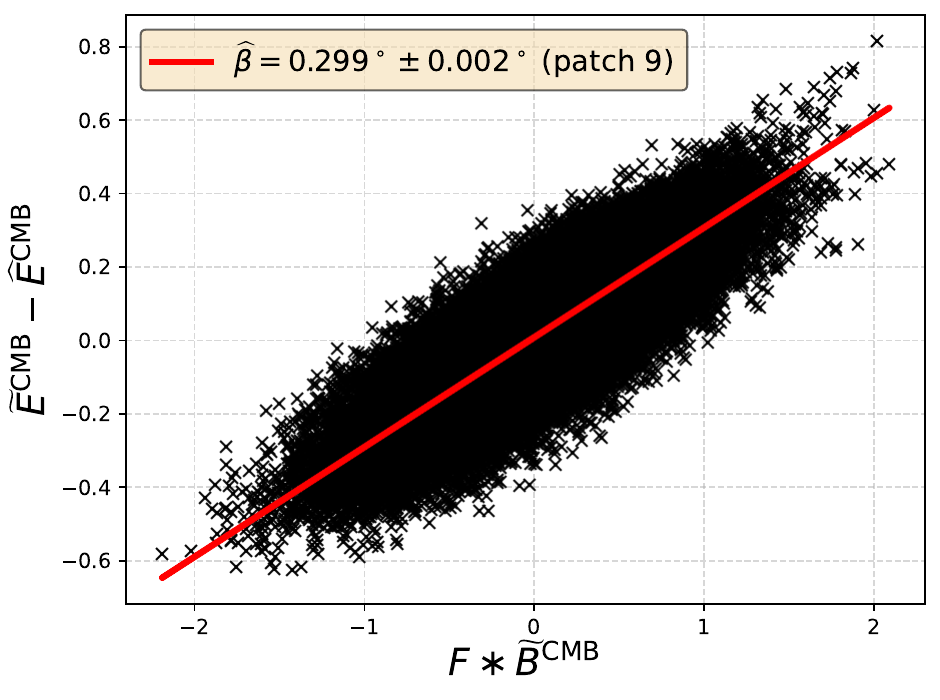}~
\includegraphics[width=0.25\textwidth,clip]{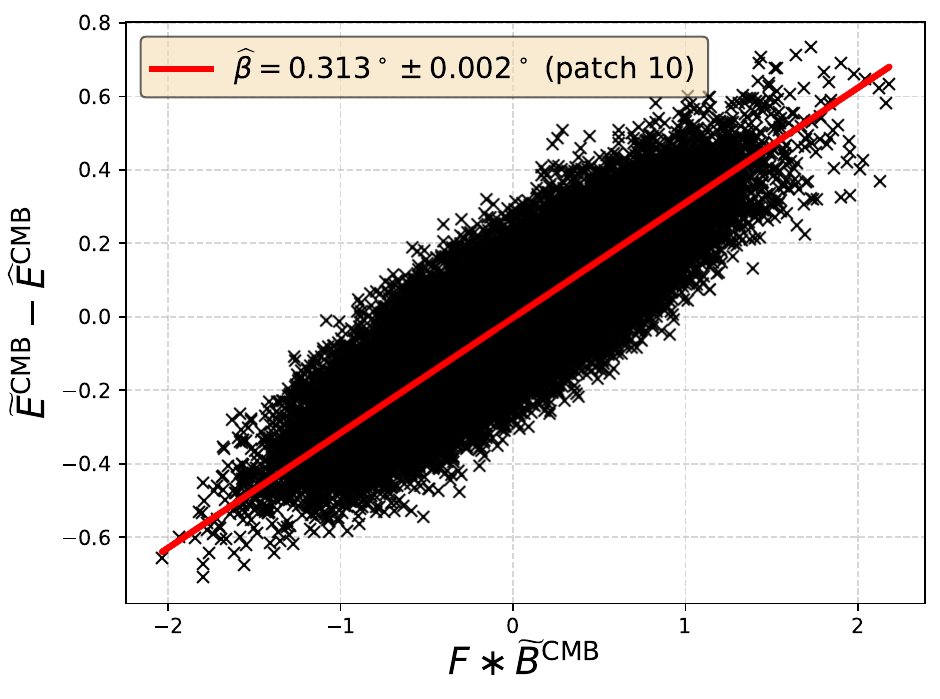}~
\includegraphics[width=0.25\textwidth,clip]{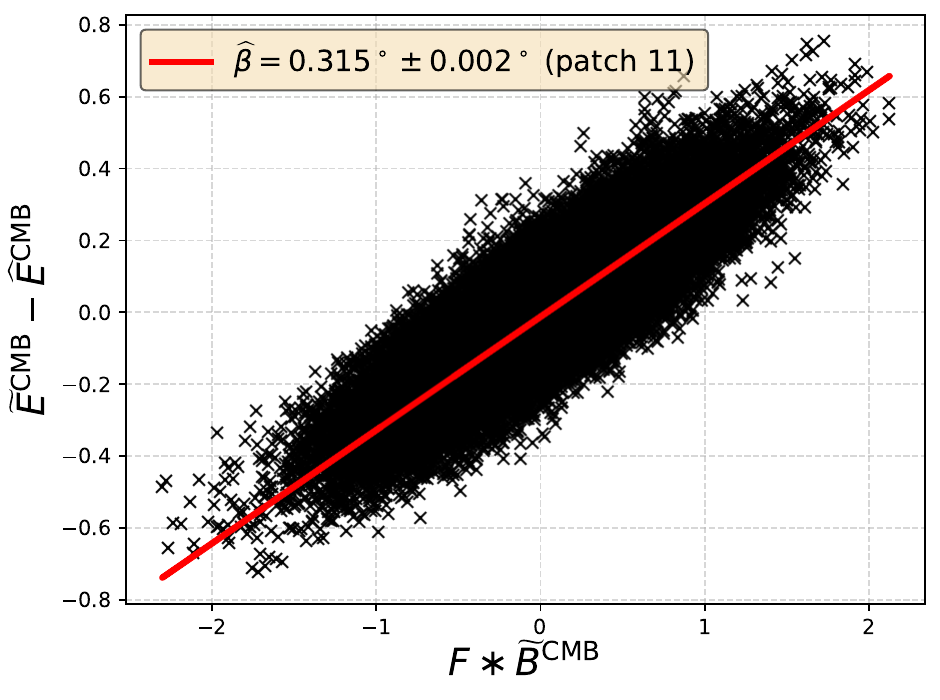}~
\includegraphics[width=0.25\textwidth,clip]{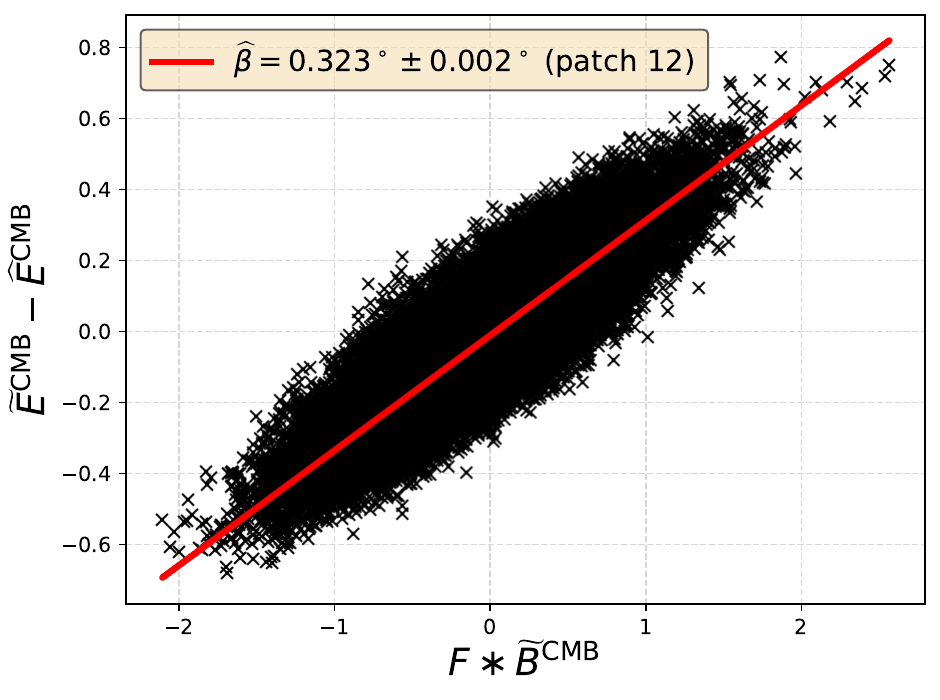}\\
\includegraphics[width=0.25\textwidth,clip]{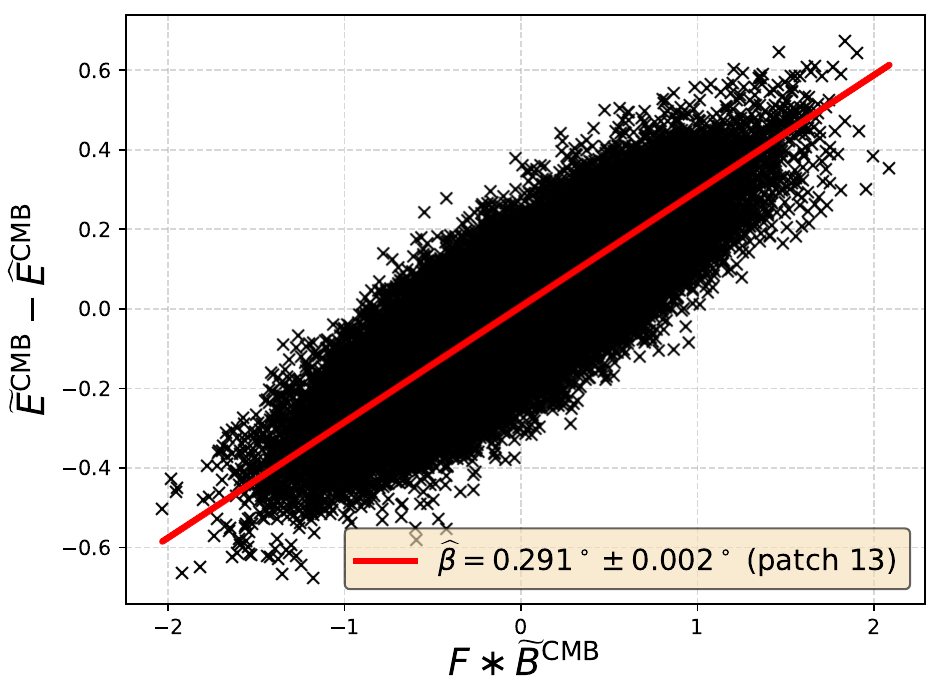}~
\includegraphics[width=0.25\textwidth,clip]{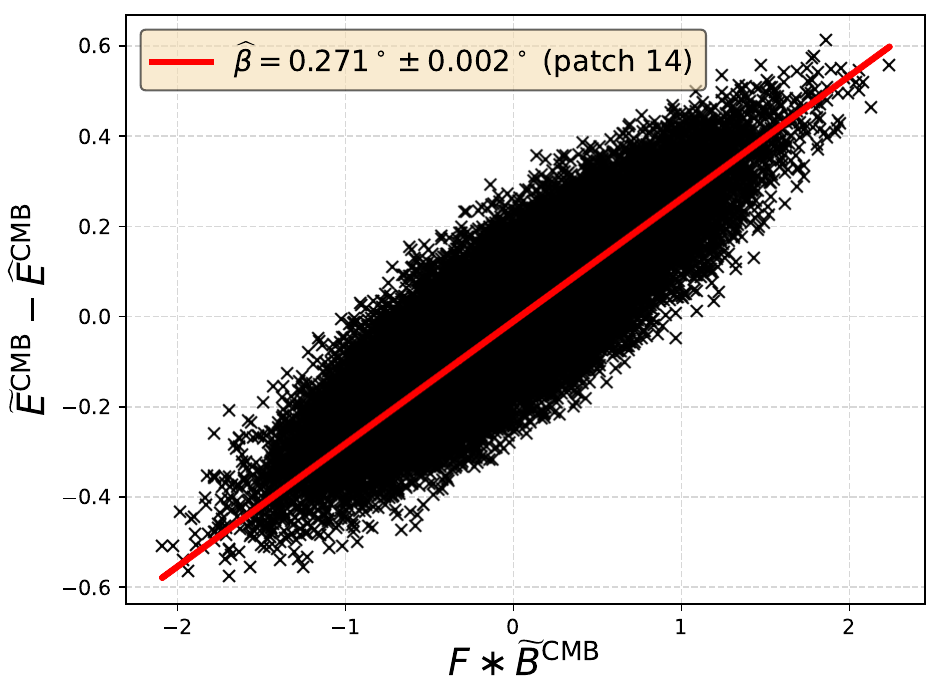}~
\includegraphics[width=0.25\textwidth,clip]{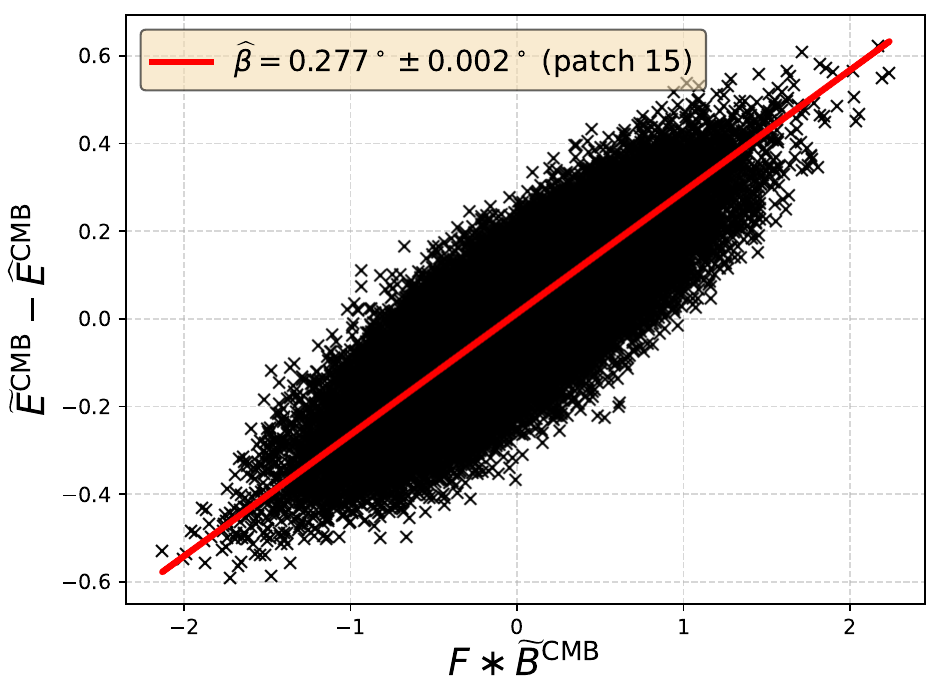}~
\includegraphics[width=0.25\textwidth,clip]{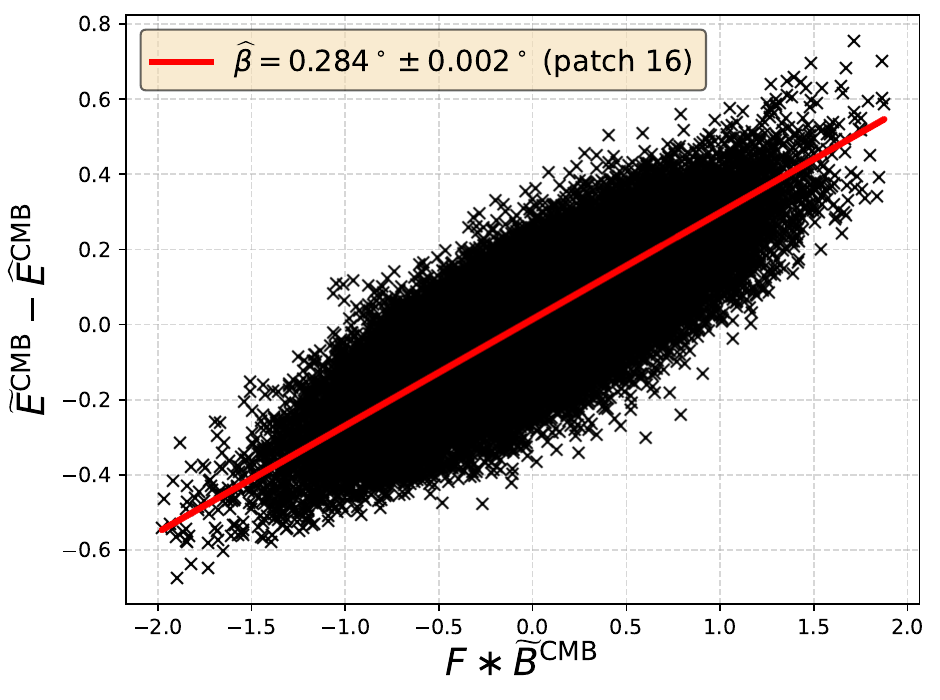}\\
\includegraphics[width=0.25\textwidth,clip]{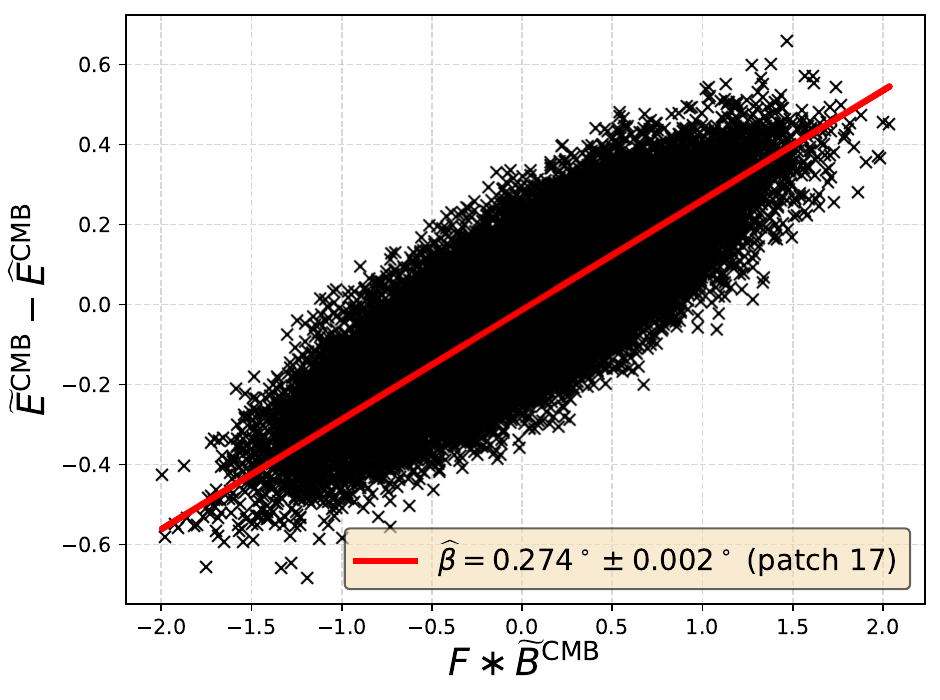}~
\includegraphics[width=0.25\textwidth,clip]{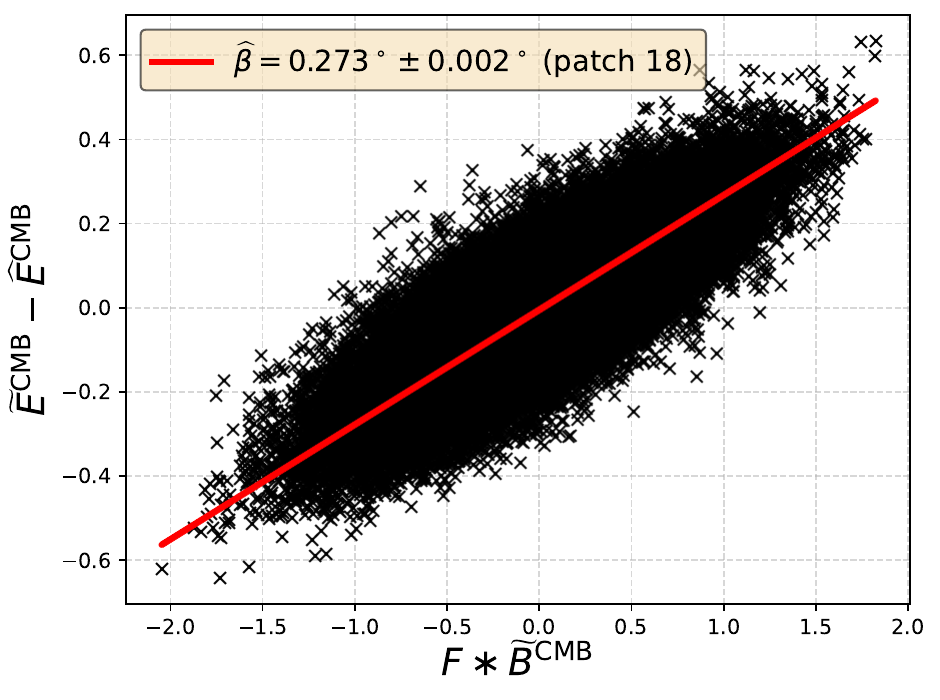}~
\includegraphics[width=0.25\textwidth,clip]{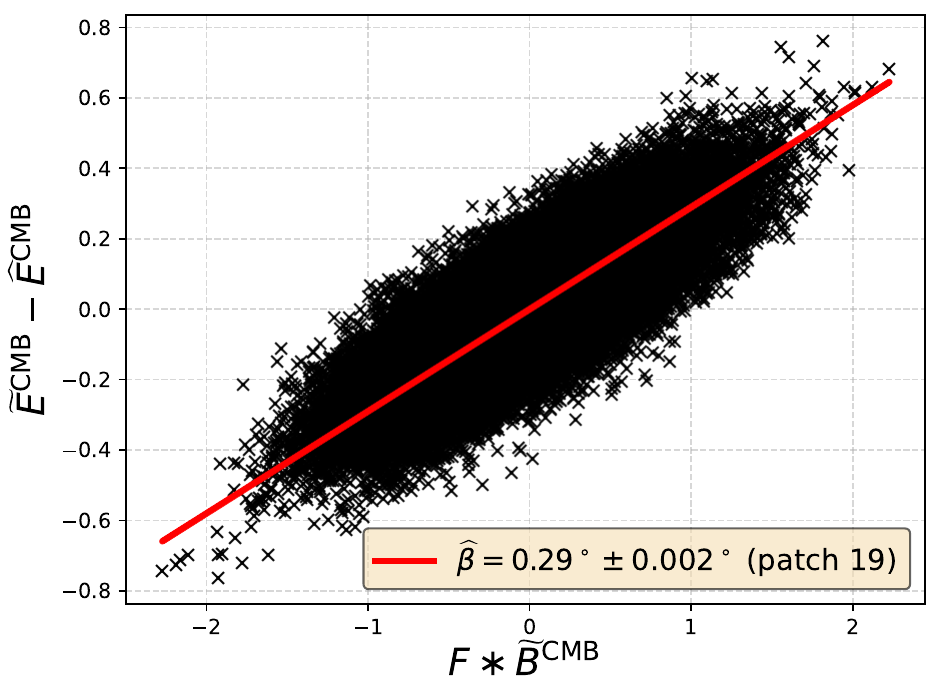}~
\includegraphics[width=0.25\textwidth,clip]{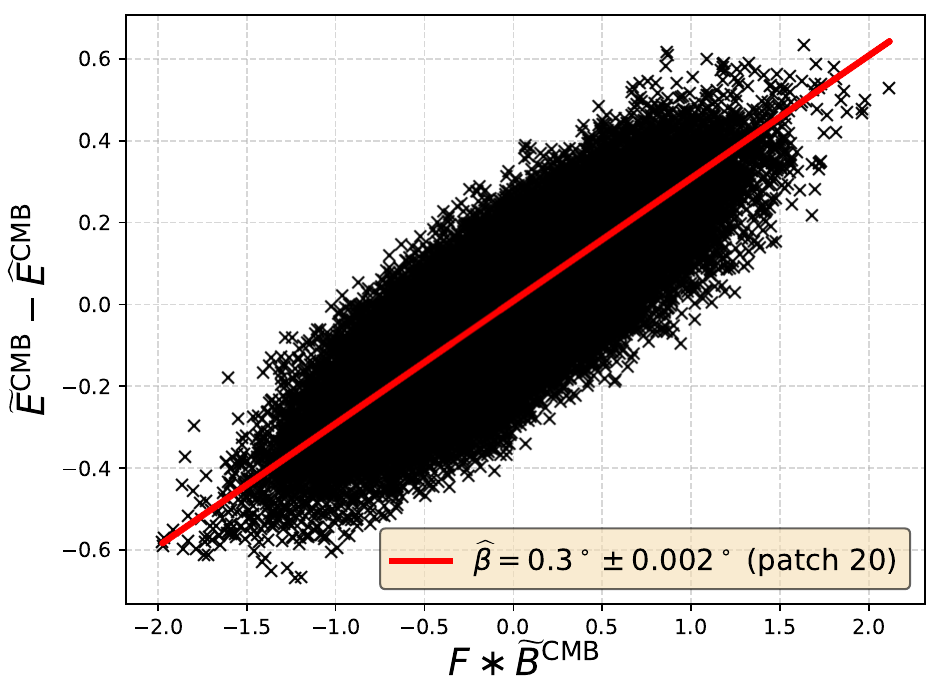}\\
\includegraphics[width=0.25\textwidth,clip]{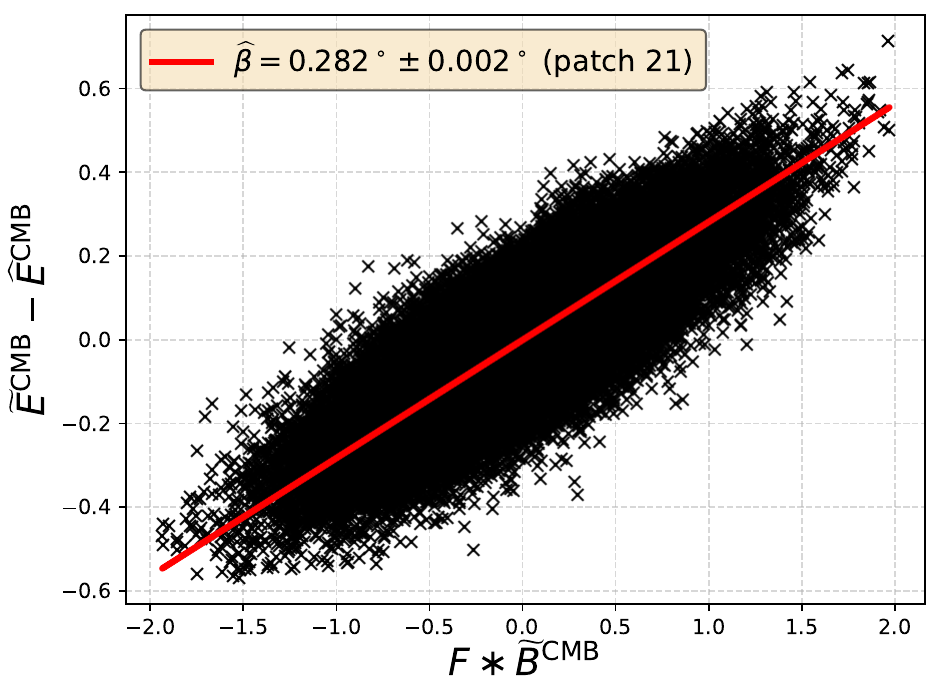}~
\includegraphics[width=0.25\textwidth,clip]{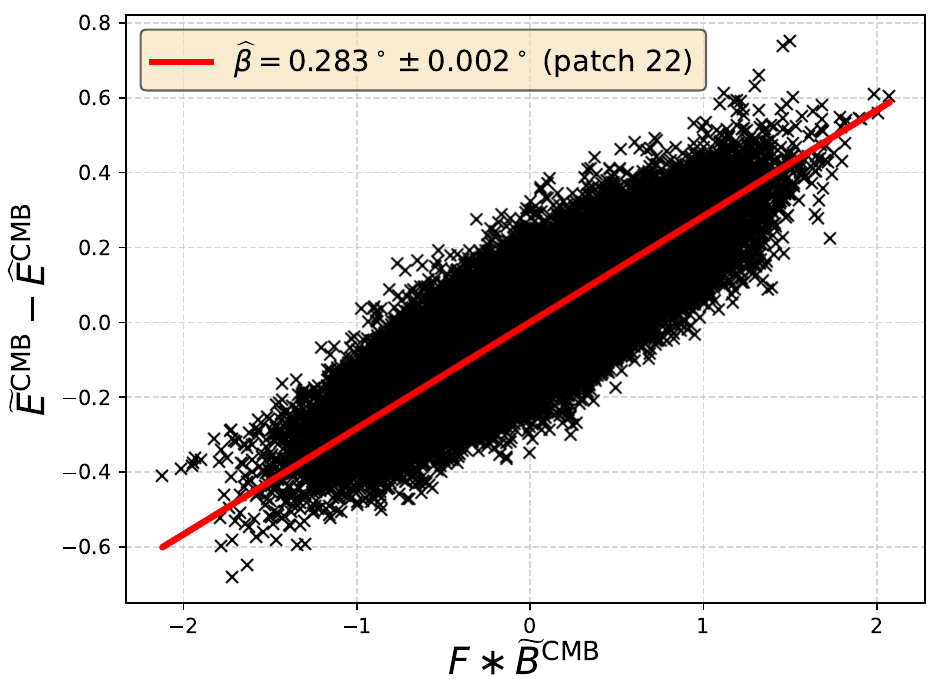}~
\includegraphics[width=0.25\textwidth,clip]{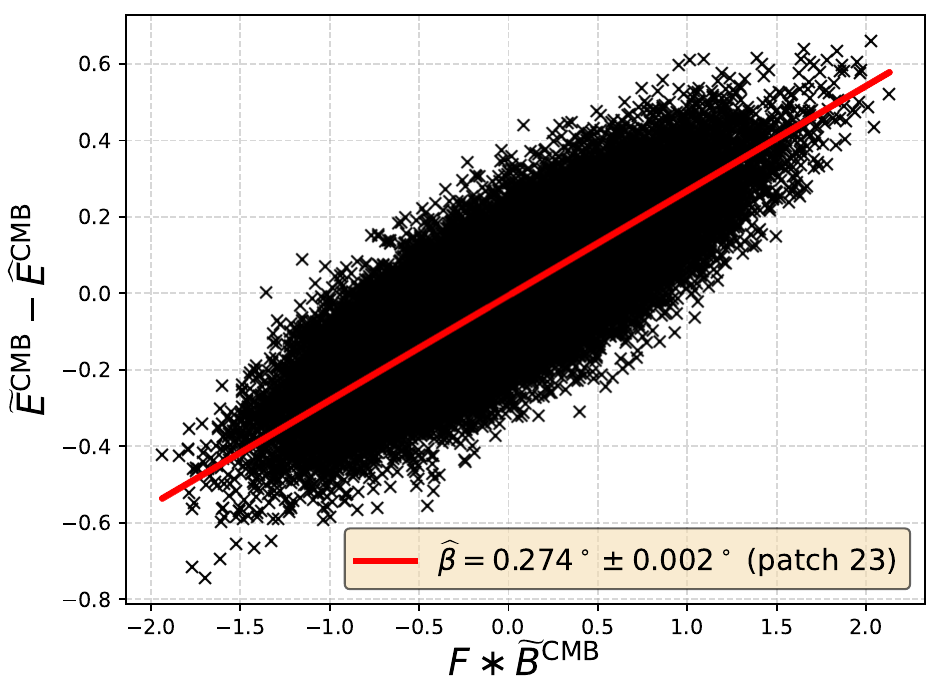}~
\includegraphics[width=0.25\textwidth,clip]{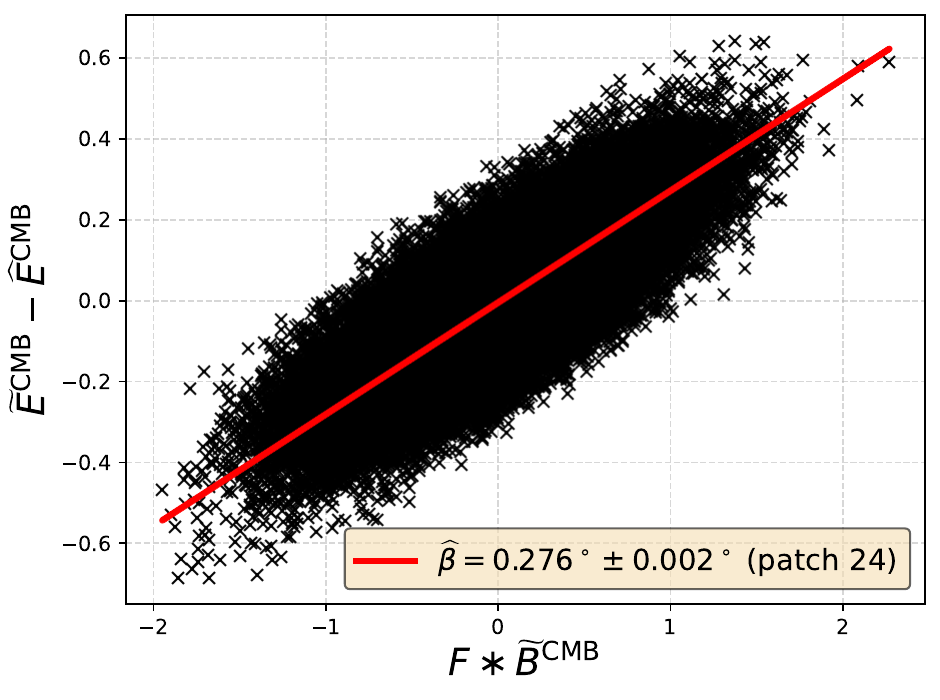}\\
\includegraphics[width=0.25\textwidth,clip]{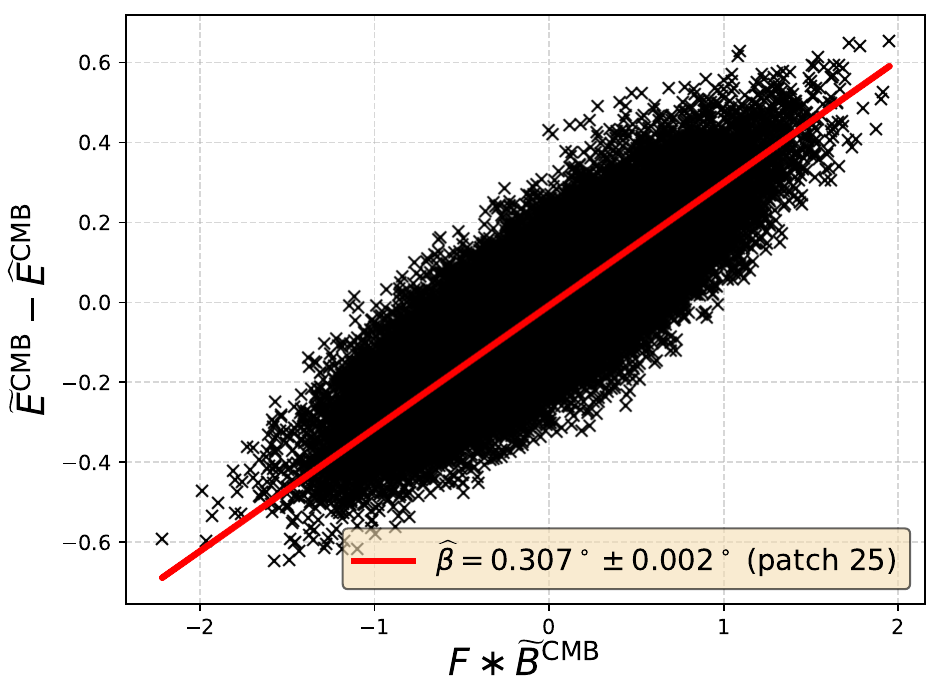}~
\includegraphics[width=0.25\textwidth,clip]{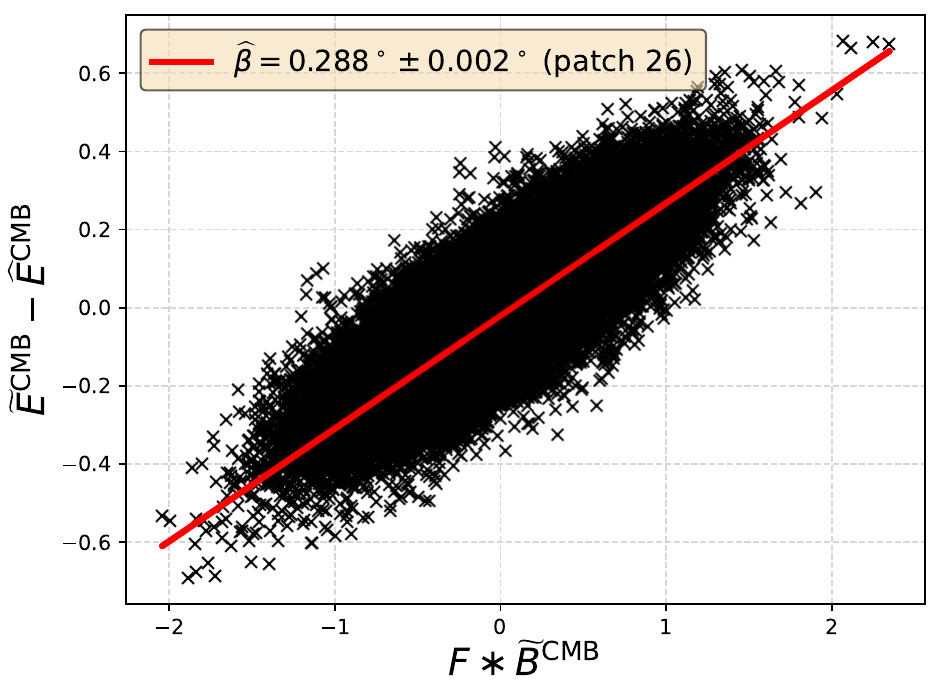}~
\hfill
\caption{\label{fig:ttplot} \lb's case $\beta = 0.3^\circ$, $\alpha \neq 0$. Estimates of the cosmic birefringence angle $\beta$ (in degrees) across the 26 sky patches shown in Figure~\ref{fig:igloo}, obtained via linear regression between the reconstructed CMB fields $\widetilde{E}^{\rm CMB} - \widehat{E}^{\rm CMB}$ and $F \ast \widetilde{B}^{\rm CMB}$ (see Equation~\ref{eq:theorem_pixel} or Equations~\ref{eq:textbook}--\ref{eq:e}). The maps are filtered in harmonic space using a top-hat bandpass filter selecting multipoles $\ell \in [250, 500]$. The average and standard deviation of the estimates across the 26 patches yield a recovered birefringence angle of $\widehat{\beta} = 0.296^\circ \pm 0.018^\circ$, corresponding to a $16\sigma$ detection of $\beta = 0.3^\circ$, with a negligible bias of less than $0.3\sigma$.}
\end{figure}

\begin{figure}[tbp]
\centering 
\includegraphics[width=0.25\textwidth,clip]{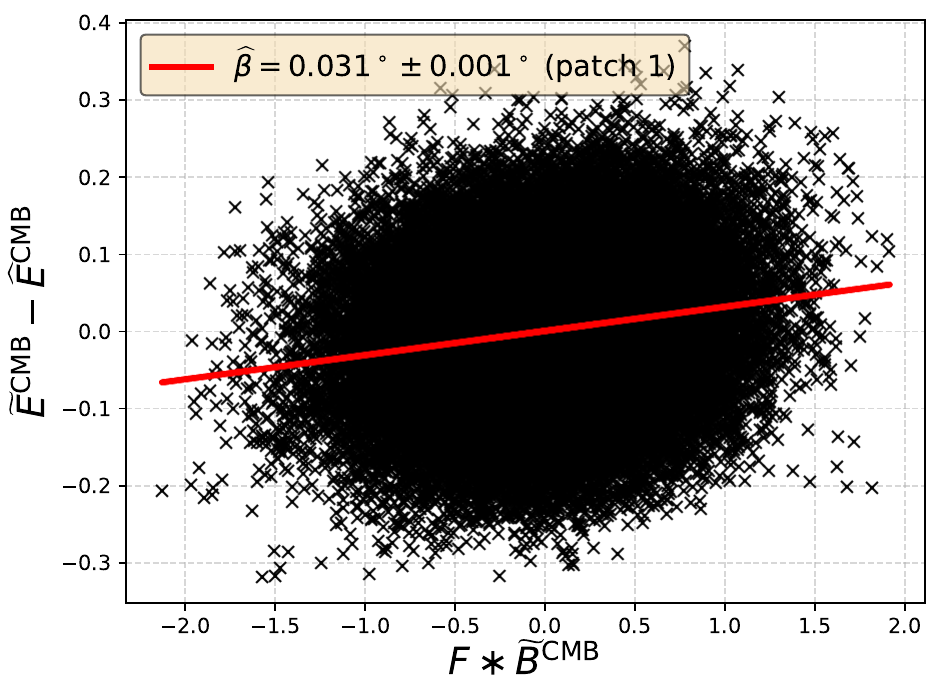}~
\includegraphics[width=0.25\textwidth,clip]{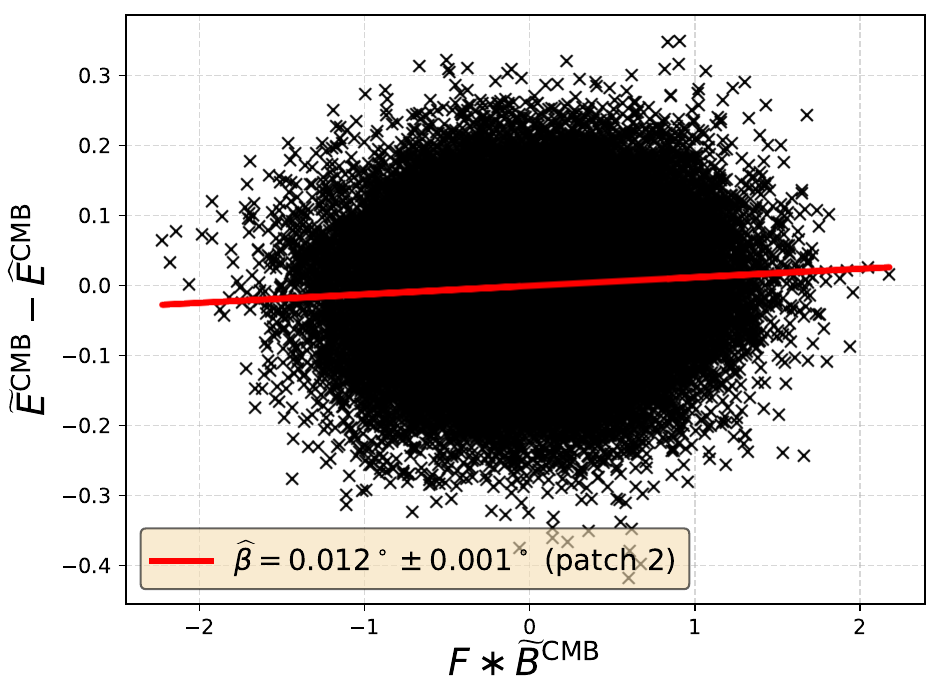}~
\includegraphics[width=0.25\textwidth,clip]{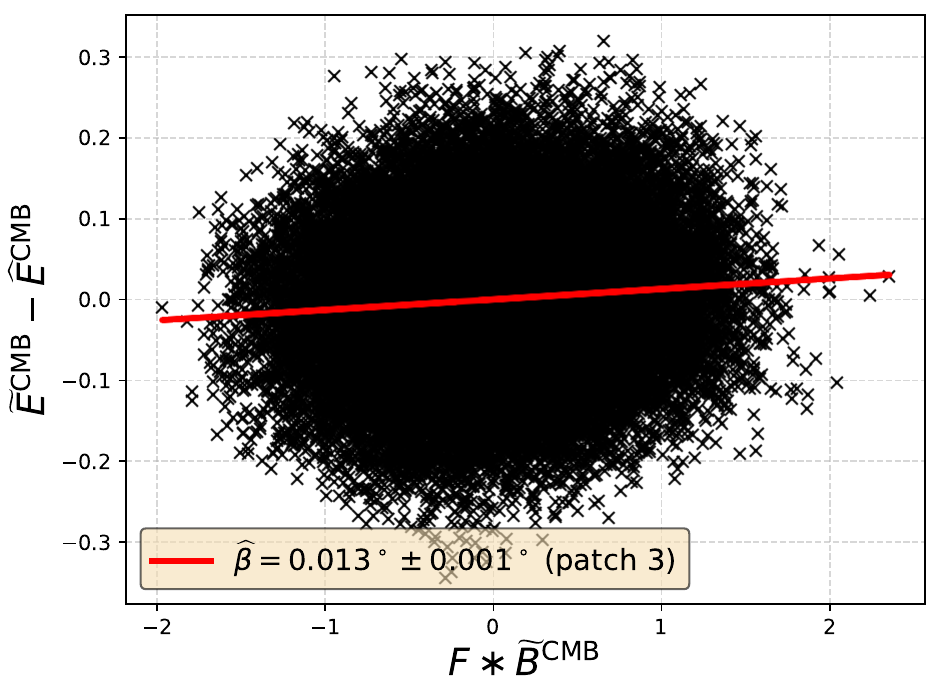}~
\includegraphics[width=0.25\textwidth,clip]{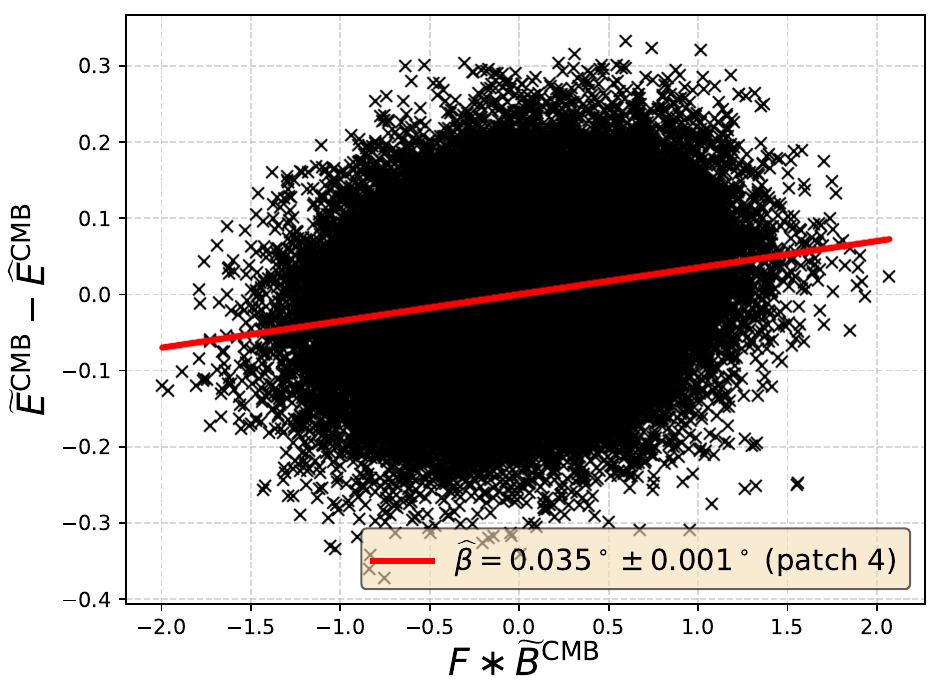}\\
\includegraphics[width=0.25\textwidth,clip]{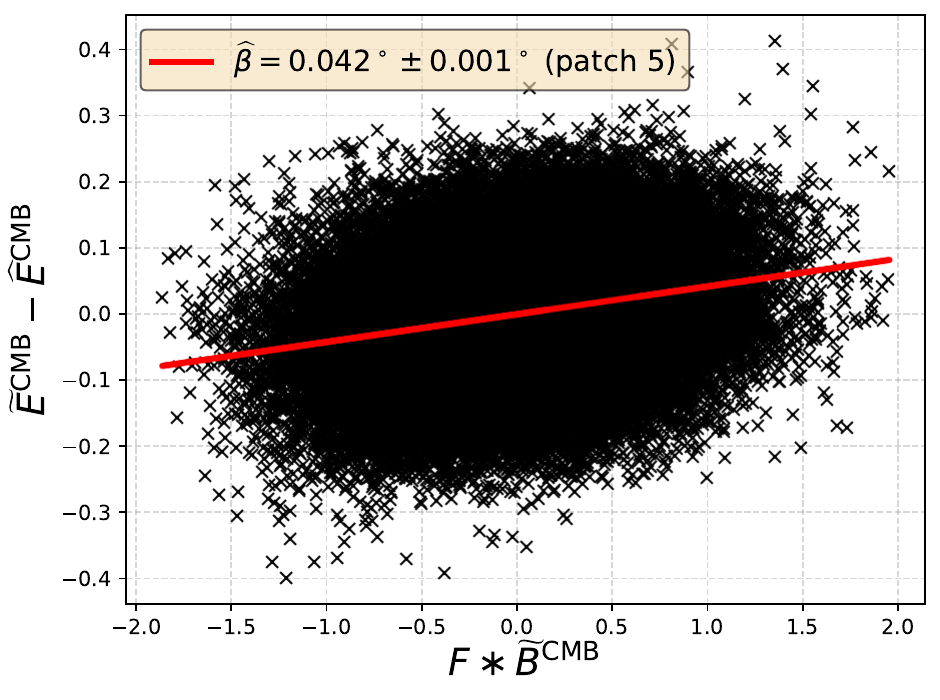}~
\includegraphics[width=0.25\textwidth,clip]{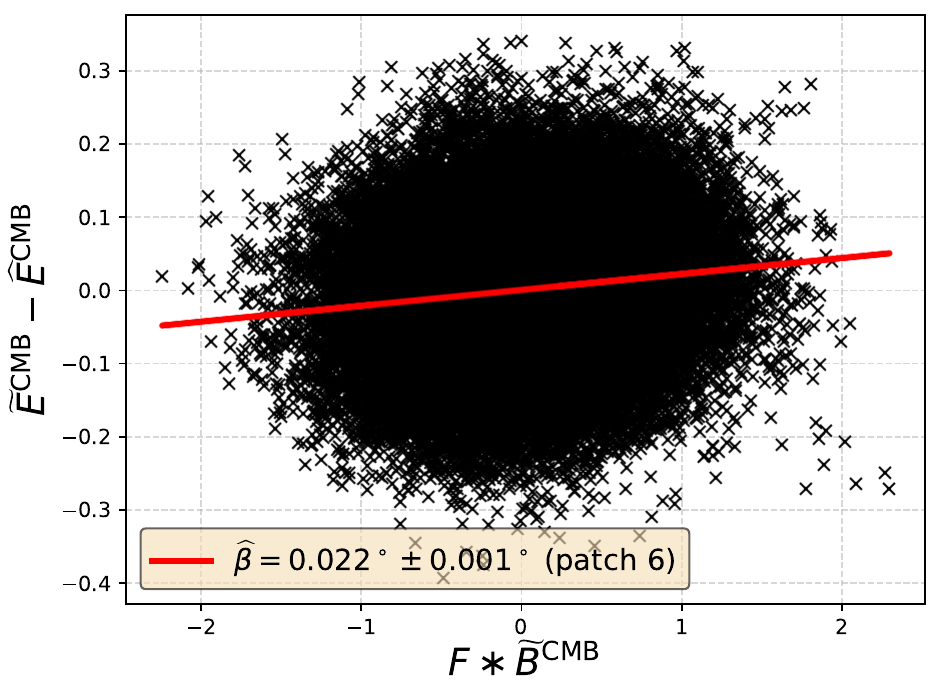}~
\includegraphics[width=0.25\textwidth,clip]{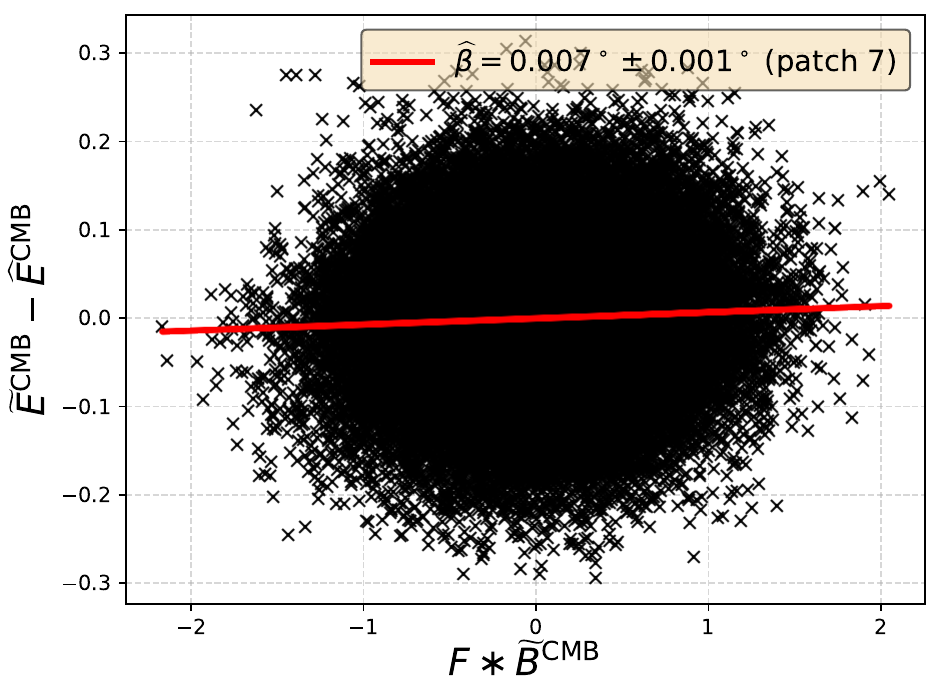}~
\includegraphics[width=0.25\textwidth,clip]{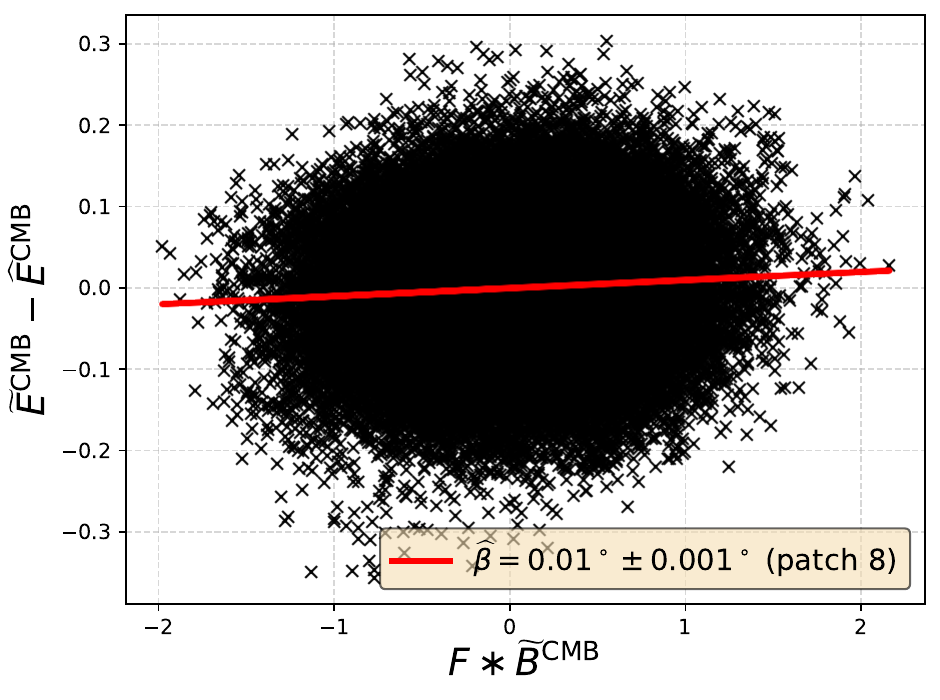}\\
\includegraphics[width=0.25\textwidth,clip]{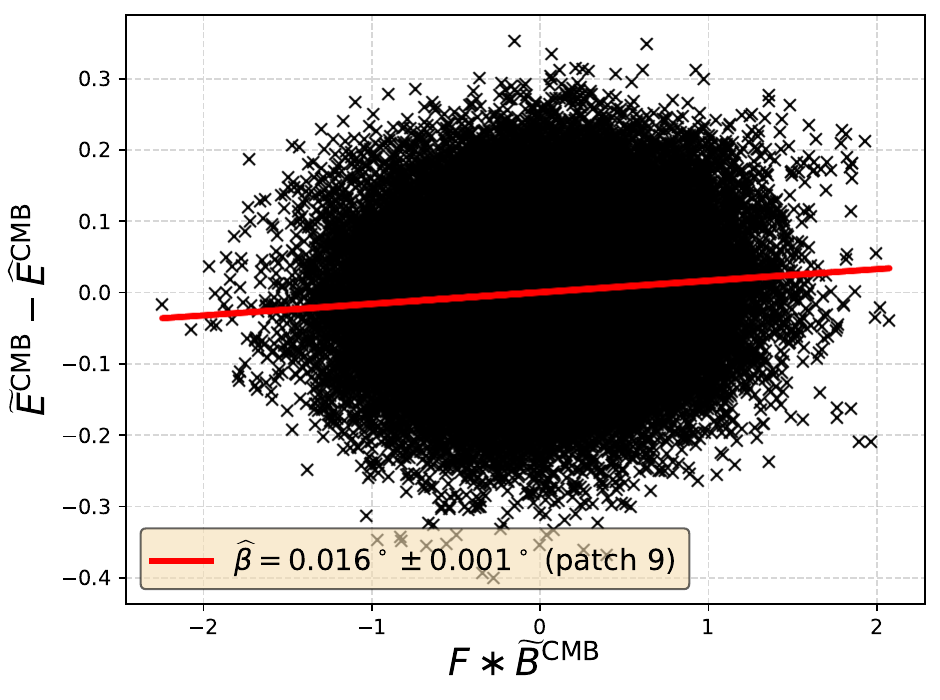}~
\includegraphics[width=0.25\textwidth,clip]{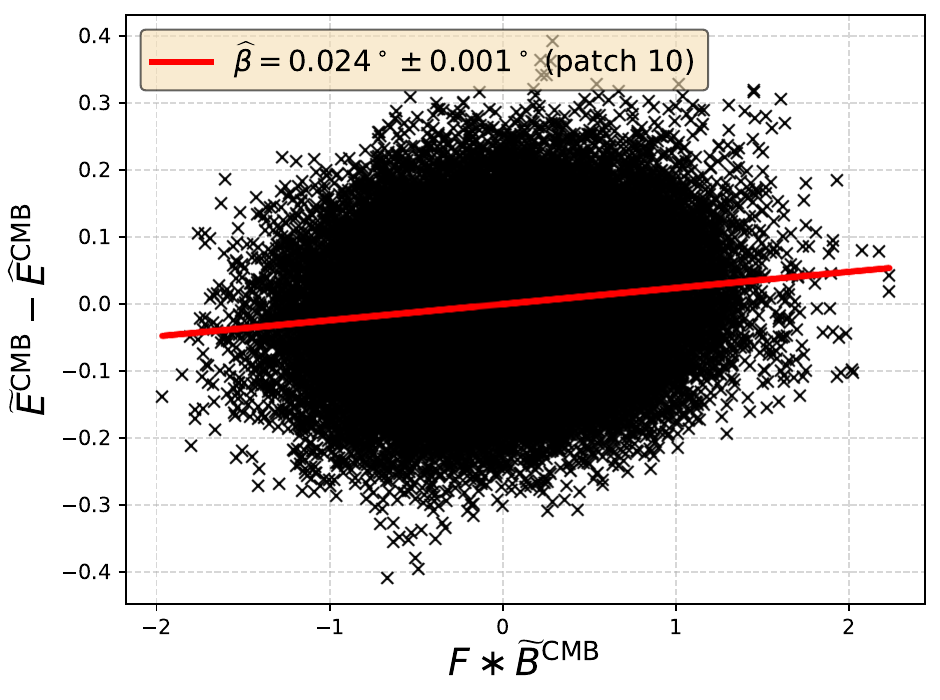}~
\includegraphics[width=0.25\textwidth,clip]{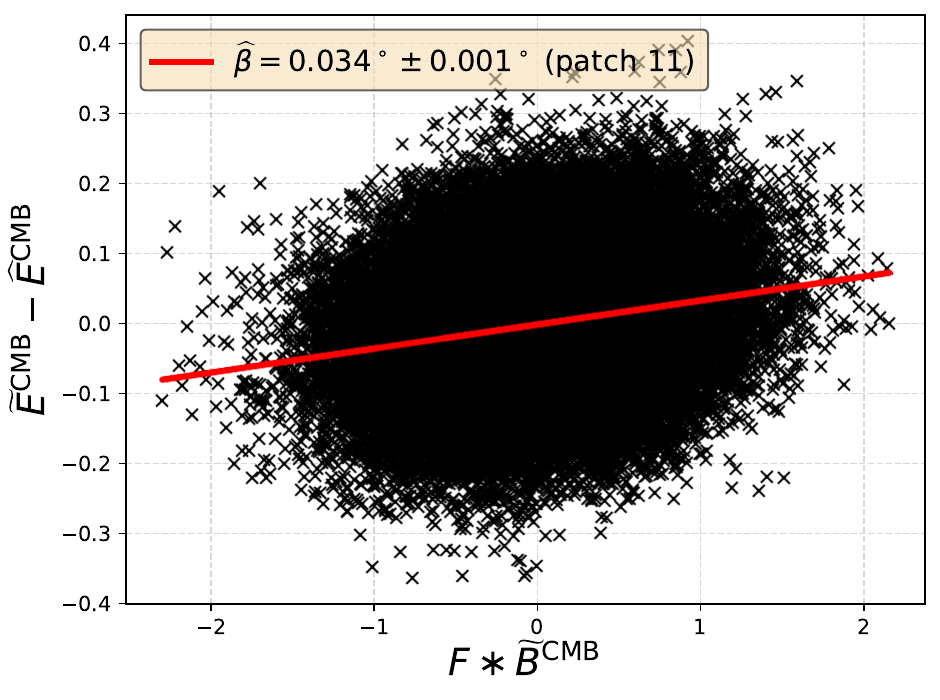}~
\includegraphics[width=0.25\textwidth,clip]{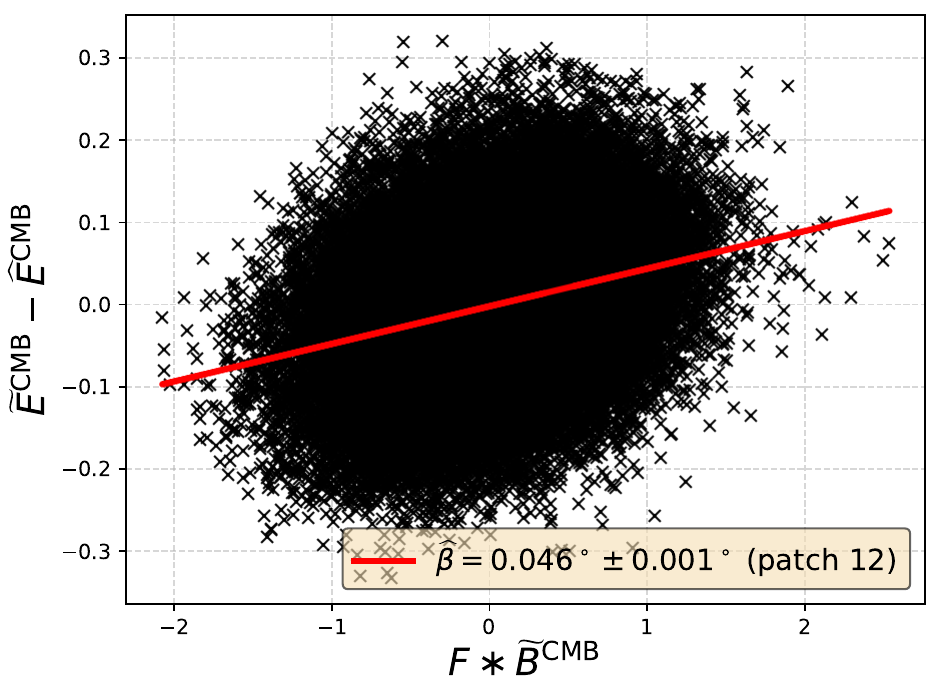}\\
\includegraphics[width=0.25\textwidth,clip]{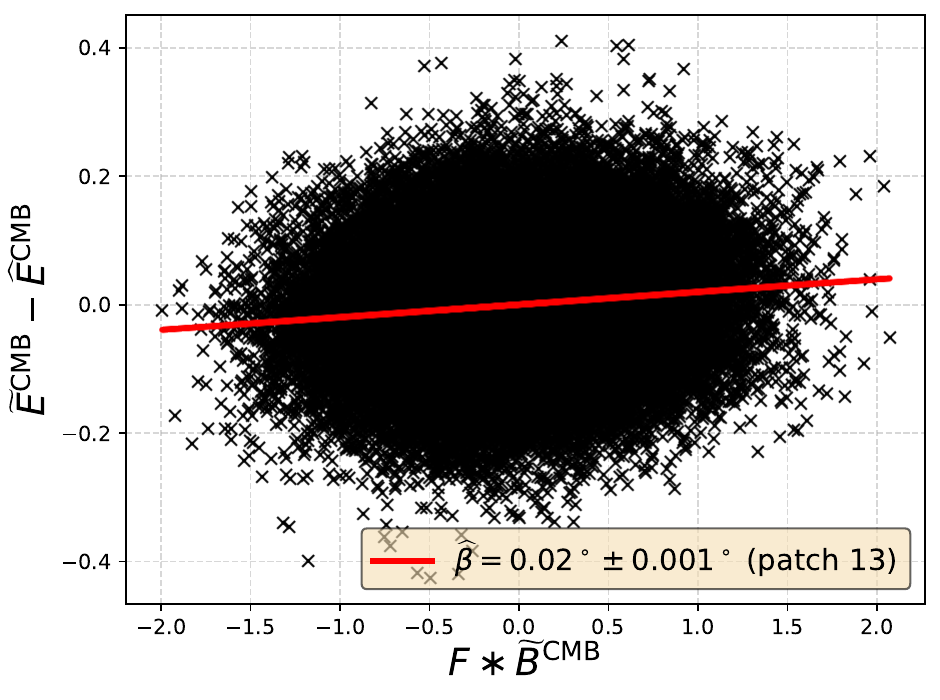}~
\includegraphics[width=0.25\textwidth,clip]{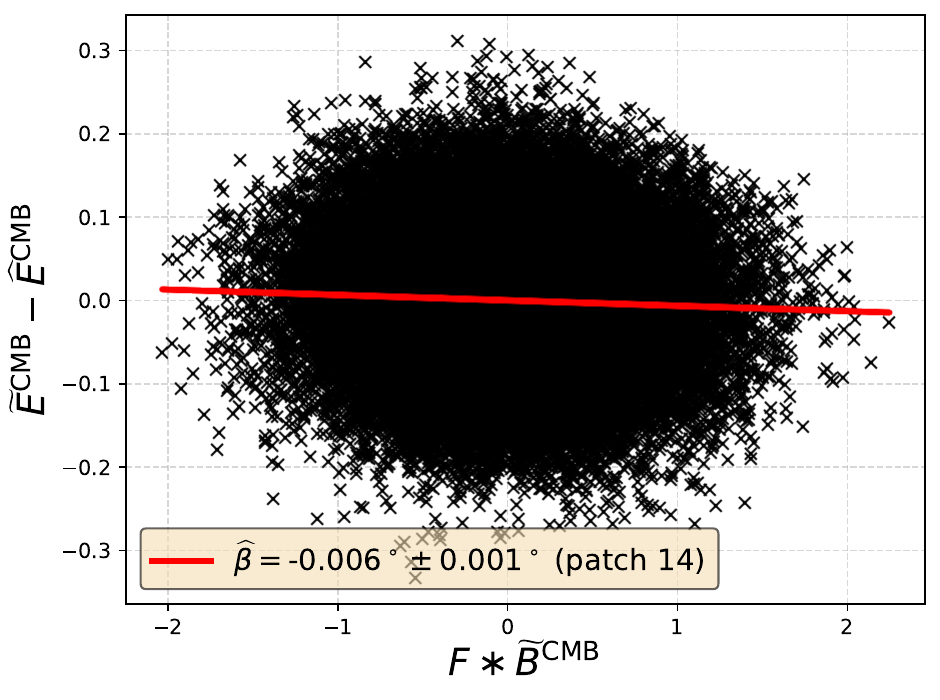}~
\includegraphics[width=0.25\textwidth,clip]{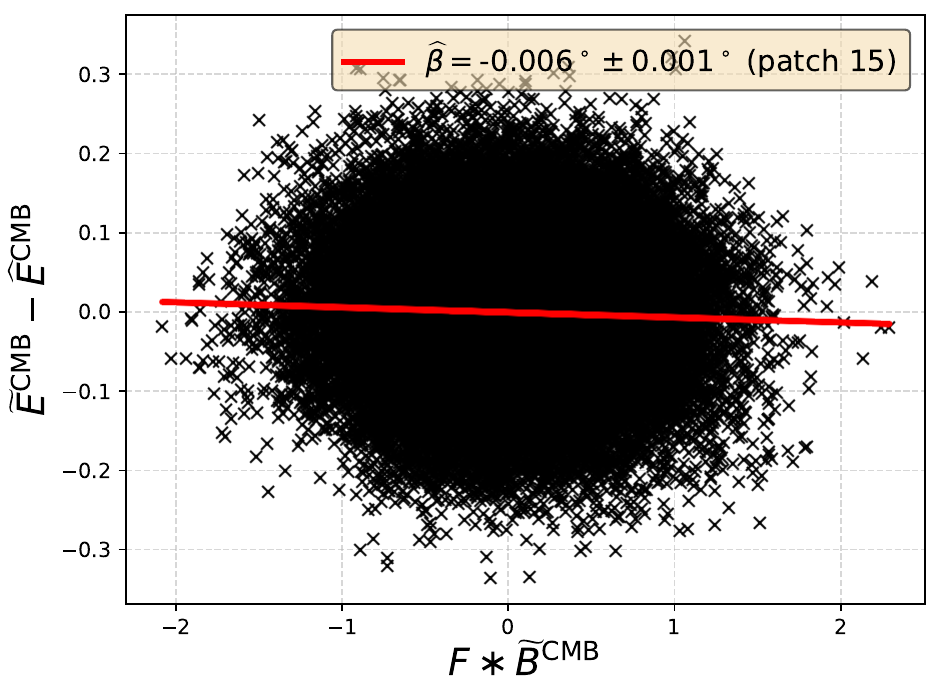}~
\includegraphics[width=0.25\textwidth,clip]{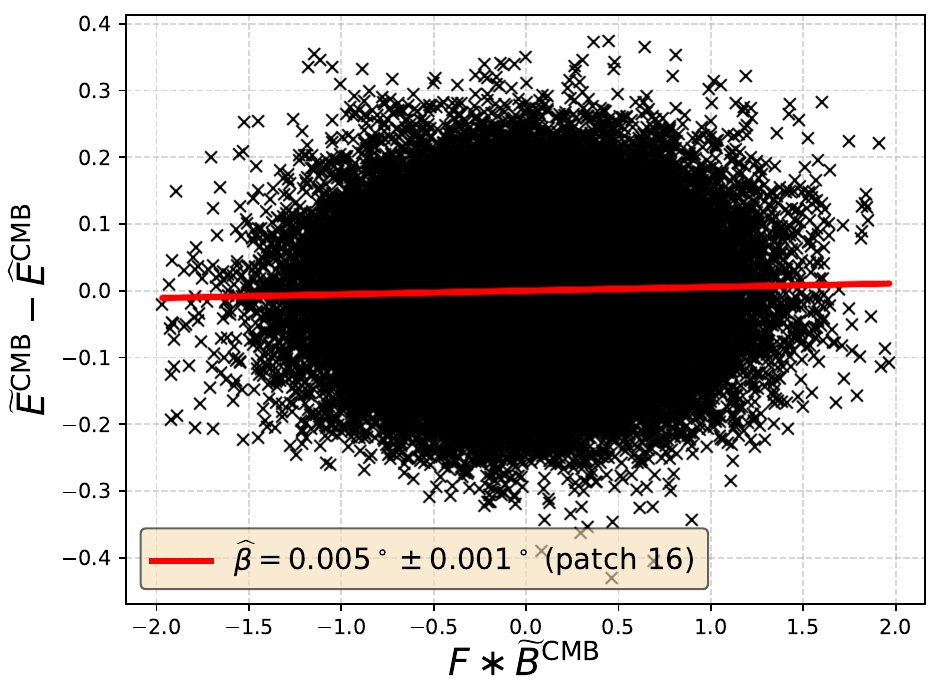}\\
\includegraphics[width=0.25\textwidth,clip]{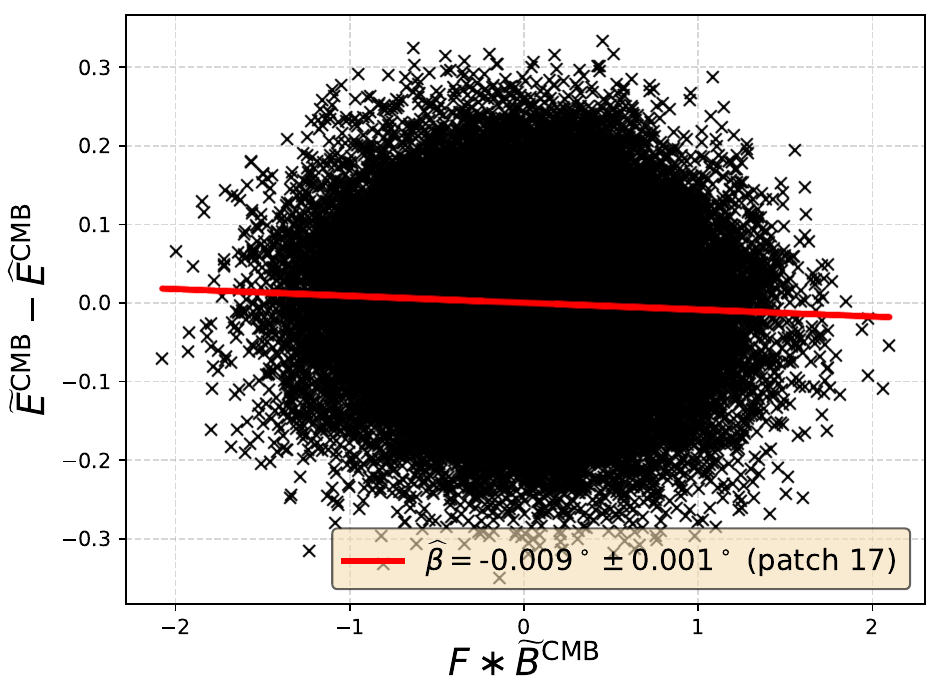}~
\includegraphics[width=0.25\textwidth,clip]{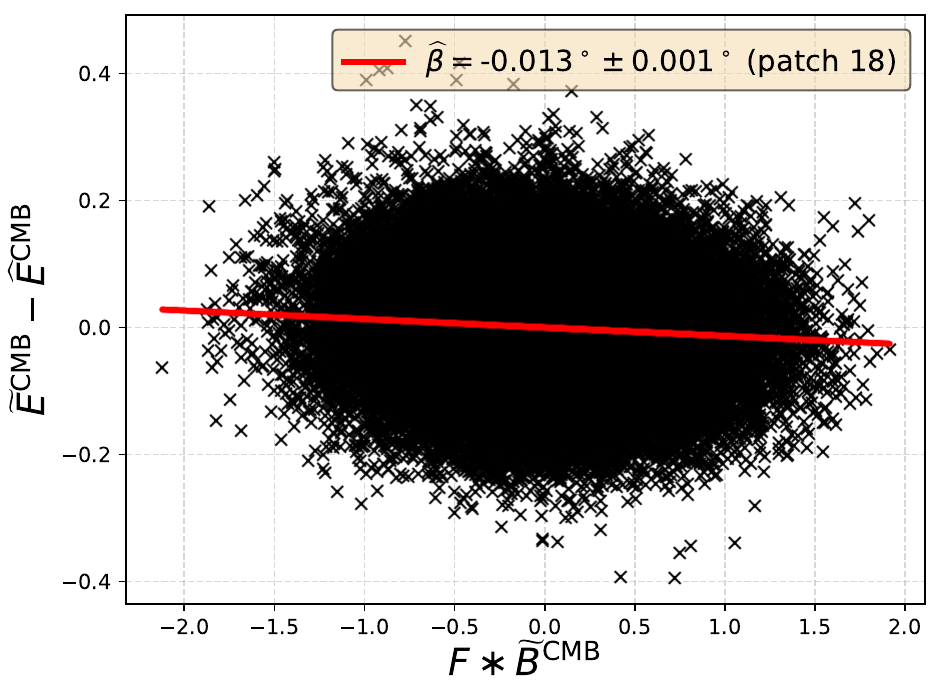}~
\includegraphics[width=0.25\textwidth,clip]{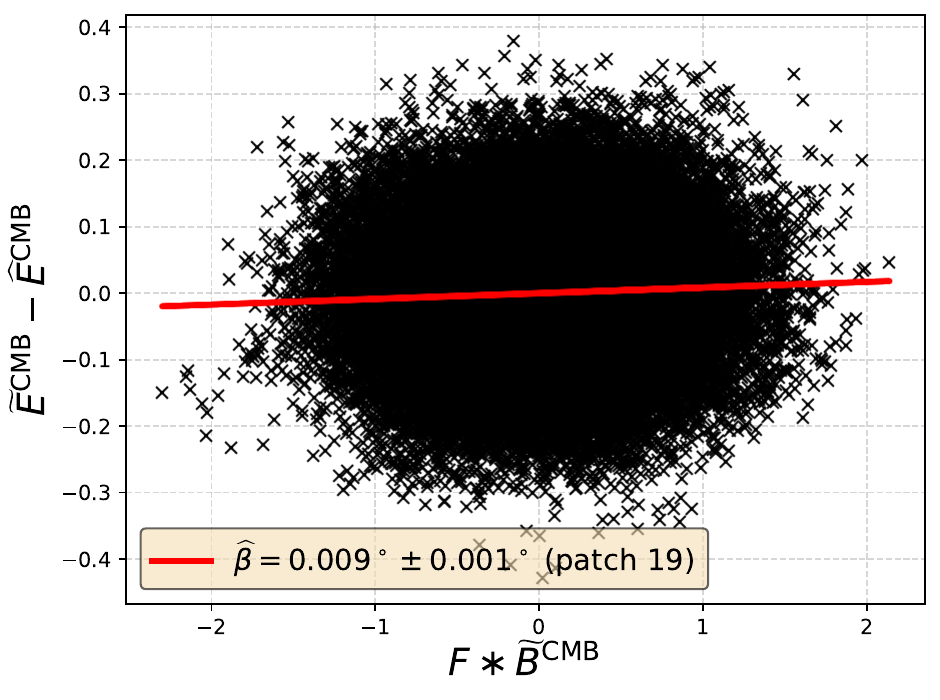}~
\includegraphics[width=0.25\textwidth,clip]{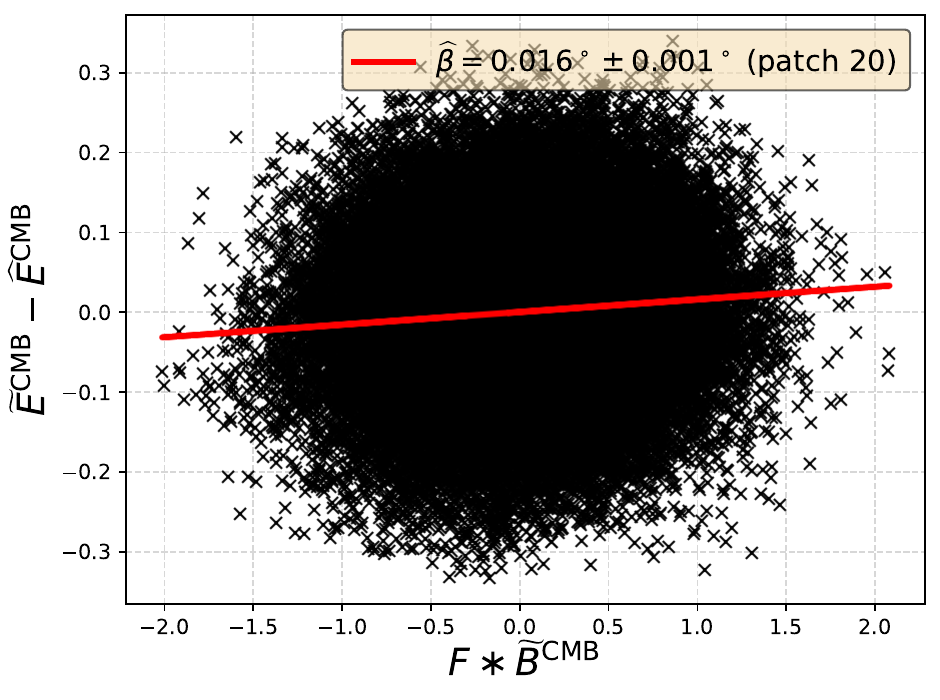}\\
\includegraphics[width=0.25\textwidth,clip]{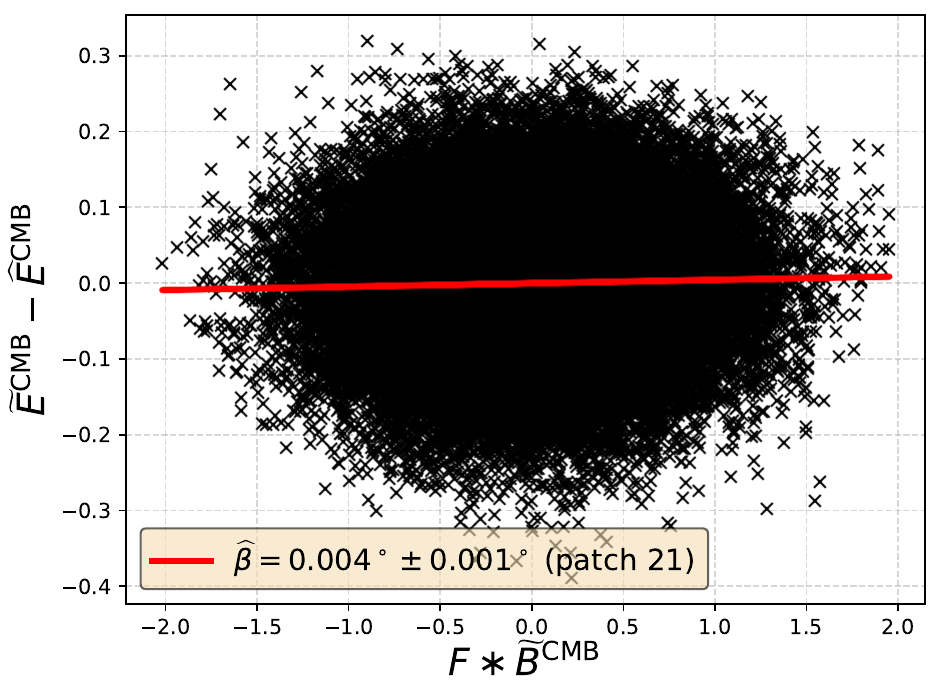}~
\includegraphics[width=0.25\textwidth,clip]{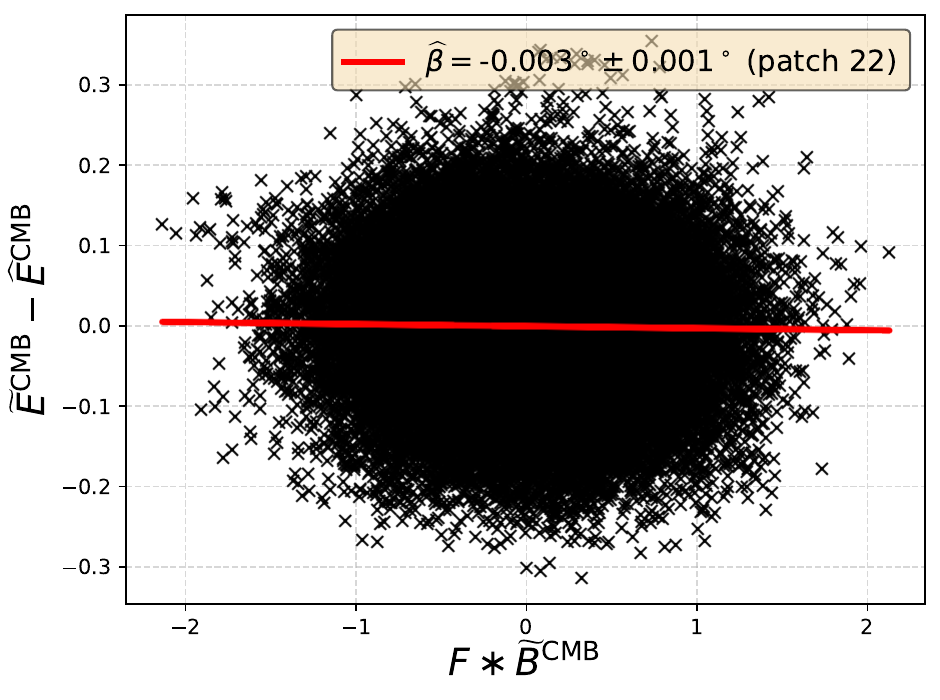}~
\includegraphics[width=0.25\textwidth,clip]{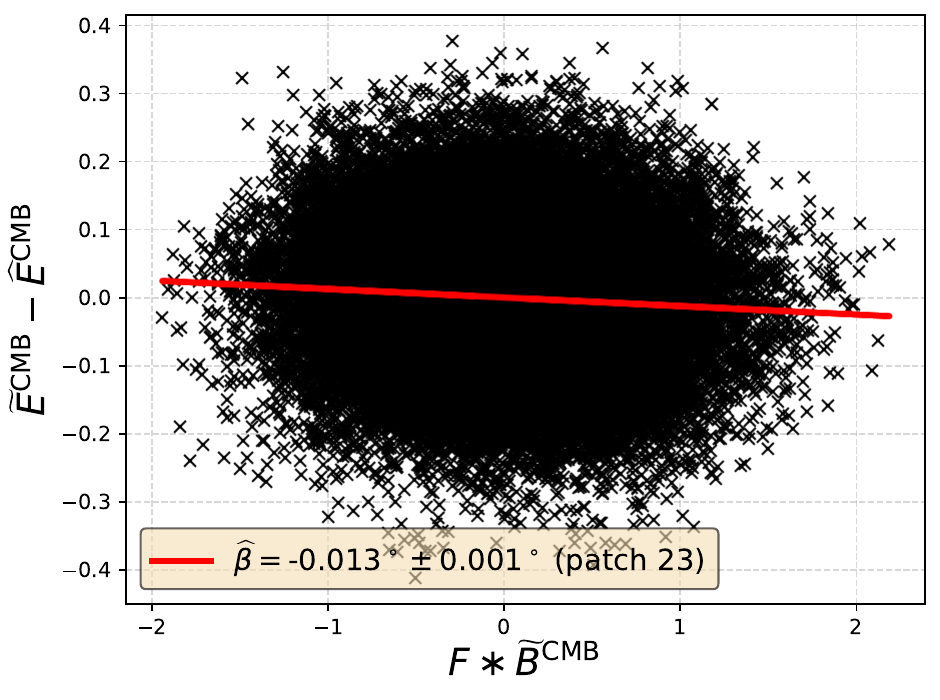}~
\includegraphics[width=0.25\textwidth,clip]{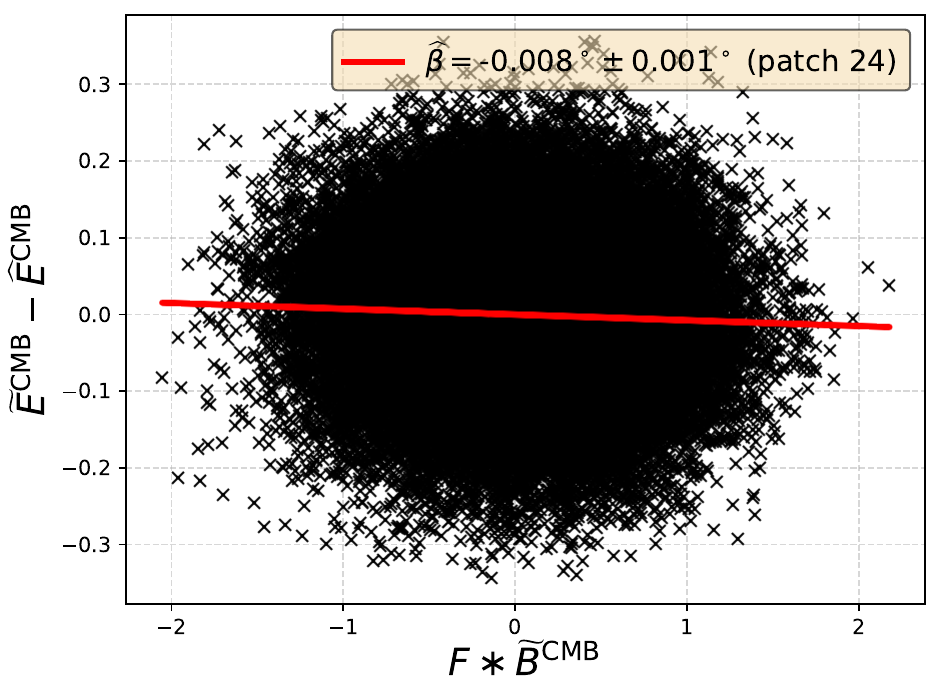}\\
\includegraphics[width=0.25\textwidth,clip]{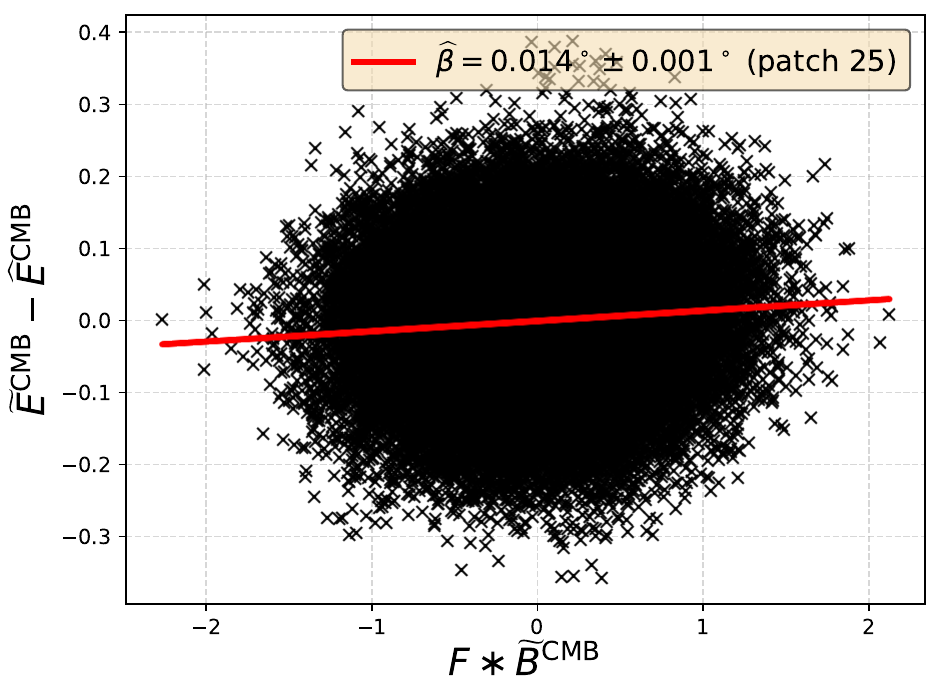}~
\includegraphics[width=0.25\textwidth,clip]{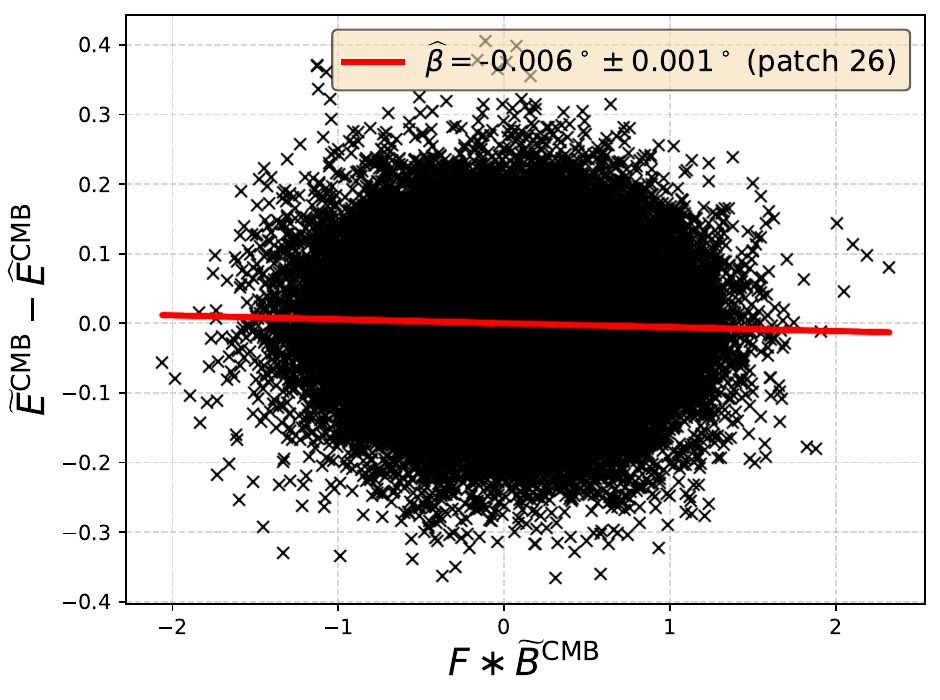}~
\hfill
\caption{\label{fig:ttplot_zerobeta_nonzeroalpha} Same as Figure~\ref{fig:ttplot}, but for the case $\beta = 0$, $\alpha \neq 0$ (i.e., no cosmic birefringence). In this scenario, the mean recovered birefringence angle is $\widehat{\beta} = 0.011^\circ \pm 0.017^\circ$, consistent with $\beta = 0$ within $1\sigma$ based on this field-level inference.}
\end{figure}

\begin{figure}[tbp]
\centering 
\includegraphics[width=0.25\textwidth,clip]{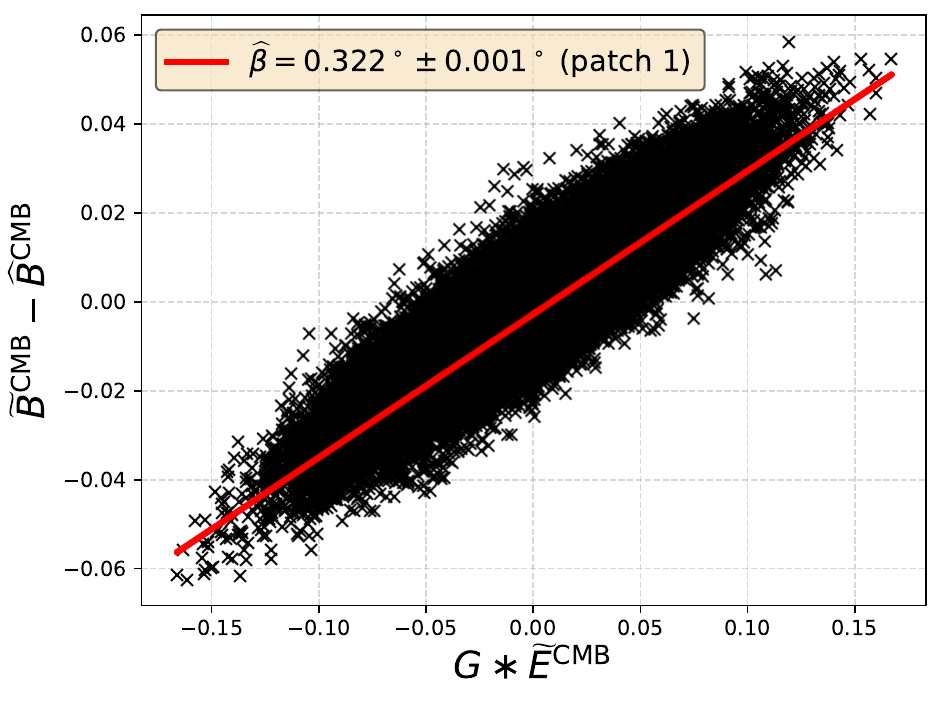}~
\includegraphics[width=0.25\textwidth,clip]{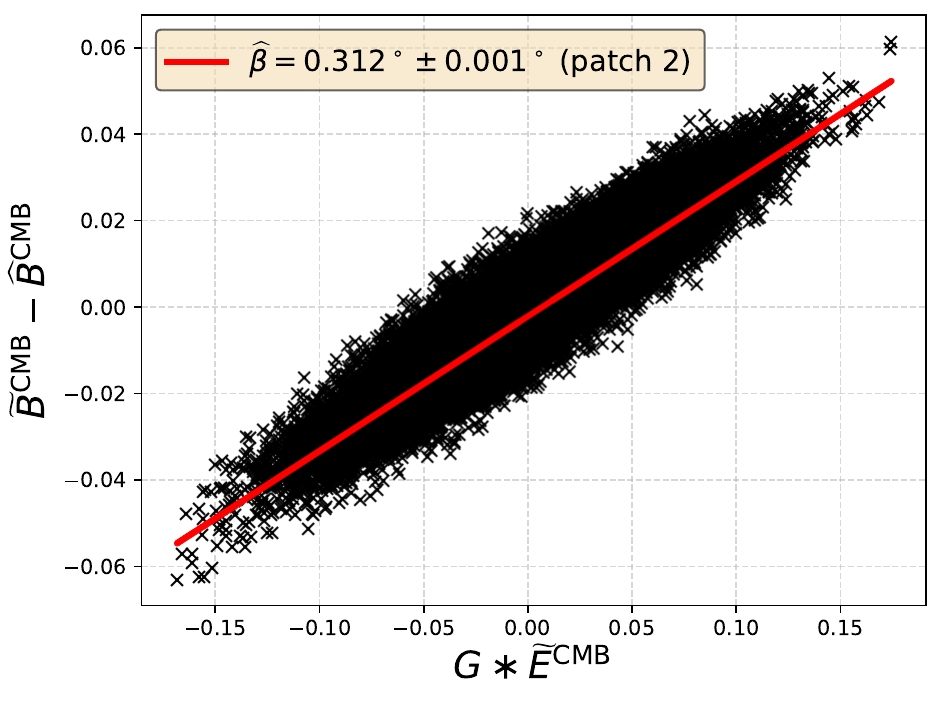}~
\includegraphics[width=0.25\textwidth,clip]{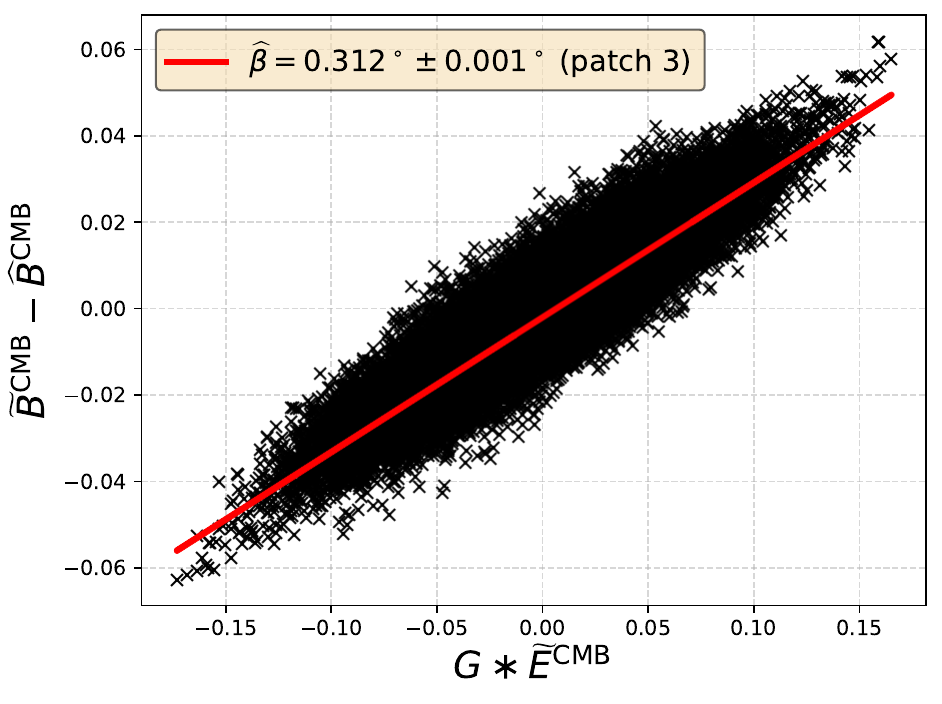}~
\includegraphics[width=0.25\textwidth,clip]{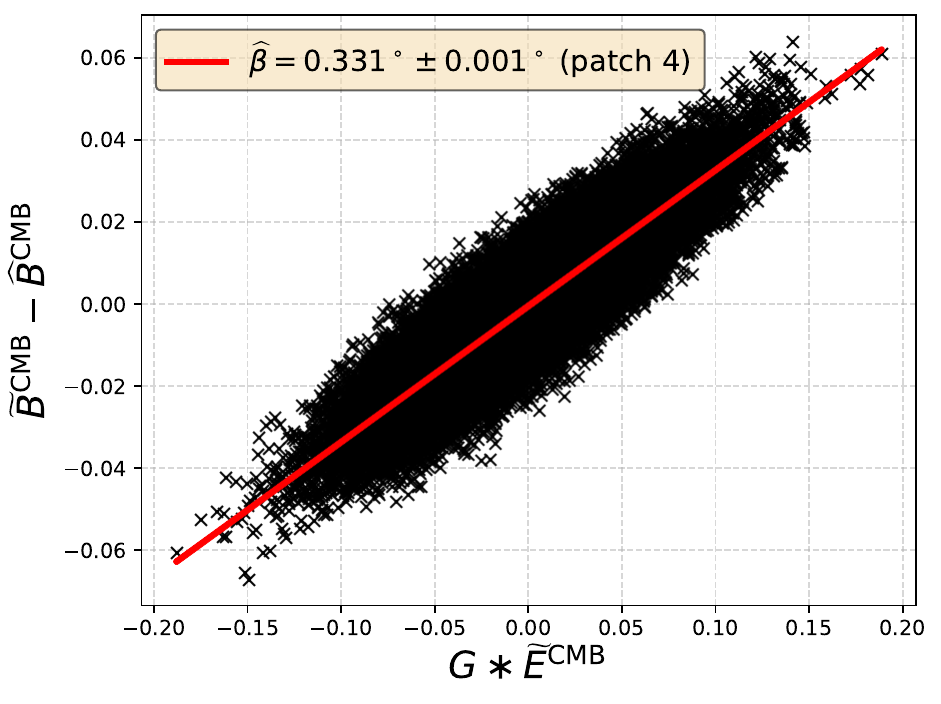}\\
\includegraphics[width=0.25\textwidth,clip]{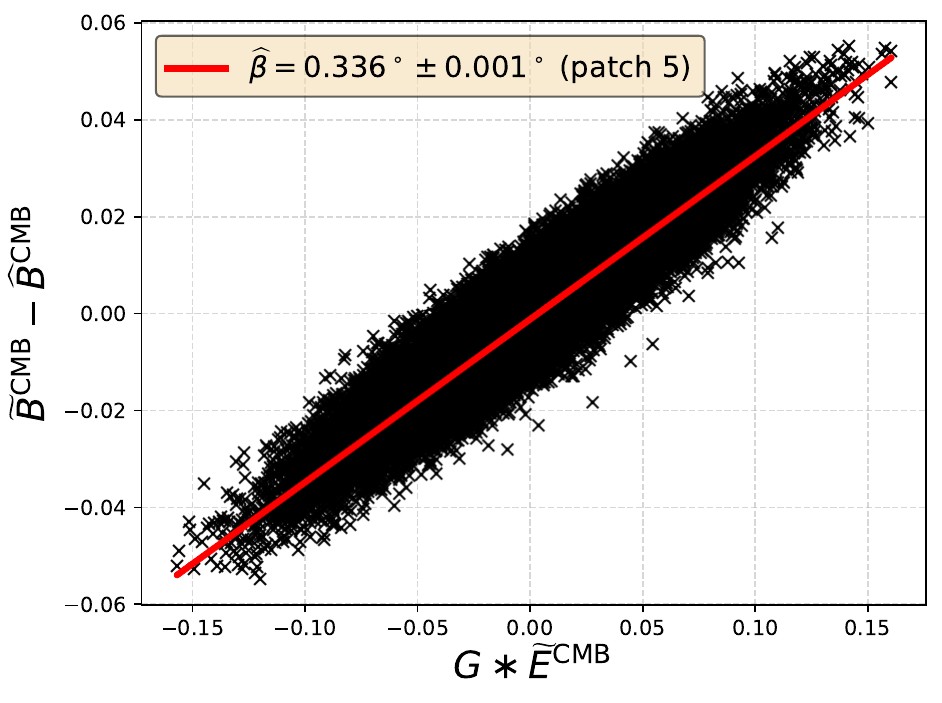}~
\includegraphics[width=0.25\textwidth,clip]{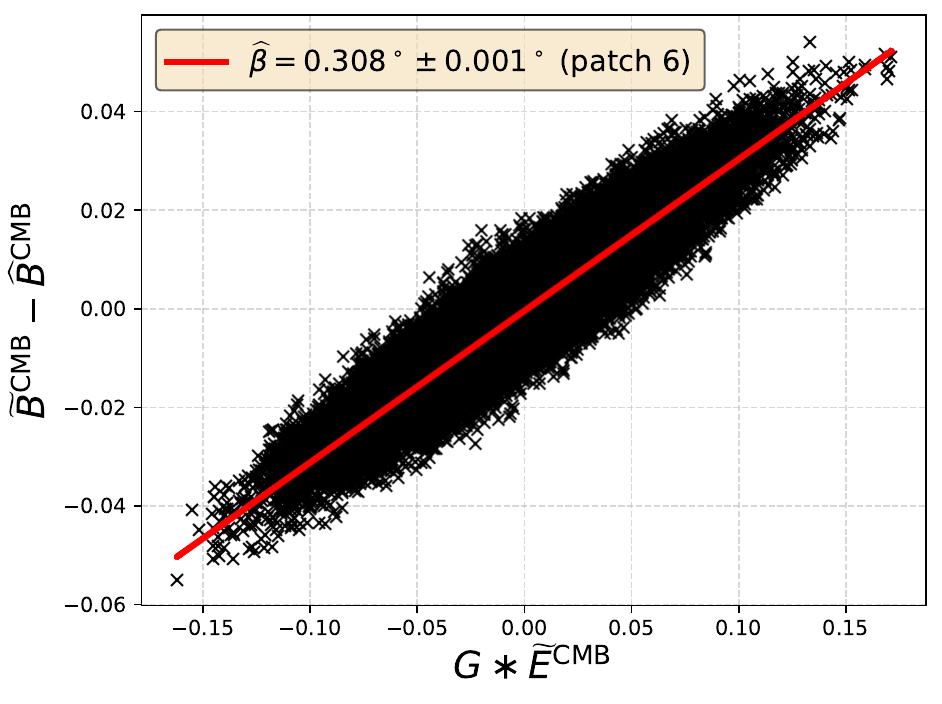}~
\includegraphics[width=0.25\textwidth,clip]{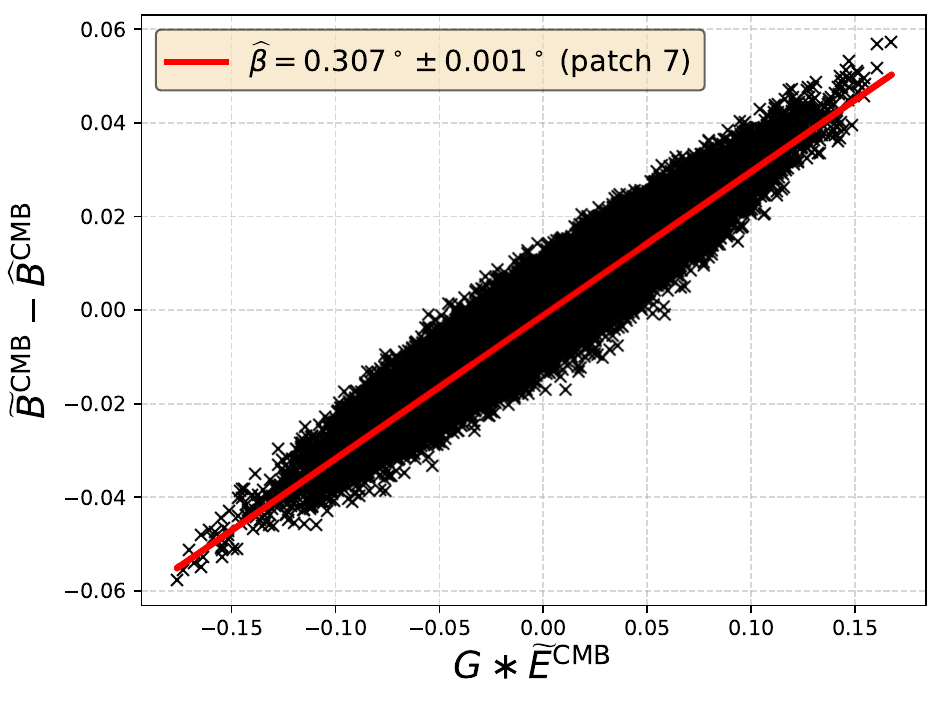}~
\includegraphics[width=0.25\textwidth,clip]{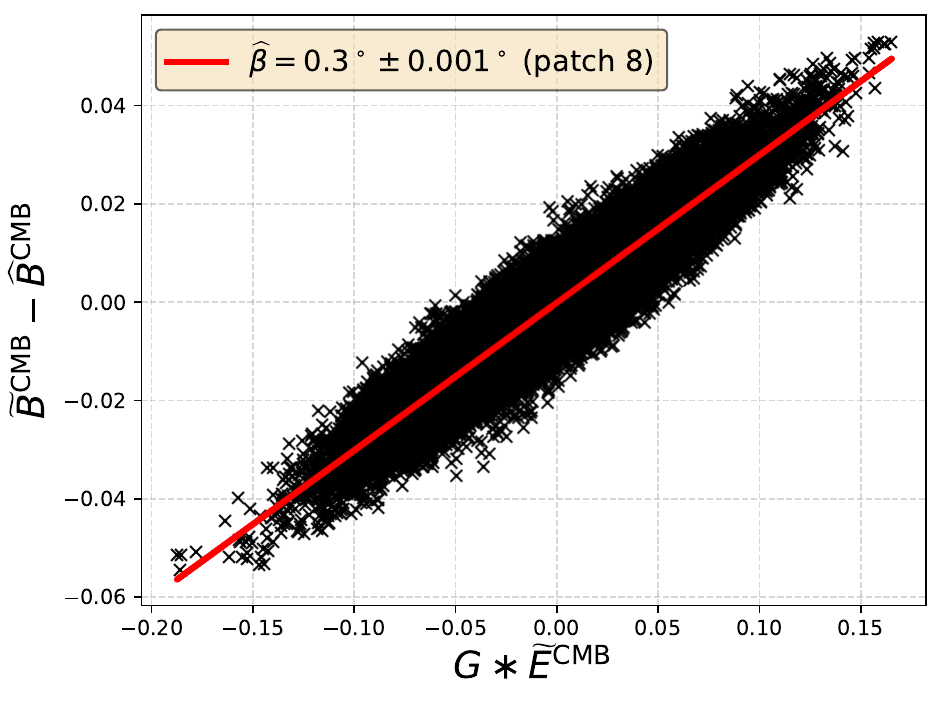}\\
\includegraphics[width=0.25\textwidth,clip]{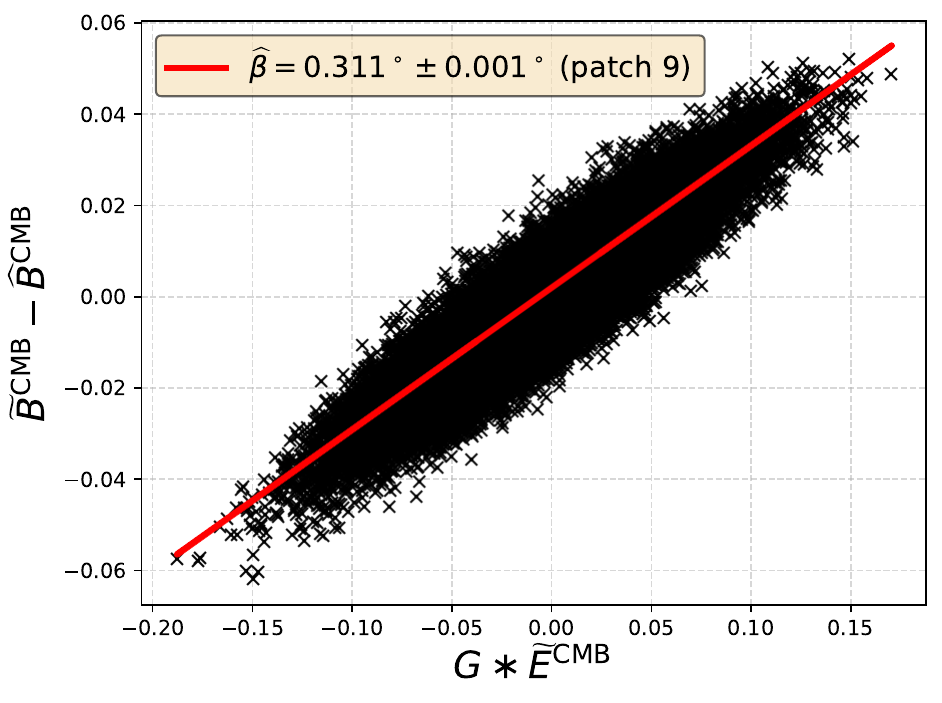}~
\includegraphics[width=0.25\textwidth,clip]{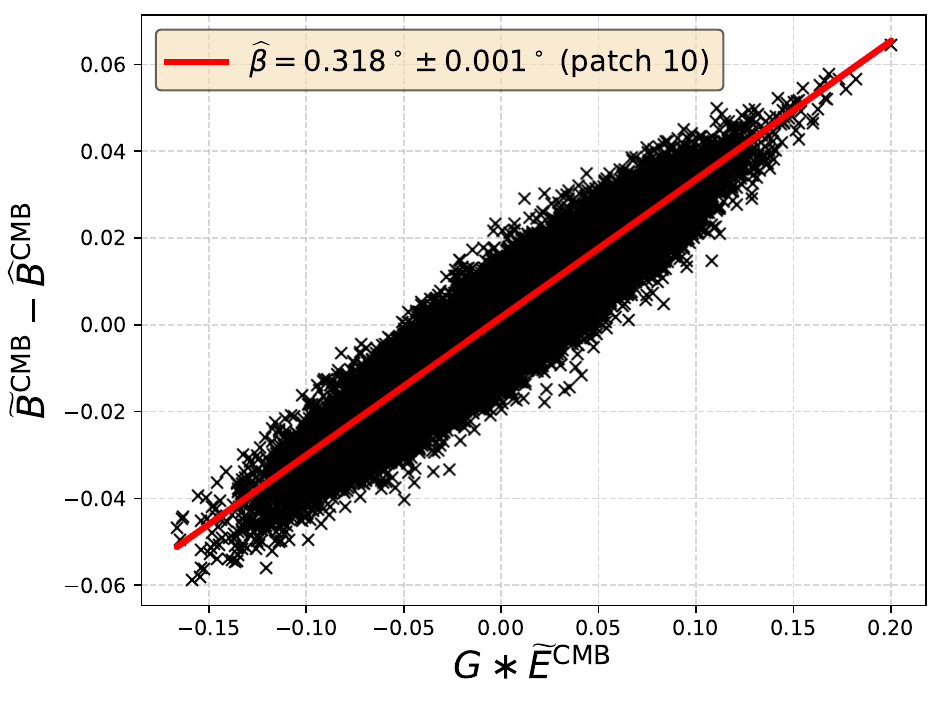}~
\includegraphics[width=0.25\textwidth,clip]{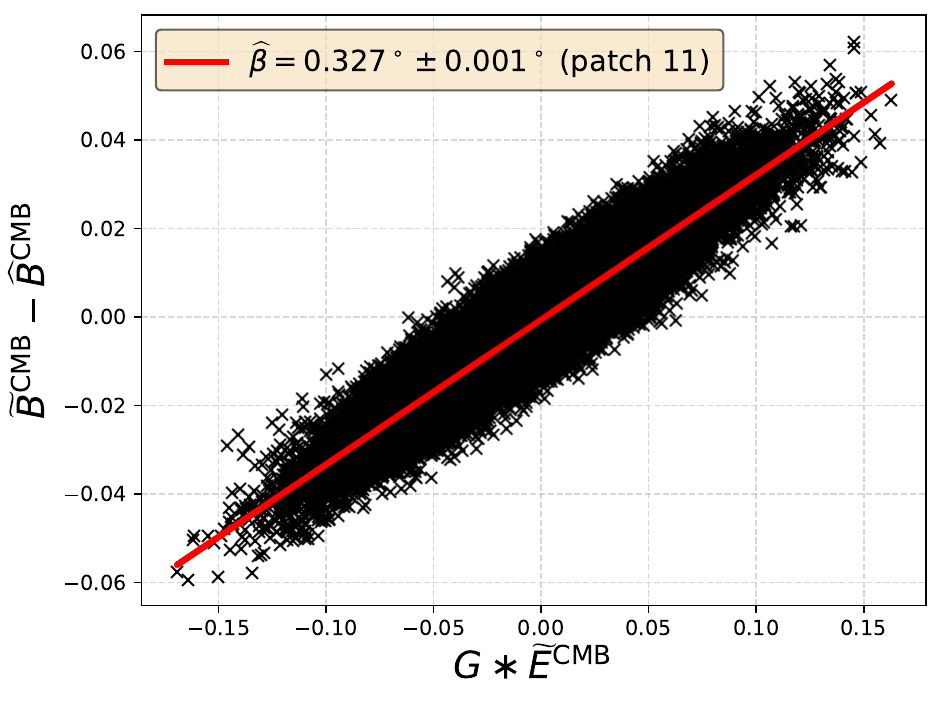}~
\includegraphics[width=0.25\textwidth,clip]{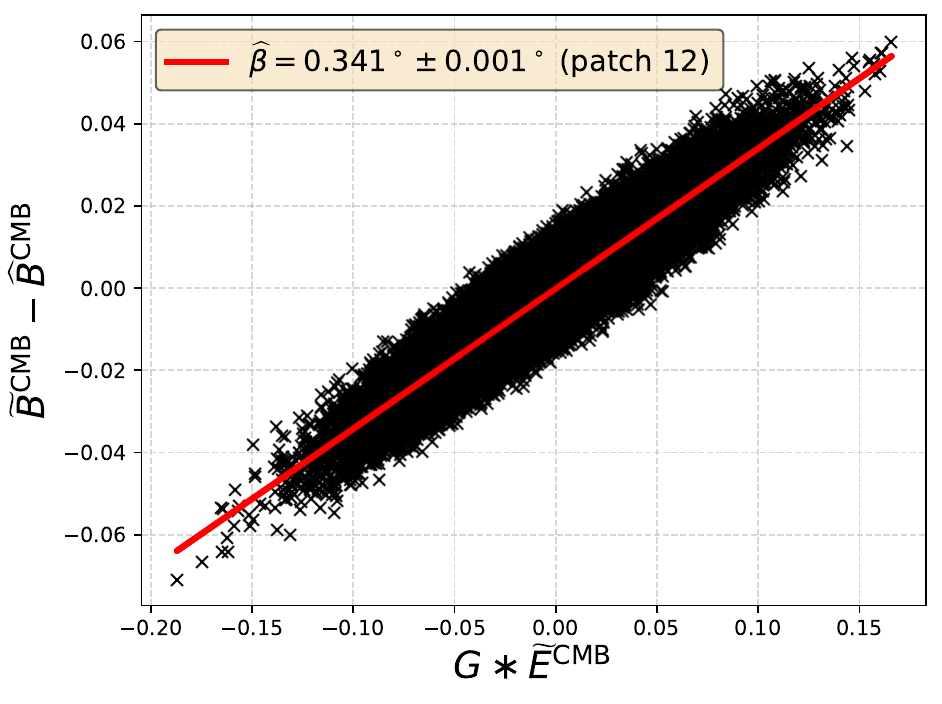}\\
\includegraphics[width=0.25\textwidth,clip]{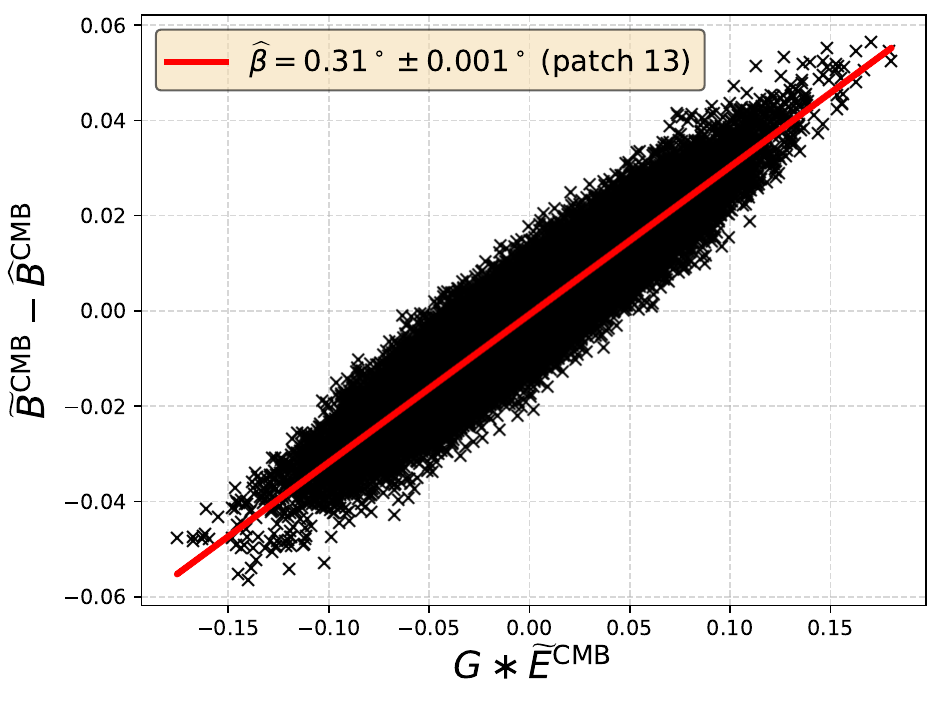}~
\includegraphics[width=0.25\textwidth,clip]{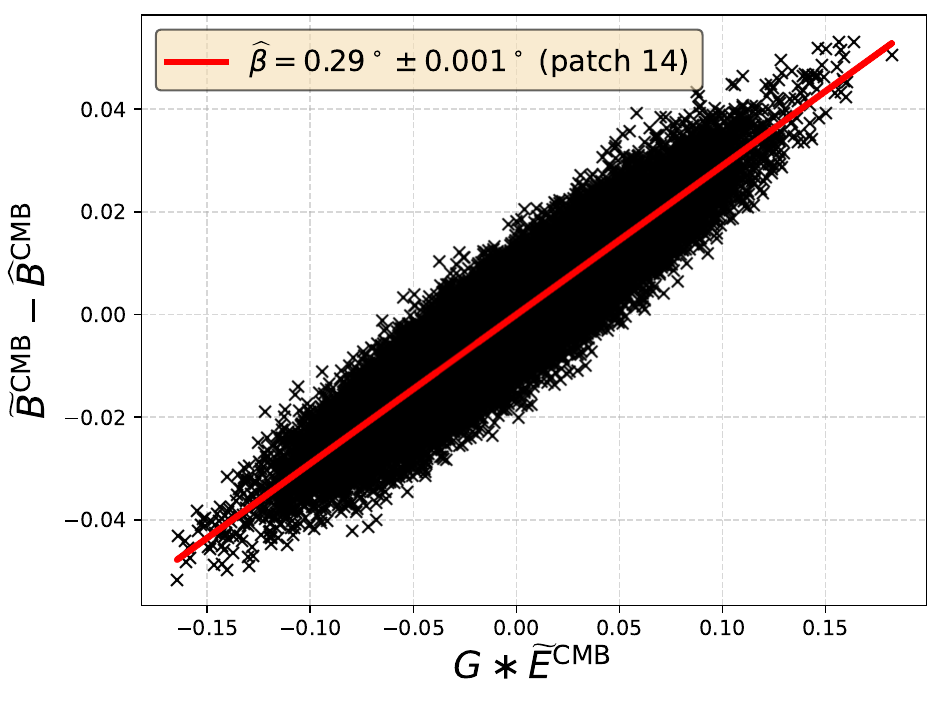}~
\includegraphics[width=0.25\textwidth,clip]{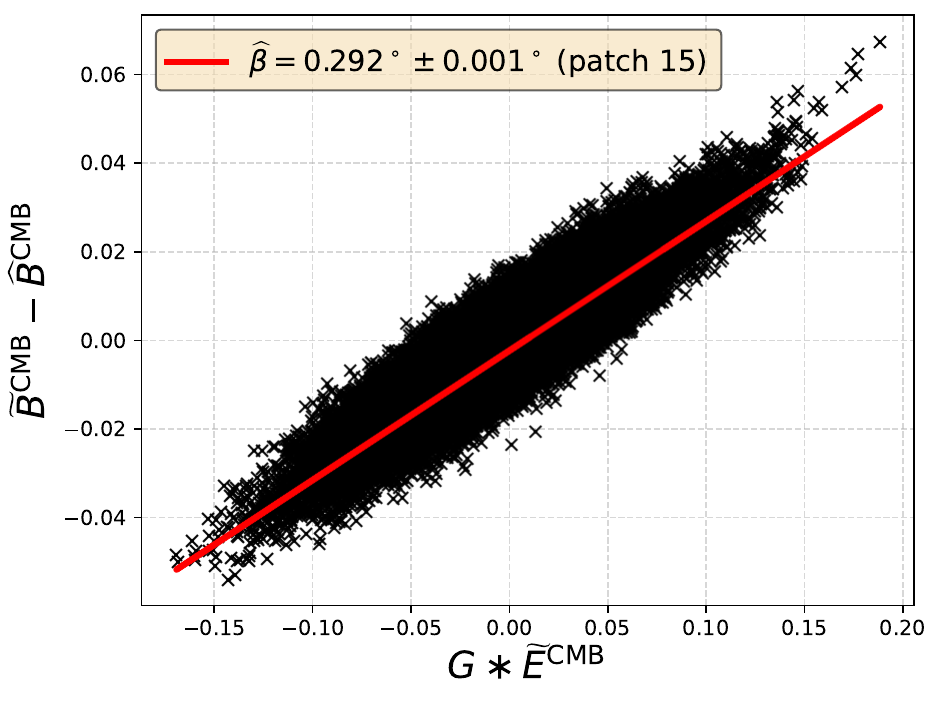}~
\includegraphics[width=0.25\textwidth,clip]{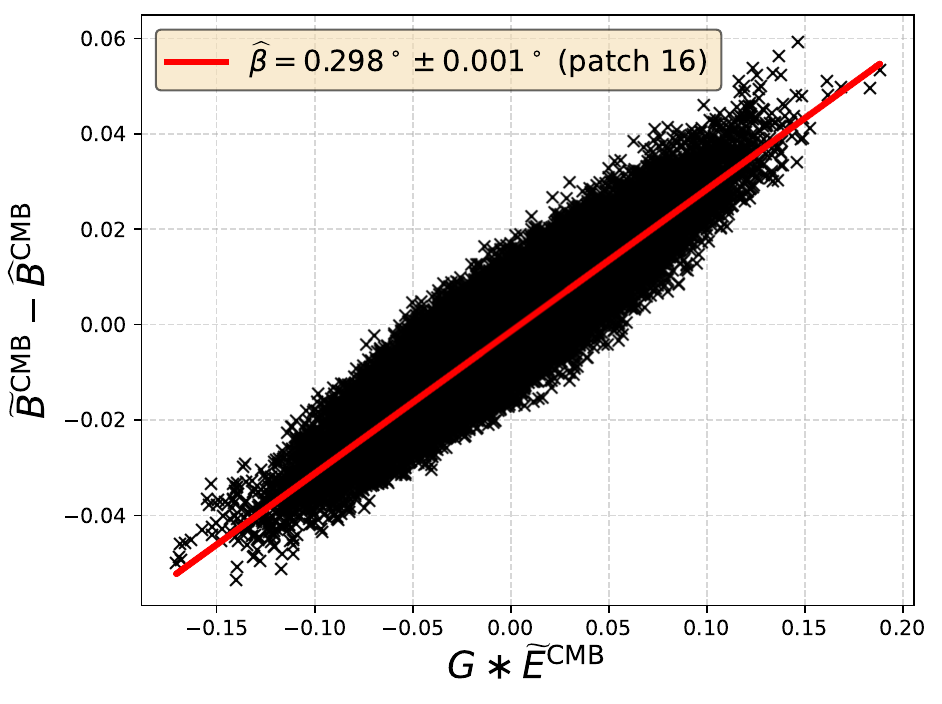}\\
\includegraphics[width=0.25\textwidth,clip]{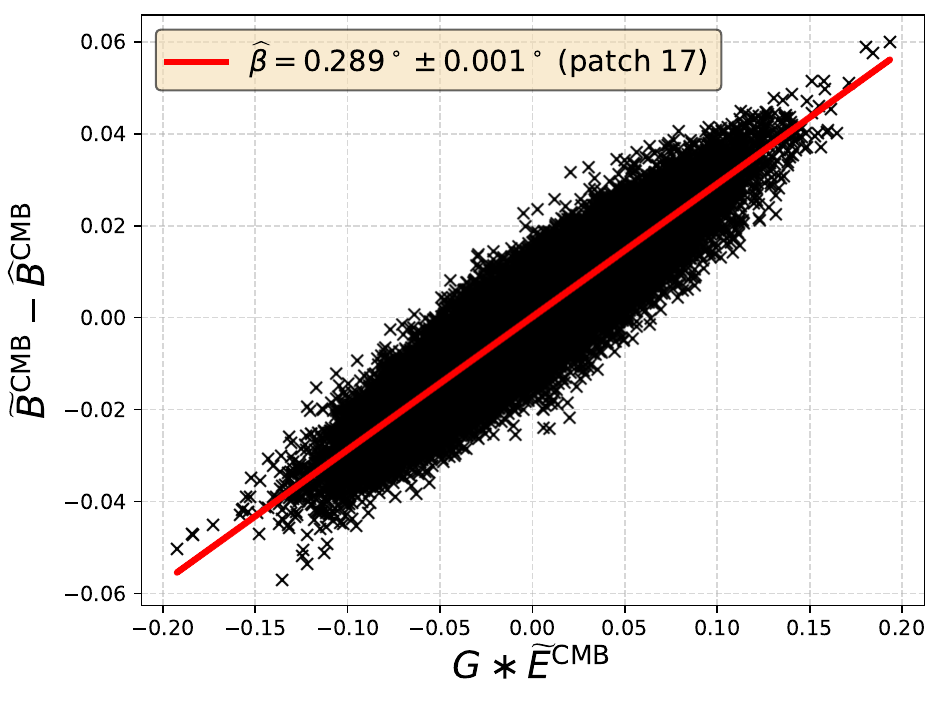}~
\includegraphics[width=0.25\textwidth,clip]{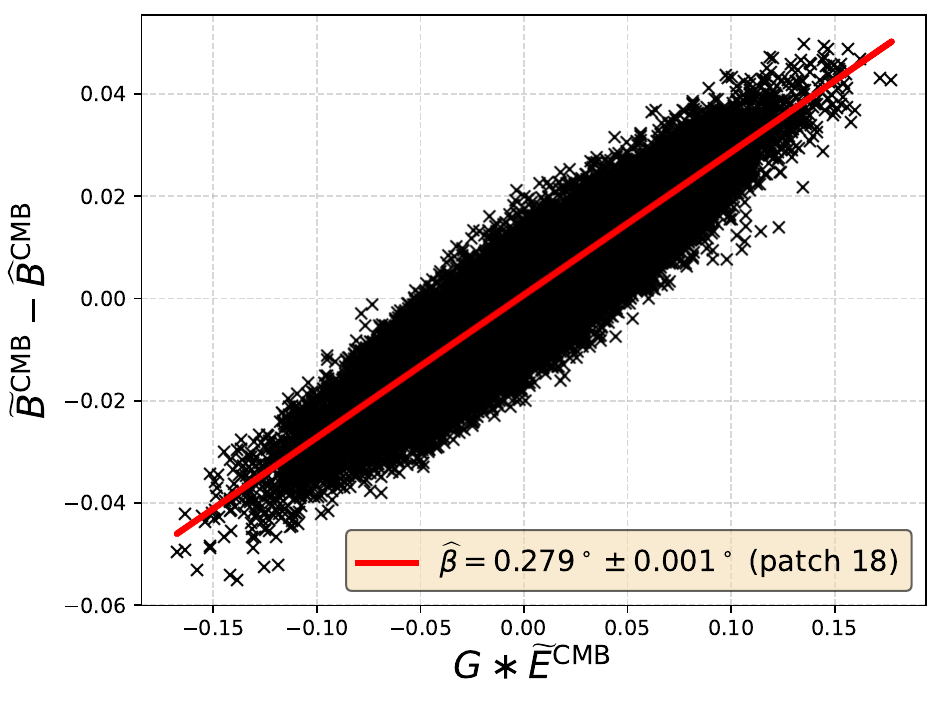}~
\includegraphics[width=0.25\textwidth,clip]{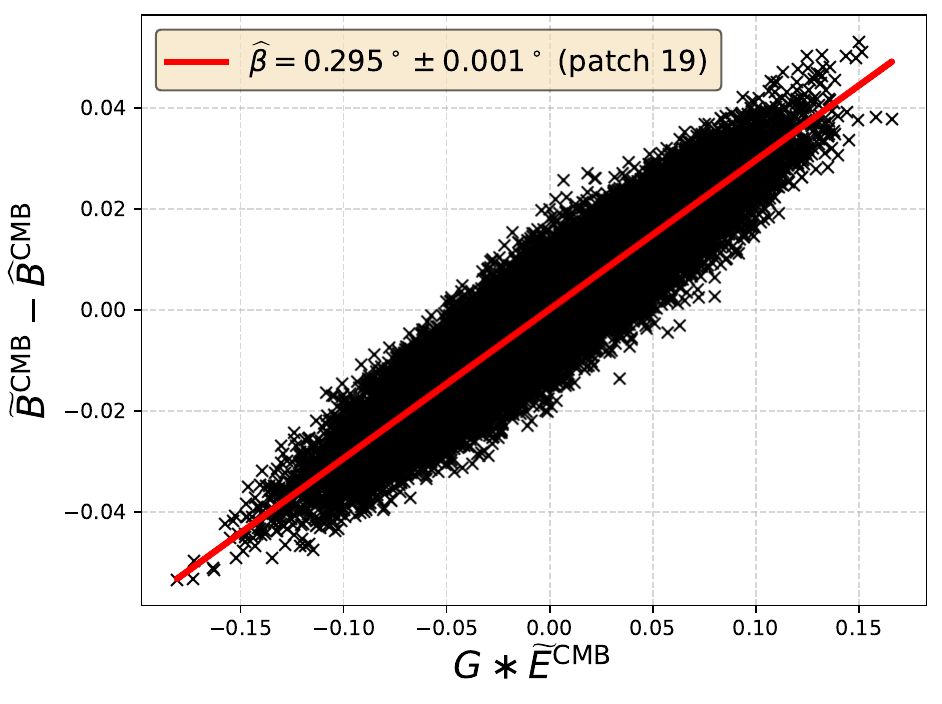}~
\includegraphics[width=0.25\textwidth,clip]{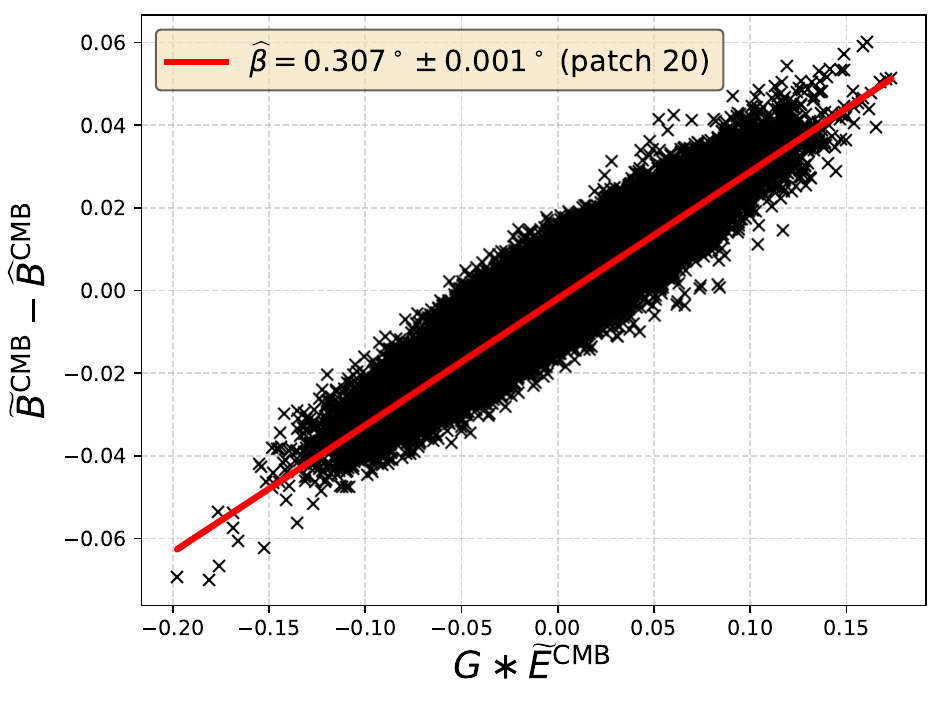}\\
\includegraphics[width=0.25\textwidth,clip]{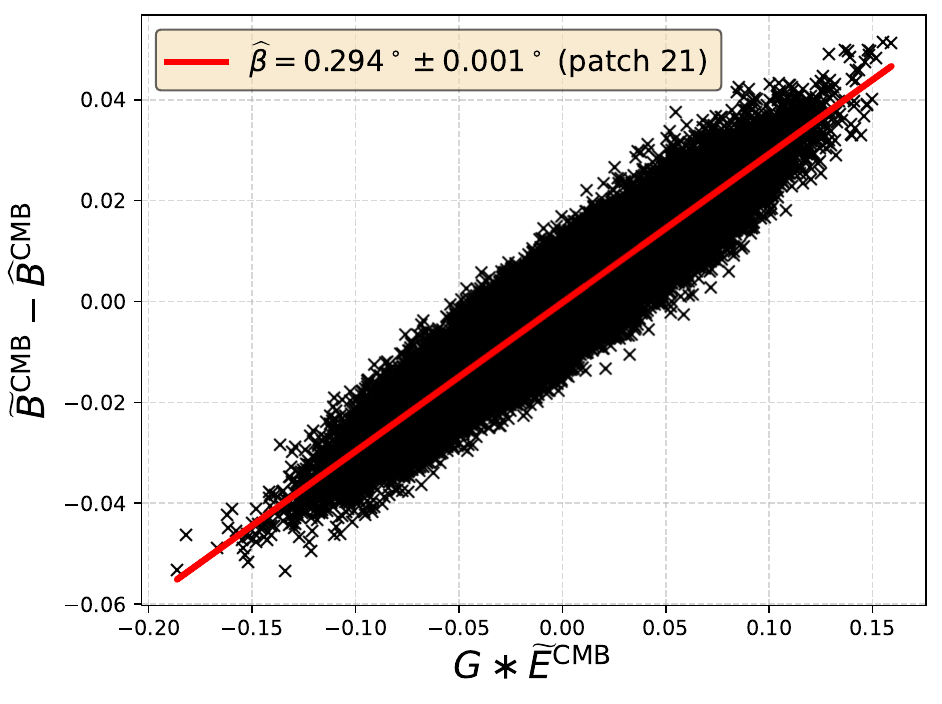}~
\includegraphics[width=0.25\textwidth,clip]{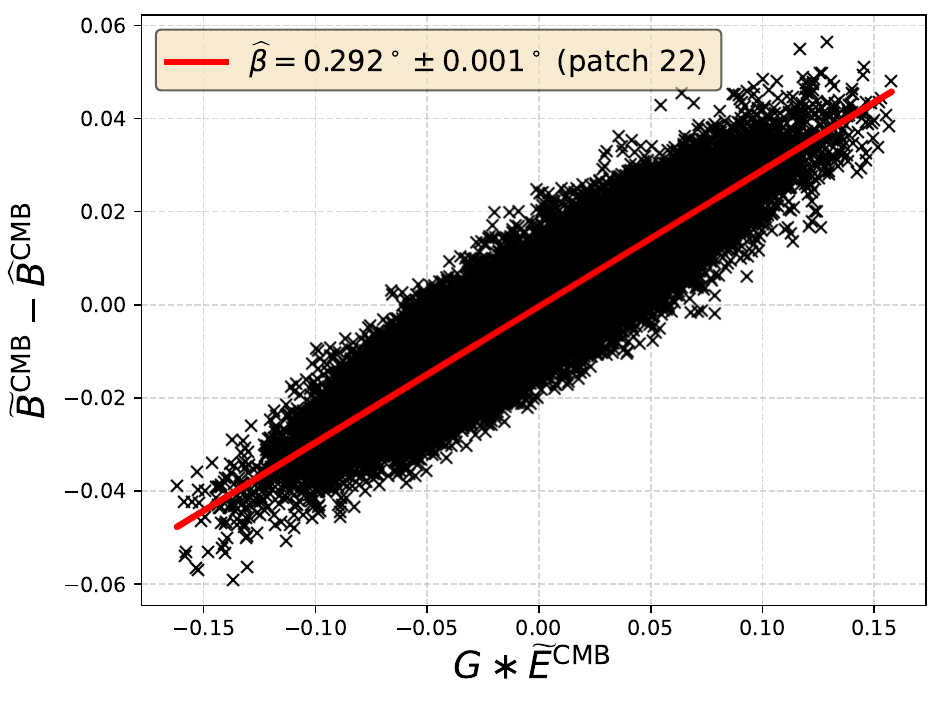}~
\includegraphics[width=0.25\textwidth,clip]{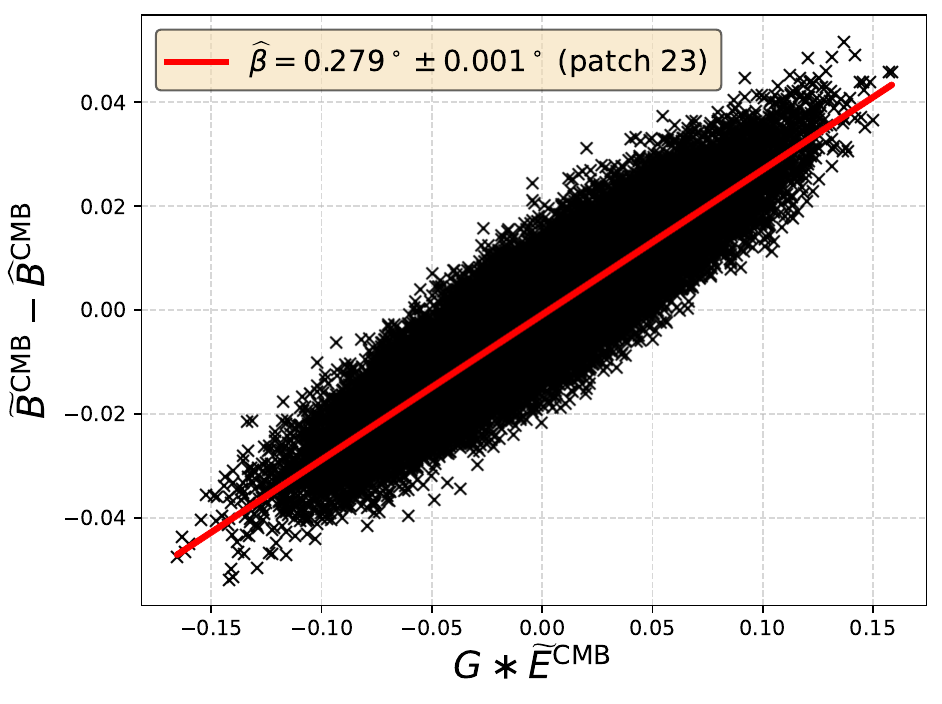}~
\includegraphics[width=0.25\textwidth,clip]{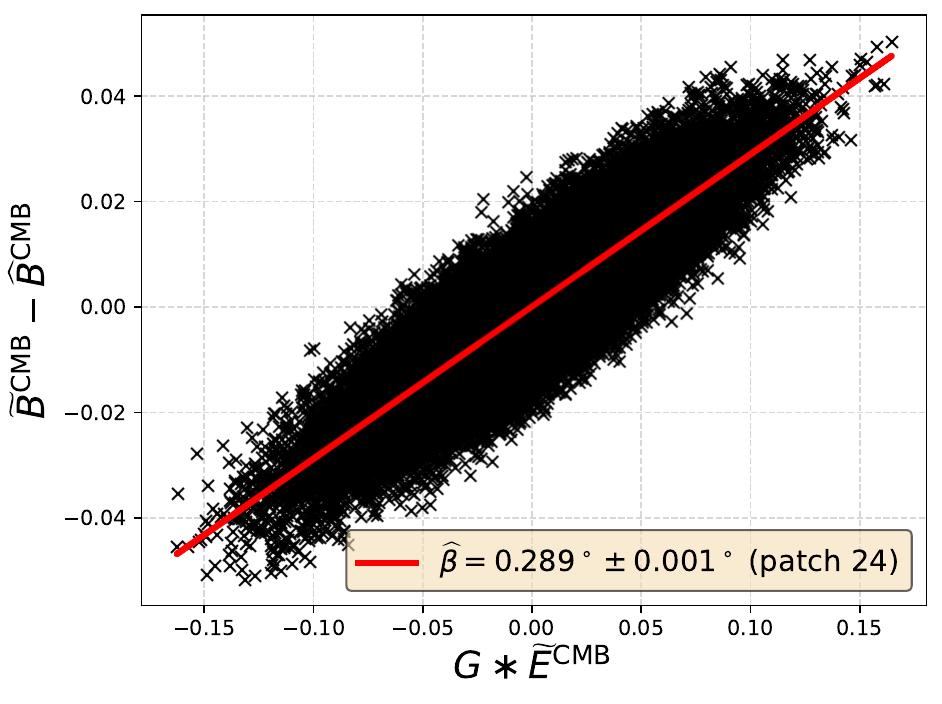}\\
\includegraphics[width=0.25\textwidth,clip]{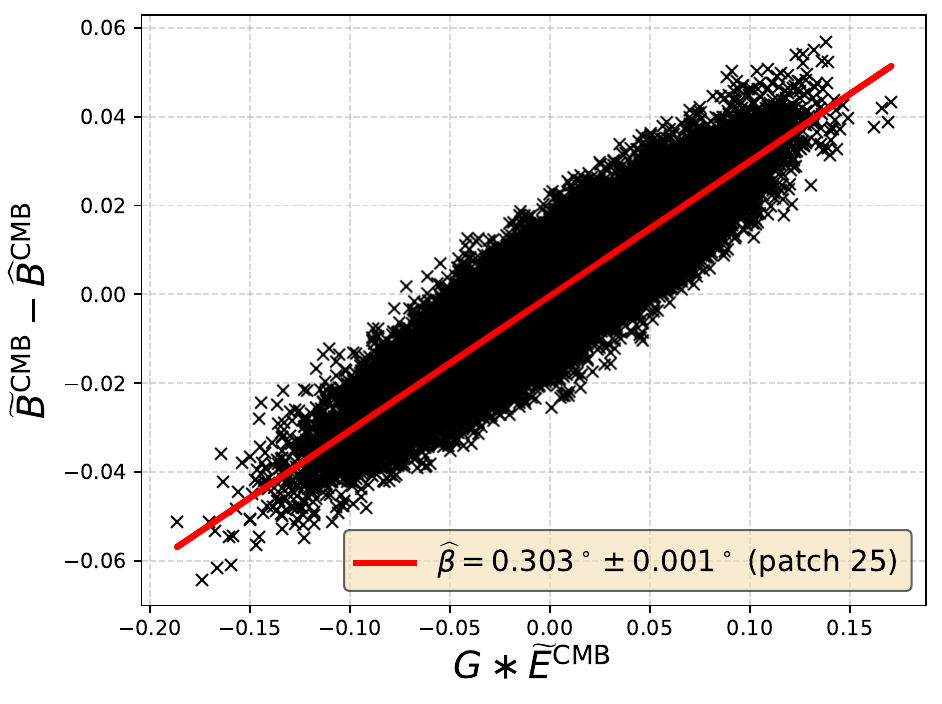}~
\includegraphics[width=0.25\textwidth,clip]{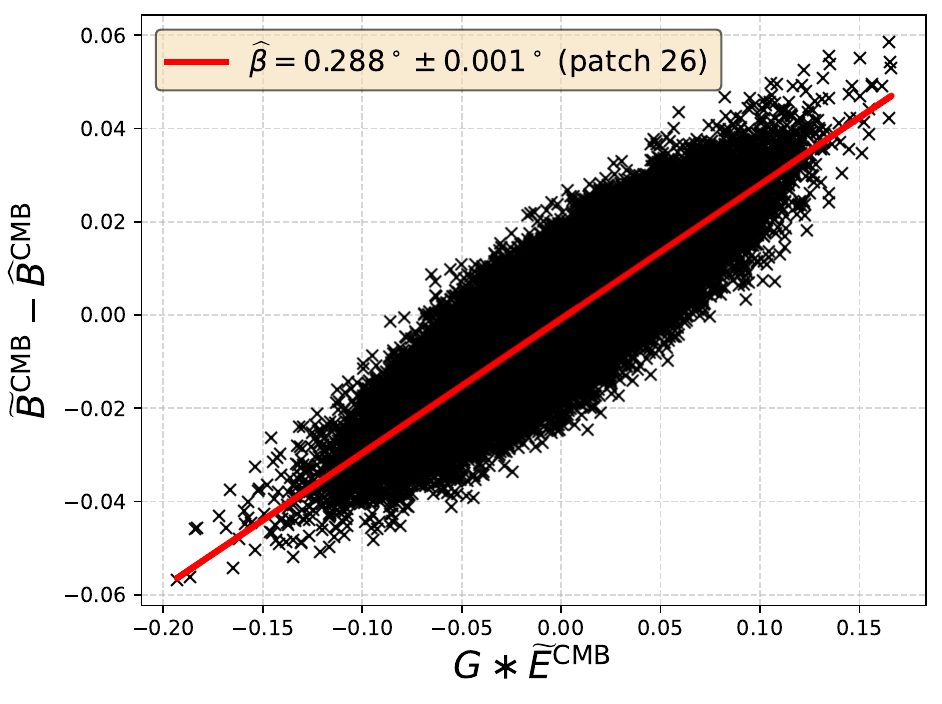}~
\hfill
\caption{\label{fig:ttplot_b} Same as Figure~\ref{fig:ttplot}, i.e., including cosmic birefringence ($\beta = 0.3^\circ$, $\alpha \neq 0$), but using linear regression between the reconstructed CMB fields $\widetilde{B}^{\rm CMB} - \widehat{B}^{\rm CMB}$ and $G \ast \widetilde{E}^{\rm CMB}$. In this case, the mean recovered birefringence angle is $\widehat{\beta} = 0.305^\circ \pm 0.016^\circ$, corresponding to a $19\sigma$ detection of $\beta = 0.3^\circ$, with a negligible bias of $0.3\sigma$.}
\end{figure}

\begin{figure}[tbp]
\centering 
\includegraphics[width=0.25\textwidth,clip]{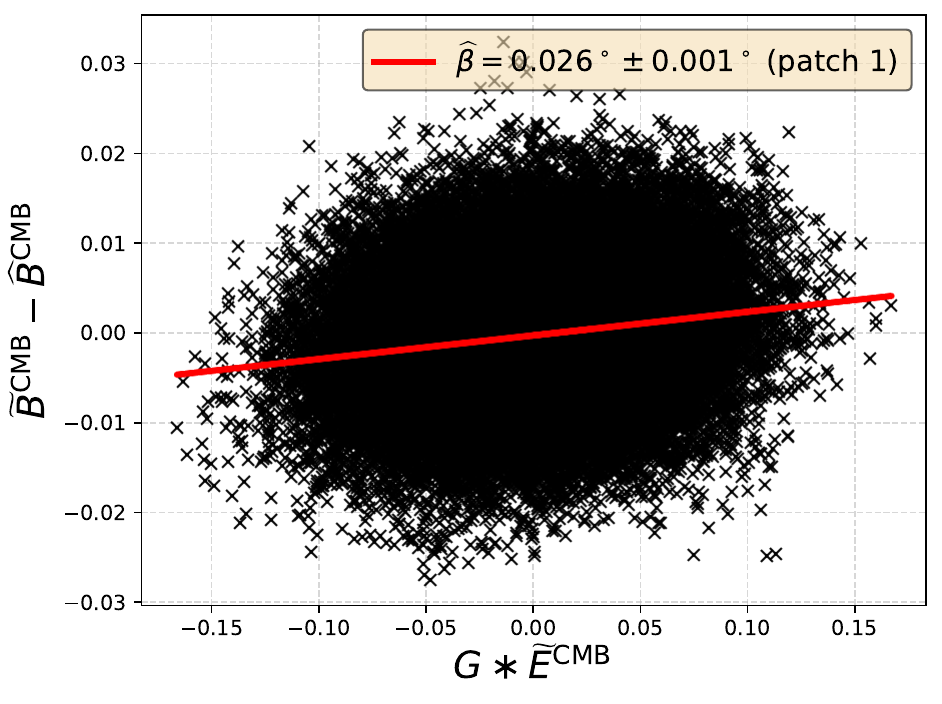}~
\includegraphics[width=0.25\textwidth,clip]{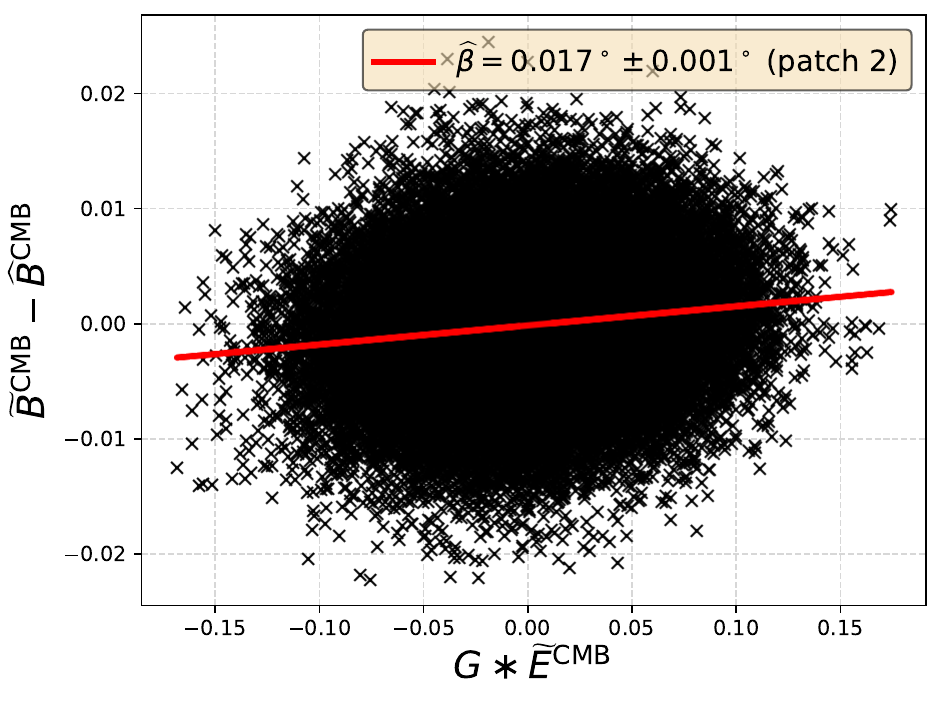}~
\includegraphics[width=0.25\textwidth,clip]{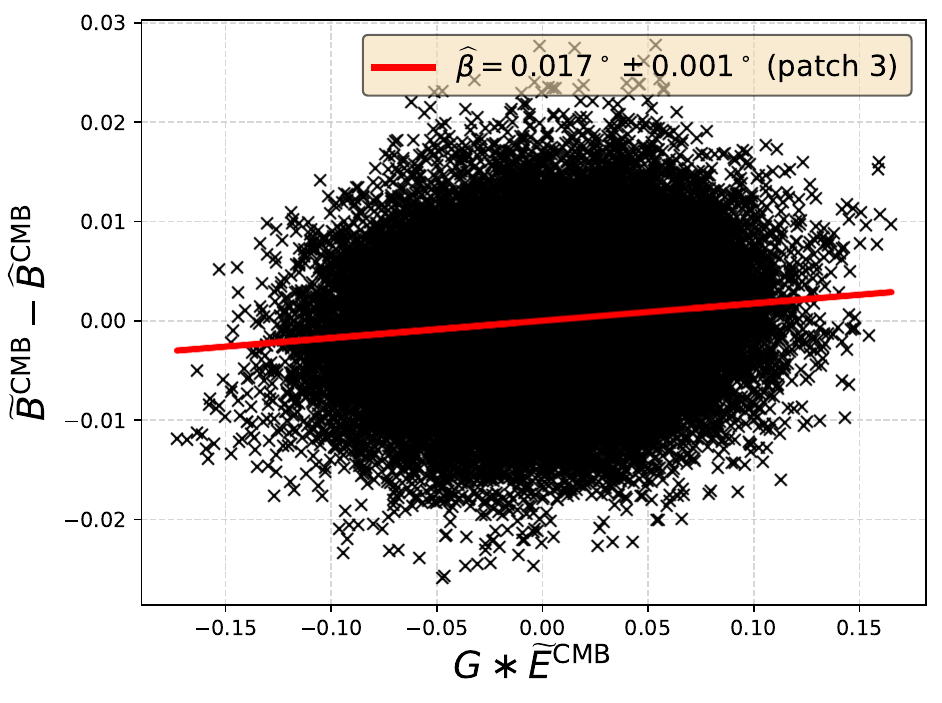}~
\includegraphics[width=0.25\textwidth,clip]{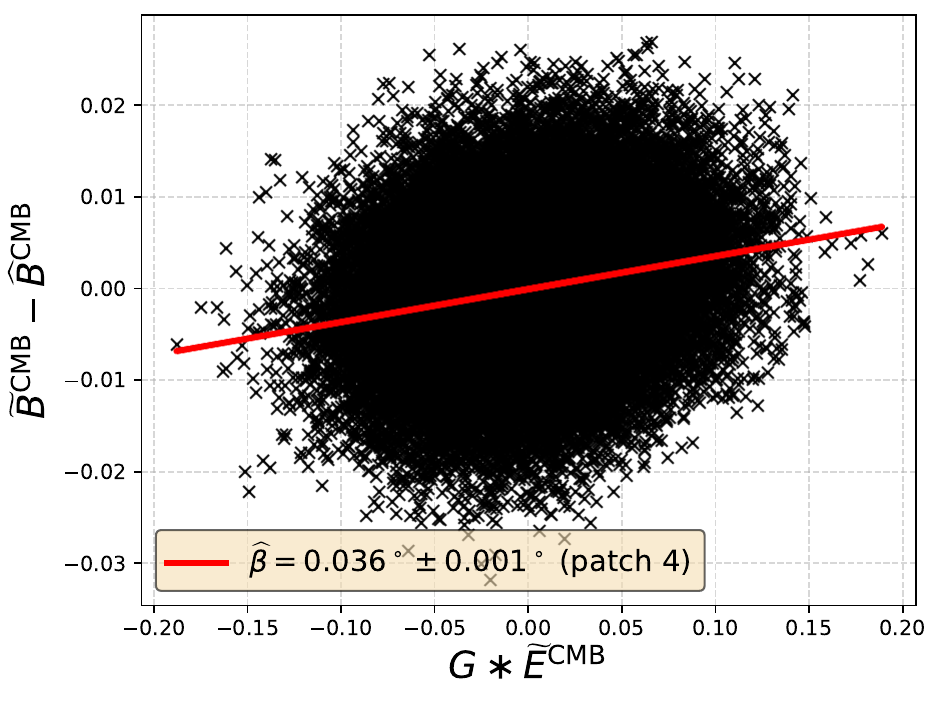}\\
\includegraphics[width=0.25\textwidth,clip]{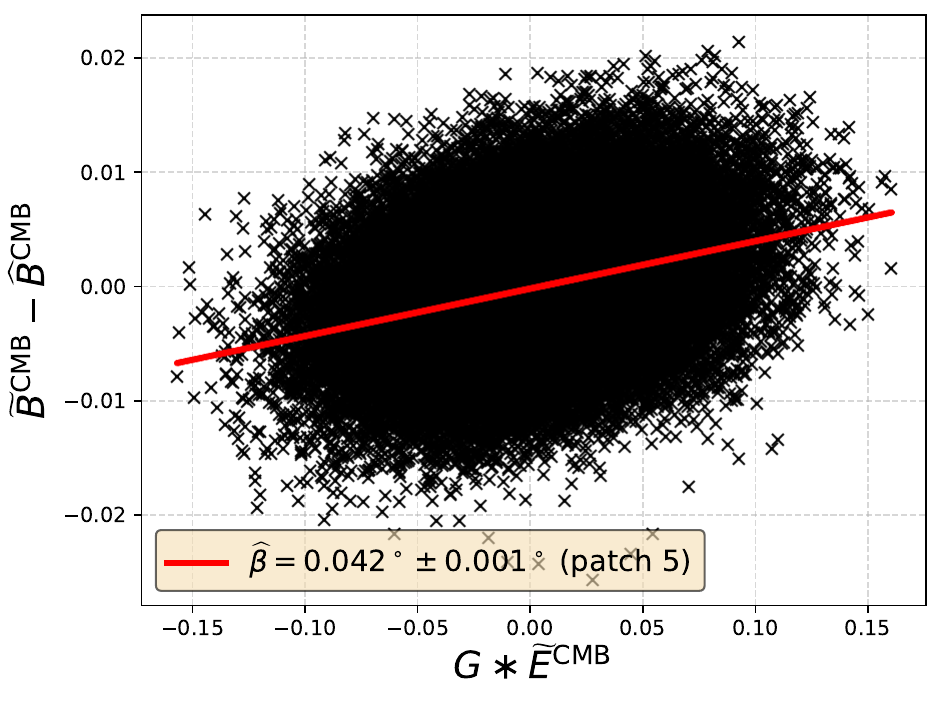}~
\includegraphics[width=0.25\textwidth,clip]{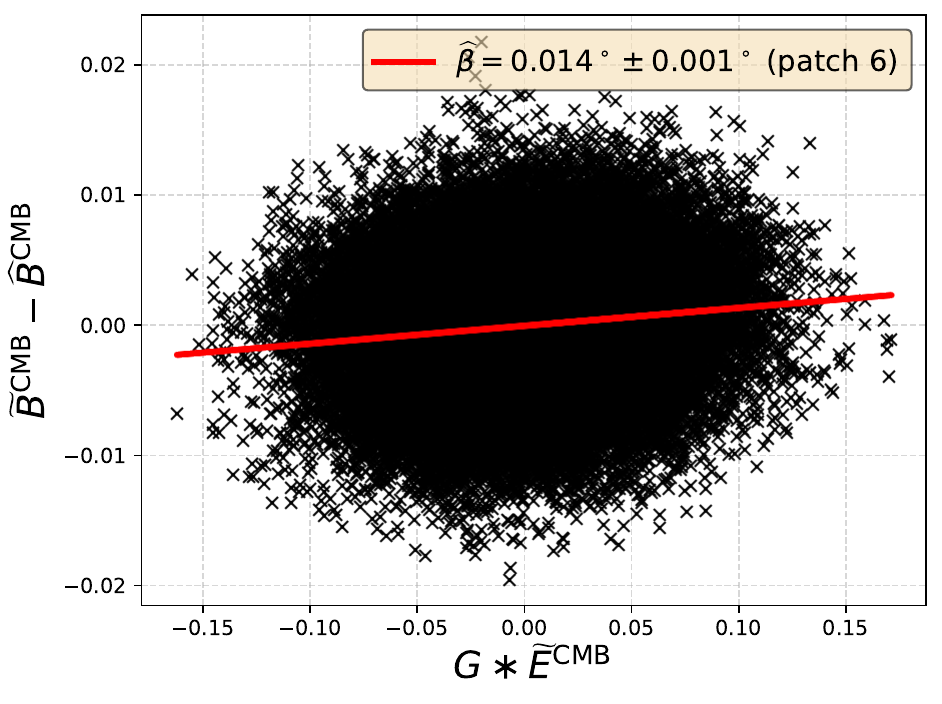}~
\includegraphics[width=0.25\textwidth,clip]{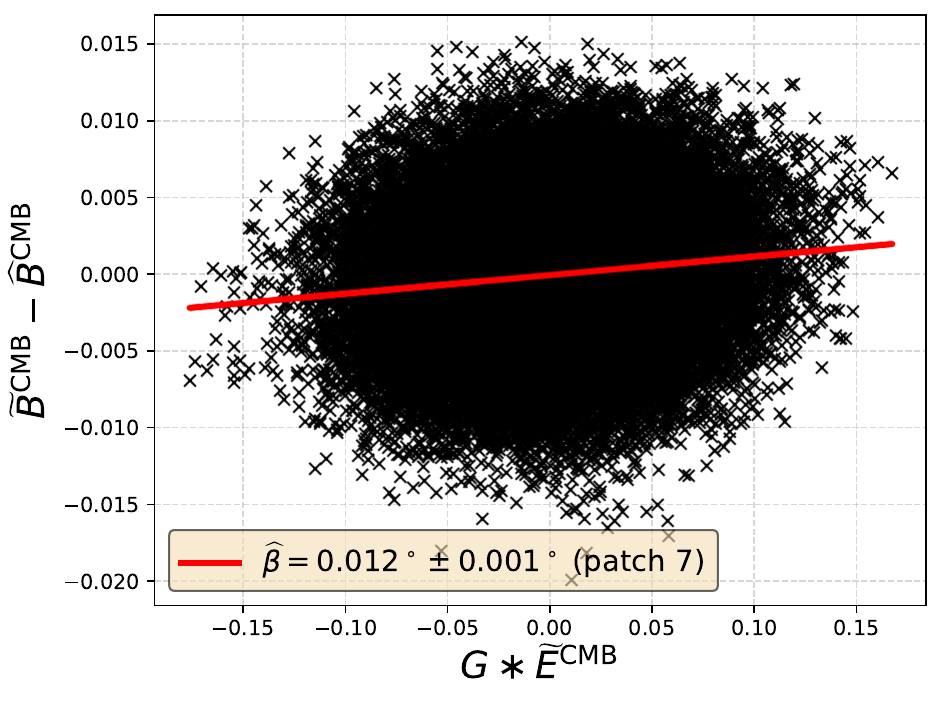}~
\includegraphics[width=0.25\textwidth,clip]{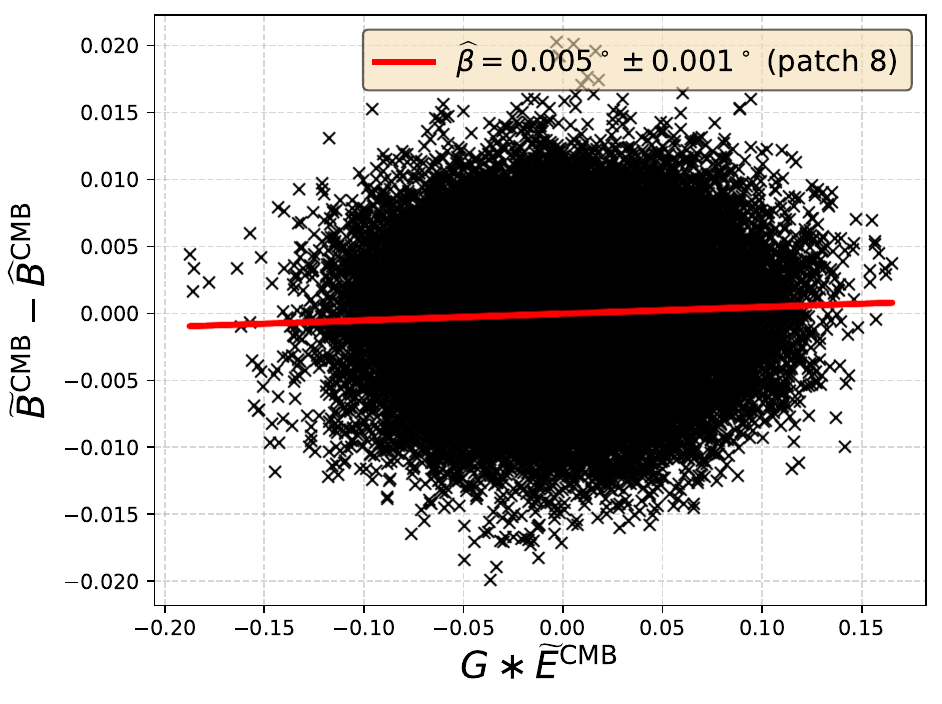}\\
\includegraphics[width=0.25\textwidth,clip]{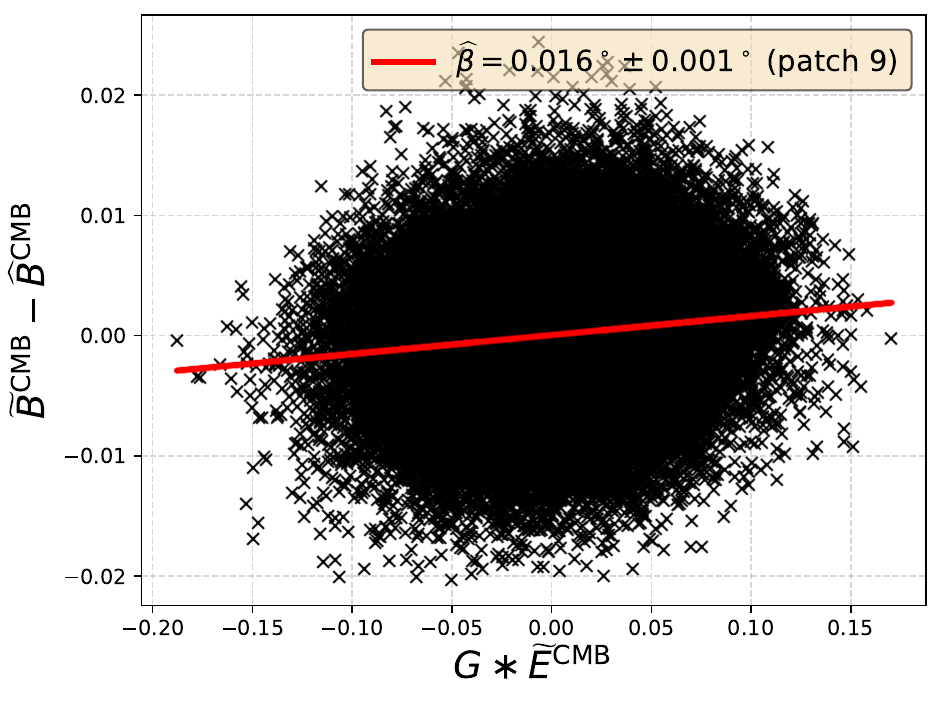}~
\includegraphics[width=0.25\textwidth,clip]{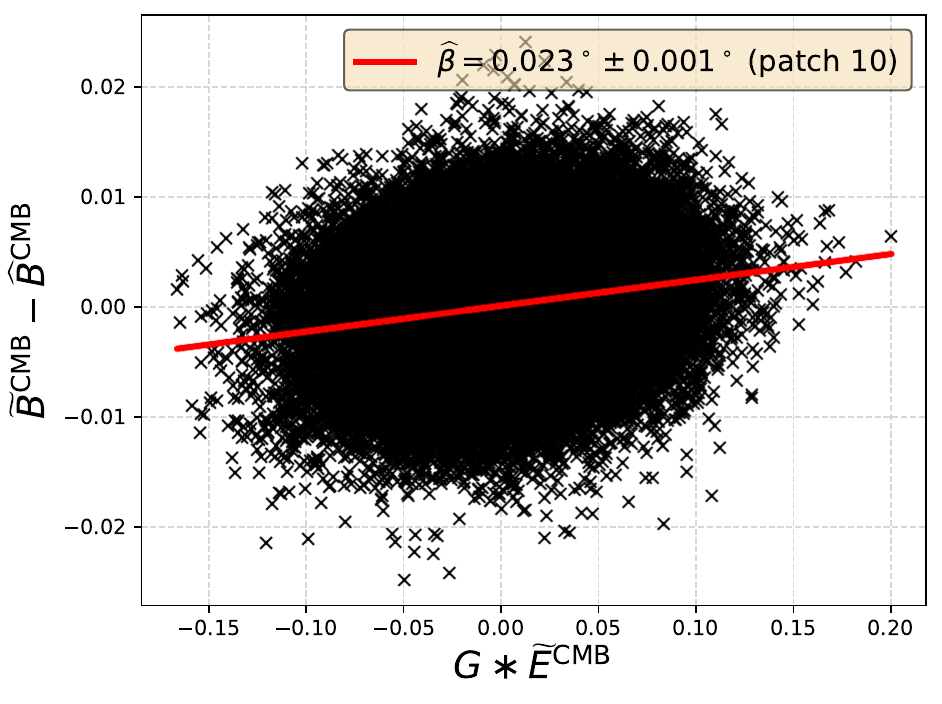}~
\includegraphics[width=0.25\textwidth,clip]{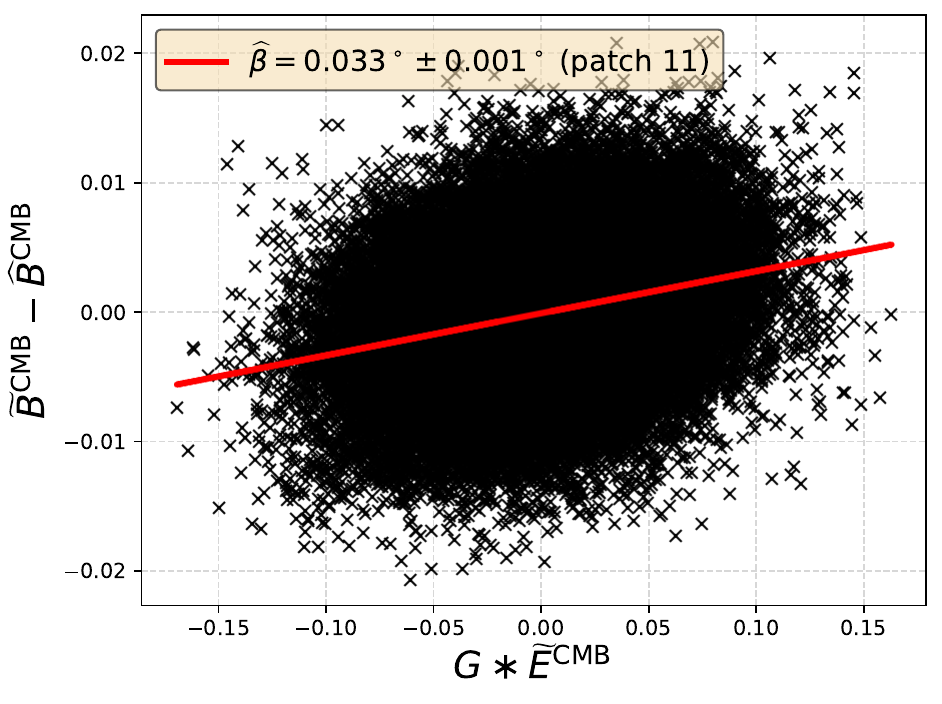}~
\includegraphics[width=0.25\textwidth,clip]{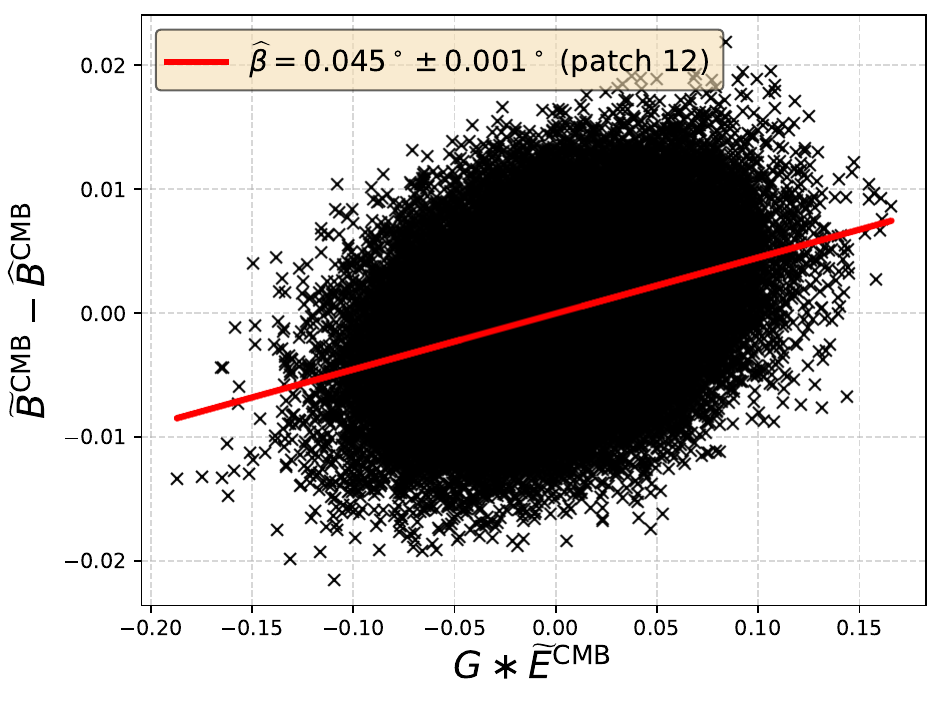}\\
\includegraphics[width=0.25\textwidth,clip]{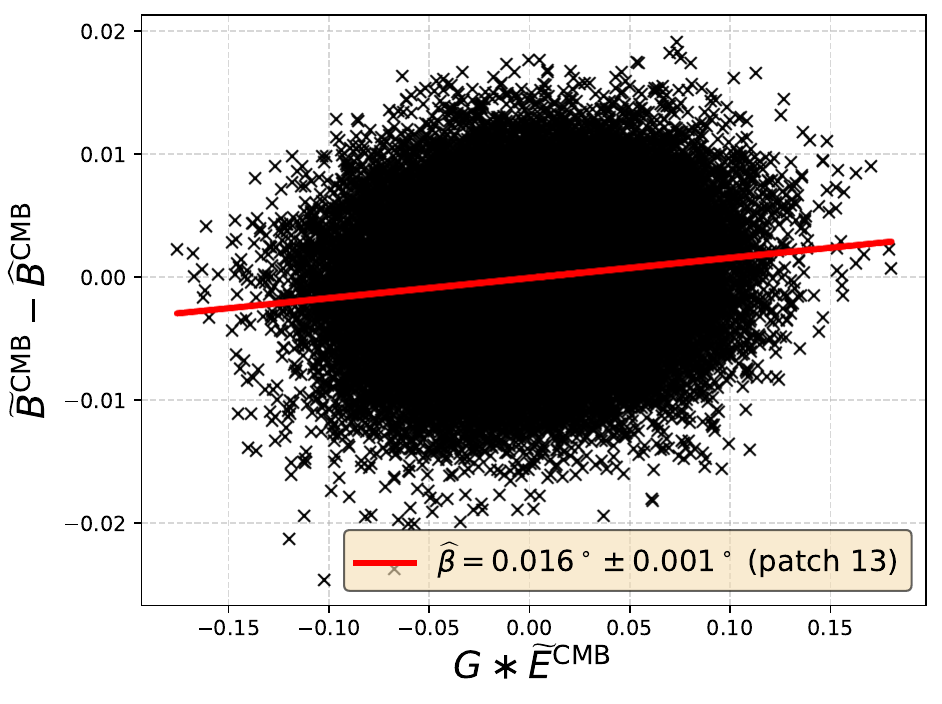}~
\includegraphics[width=0.25\textwidth,clip]{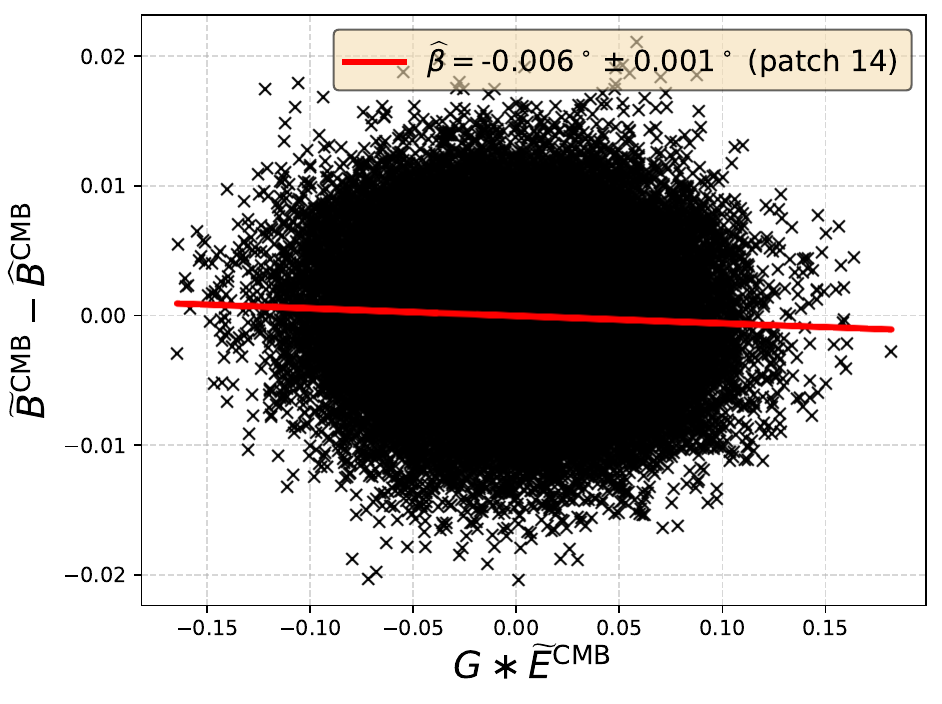}~
\includegraphics[width=0.25\textwidth,clip]{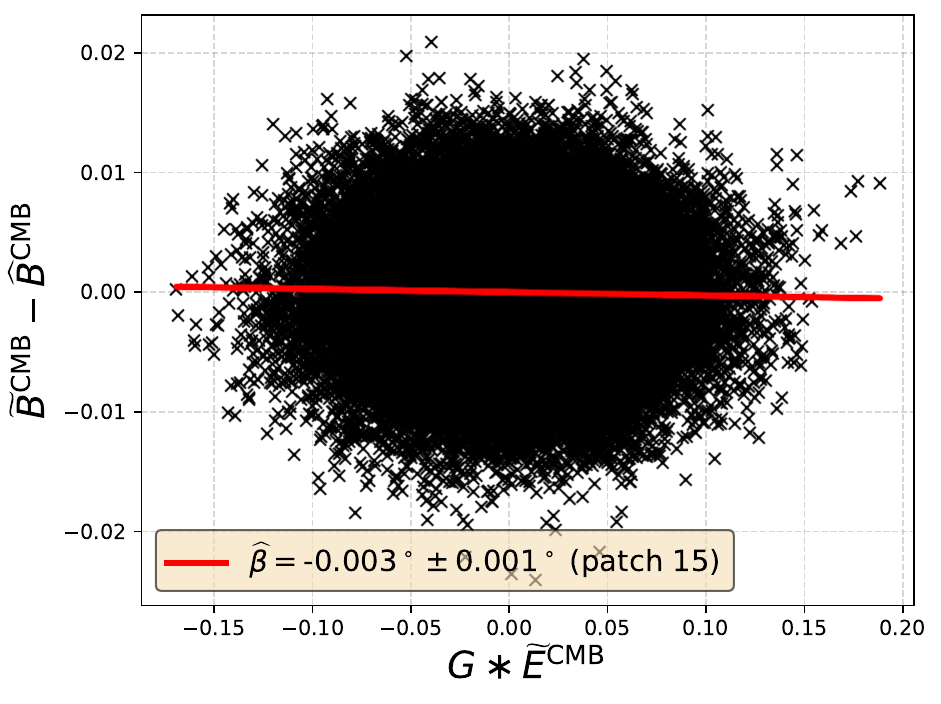}~
\includegraphics[width=0.25\textwidth,clip]{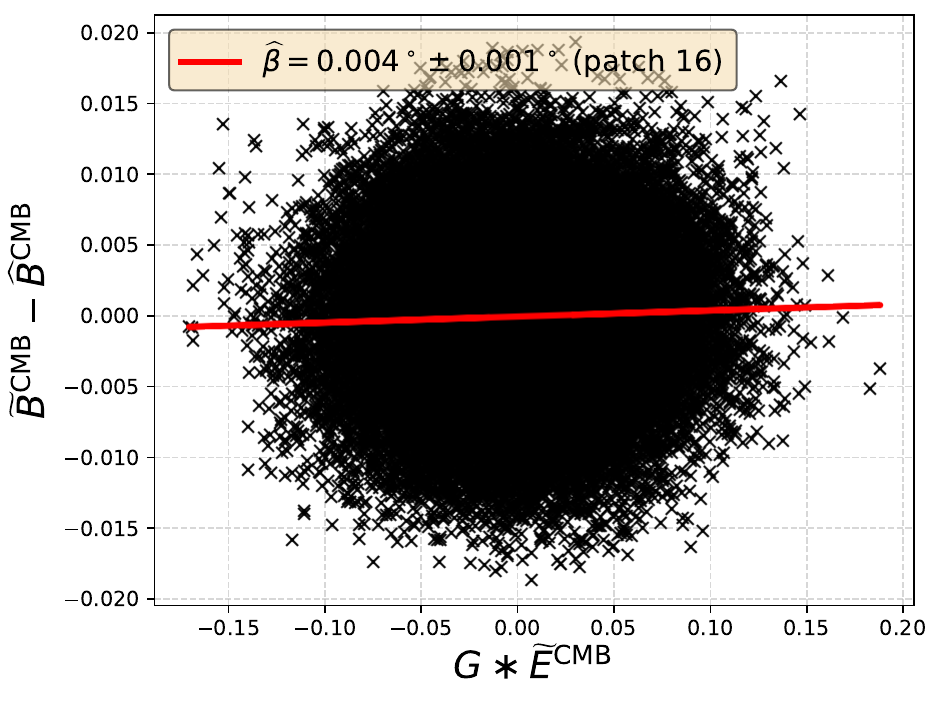}\\
\includegraphics[width=0.25\textwidth,clip]{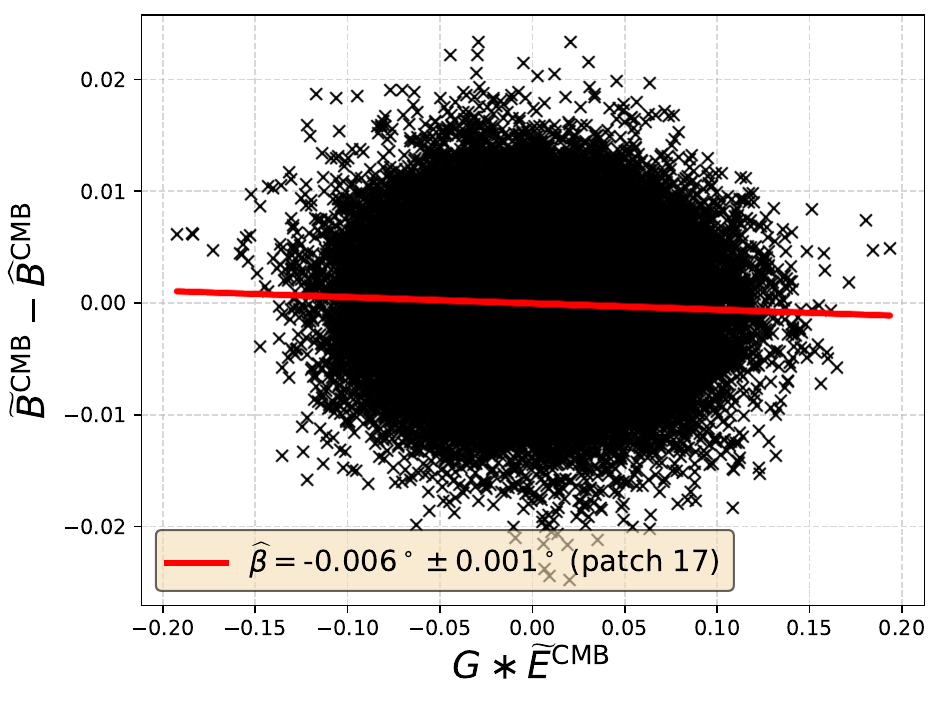}~
\includegraphics[width=0.25\textwidth,clip]{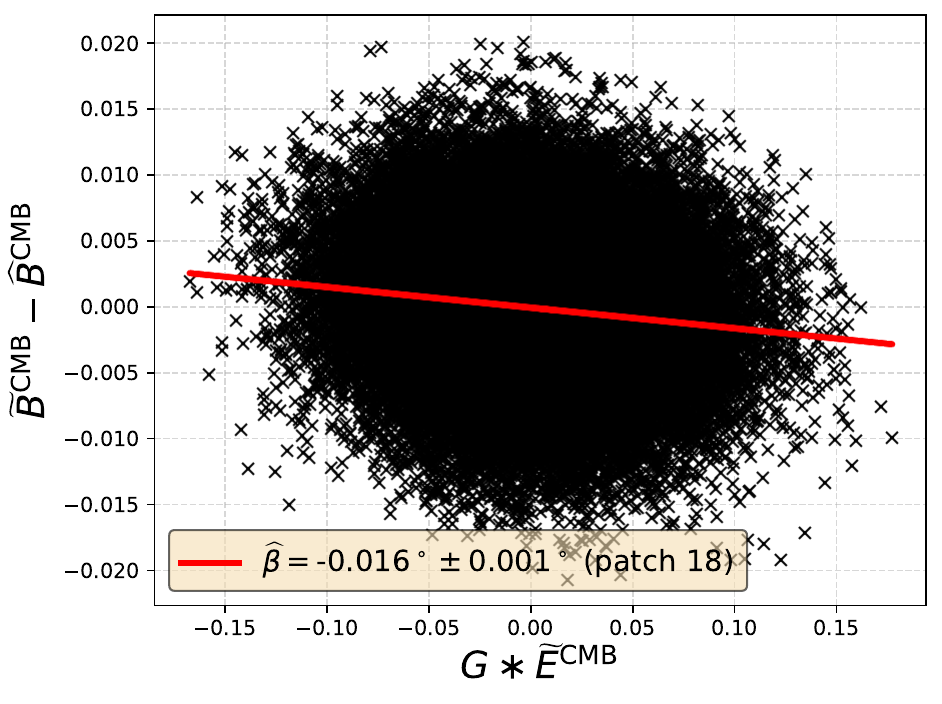}~
\includegraphics[width=0.25\textwidth,clip]{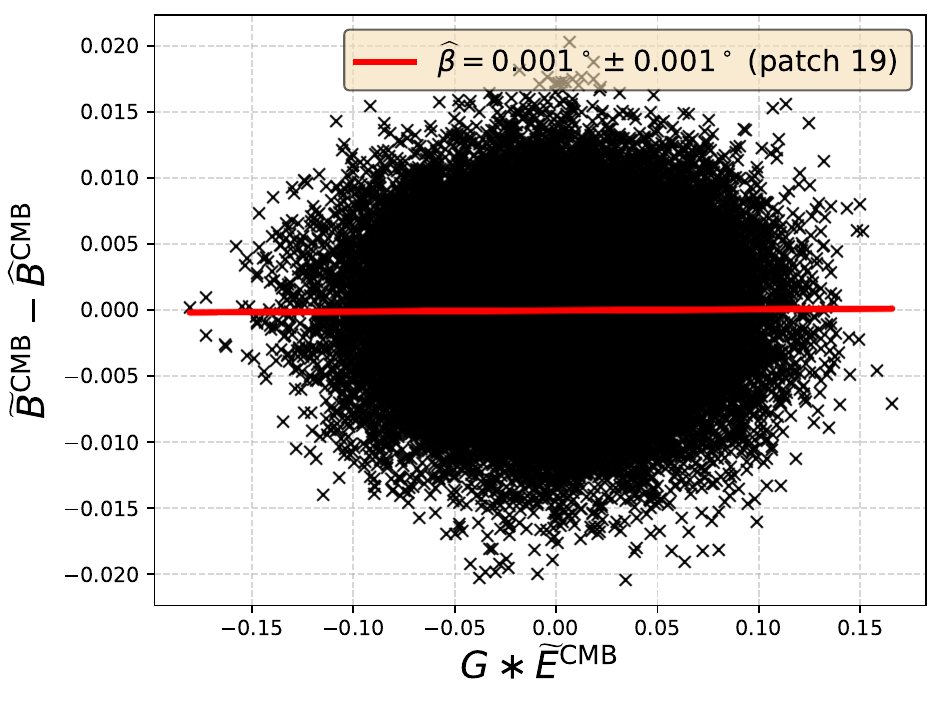}~
\includegraphics[width=0.25\textwidth,clip]{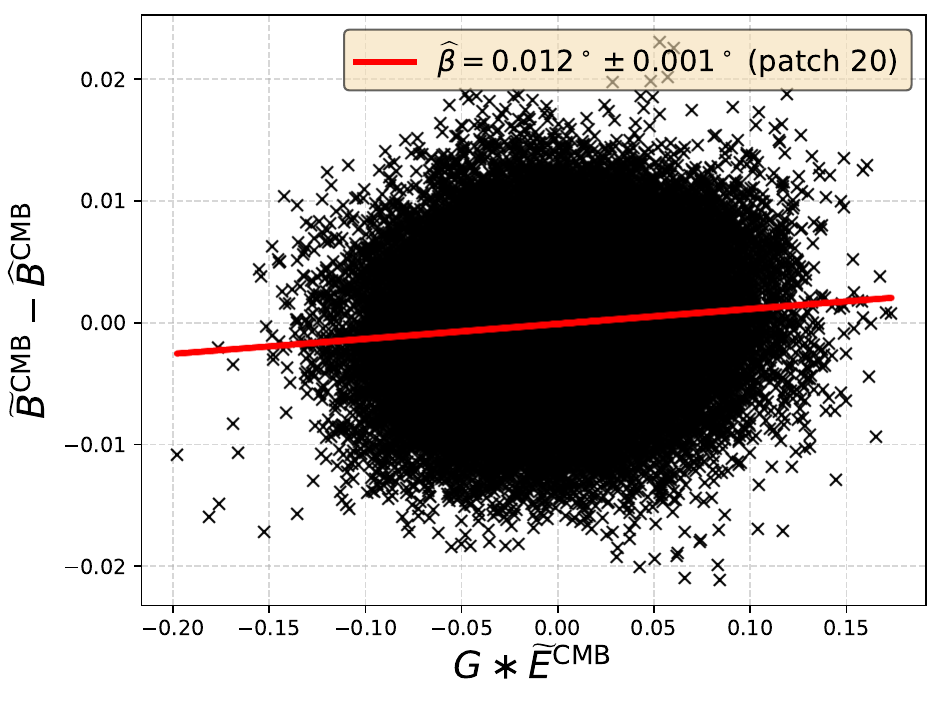}\\
\includegraphics[width=0.25\textwidth,clip]{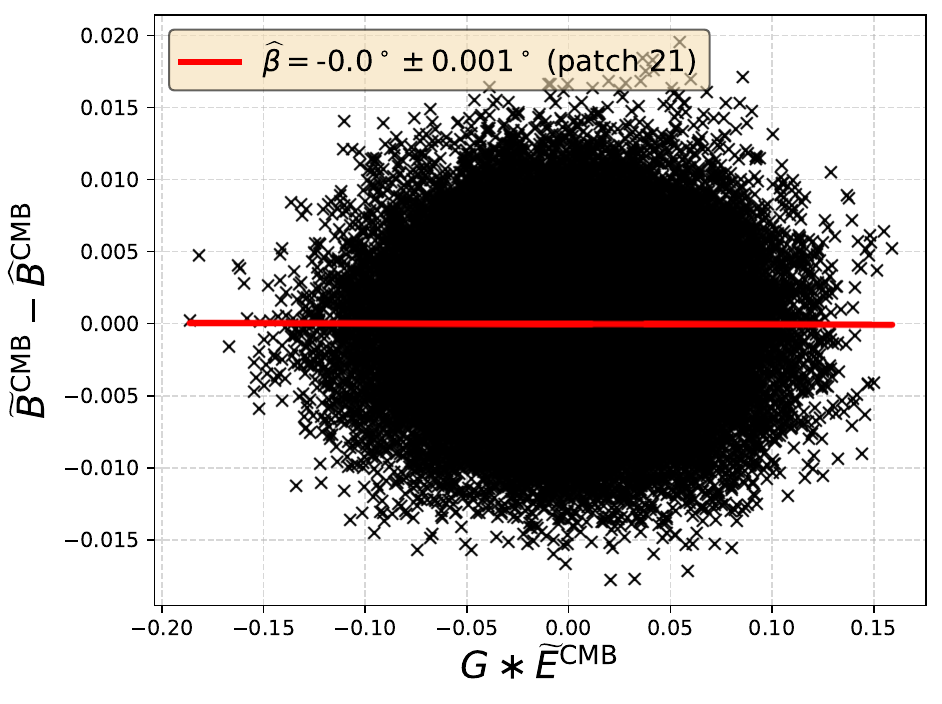}~
\includegraphics[width=0.25\textwidth,clip]{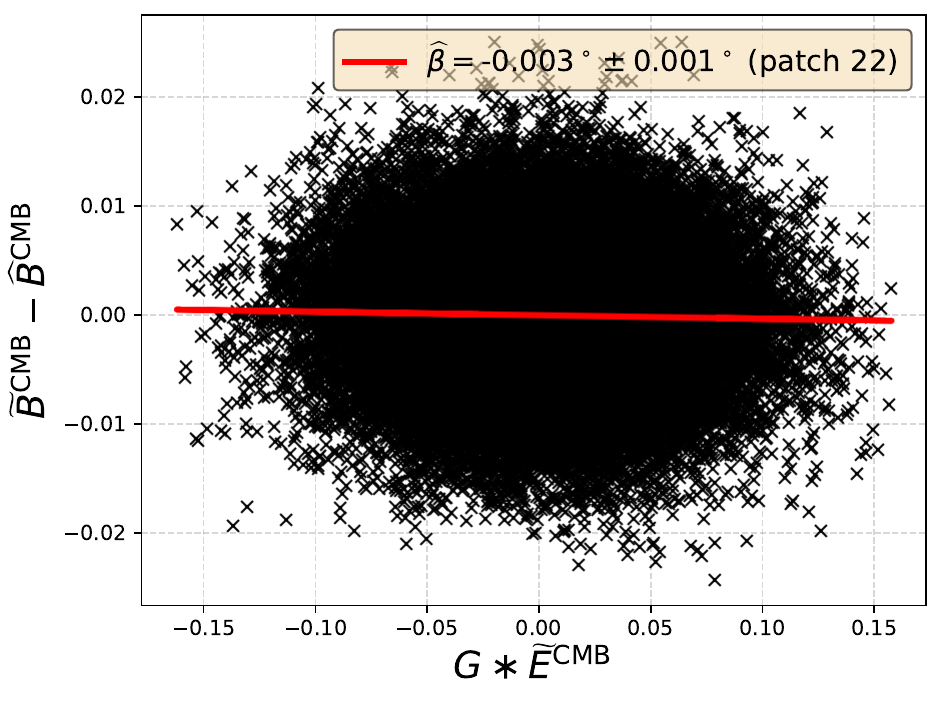}~
\includegraphics[width=0.25\textwidth,clip]{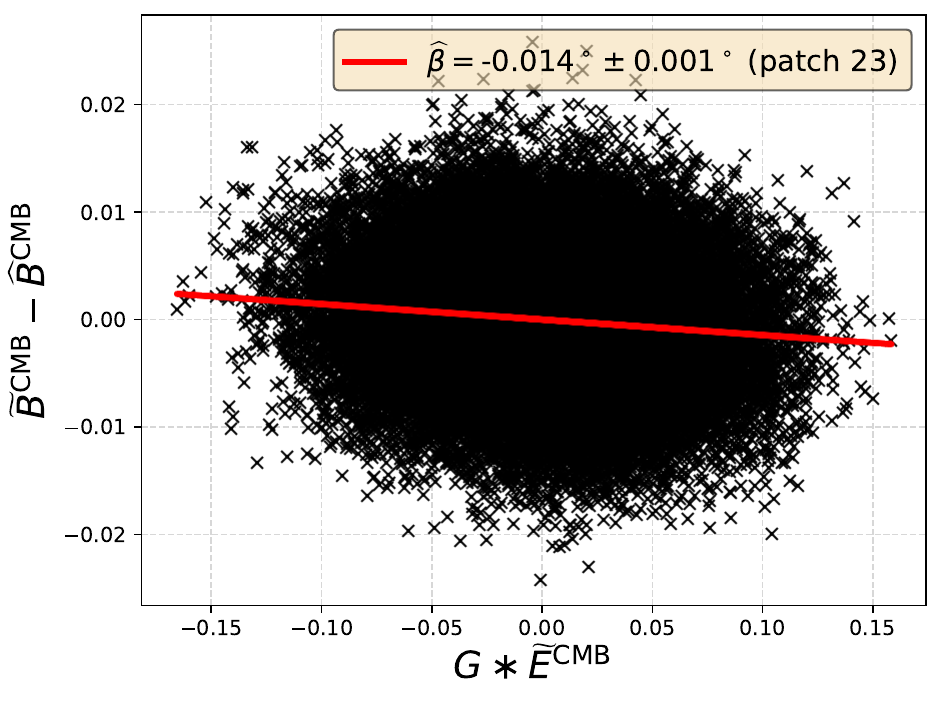}~
\includegraphics[width=0.25\textwidth,clip]{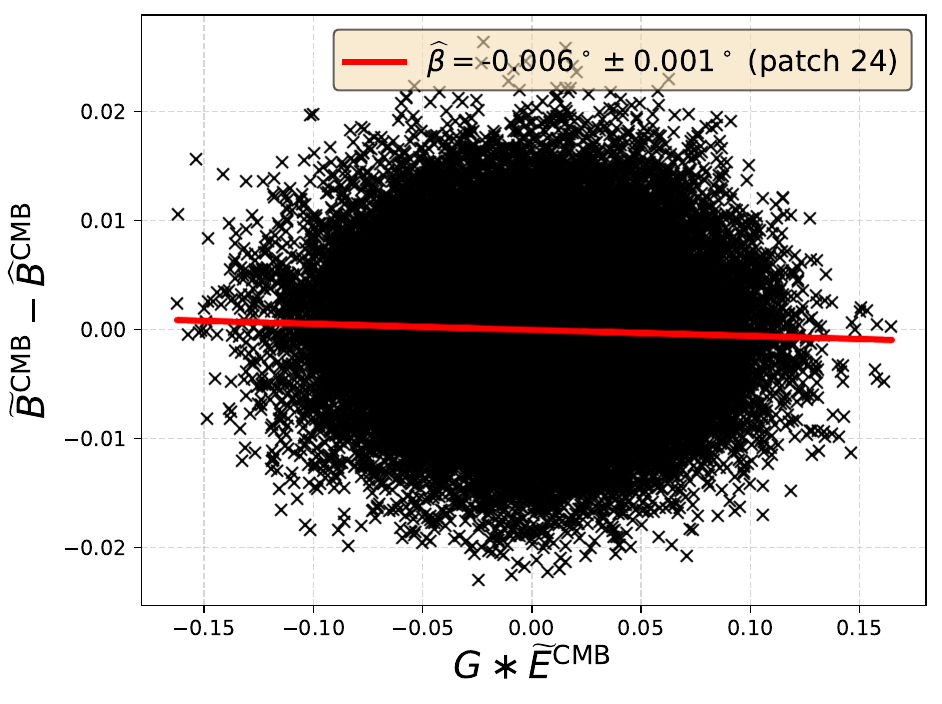}\\
\includegraphics[width=0.25\textwidth,clip]{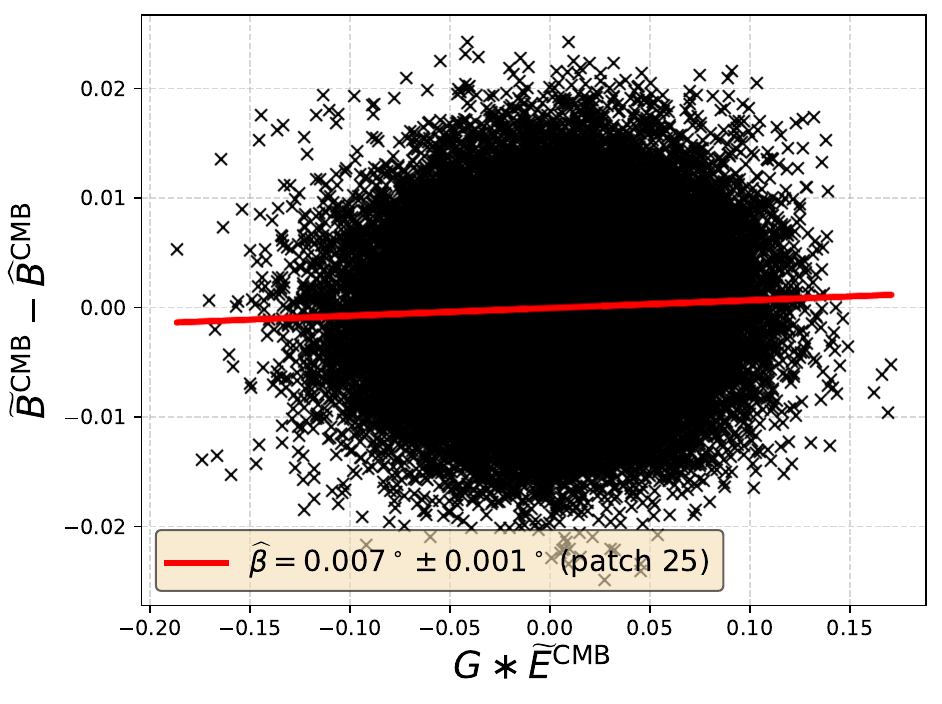}~
\includegraphics[width=0.25\textwidth,clip]{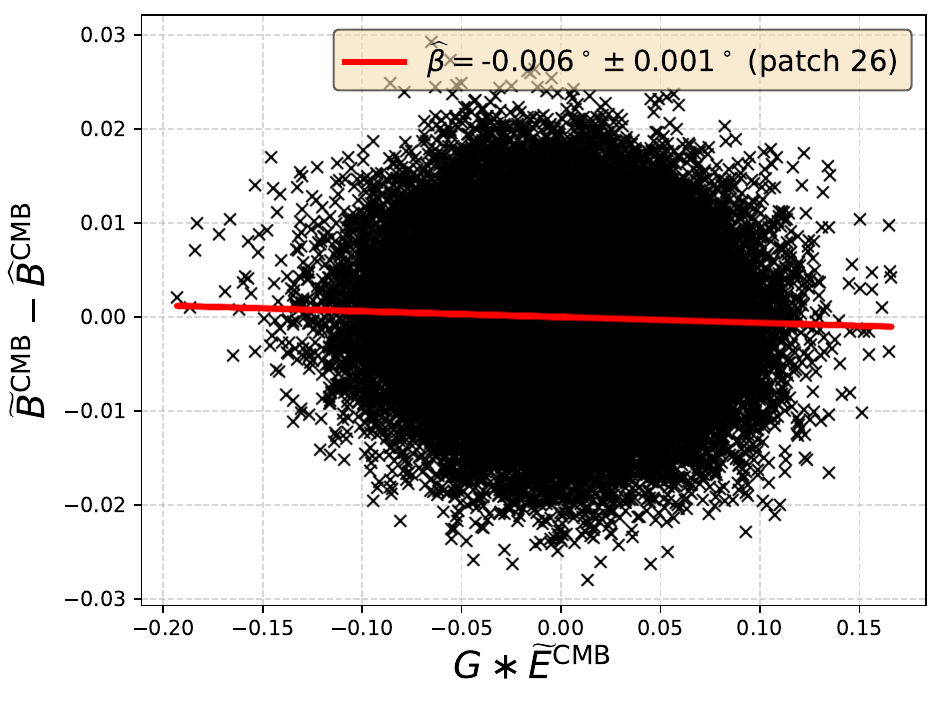}~
\hfill
\caption{\label{fig:ttplot_zerobeta_nonzeroalpha_b} Same as Figure~\ref{fig:ttplot_b}, i.e., regression of $\widetilde{B}^{\rm CMB} - \widehat{B}^{\rm CMB}$ on $G \ast \widetilde{E}^{\rm CMB}$, but for the case $\beta = 0$, $\alpha \neq 0$ (no cosmic birefringence). The mean recovered birefringence angle is $\widehat{\beta} = 0.010^\circ \pm 0.016^\circ$, consistent with $\beta = 0$ within $1\sigma$.}
\end{figure}

To streamline the main text, this appendix presents the correlation plots of the reconstructed \lb\ CMB fields across the 26 sky patches shown in Figure~\ref{fig:igloo}. Specifically, we display the linear regressions between the reconstructed fields $\widetilde{E}^{\rm CMB}(\hat{n}) - \widehat{E}^{\rm CMB}(\hat{n})$ and $(F \ast \widetilde{B}^{\rm CMB})(\hat{n})$ (Figures~\ref{fig:ttplot} and~\ref{fig:ttplot_zerobeta_nonzeroalpha}), as well as between $\widetilde{B}^{\rm CMB}(\hat{n}) - \widehat{B}^{\rm CMB}(\hat{n})$ and $(G \ast \widetilde{E}^{\rm CMB})(\hat{n})$ (Figures~\ref{fig:ttplot_b} and~\ref{fig:ttplot_zerobeta_nonzeroalpha_b}), under scenarios with and without cosmic birefringence. See Sections~\ref{subsec:formula} and~\ref{subsec:beta_linreg} of the main text for further details and discussion.

\end{document}